\documentclass[a4paper,UKenglish,cleveref, autoref, thm-restate]{lipics-v2021}

\pdfoutput=1 
\hideLIPIcs  

\graphicspath{{./results/}}

\title{Same Quality Metrics, Different Graph Drawings} 

\author{Simon {van Wageningen}}{Utrecht University Department of Information and Computing Sciences, The Netherlands \and \url{https://www.uu.nl/staff/SvanWageningen} }{s.vanwageningen@uu.nl}{https://orcid.org/0000-0002-0346-5597}{}

\author{Tamara Mchedlidze}{Utrecht University Department of Information and Computing Sciences, The Netherlands \and \url{https://www.uu.nl/staff/TMtsentlintze1/} }{ t.mtsentlintze@uu.nl}{https://orcid.org/0000-0001-6249-3419}{}

\author{Alexandru C. Telea}{Utrecht University Department of Information and Computing Sciences, The Netherlands \and \url{https://webspace.science.uu.nl/~telea001/} }{a.c.telea@uu.nl}{https://orcid.org/0000-0003-0750-0502}{}

\authorrunning{S. van Wageningen, T. Mchedlidze, and A.\,C. Telea } 

\Copyright{Simon van Wageningen, Tamara Mchedlidze, and Alexandru C. Telea} 

\ccsdesc[500]{Human-centered computing~Graph drawings} 

\keywords{graph drawing, quality metrics, assumptions, fooling} 

\supplement{https://github.com/simonvw95/Same-Quality-Metric-{}-Different-Graph-Drawings}

\nolinenumbers 

\usepackage{xcolor}
\usepackage{amsmath}
\usepackage{microtype}
\usepackage{colortbl}
\usepackage{algorithm}
\usepackage{algorithmicx}
\usepackage{algpseudocode}
\usepackage{amsfonts}
\usepackage{hyperref}
\usepackage{xcolor}

\EventEditors{Vida Dujmovi\'c and Fabrizio Montecchiani}
\EventNoEds{2}
\EventLongTitle{33rd International Symposium on Graph Drawing and Network Visualization (GD 2025)}
\EventShortTitle{GD 2025}
\EventAcronym{GD}
\EventYear{2025}
\EventDate{September 24--26, 2025}
\EventLocation{Norrk\"{o}ping, Sweden}
\EventLogo{}
\SeriesVolume{357}
\ArticleNo{5}

\begin{document}

\maketitle

\begin{figure}[!ht]
\centering
\includegraphics[width=\linewidth]{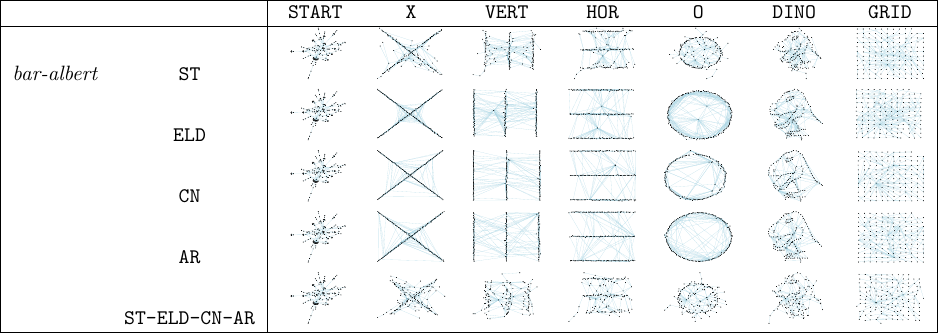}
\caption{Morphing a graph drawing (\texttt{START}) into six different target shapes (\texttt{X,VERT,HOR,O,DINO,GRID}) while keeping one or more quality metric(s) nearly constant (\texttt{ST,ELD,CN,AR}). Values for \texttt{ST,ELD,AR} are within $\pm\epsilon=0.0025$ and within $\pm\epsilon=\texttt{CN}(\Gamma)*0.05$ for 
\texttt{CN}, so the morphed drawings have very similar quality to the starting ones.}

\label{fig:teaser}
\end{figure}

\begin{abstract}
Graph drawings are commonly used to visualize relational data. User understanding and performance are linked to the quality of such drawings, which is measured by quality metrics. 
The tacit knowledge in the graph drawing community about these quality metrics is that they are not always able to \emph{accurately} capture the quality of graph drawings. In particular, such metrics may rate drawings with very poor quality as very good. In this work we make this tacit knowledge explicit by showing that we can modify existing graph drawings into arbitrary target shapes while keeping one or more quality metrics almost identical. This supports the claim that more advanced quality metrics are needed to capture the `goodness' of a graph drawing and that we cannot confidently rely  on the value of a single (or several) certain quality metrics.
\end{abstract}


\section{Introduction}

Relational data is typically visualized using node-link diagrams, commonly known as \emph{graph drawings} (GD). High-quality graph drawings are key to user understanding and performance~\cite{Purchase1996_aesthetics}. In turn, such quality is measured by various so-called \emph{quality metrics} (QM)~\cite{Purchase_2002_metrics}. At a high level QMs capture two different aspects of the goodness of a GD: 1. how readable the elements of the drawing are, e.g., few crossings, few node overlaps -- we call these \emph{readability} metrics; and 2. how well the graph structure is represented by the drawing -- \emph{faithfulness metrics}. For instance, nodes lie at distances similar to their graph theoretical distances (measured by \emph{stress}) or clusters are clearly depicted~\cite{Nguyen_Eades_Hong_2013}

Quality metrics tend to correlate -- often by design -- with the aesthetic preferences of users~\cite{Huang07,Purchase1996_aesthetics} and task performance~\cite{Purchase_Performance_1997}. As such, GD algorithms aim to (in)directly optimize them in order to produce pleasing drawings. For instance, spring-based techniques like Fruchterman-Reingold~\cite{Fruchterman_Reingold_fd_1991} and ForceAtlas2~\cite{Jacomy_Venturini_Heymann_Bastian__FA2_2014} use forces on their nodes (and edges) in various ways, leading to drawings with `good' scores on most GD quality metrics. Other techniques like Stress Majorization~\cite{Gansner_Koren_North_SM_2005}, Stress-Plus-X~\cite{stressplusx}, DeepGD~\cite{deepgd}, \emph{SGD}$^2$~\cite{Ahmed_Luca_Devkota_Kobourov_Li_2022_sgd}, and Core-GD~\cite{groetschla2024coregd} directly optimize one or more metrics to produce aesthetically pleasing drawings. 


Two underlying and often implicit assumptions on quality metrics are that a `high' value indicates a `good' drawing; and that drawings of the same graph with similar metric values have similar quality. It is, however, widely known by Graph Drawing researchers, and has been exemplified multiple times by the Graph Drawing live challenge~\cite{gdnvcontest}, that these assumptions do not always hold. In particular, over the years, GD challenge participants have presented drawings with very high QMs values that had an obfuscating appearance and did not reveal the graph structure.

We devote this work to make this tacit knowledge explicit and show that we cannot confidently rely on the value of a single quality metric nor on combinations of certain quality metrics. For this, we \emph{`fool'} the quality metrics by morphing given, high-quality graph drawings into a given target shape (which can be any arbitrary shape) using a simulated annealing process while keeping one or more given quality metric values relatively fixed. 
Thus we are able to obtain drawings that obfuscate the graph structure and have very high values of quality metrics. While for readability metrics this is rather a confirmation of the tacit knowledge we already possess, it is very surprising to achieve this for a faithfulness quality metric such as stress as well as for combinations of certain QMs. We also show the limitations of our approach -- certain combinations of metrics can not be fooled by it. Finally, we experiment with graphs of varying structures to illustrate how the graph’s structure affects the difficulty of fooling quality metrics.

\section{Related Work}
\label{sec:related_work}
The assumption that quality metrics directly correlate with human understanding, perception, and preference has been researched by many previous (user) studies~\cite{Burch_2021_usereval}. The authors in \,\cite{Purchase1996_aesthetics} showed that larger numbers of edge crossings and edge bends are negatively correlated with human perception. Huang\,\cite{Huang07} refined such findings by using eye tracking to show that small edge angles lead to poorer performance in path search tasks.

The number of edge crossings in a drawing has been shown to correlate with user preferences\,\cite{Chimani_etal2014_stressfavor,Purchase1996_aesthetics}. The perception of how well nodes are placed according to their shortest path, also called \emph{stress}, was researched in a 2024 study \cite{Mooney_Purchase_Wybrow_Kobourov_Miller_2024_stressperception}, which found that users are able to perceive different values of stress in different drawings. Furthermore, stress also tends to correlate with user preferences\,\cite{Chimani_etal2014_stressfavor}.

Several types of visualizations beyond GD have been put to the test to try and fool statistical properties. Many works\,\cite{Anscombe_1973_stats,Boger_et_al_2021_stats, Chatterjee_Aykut_2007_stats,Matejka_Fitzmaurice_2017_stats_sim_anneal} have shown that it is possible to control the visual appearance of 2D scatterplots, by iteratively morphing these into target shapes, while  keeping summary statistical metrics such as means, standard deviations and correlation similar. Different DR algorithms can create wildly different projections of the same dataset which have very similar quality metric values\,\cite{espadoto19}. Separately, Machado, Telea, and Behrisch showed that one can control the shapes of clusters in projections and only marginally affect quality metrics\,\cite{sharp}, as well as completely distort the visual appearance of projections without affecting multiple quality metrics\,\cite{machado2025_metrics}.

Similarly, Chari and Pachter~\cite{Chari_Pachter_2023_elephant} showed that they could transform projections of the same dataset into any specified shape without heavily adjusting the quality metric values. While such work did not aim to technically `fool' quality metrics, it provides strong evidence that metrics and projection appearance are not inherently correlated -- an effect that we also show, in our work, to hold for graph drawings.

Regarding graphs (but not their drawings), Chen et al.~\cite{Chen_2021_graphstats} showed that one can generate graphs with the exact same graph statistics, such as the number of nodes, edges, and triangles. Di Bartolomeo, Lang and Dunne\,\cite{bartolomeo_Lang_Dunne_2022_worst} (potentially a satirical paper) suggested as future research directions the idea of `rocketshipness', in which a graph drawing is optimized to resemble a rocket ship while maintaining one or more GD quality metrics. However, this idea was not further explored.

\section{Fooling GD metrics: Experimental setup}
\label{sec:setup}
\subsection{Preliminaries}
We first introduce a few notations and concepts. An \emph{undirected graph} $G=(V,E)$ is a set of nodes $V = \{v_{1}, \ldots, v_{n}\}$ and a set of undirected unweighted edges $E = \{e_{1},\ldots,e_{m}\} \subseteq V \times V$. A straight-line \emph{graph drawing} $\Gamma$ of $G$ assigns $2$-dimensional coordinates $X_i$ to nodes $v_i \in V$ and line segments $L_i$ to edges $e_i \in E$, so $\Gamma$ can be represented as a matrix $X^2 \in \mathbb{R}^{n \times 2}$ with rows $X_i\in \mathbb{R}^{2}$. Let $D \in \mathbb{R}^{n \times n}$ denote the \emph{shortest path matrix} of graph-theoretic distances $d_{ij}$ between all node-pairs $(v_i,v_j) \in V \times V$. We next refer to the Euclidean distance $\|X_{i} - X_{j}\|$  between two nodes $v_i$ and $v_j$ simply as \emph{distance}. Let $deg(v)$ be the degree of node $v$, \emph{i.e.}, the number of edges incident to $v$. A \emph{quality metric} is a function $Q(\Gamma) \in [0,1]$ which assigns a value to $\Gamma$, with low (resp. high) values denoting better (resp. poorer) drawings.

\subsection{Metrics \label{sec:experimentalsetup_metrics}}
We consider in our work the following GD quality metrics and their combinations:

\subparagraph*{Stress:} Measures the difference between the Euclidean distances of all node-pairs and their  shortest path distances\,\cite{KamadaKawai1989}  as $\texttt{ST}(\Gamma) = \frac{1}{n(n-1)/2}\sum^{n}_{i=0}\sum^{n}_{j=i+1} \frac{(\|Z_{i}-Z_{j}\|-d_{ij})^{2}}{d_{ij}^2}$, where $Z_i$ scales the coordinate $X_i$ by the shortest path distances as $Z_i =  \frac{\sum_{i \neq j}{\|X_{i} - X_{j}\|} / d_{ij}} {\sum_{i \neq j}{\|X_{i} - X_{j}\|^2} / d_{ij}^2} X_i.$

$\texttt{ST}(\Gamma) = \frac{1}{n(n-1)/2}\sum^{n}_{i=0}\sum^{i}_{j=0} \frac{(\|Z_{i}-Z_{j}\|-d_{ij})^{2}}{d_{ij}^2}$,

We expect the \texttt{ST} metric to be the most difficult metric to fool in drawings, as it 
explicitly evaluates how well graph structure is depicted in a drawing, i.e., is a faithfulness QM. 

\subparagraph*{Edge length deviation:} Measures the average deviation of edge lengths~\cite{Mooney_Purchase_Wybrow_Kobourov_2024_metriclandscape} from the mean $\mu$ of all lengths in a drawing as $\texttt{ELD}(\Gamma) = \sqrt{ \frac{1}{m}\sum_{i=1}^{m}{\|L_i\| - \mu)^{2}}}$. We expect this metric to be relatively easy to fool as it \emph{solely} focuses on the uniformity of edge lengths.

\subparagraph*{Crossing number:} We capture this by the number of intersections of all edge segments $L_i$, \emph{i.e.} $\texttt{CN}(\Gamma) = \sum^{m}_{i=0}\sum^{m}_{j=i+1} \mathbf{1} (L_i \cap L_j \neq \emptyset)$. We expect that \texttt{CN} is hard to fool, as making subtle changes to any drawing tends to have a strong impact on how its lines cross.

\subparagraph*{Angular resolution} We measure the frequency of small angles occurrences in a graph drawing by computing the average deviation of angles~\cite{Mooney_Purchase_Wybrow_Kobourov_2024_metriclandscape} of adjacent edges \emph{vs} the best possible angle as $\texttt{AR} = \frac{1}{n'} \sum^{n'}_{i=1} \left\vert \frac{\Theta_{i}-\theta_{i}}{\Theta_{i}}\right\vert,$ where $n'$ is the number of nodes with degree $\geq 2$; $\theta_{i}$ is the smallest measured angle between consecutive adjacent edges of node $i$ and $\Theta_{i} = 2\pi/deg(v_{i})$ is the best possible angle between consecutive edges for node $i$. We speculate that this metric is easy to fool as it is easy to keep similar angles while making large visual changes.

\subsection{Simulated Annealing Approach}
\label{sec:approach}

We follow the approach in~\cite{Matejka_Fitzmaurice_2017_stats_sim_anneal} to slowly morph any given graph drawing $\Gamma$ with node coordinates $X$ into a drawing $\Gamma'$ with node coordinates $X'$, while keeping one or more quality metrics values ${QM}^i(\Gamma')$, $i\in\{a,b,\dots\}$ very close to their original values $QM^i(\Gamma)$ and moving $X'$ closer to a target shape $Y$. Here, $Y$ can be \emph{any} set of node coordinates, as long as $|Y|=|X|$, as this simplifies our morphing implementation.

\begin{algorithm}
\caption{Simulated Annealing Morphing
}\label{alg:sim_anneal}

\begin{algorithmic}[1]
\State $DIFF_{curr} \gets Sim(X, Y)$
\State ${\mathbf{qm}} \gets [QM^a(\Gamma), QM^b(\Gamma), \dots]$
\For{$i = 1$ to $N_{max}$}
    \State $(X', DIFF_{test})  \gets Jit(X, T_{i}, DIFF_{curr}, Y)$
    \State $\mathbf{qm}' \gets [QM^a(\Gamma'), QM^b(\Gamma'), \dots]$
    \If{$\forall j,\ |\mathbf{qm}'[j] - \mathbf{qm}[j]| \leq \mathbf{\epsilon_j}$}
        \State $X \gets X'$
        \State $DIFF_{curr} \gets DIFF_{test}$
    \EndIf
\EndFor
\end{algorithmic}
\end{algorithm}

Algorithm~\ref{alg:sim_anneal} details our morphing: We start by computing the similarity of our graph drawing to the target shape and its quality metric value (lines 1-2). The $Jit$ function jitters a random selection of nodes $X^s \subset X, 1 \leq |X|/15$ by a random value drawn from a normal distribution [$-0.5$, $0.5$]$/25$ while increasing the similarity of $X'$ to $Y$ (line 4, see also Alg.~\ref{alg:jitter}). We accept $X'$ if the quality $\mathbf{qm'}$ of its drawing $\Gamma'$ is within a small range $\mathbf{\epsilon}$ of the initial quality $\mathbf{qm}$ (lines 5-8). We repeat this process for $N_{max}=30000$ iterations.
To escape local minima, we also use simulated annealing: If a randomly generated value is below the current temperature $T_{i}$, we accept the node coordinates $X_{jit}$, irrespective of its (dis)similarity to $Y$. $T_i$ varies over iterations: We start with $T_{init} = 0.4$ and quadratically decrease it to a final value $T_{N_{max}} = 0.001$. After exploring the practical ranges of the metrics in\,\cite{Mooney_Purchase_Wybrow_Kobourov_2024_metriclandscape} and their variations discussed in\,\cite{vanWageningen_Mchedlidze_Telea_2024}, we set $\epsilon = 0.0025$ for \texttt{ST,ELD,AR} and $\epsilon = \texttt{CN}(\Gamma)*0.05$ for \texttt{CN}.

\begin{table}[b!]
\centering
\small
\begin{tabular}{||l||c|c|c|c||}
\hline
 \emph{Graph} & $n$ & $m$ & $deg(V)_{avg}$ & \emph{source}\\
\hline
\hline
polbooks & 105 & 441 &  8.40 & ~\cite{networkrepository}\\
lnsp\_131 & 123 & 275 &  4.47 & ~\cite{networkrepository}\\
bar-albert & 142 & 175 &  2.46 & generated\\
gams10am & 171 & 298 &  3.49 & ~\cite{networkrepository}\\
dwt\_307 & 307 & 1108 &  7.22 & ~\cite{networkrepository}\\
\hline
\end{tabular}
\caption{Descriptive statistics of graphs used in our experiments.\label{tab:dataset_description}}
\end{table}

\begin{algorithm}
\caption{Jitter function $Jit(X,T,DIFF,Y)$
}\label{alg:jitter}

\begin{algorithmic}[1]
\While{}
    \State $X^s \subset X, 1 \leq |X|/15$ 
    \State $X_{jit} \gets X + (X^s + \frac{\text{rand}(\mathcal{N}(-0.5, 0.5))}{25})$
    \If{$Sim(X_{jit},Y) < DIFF$ $\lor $ $T > \text{rand}(\mathcal{U}(0, 1))$}
        \State \textbf{return} $(X_{jit}, Sim(X_{jit}, Y))$
    \EndIf
\EndWhile
\end{algorithmic}
\end{algorithm}

For the similarity $Sim(X,Y)$ (see Alg.\,\ref{alg:similarity}), one can use Mean Squared Error $MSE =\frac{1}{n} \sum_{i=1}^{n} (X_i - Y_i)^2$ or the Procrustes Statistic\,\cite{deepdrawing} (similar to \emph{MSE} but invariant to rotation, translation and scaling). We also briefly experimented with the \emph{Wasserstein Distance}\,\cite{Villani2009_wasserstein} and the \emph{Sinkhorn}\,\cite{Cuturi_2013_sinkhorn} algorithm. However, these approaches yielded slightly worse results and were thus discarded.

The authors of \cite{Matejka_Fitzmaurice_2017_stats_sim_anneal} used the average distance of all points in $X$ to the closest point in $Y$, which gives more morphing freedom. Our $Sim$ design slightly adapts this idea. Algorithm~\ref{alg:similarity} details our Similarity function: We start by computing the closest distances from two sets of node coordinates $X$ and $Y$. We compute the distance of node $X_0$ (the \emph{current} first node) in coordinate set $X$ to all nodes in coordinate set $Y$. The distance to the closest node $Y_{idx}$ in $Y$ is then added to the total loss after which both $X_0$ and $Y_{idx}$ are removed from $X$ and $Y$, respectively. This iterative process is then repeated until all nodes are removed from the coordinate set and ensures no repetition of nodes in any pairs. To compare similarity values across different graphs, we express the similarity in a percentage, where we consider perfect similarity to the target shape ($Sim(Y,Y)=0$) as $100\%$ and perfect similarity to the starting shape as $0\%$, i.e.: $100 - (Sim(X',Y) / Sim(X,Y) *100)$ where $Sim(X,Y) \to \mathbb{R}$ is equal to the starting similarity from Alg~\ref{alg:sim_anneal} (line 1).

\begin{algorithm}[htb]
\caption{Similarity function $Sim(X,Y)$
}\label{alg:similarity}

\begin{algorithmic}[1]
\State $loss \gets 0$
\While{$n > 0$}
    \State $DIS \gets \left[ \|X_{0} - Y\| \right]_{i=0}^{n}$
    \State ($idx,dis) \gets (argmin(DIS),min(DIS)$)
    \State $X \gets RemoveRow(X, 0)$
    \State $Y \gets RemoveRow(Y, idx)$
    \State $loss \gets loss + dis$
    \State $n \gets n - 1$
\EndWhile\\
\textbf{return} $loss$

\end{algorithmic}
\end{algorithm}

\subsection{Datasets}
\label{sec:datasets}
\begin{figure}[!ht]
\centering
\includegraphics[width=0.75\linewidth]{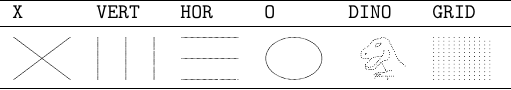}
\caption{Target shapes used in our experiments.\label{fig:targets}}
\end{figure}

We perform experiments on the five graphs in Tab.~\ref{tab:dataset_description}. Four come from the Network Repository\,\cite{networkrepository} due to their interesting varying structures. We created the fifth one using the dual-Barabasi-Albert algorithm\,\cite{Moshiri_2018_dual_barabasi} to control the node count $n$. For each graph we create initial drawings (\texttt{START}) using  ForceAtlas2~\cite{Jacomy_Venturini_Heymann_Bastian__FA2_2014} with 5000 iterations and default parameters. We translate and scale all drawings to have coordinates in $[0, 1]$ so that \emph{Jit} becomes scale invariant.
We use six target shapes $Y$ from \cite{Matejka_Fitzmaurice_2017_stats_sim_anneal} including the Dinosaur shape from \cite{Cairo_2016_Datausaurus}, see Fig.~\ref{fig:targets}. For all shapes we create simple algorithms that generate $Y$ for any point count $n$ so we can next easily enforce $|X|=|Y|$ (see Sec.~\ref{sec:approach}). Our data and code are available on GitHub\,\cite{githubpage}. 

\begin{figure}[!b]
\centering
\begin{tabular}{|l|cccc|}
  \hline
  & \emph{frame 1} & \emph{frame 2} & \emph{frame 3} & \emph{frame 4}\\
\hline
  \emph{targets} &
  \includegraphics[width=22.5mm]{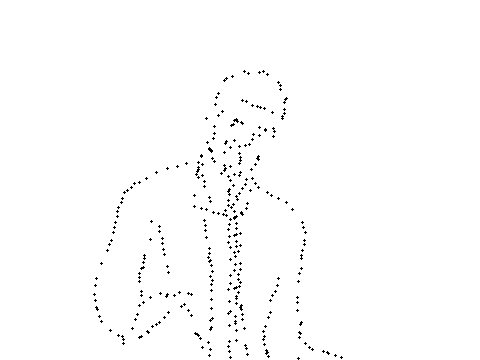} & \includegraphics[width=22.5mm]{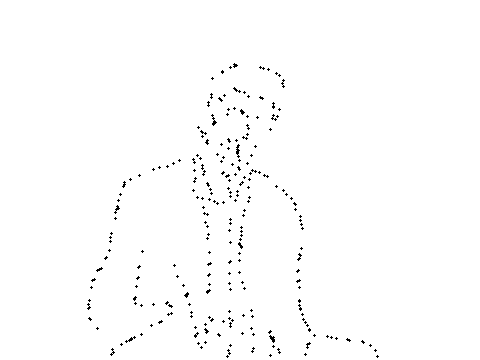} & \includegraphics[width=22.5mm]{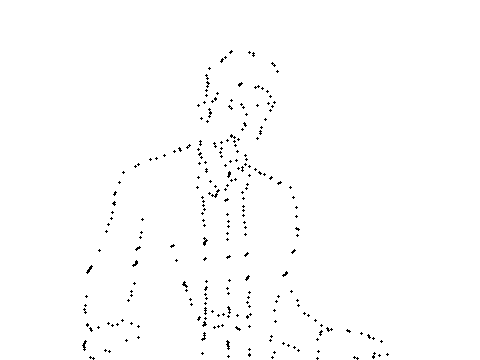} & \includegraphics[width=22.5mm]{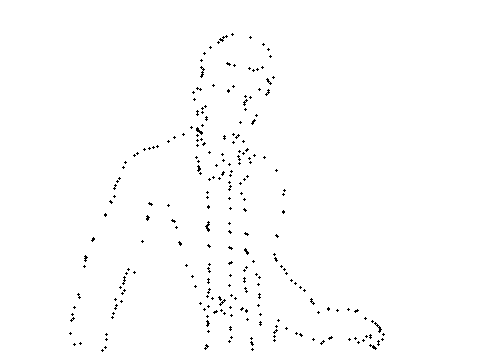}\\
  
  \emph{drawings} &
  \includegraphics[width=22.5mm, angle=270, origin=c]{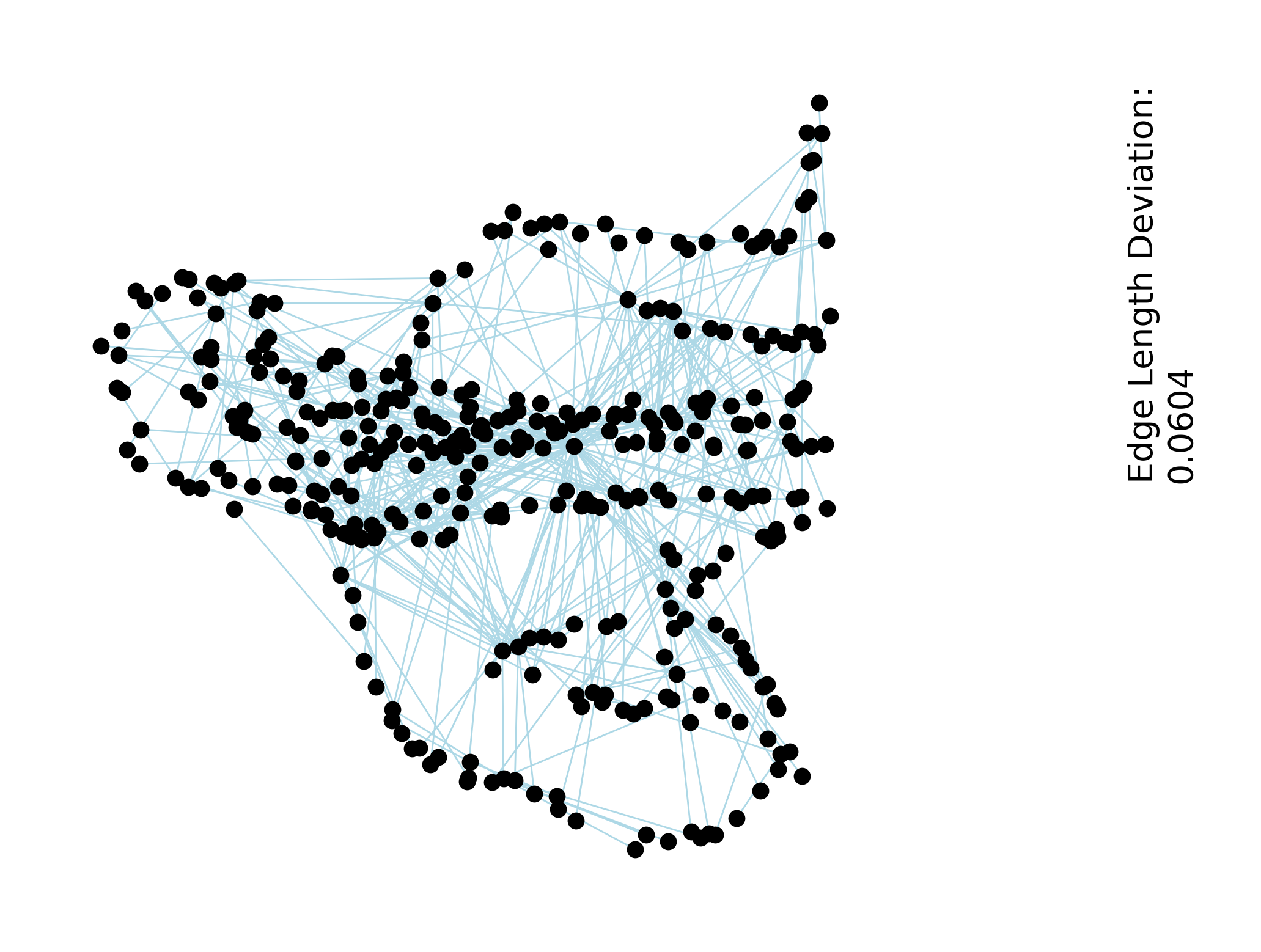} & \includegraphics[width=22.5mm, angle=270, origin=c]{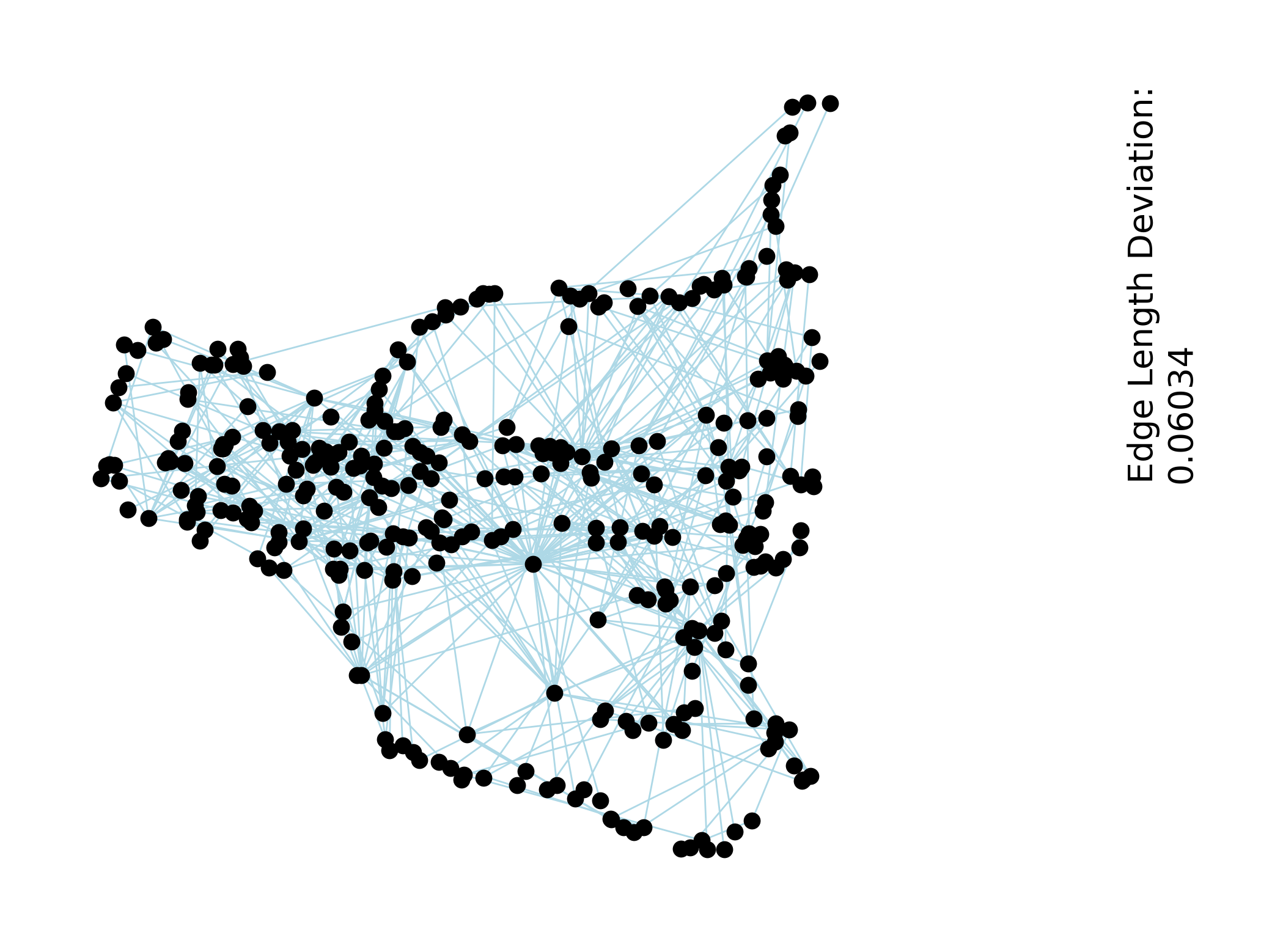} & \includegraphics[width=22.5mm, angle=270, origin=c]{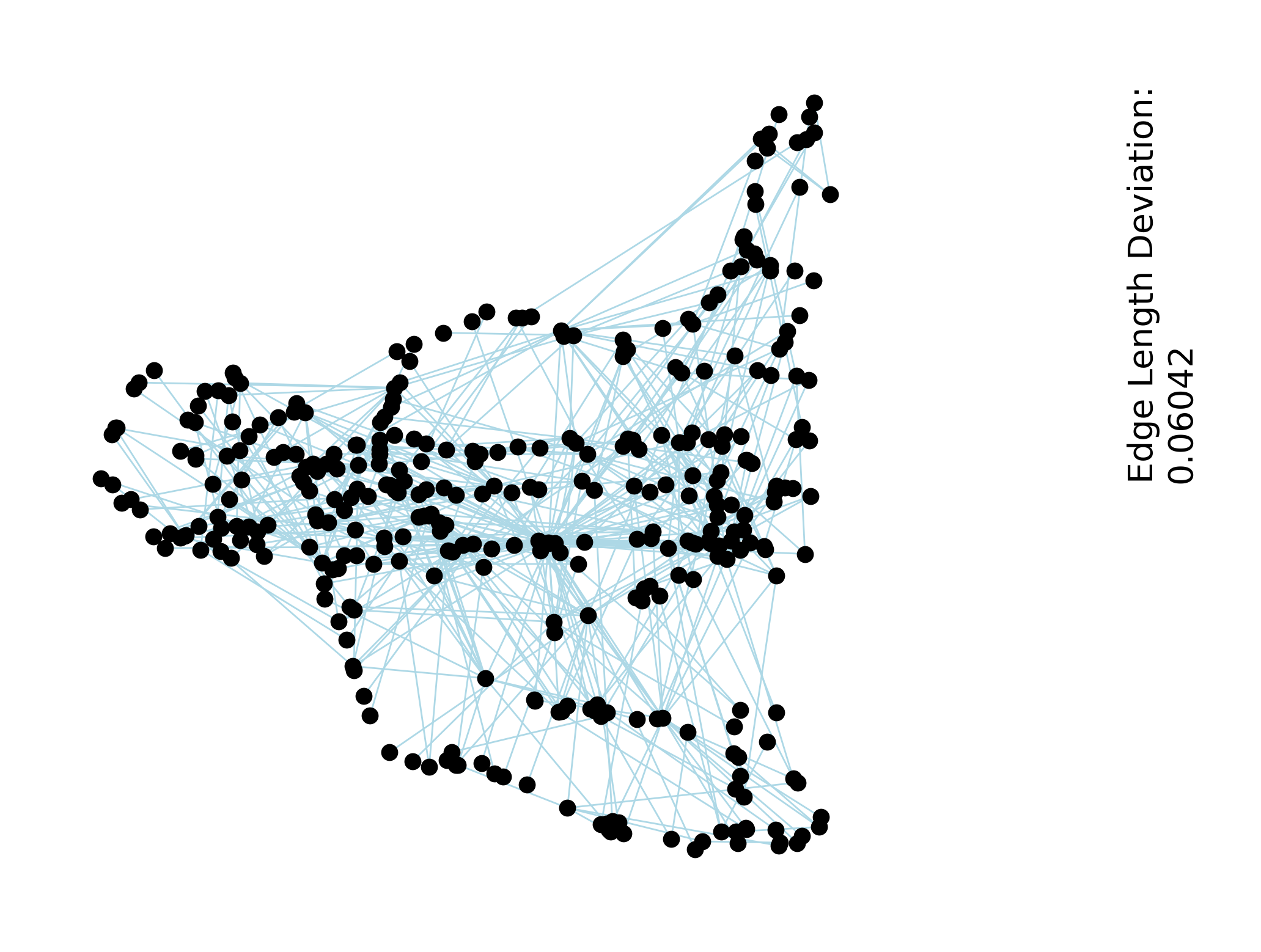} & \includegraphics[width=22.5mm, angle=270, origin=c]{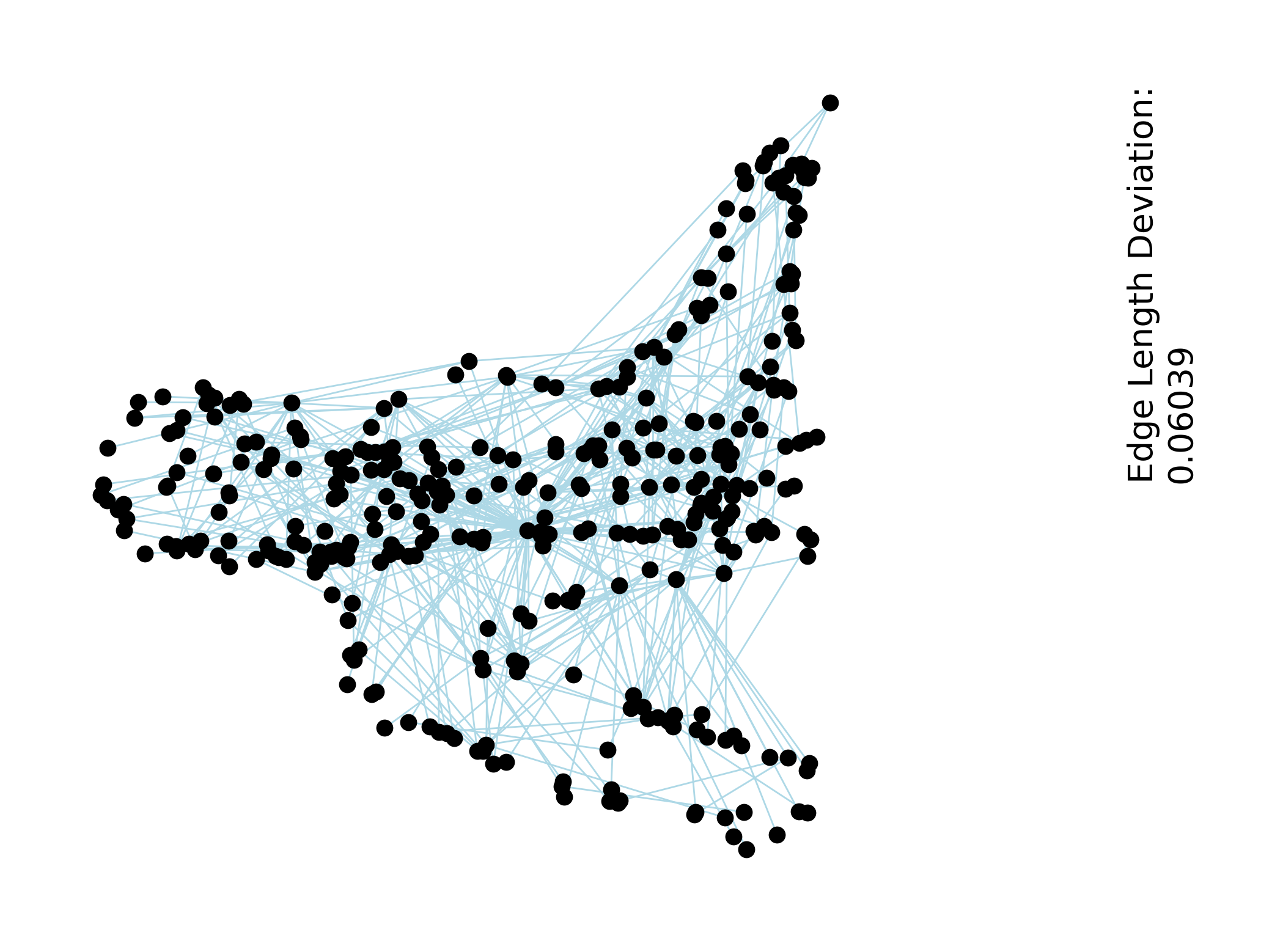}\\
  \hline
\end{tabular}
\caption{Four sequential target frames from the video and the resulting sequential drawings from attempting to fool a quality metric (\texttt{ELD}) for the target frames.}

\label{fig:rickroll}
\end{figure}

\begin{figure*}[!hbt]
\centering
\includegraphics[width=\linewidth]{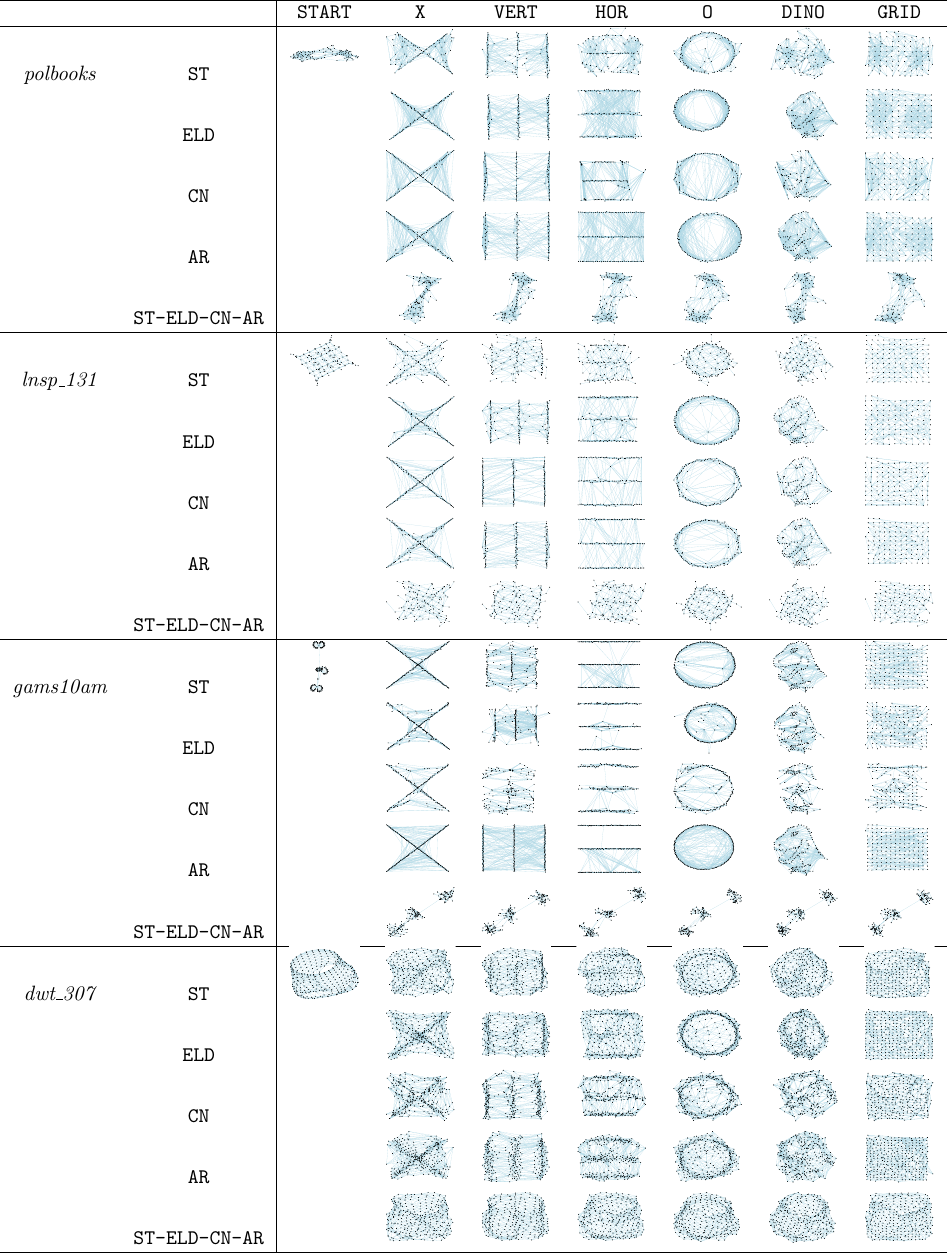}
\caption{Collection of different graph drawings. The \texttt{START} column indicates the starting drawing of various graphs. The results of applying Alg.~\ref{alg:sim_anneal}  to \texttt{START} to six target shapes (\texttt{X}, \texttt{VERT}, \texttt{HOR}, \texttt{O}, \texttt{DINO}, \texttt{GRID}) are shown in their respective columns. The rows indicate the (combinations of) metrics that have $\pm\epsilon=0.0025$ for \texttt{ST,ELD,AR} and $\pm\epsilon=\texttt{CN}(\Gamma)*0.05$ for \texttt{CN}.}
\label{fig:results}
\end{figure*}

\section{Results and Discussion}
\label{sec:results}

We next refer to results from the figures and tables in the Appendix with the $\star$ notation. These extra results include the drawings of all pairs and triples of metric combinations, distribution of similarity percentages, and results of statistical testing. We also refer the reader to our {\color{blue}\href{https://github.com/simonvw95/Same-Quality-Metric--Different-Graph-Drawings/blob/main/example_figures/rr-ba_rr-ELD.gif}{video material}}, in which we replicate a dancing person while keeping a quality metric value the same. An example of four frames of this video can be seen in Fig~\ref{fig:rickroll}. The Appendix and our GitHub page\,\cite{githubpage} cover extra details on how the videos were created and highlight extra video results, respectively.

\subparagraph*{Individual metrics:} Figures~\ref{fig:teaser} and~\ref{fig:results} show the resulting drawings of fooling individual metrics and all four metrics at once. We see that not all individual quality metrics can be fooled equally well. As speculated, fooling \texttt{ST} proves to be the most difficult. For example, the \emph{bar-albert} graph nearly perfectly morphs to all six targets for \texttt{ELD,CN,AR} but not for \texttt{ST}. We see similar results for \emph{polbooks}, \emph{lnsp\_131} and \emph{dwt\_307}. For the metrics \texttt{ELD} and \texttt{AR}, we observe easy fooling for all graphs except \emph{dwt\_307}. We also see that fooling the number of crossings (\texttt{CN}) is easier than expected -- see the results for all graphs except \emph{gams01am} and \emph{dwt\_307}. Aggregating the results over all graphs and target shapes (see Fig.~\ref{fig:sig_tests_metrics}$\star$) shows that \texttt{ELD} is much easier to fool than \texttt{ST}; and \texttt{AR} is much easier to fool than \texttt{ST} and \texttt{CN}.

\begin{figure*}[!ht]
\centering
\begin{tabular}{cc|ccccccc}
  \hline
  & & \texttt{START} & \texttt{X} & \texttt{VERT} & \texttt{HOR} & \texttt{O} & \texttt{DINO} & \texttt{GRID}\\
\hline
 
    \emph{dwt\_307} & \texttt{ST-ELD} & \includegraphics[width=9.5mm]{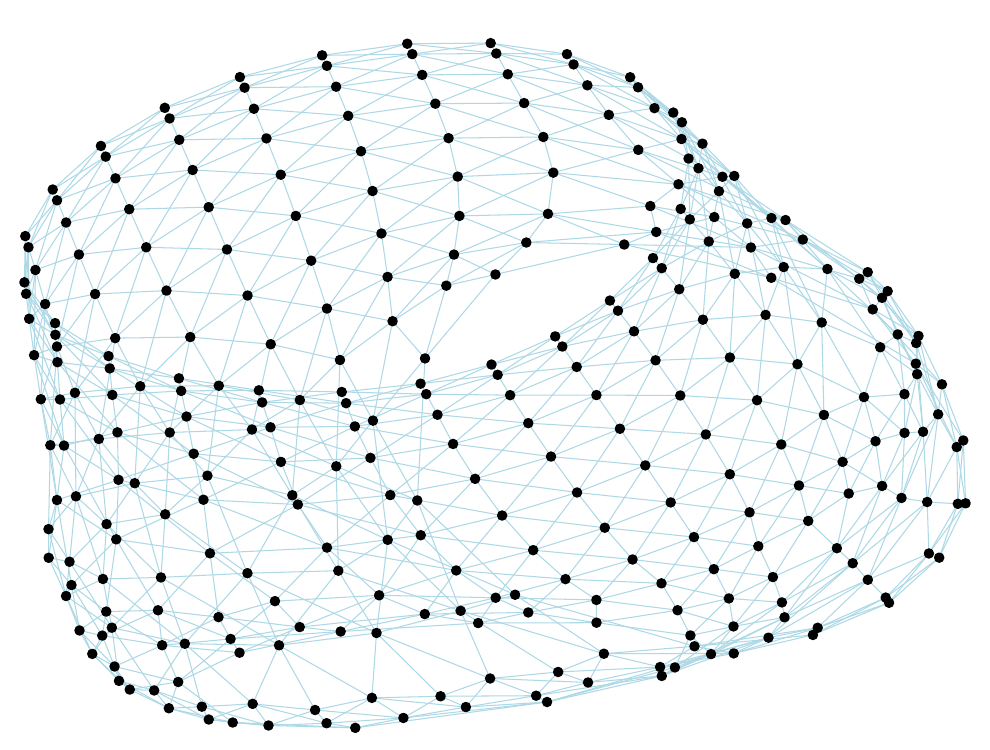} & \includegraphics[width=9.5mm]{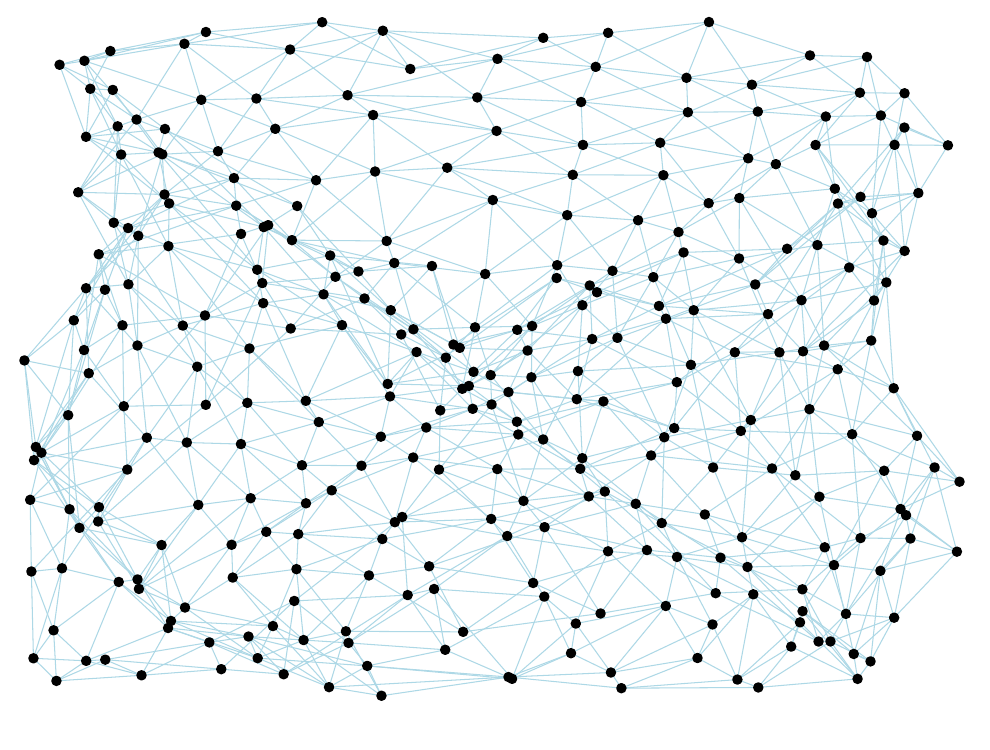} &
  \includegraphics[width=9.5mm]{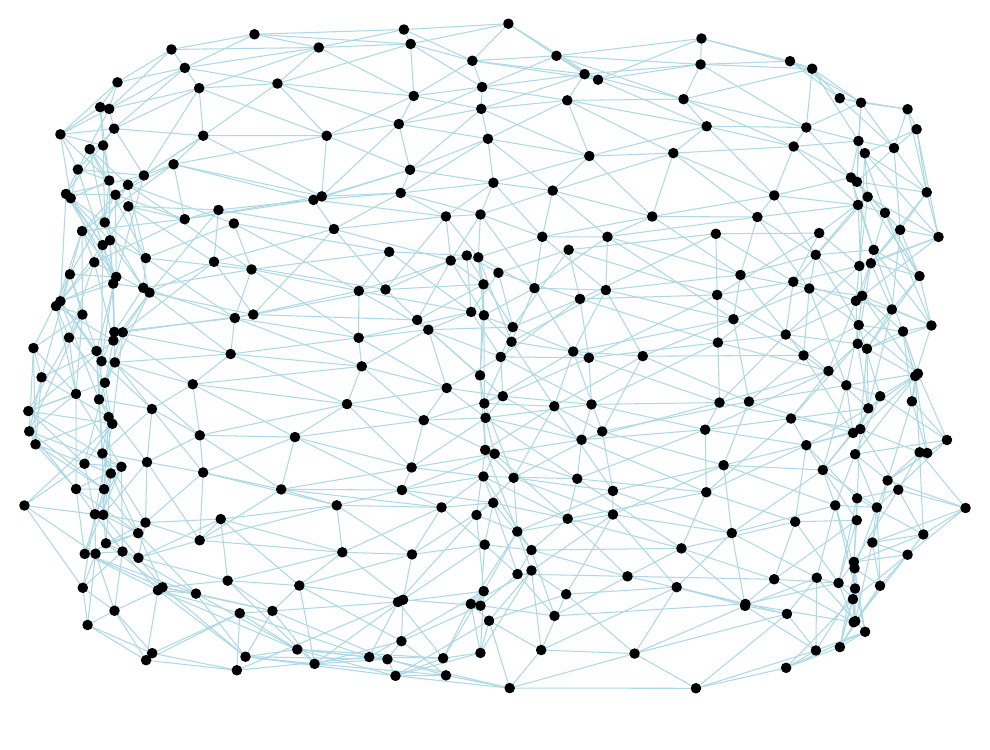} &
  \includegraphics[width=9.5mm]{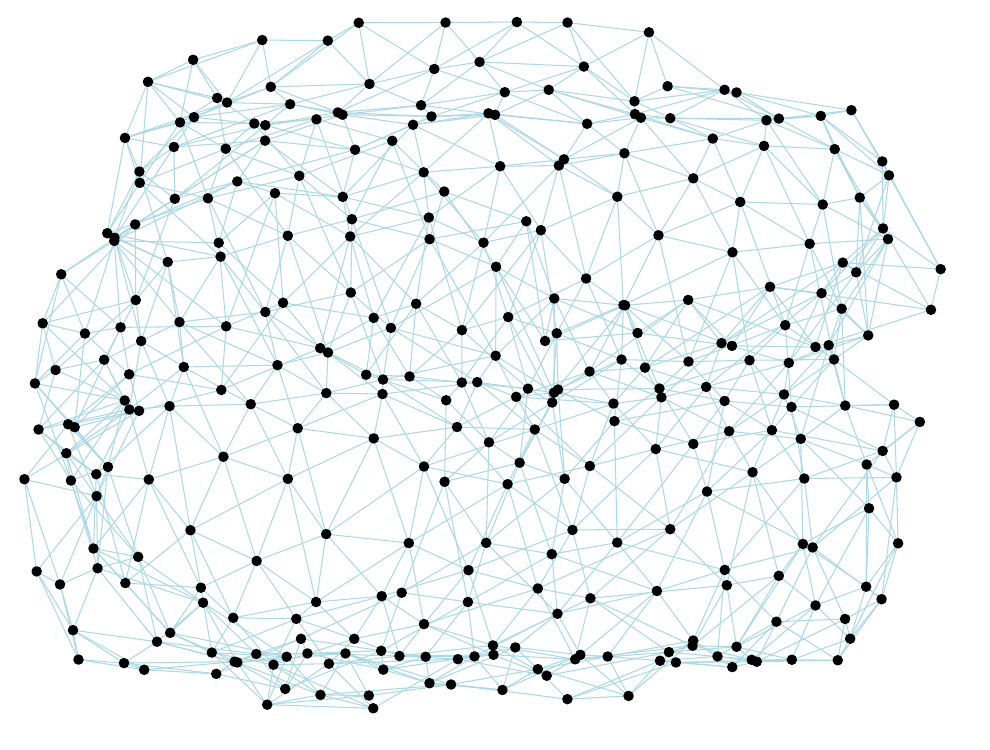} &
  \includegraphics[width=9.5mm]{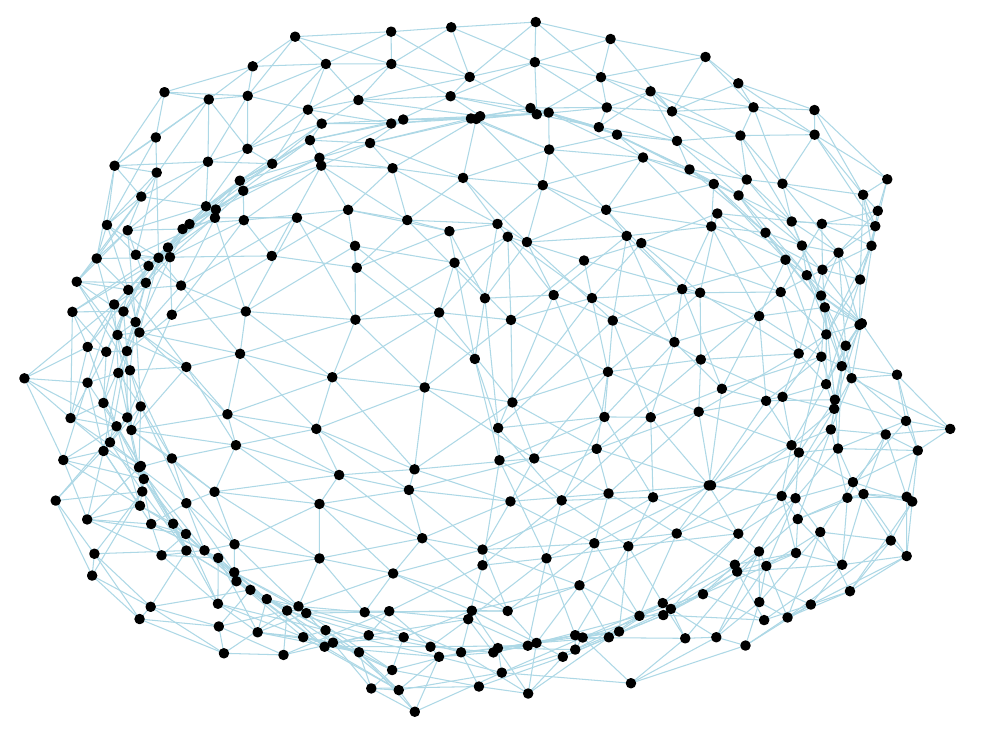} &
  \includegraphics[width=9.5mm]{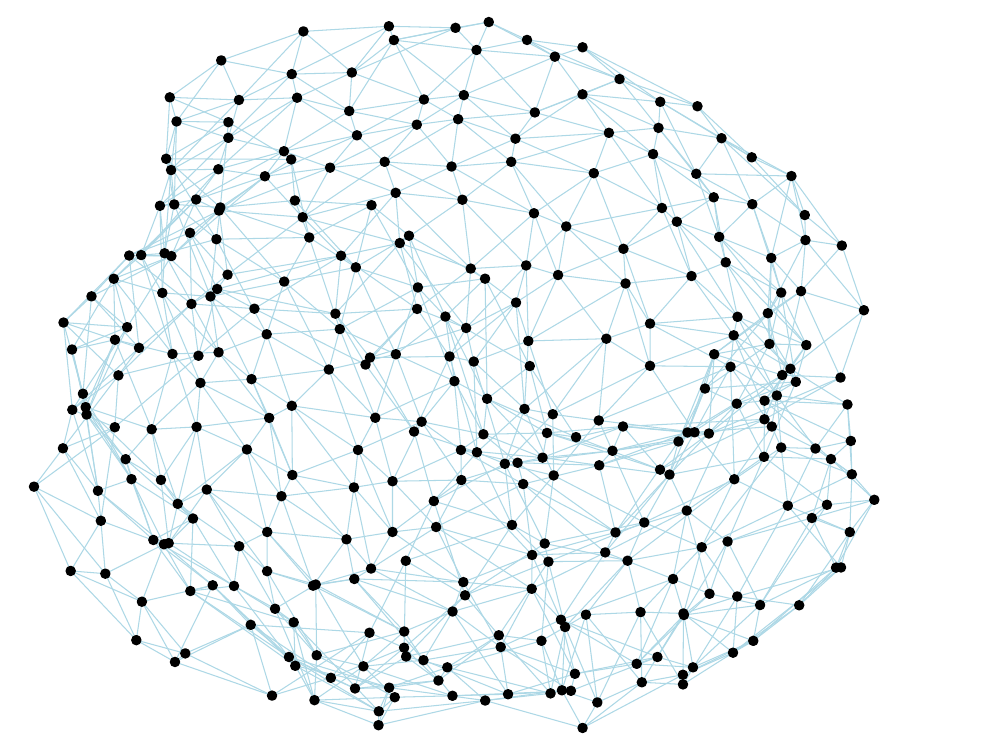} &
  \includegraphics[width=9.5mm]{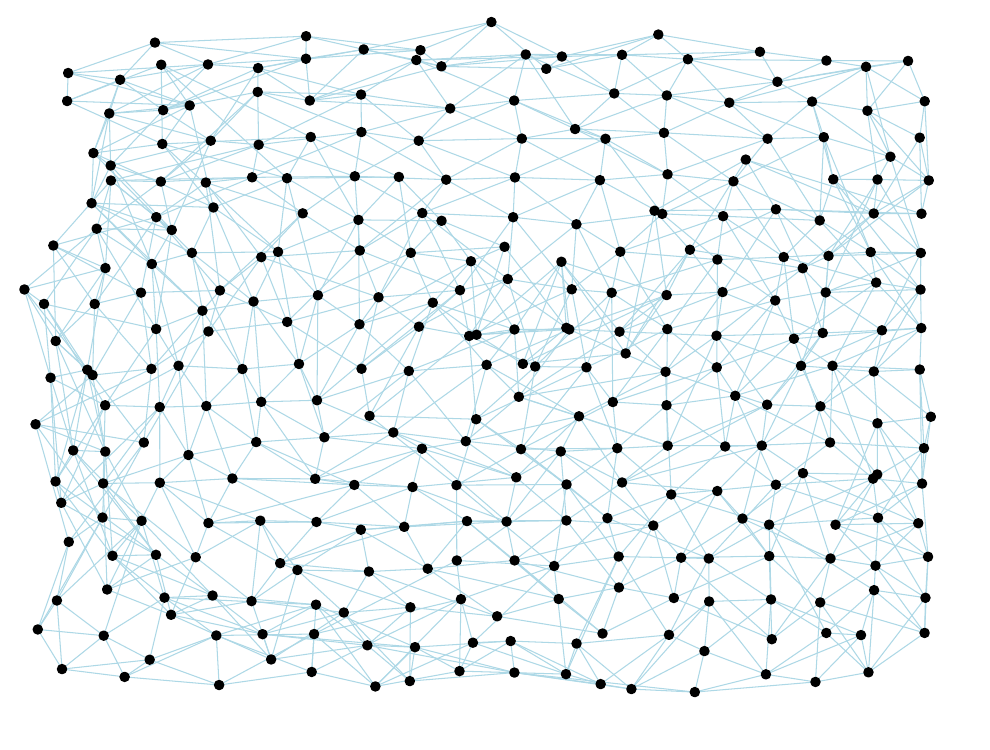}
  \\
     & \texttt{ST-CN} & & \includegraphics[width=9.5mm]{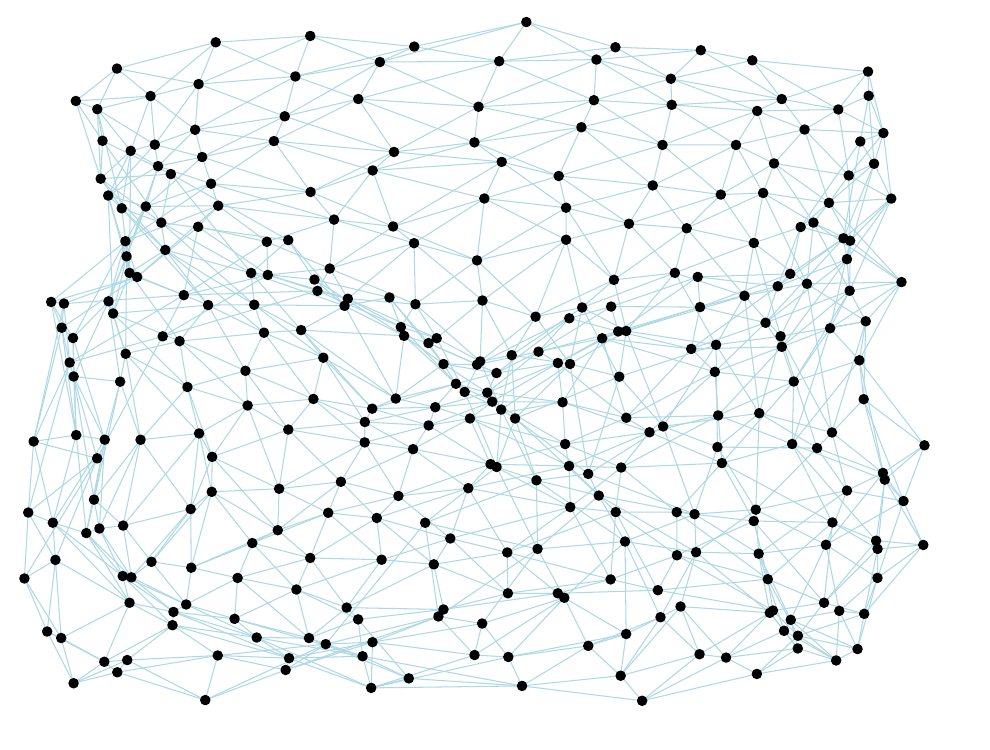} &
  \includegraphics[width=9.5mm]{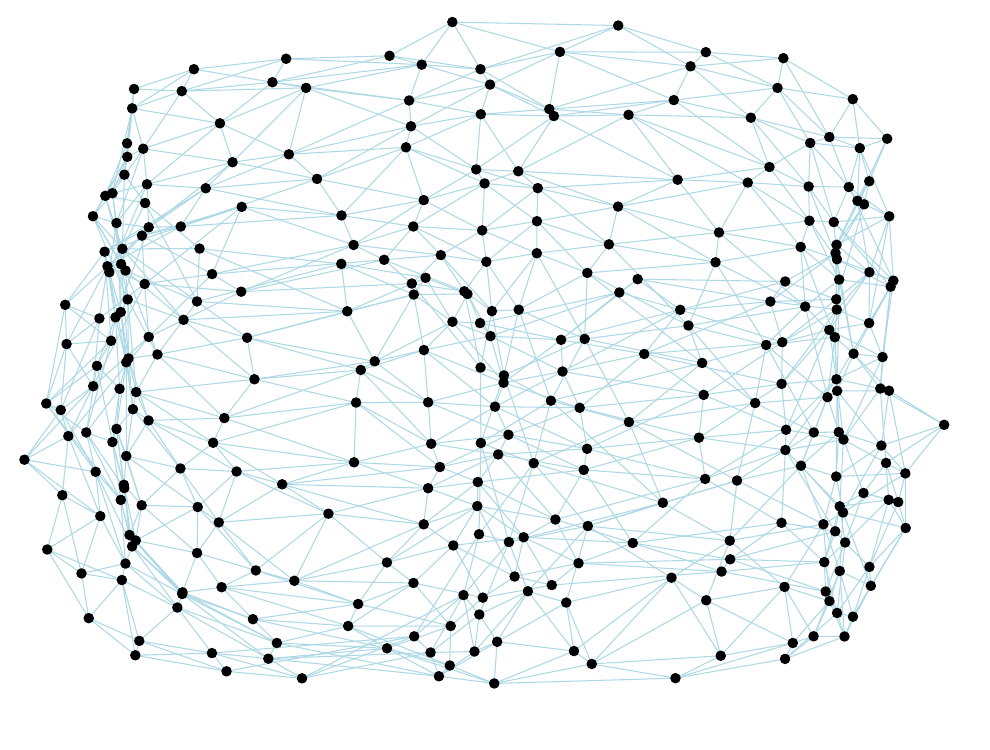} &
  \includegraphics[width=9.5mm]{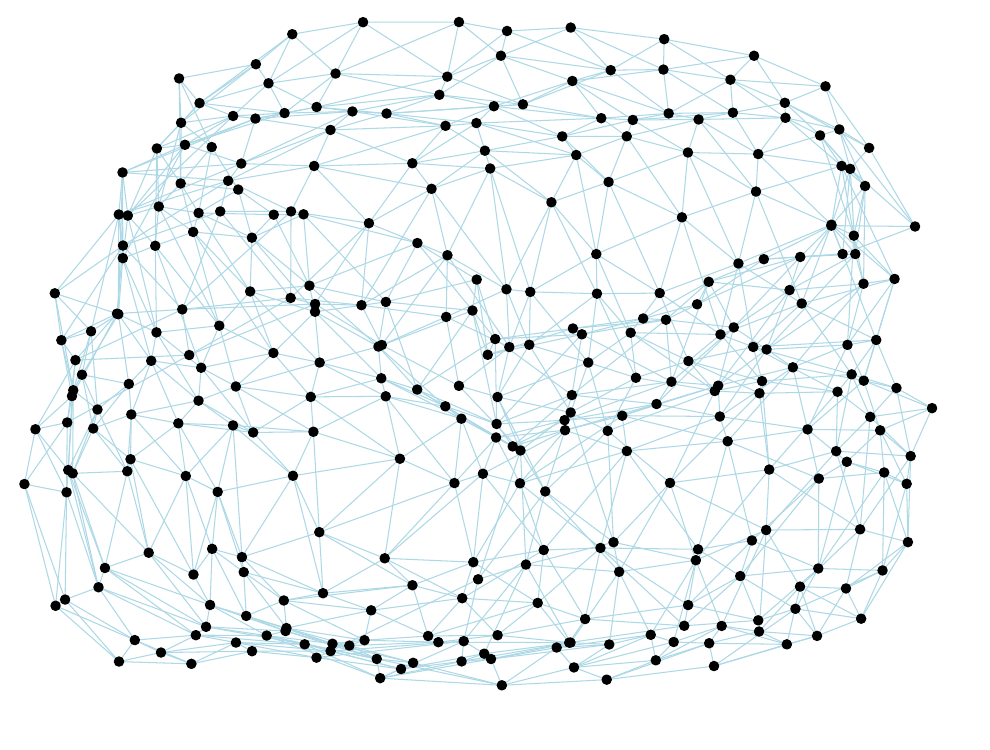} &
  \includegraphics[width=9.5mm]{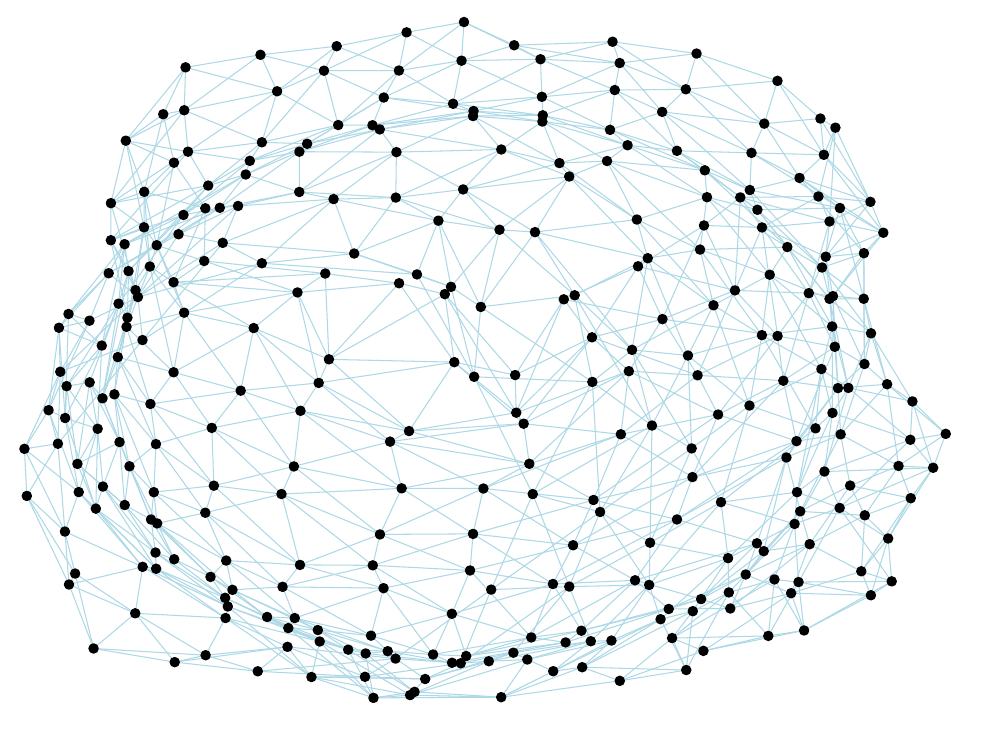} &
  \includegraphics[width=9.5mm]{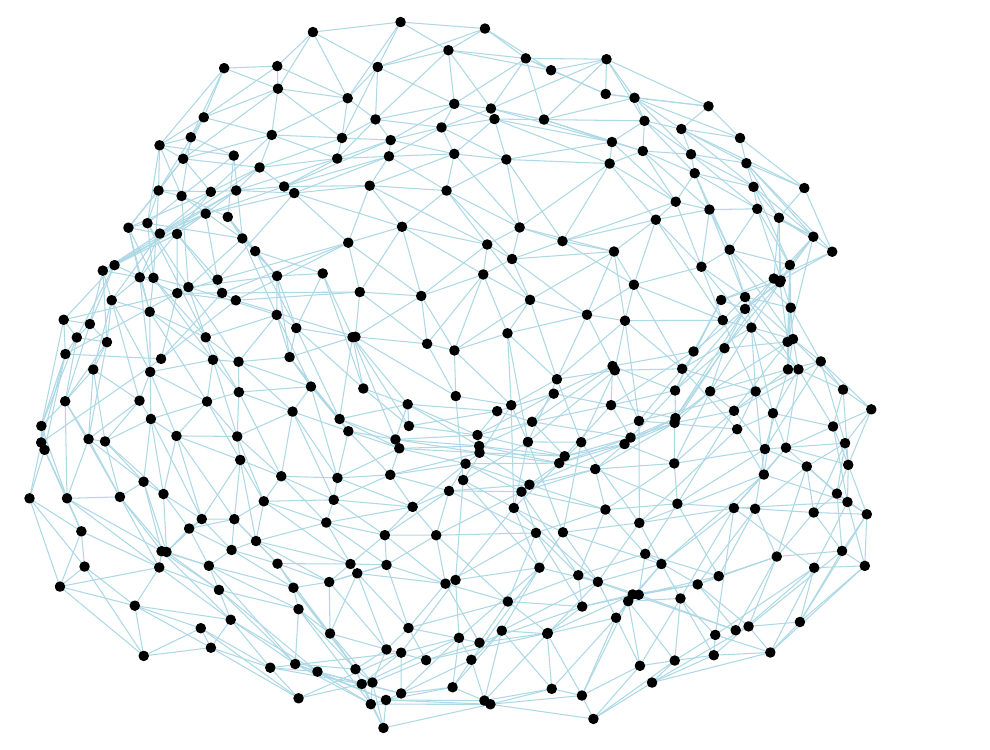} &
  \includegraphics[width=9.5mm]{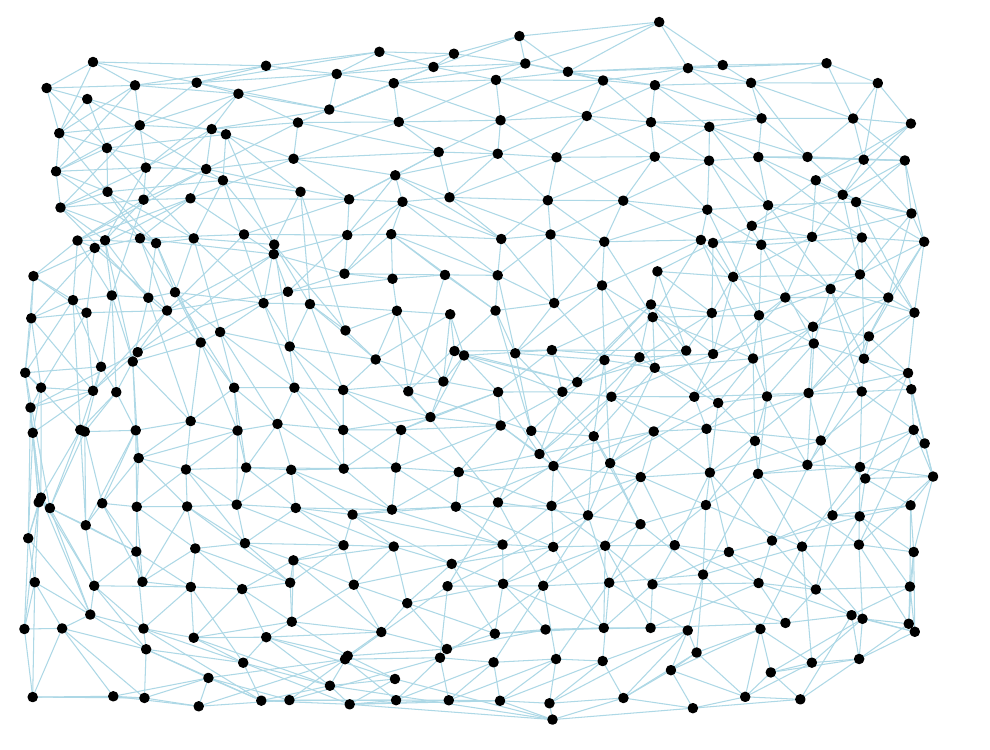}
  \\
  & \texttt{ST-AR} &
  & \includegraphics[width=9.5mm]{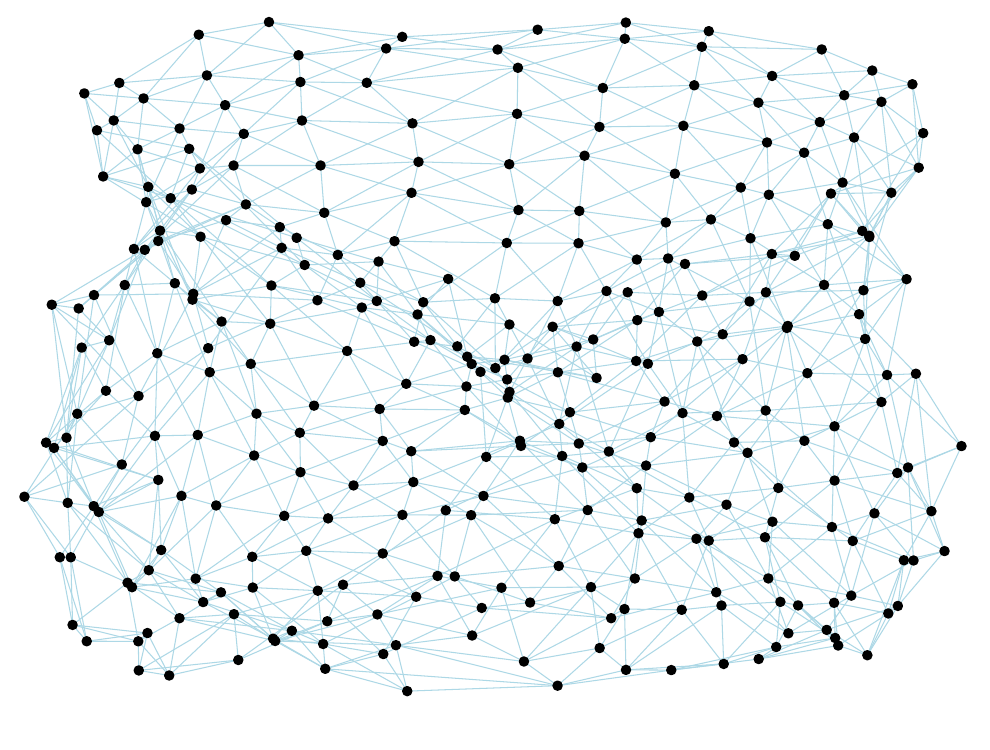} &
  \includegraphics[width=9.5mm]{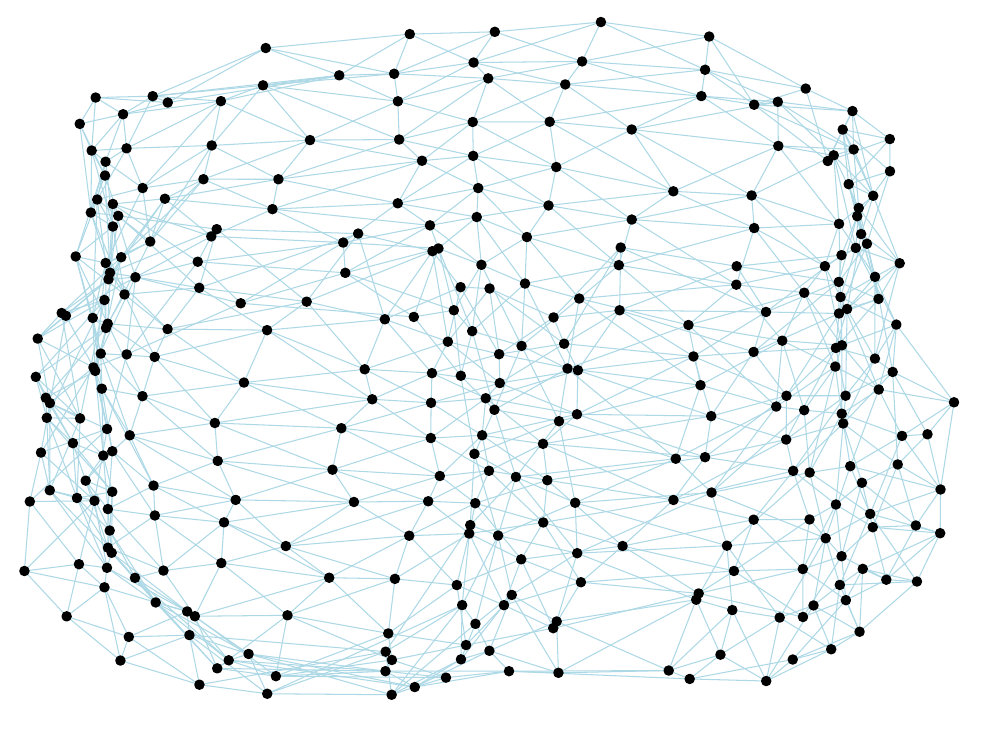} &
  \includegraphics[width=9.5mm]{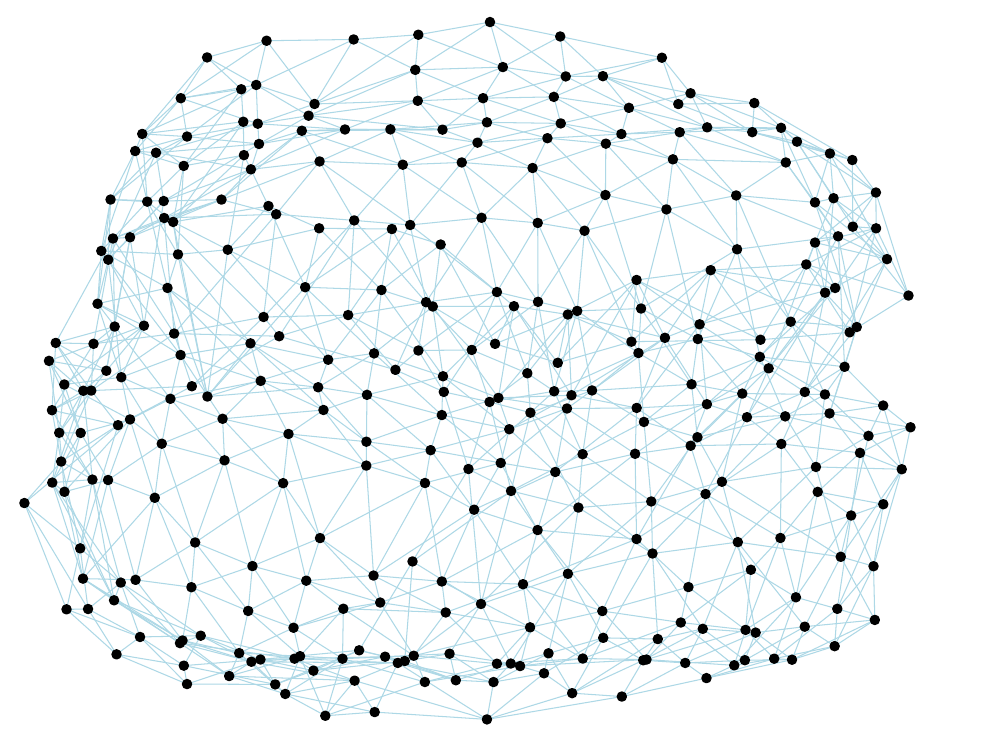} &
  \includegraphics[width=9.5mm]{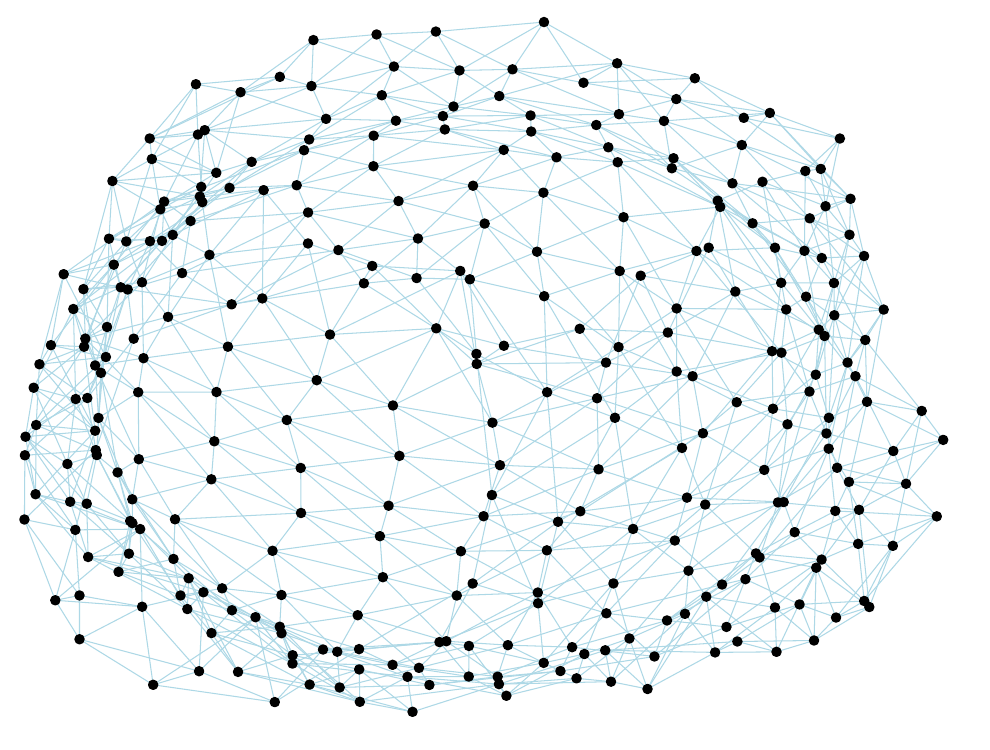} &
  \includegraphics[width=9.5mm]{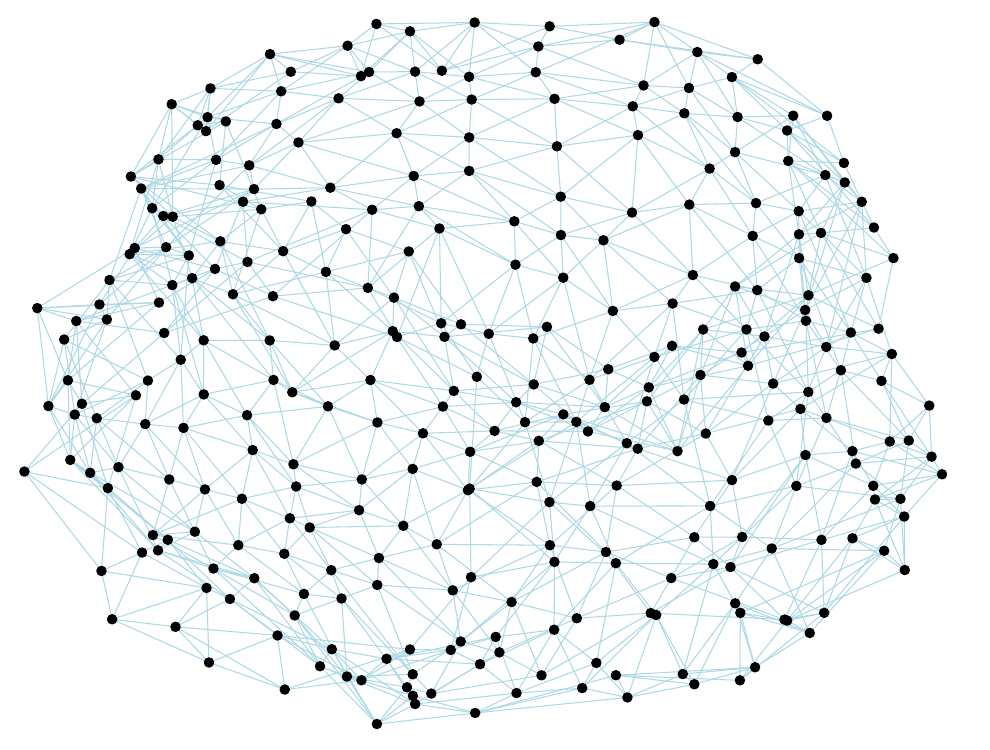} &
  \includegraphics[width=9.5mm]{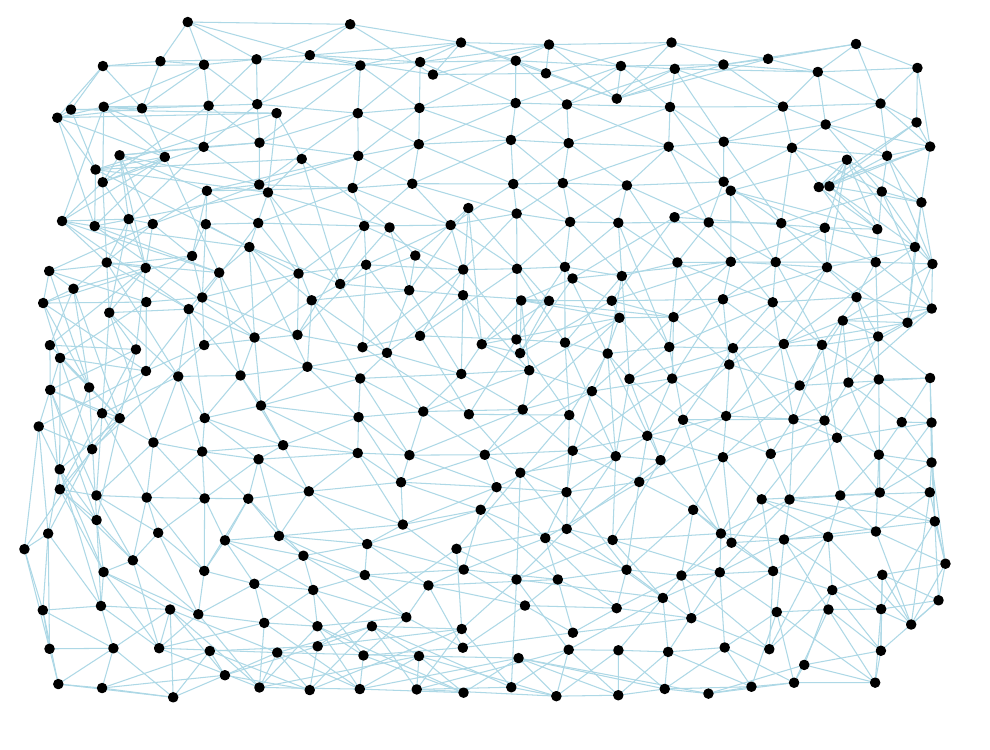}
  \\

     & \texttt{ELD-CN} & & \includegraphics[width=9.5mm]{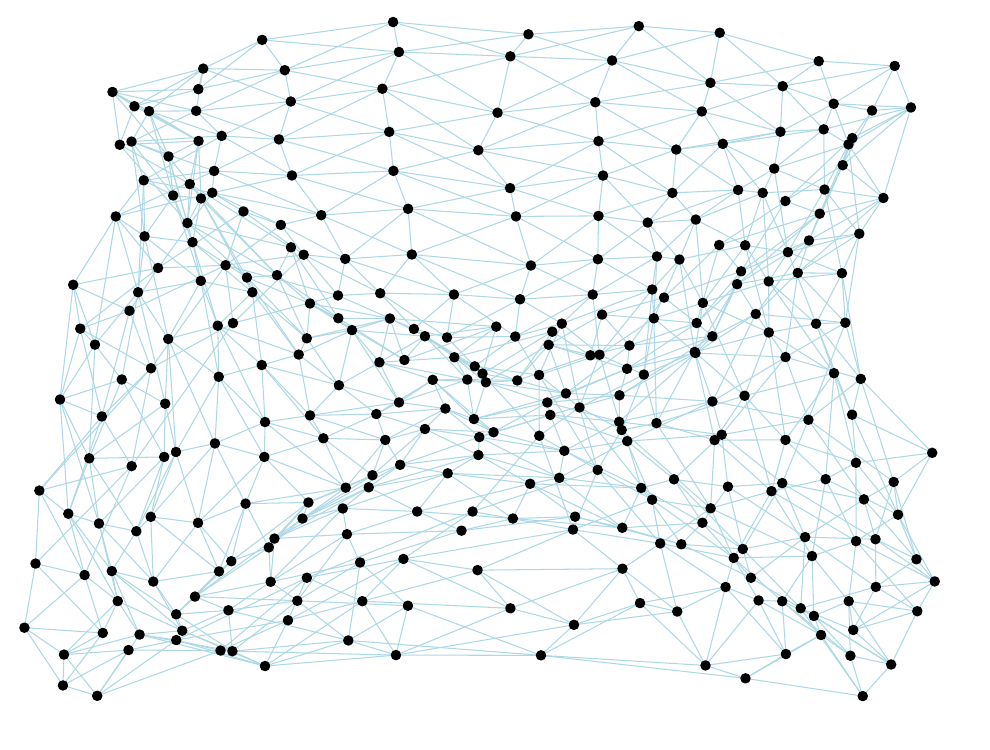} &
  \includegraphics[width=9.5mm]{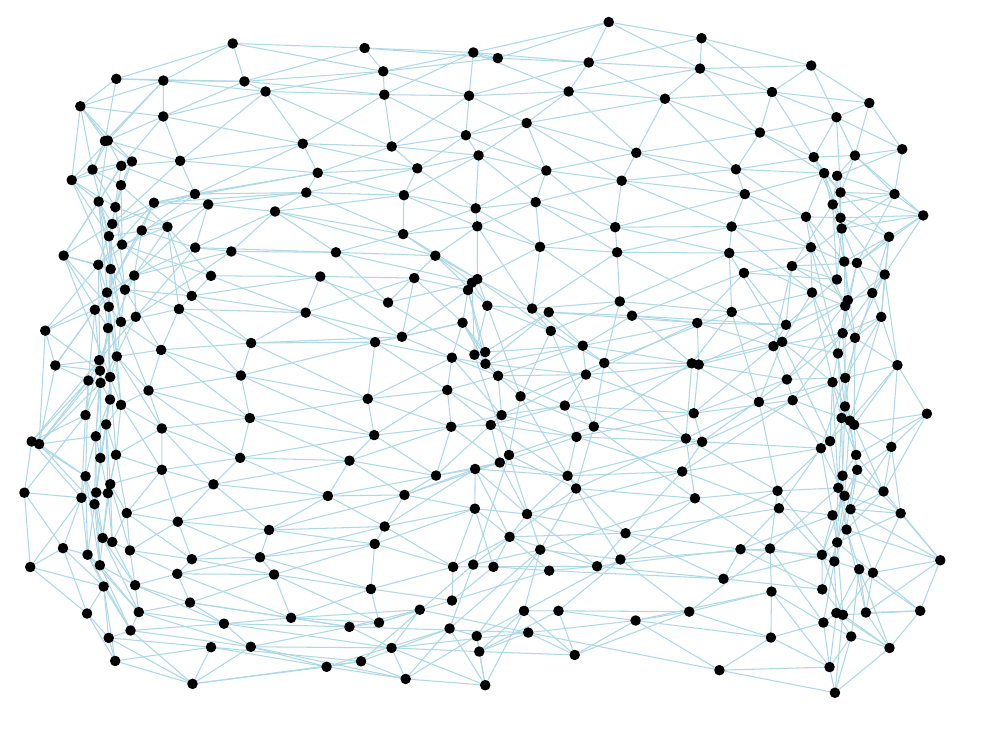} &
  \includegraphics[width=9.5mm]{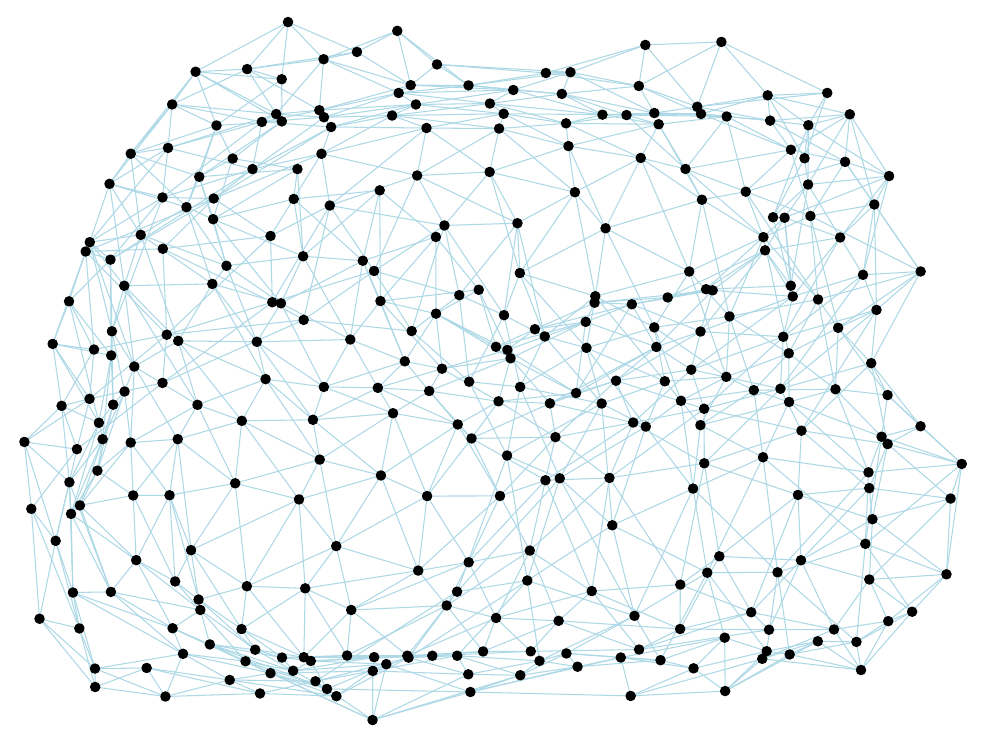} &
  \includegraphics[width=9.5mm]{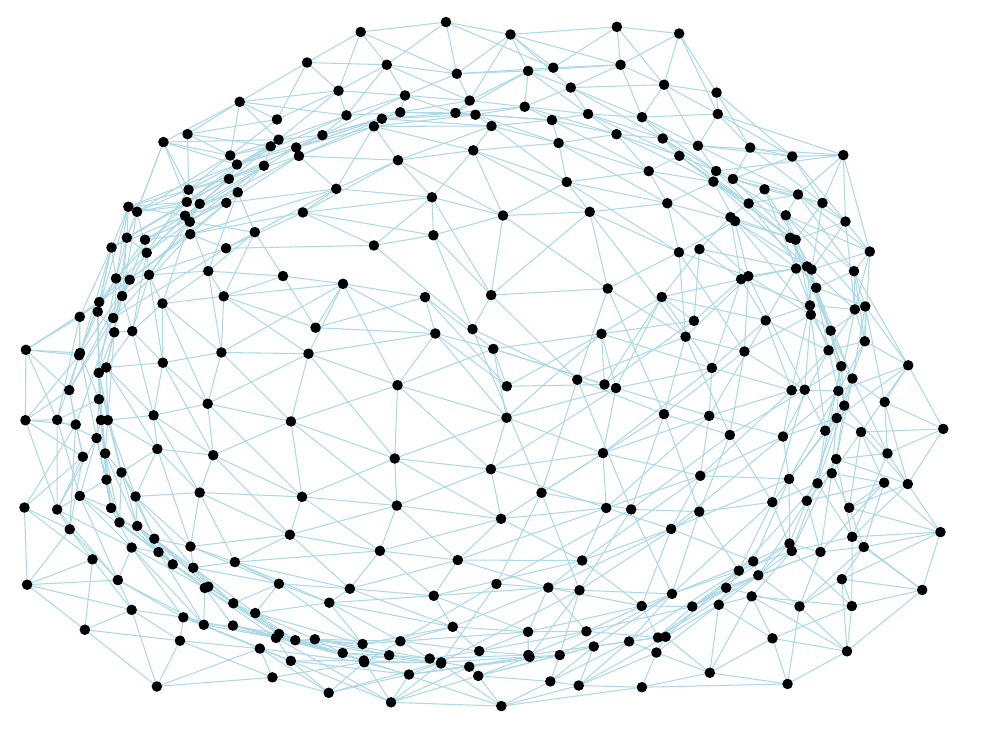} &
  \includegraphics[width=9.5mm]{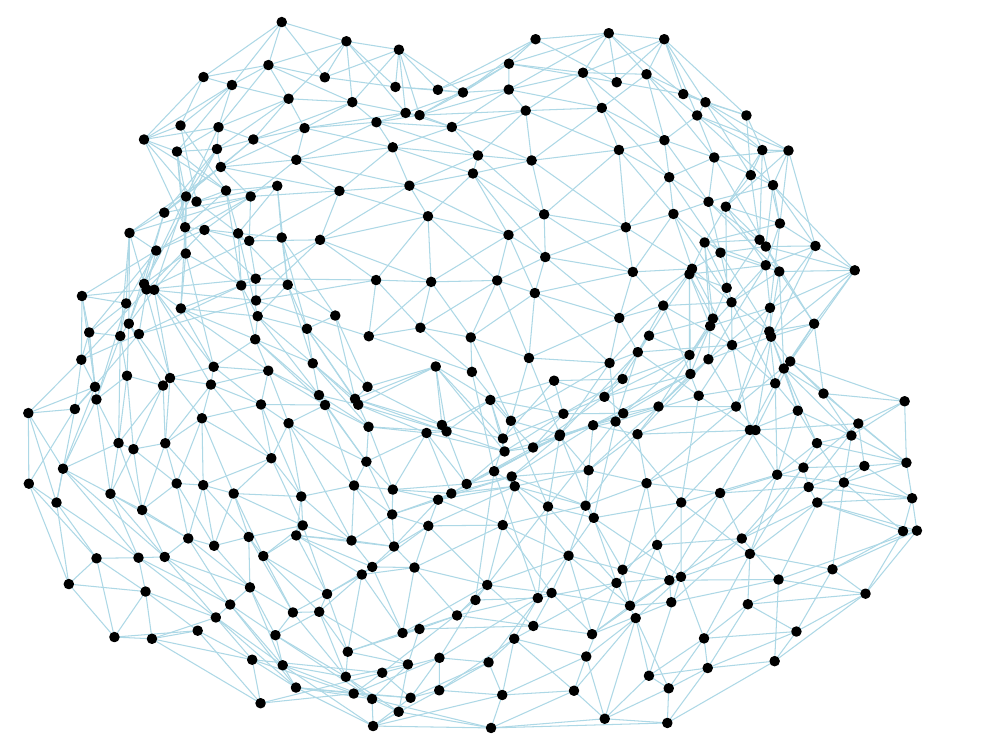} &
  \includegraphics[width=9.5mm]{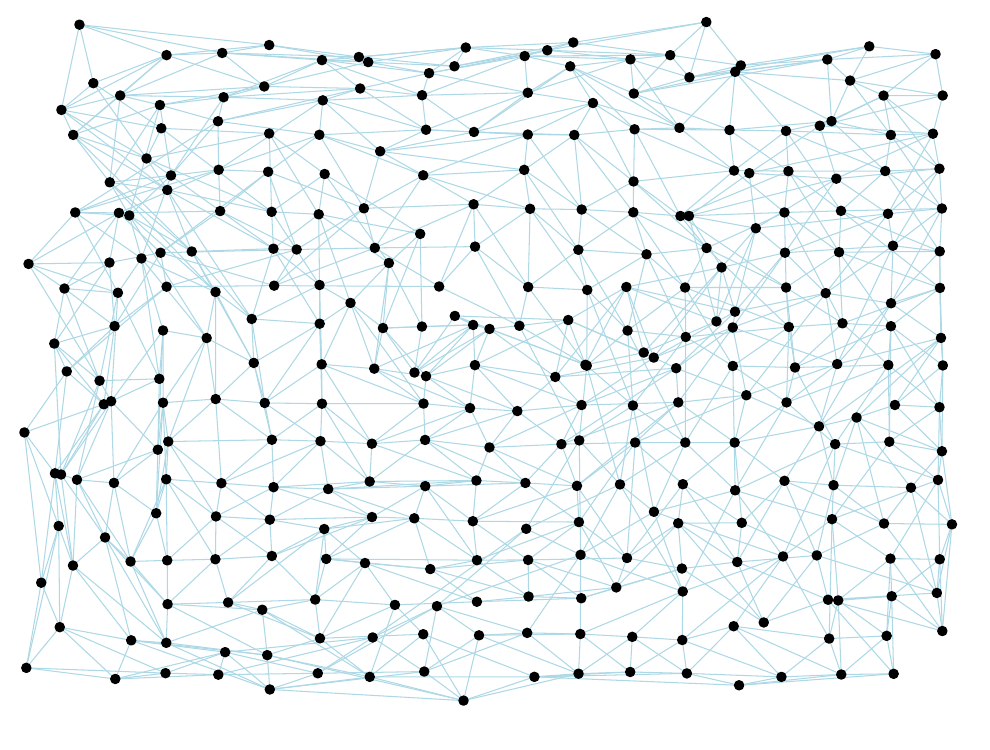}
  \\
       & \texttt{ELD-AR} & & \includegraphics[width=9.5mm]{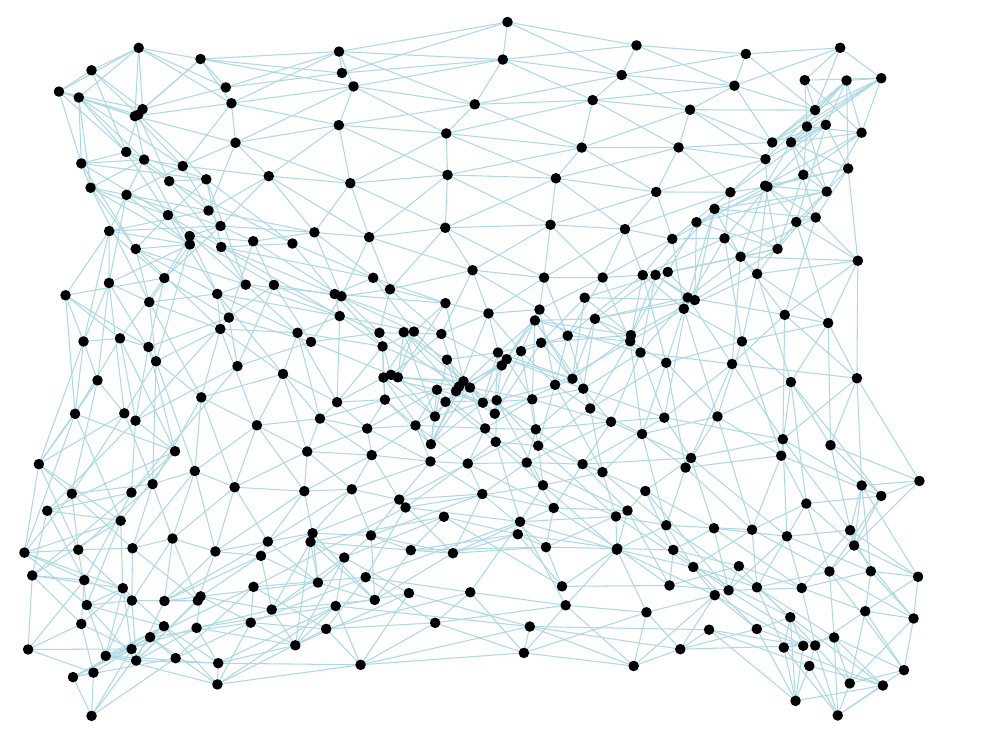} &
  \includegraphics[width=9.5mm]{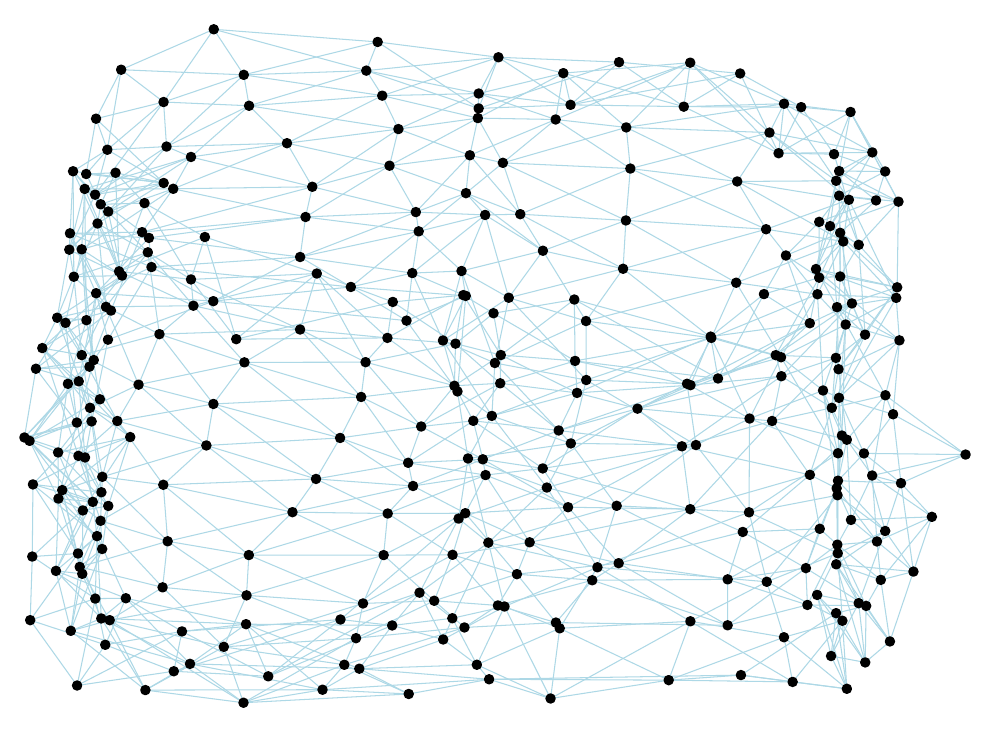} &
  \includegraphics[width=9.5mm]{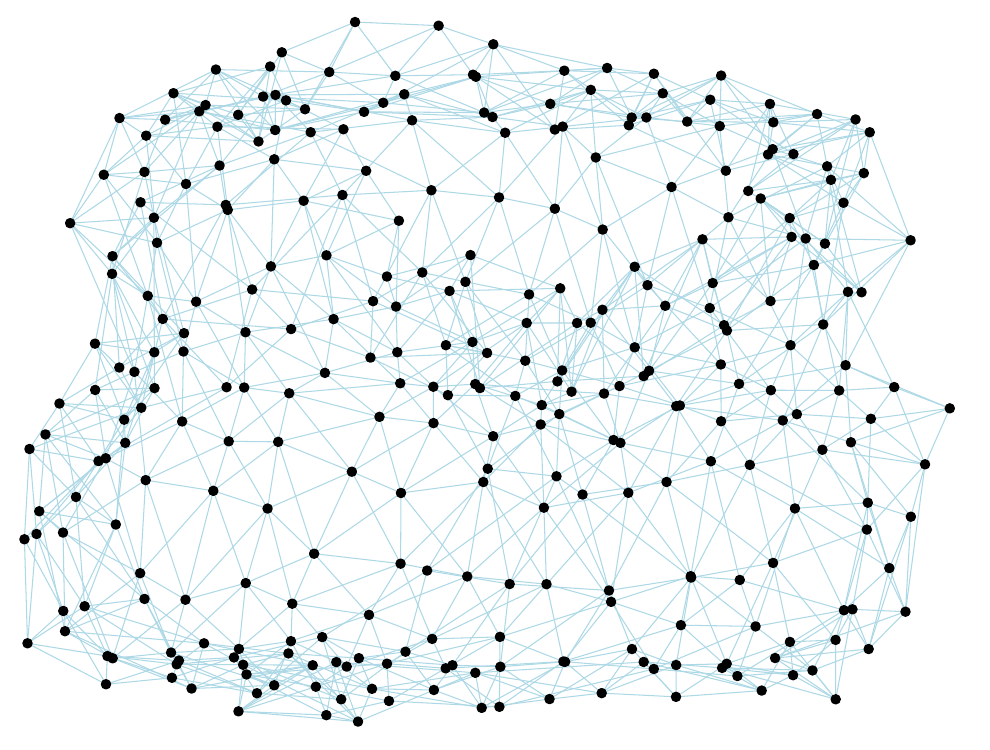} &
  \includegraphics[width=9.5mm]{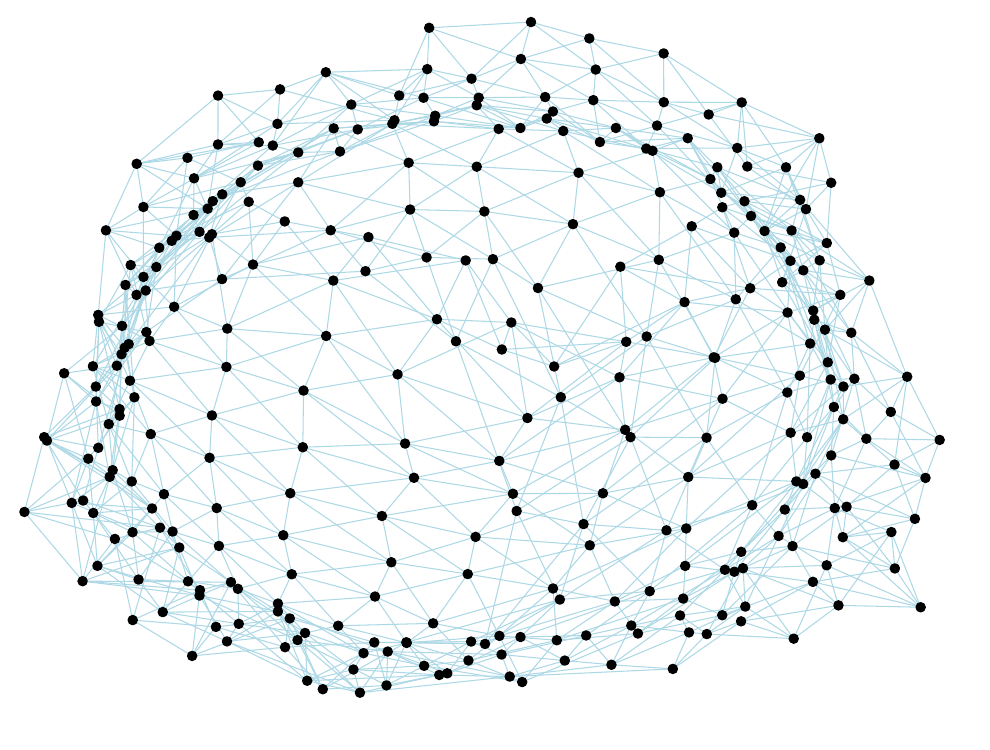} &
  \includegraphics[width=9.5mm]{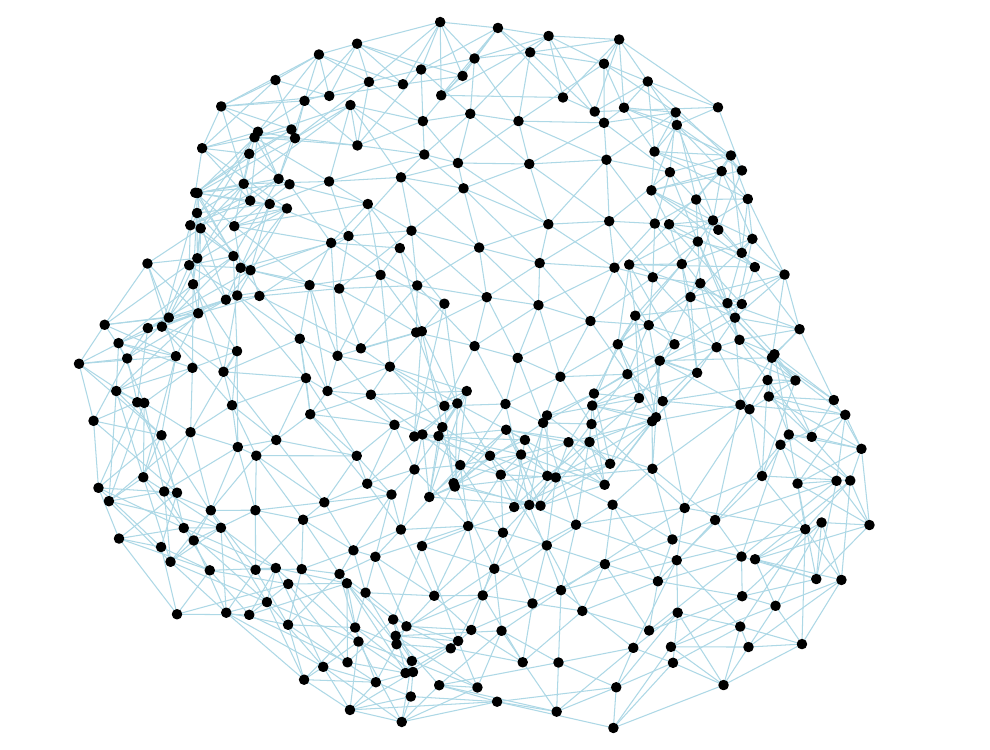} &
  \includegraphics[width=9.5mm]{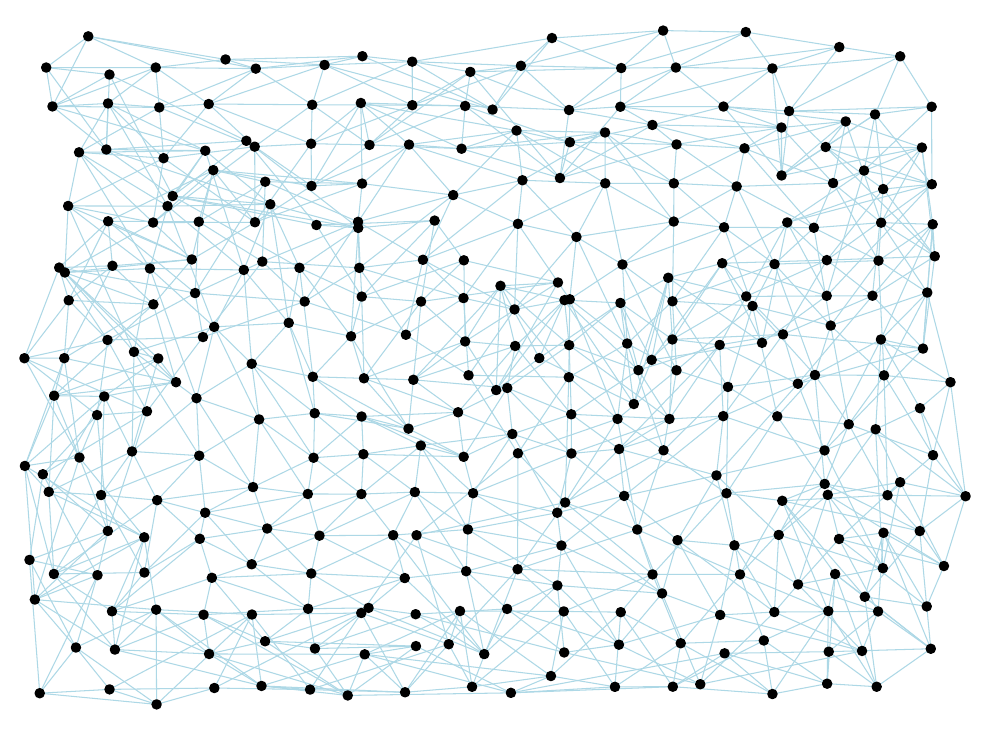}
  \\
       & \texttt{CN-AR} & & \includegraphics[width=9.5mm]{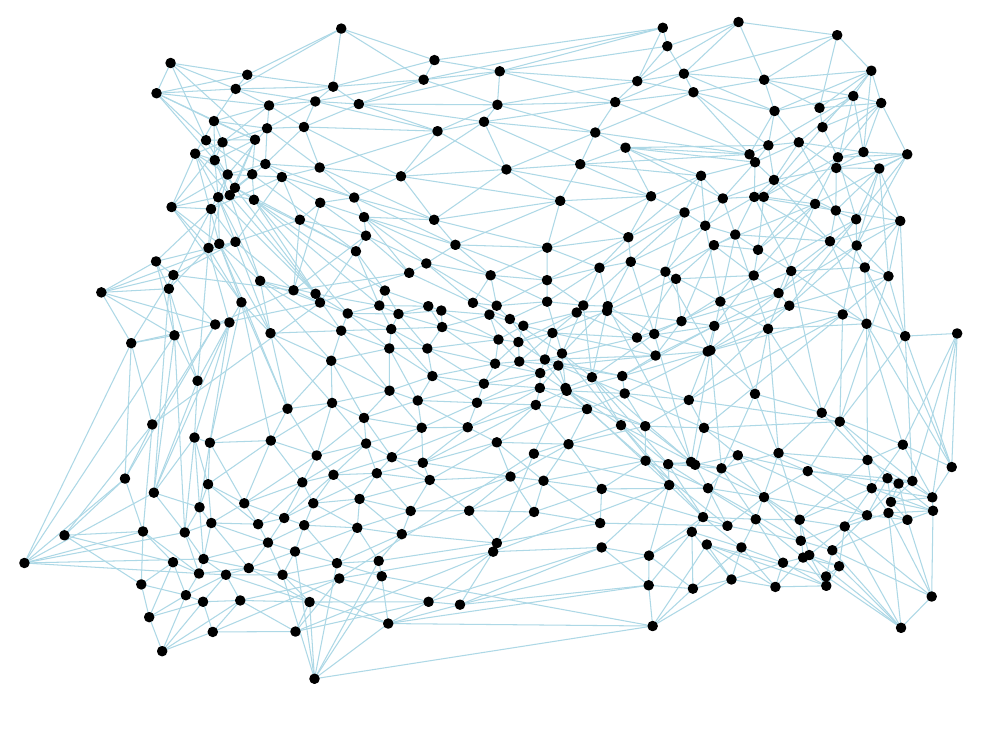} &
  \includegraphics[width=9.5mm]{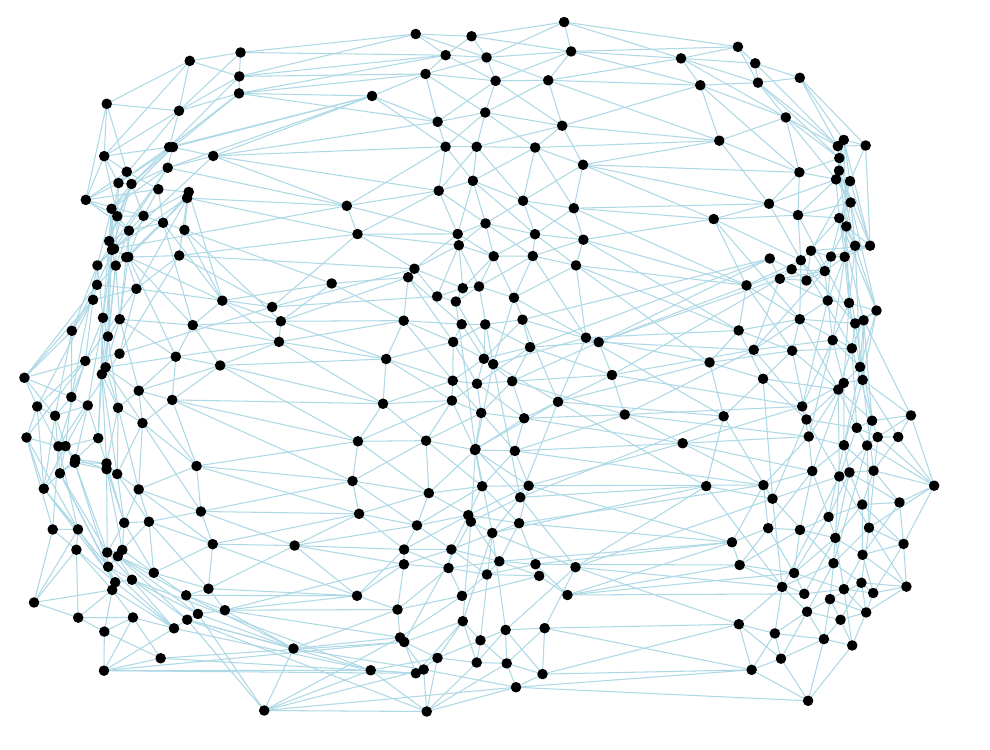} &
  \includegraphics[width=9.5mm]{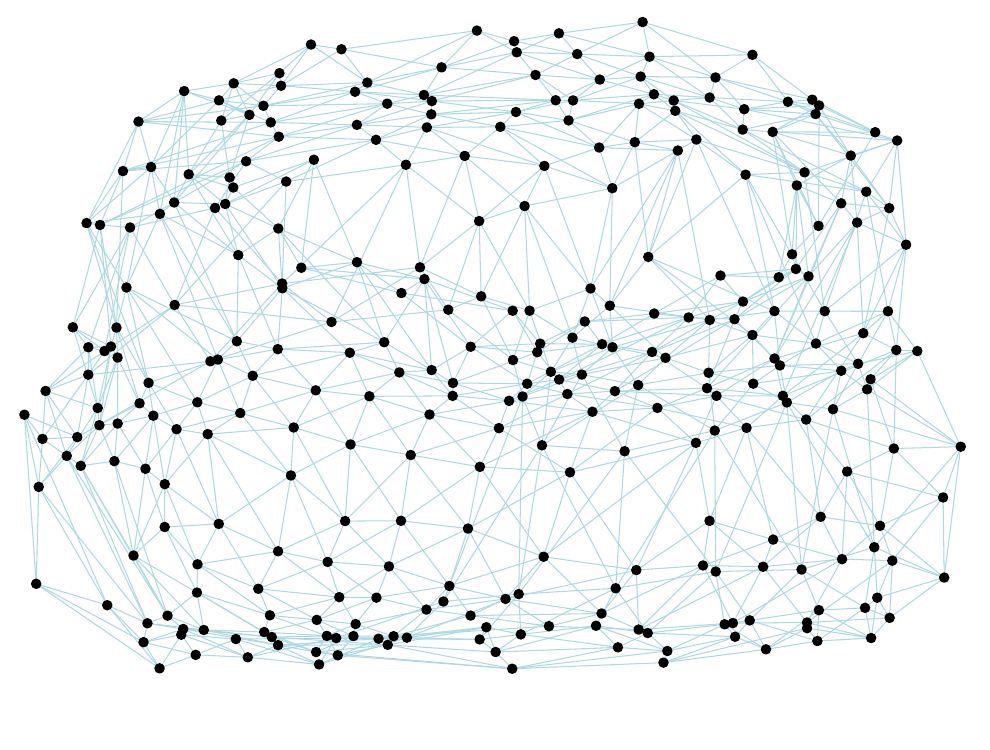} &
  \includegraphics[width=9.5mm]{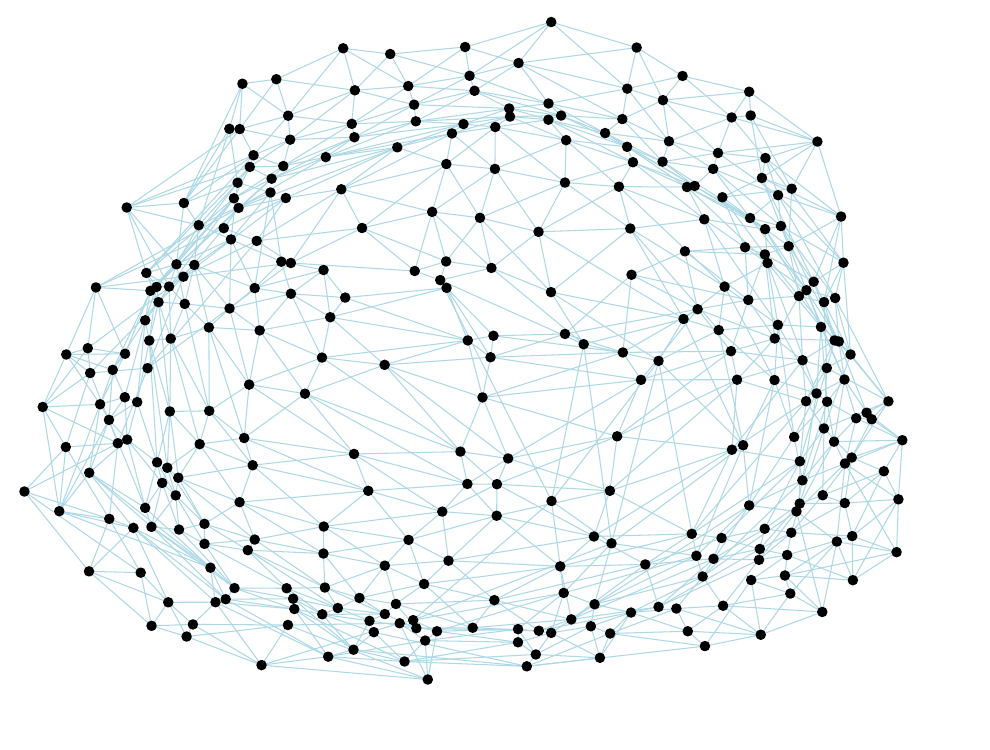} &
  \includegraphics[width=9.5mm]{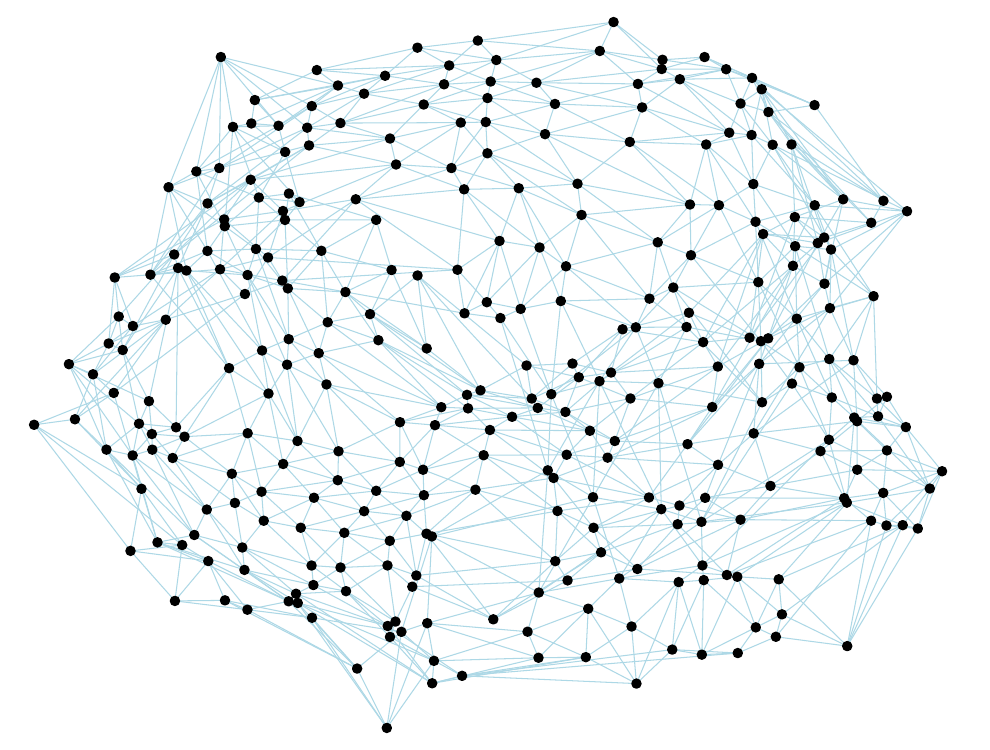} &
  \includegraphics[width=9.5mm]{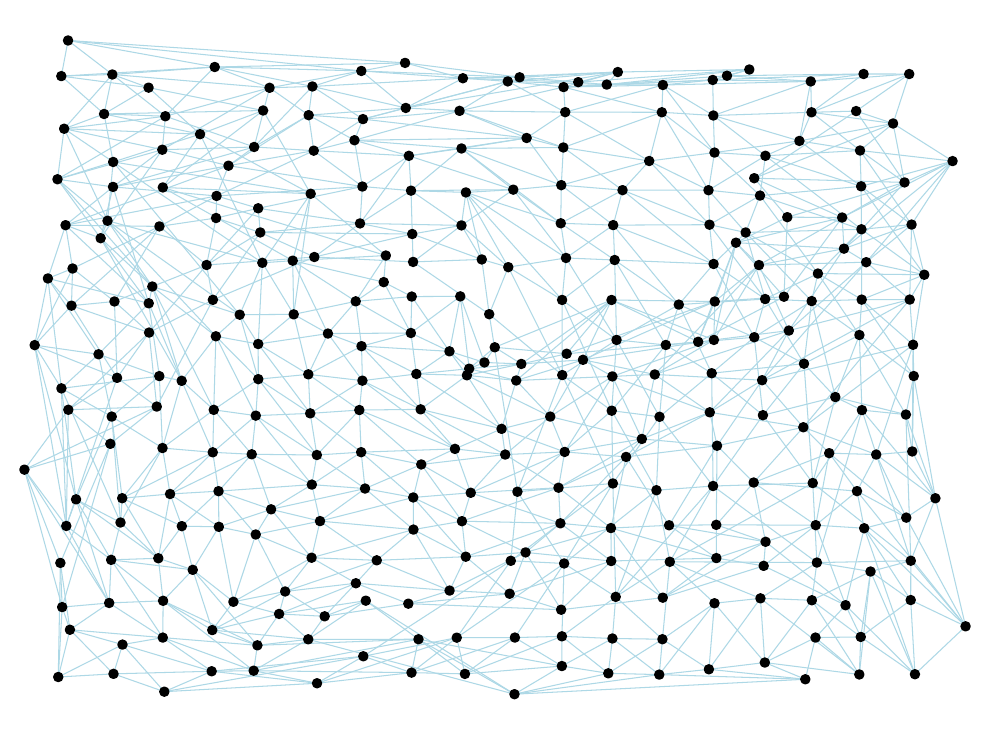}
  \\
       & \texttt{ST-ELD-CN} & &\includegraphics[width=9.5mm]{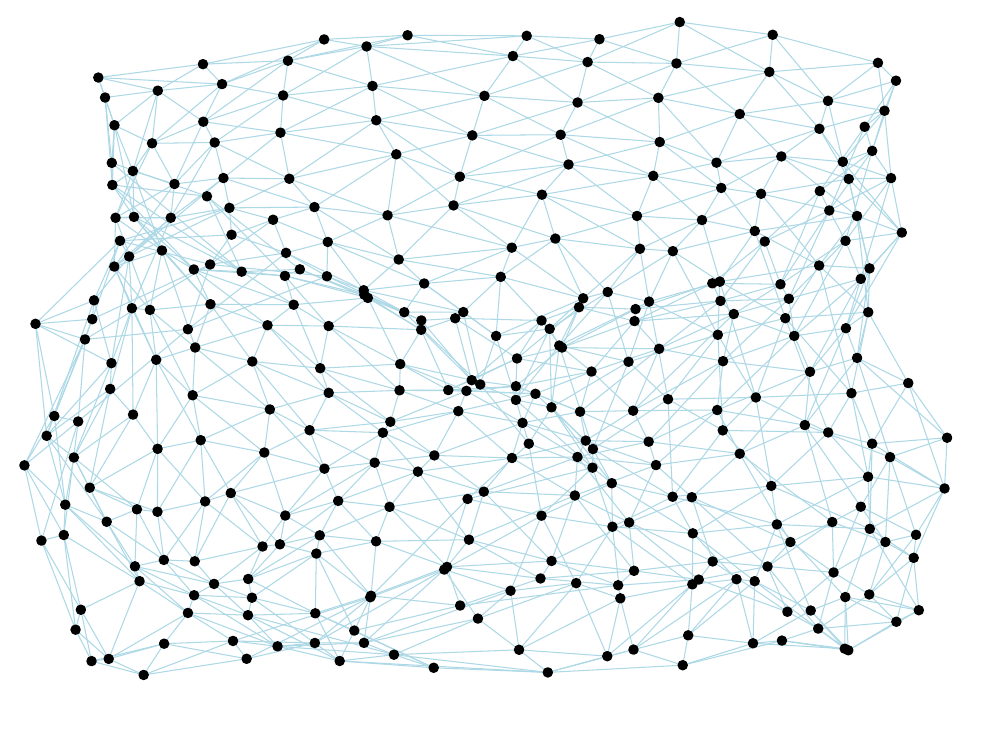} &
  \includegraphics[width=9.5mm]{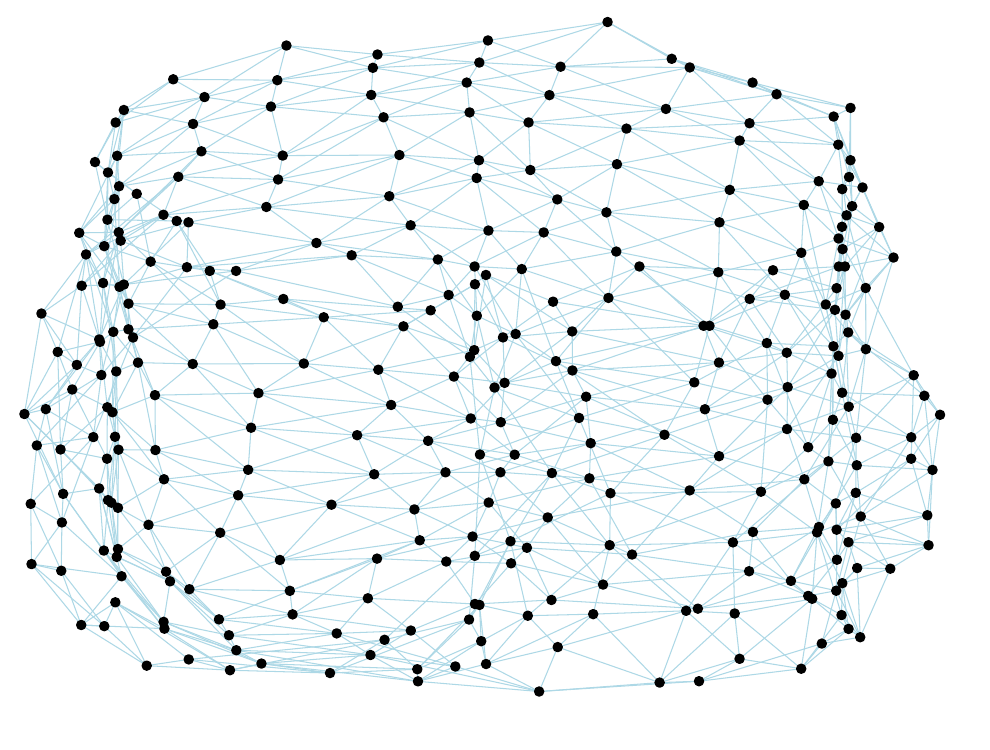} &
  \includegraphics[width=9.5mm]{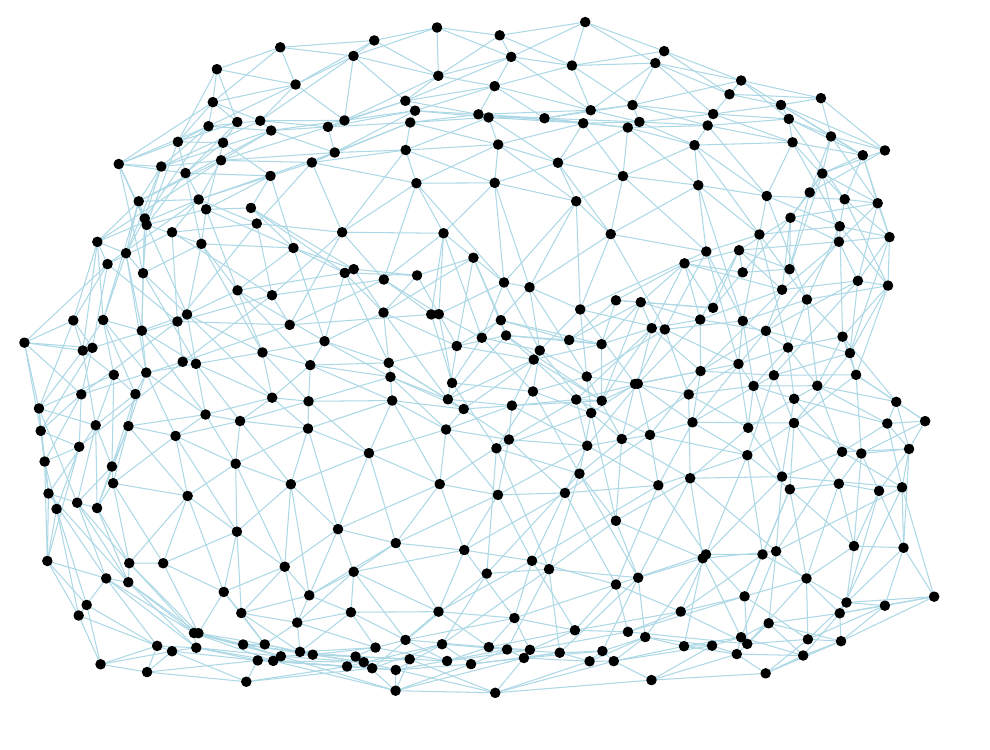} &
  \includegraphics[width=9.5mm]{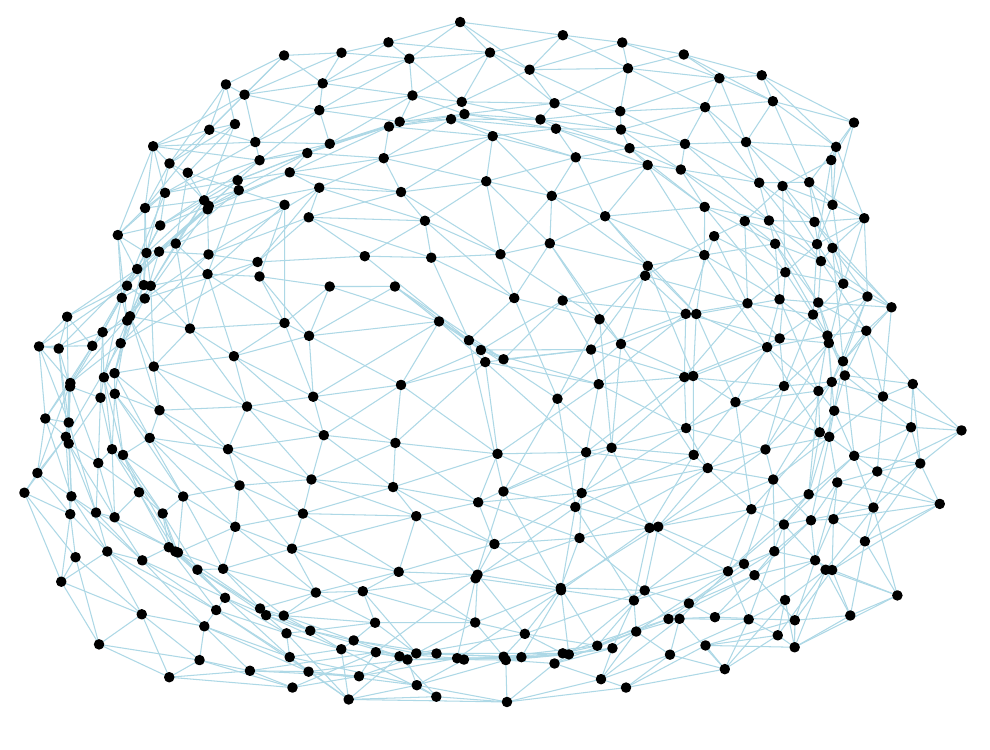} &
  \includegraphics[width=9.5mm]{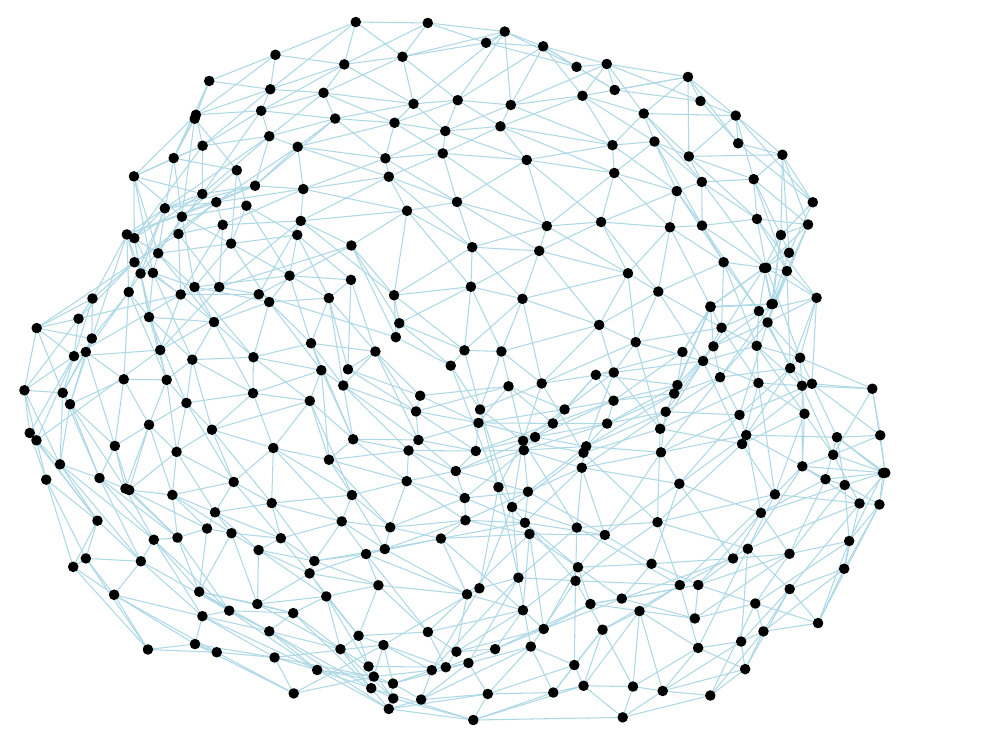} &
  \includegraphics[width=9.5mm]{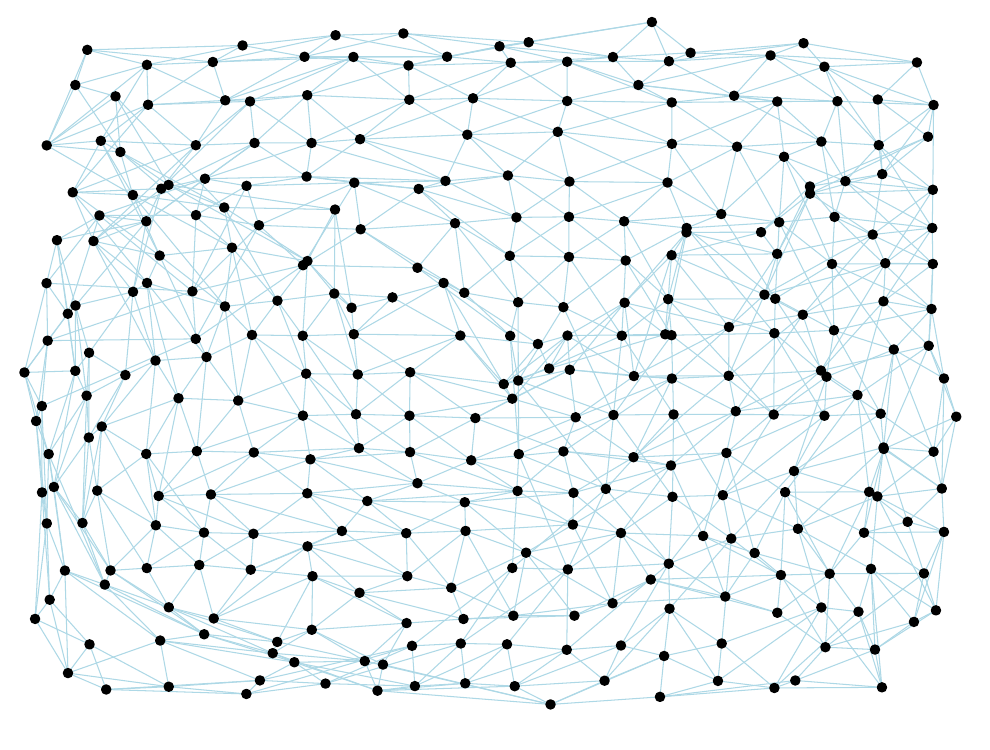}
  \\
         & \texttt{ST-ELD-AR} & &\includegraphics[width=9.5mm]{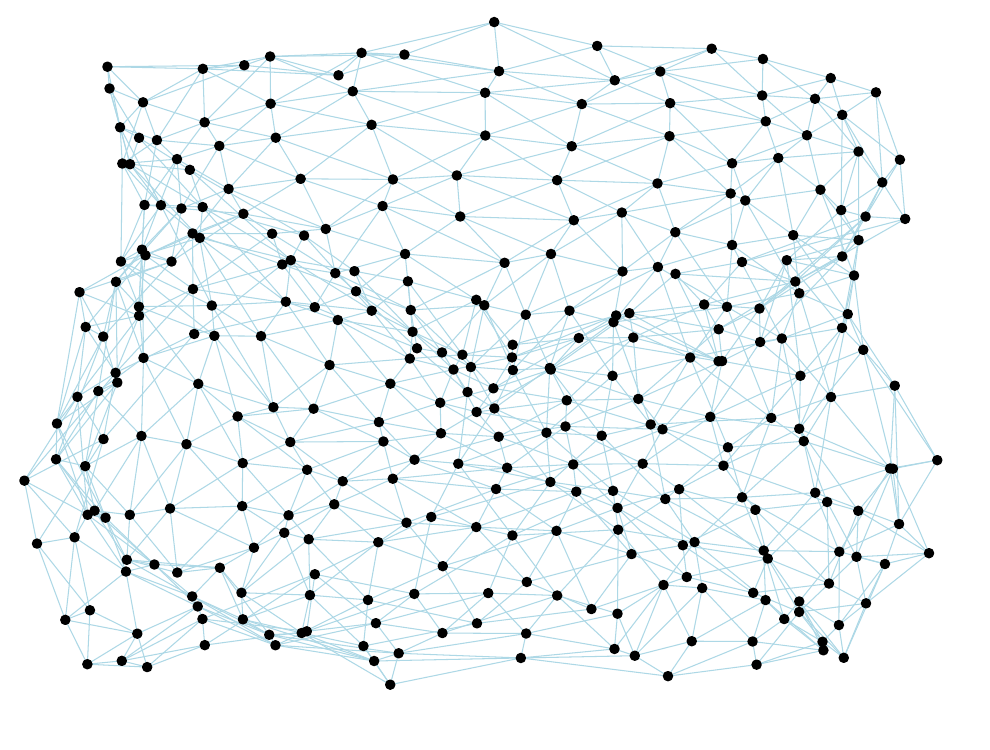} &
  \includegraphics[width=9.5mm]{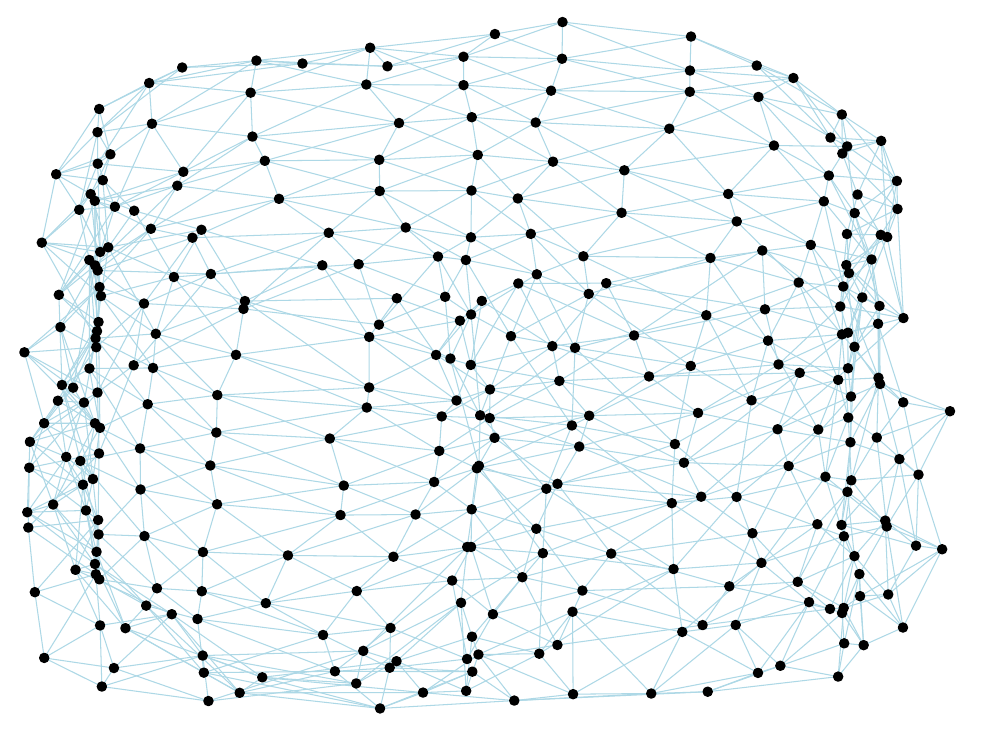} &
  \includegraphics[width=9.5mm]{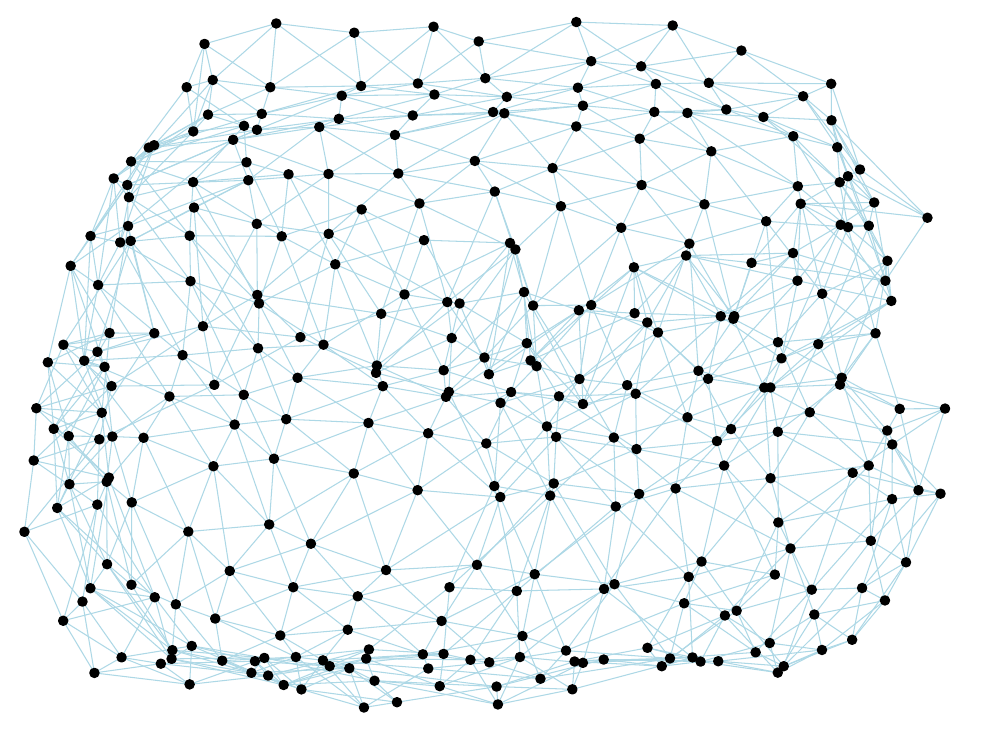} &
  \includegraphics[width=9.5mm]{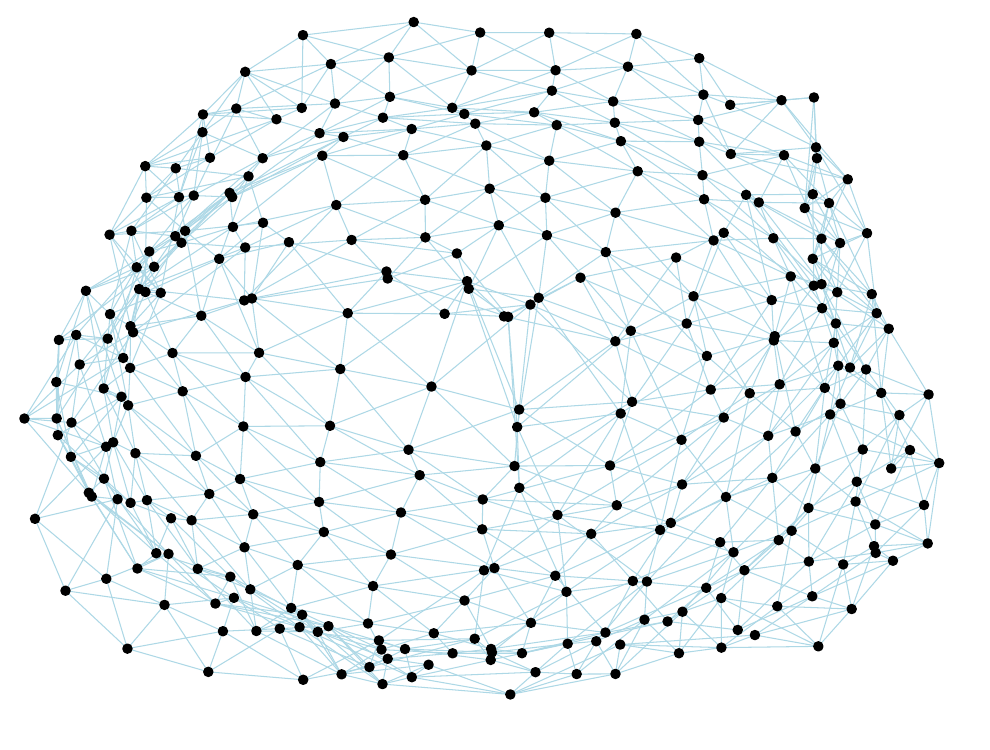} &
  \includegraphics[width=9.5mm]{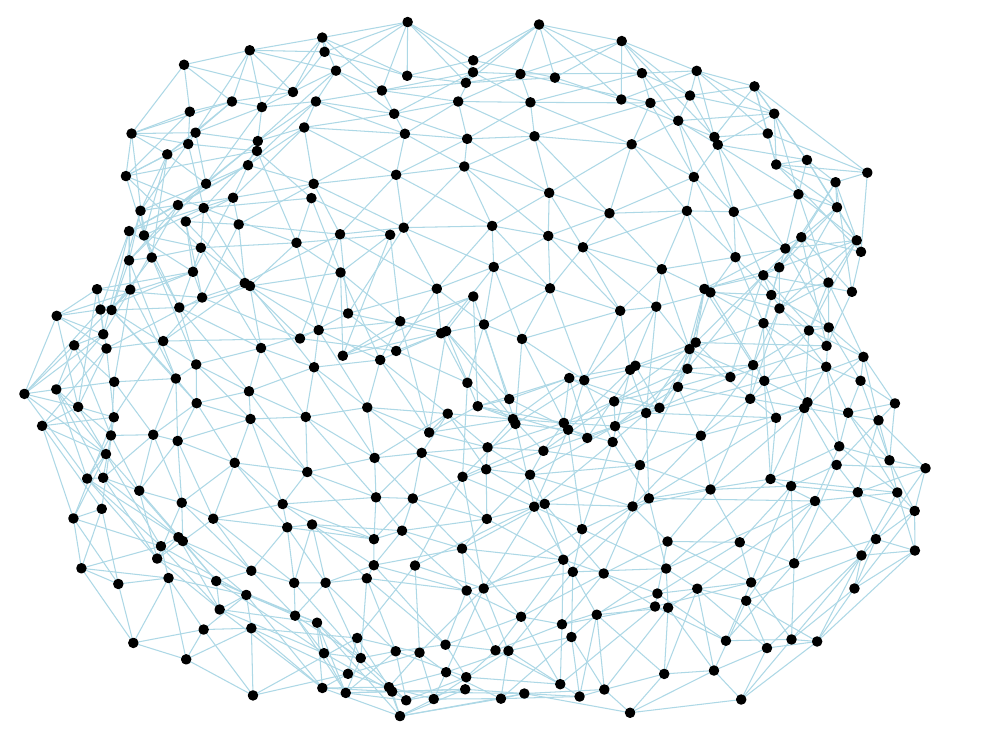} &
  \includegraphics[width=9.5mm]{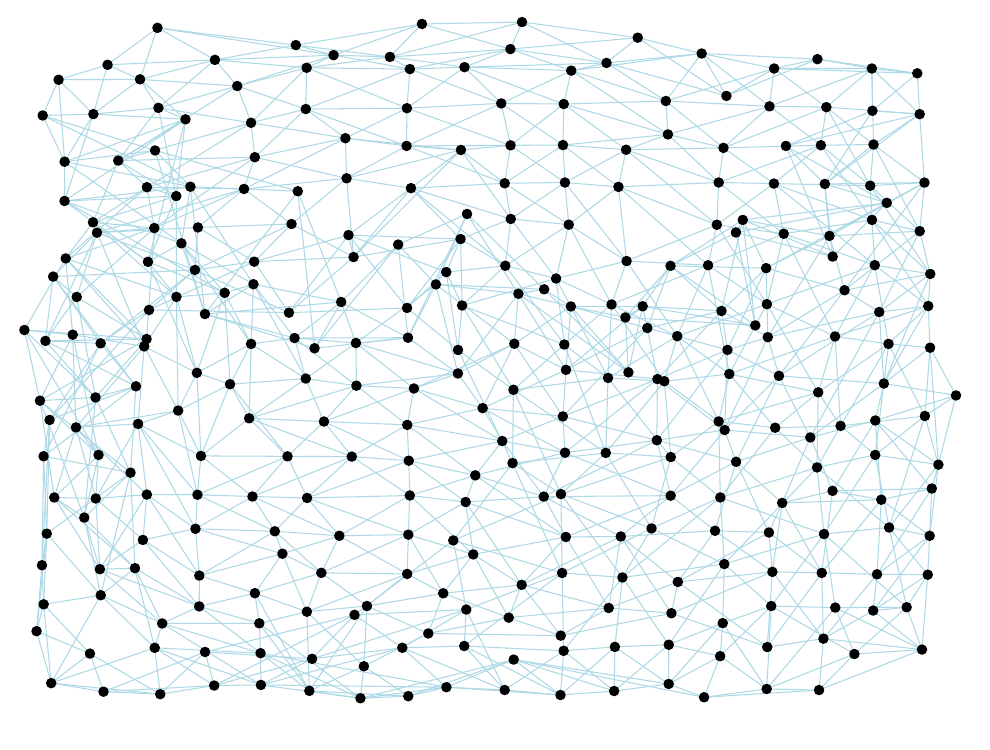}
  \\
         & \texttt{ST-CN-AR} & &\includegraphics[width=9.5mm]{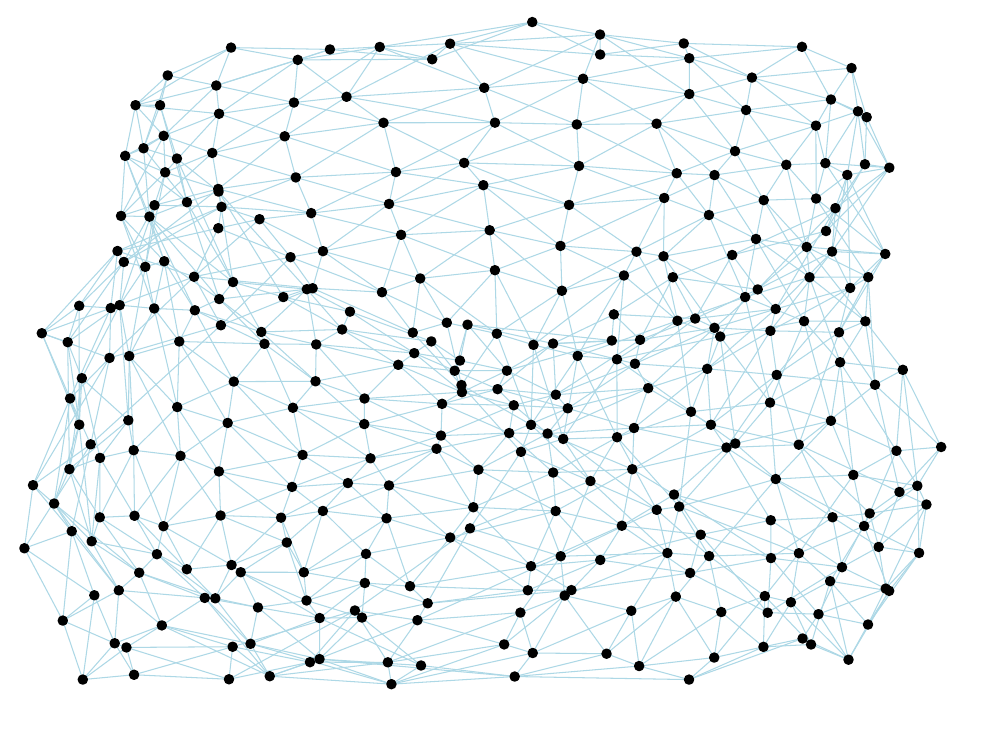} &
  \includegraphics[width=9.5mm]{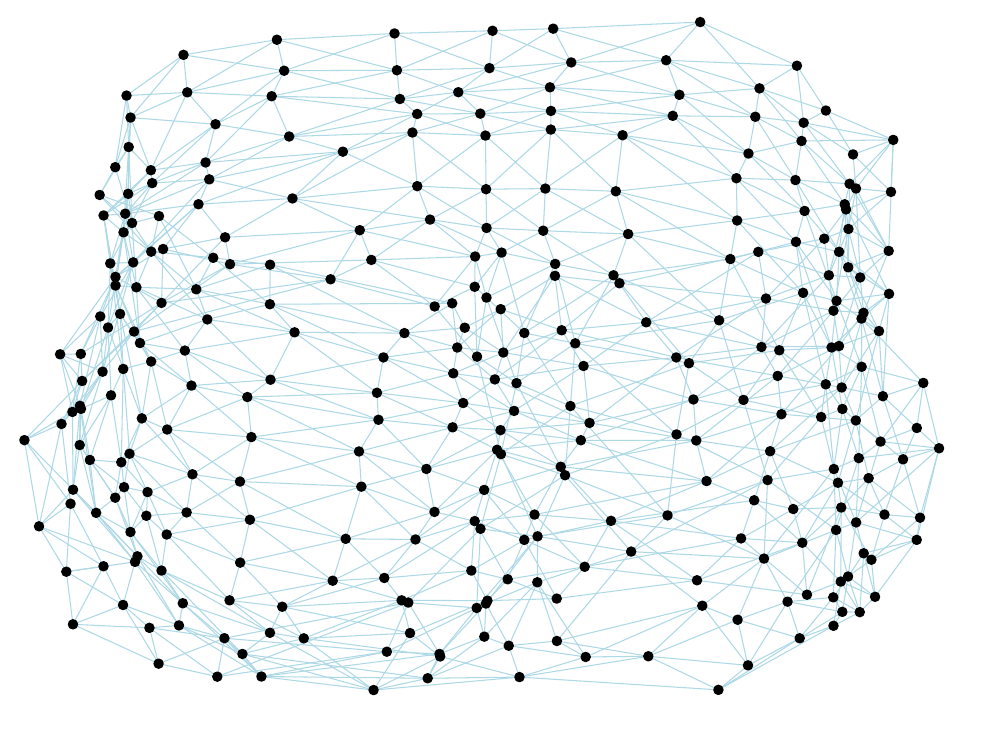} &
  \includegraphics[width=9.5mm]{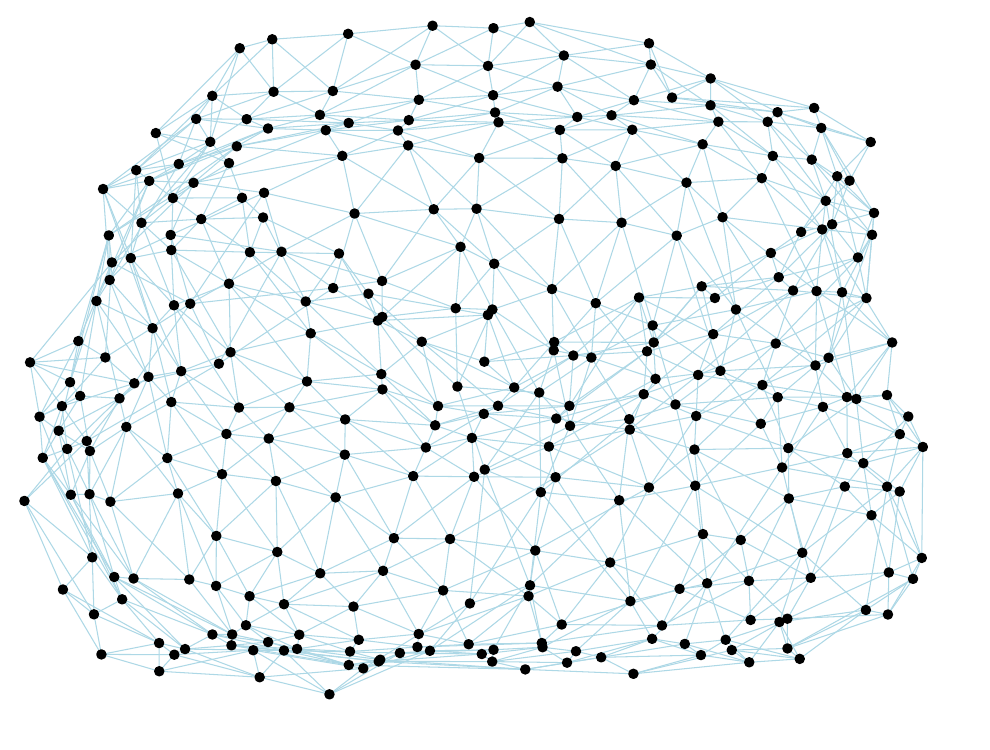} &
  \includegraphics[width=9.5mm]{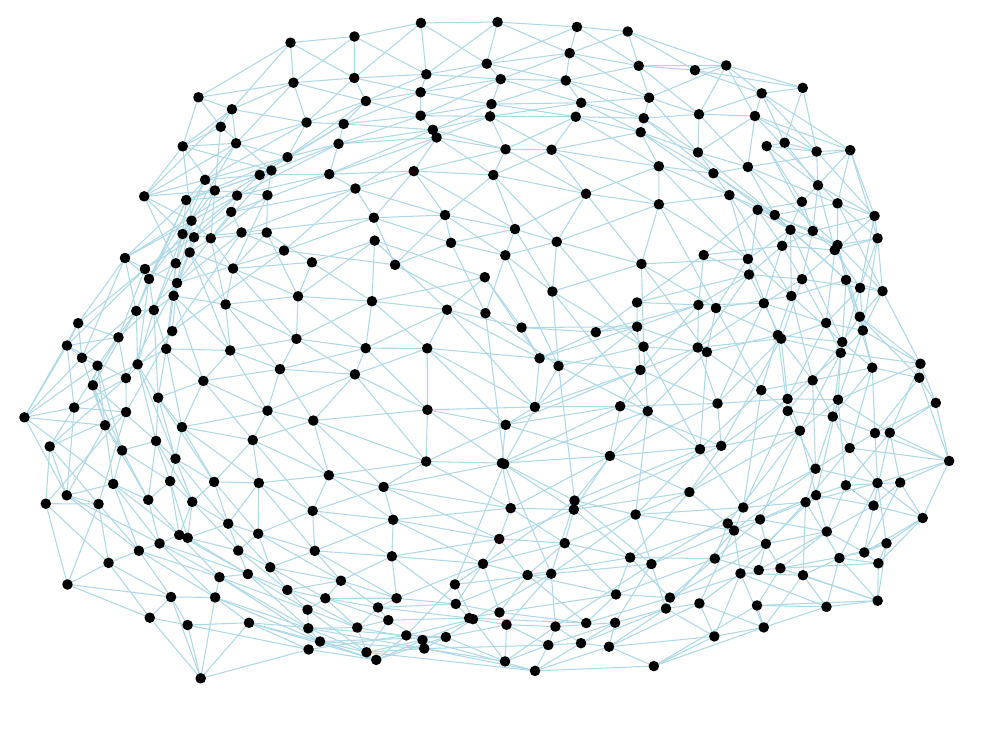} &
  \includegraphics[width=9.5mm]{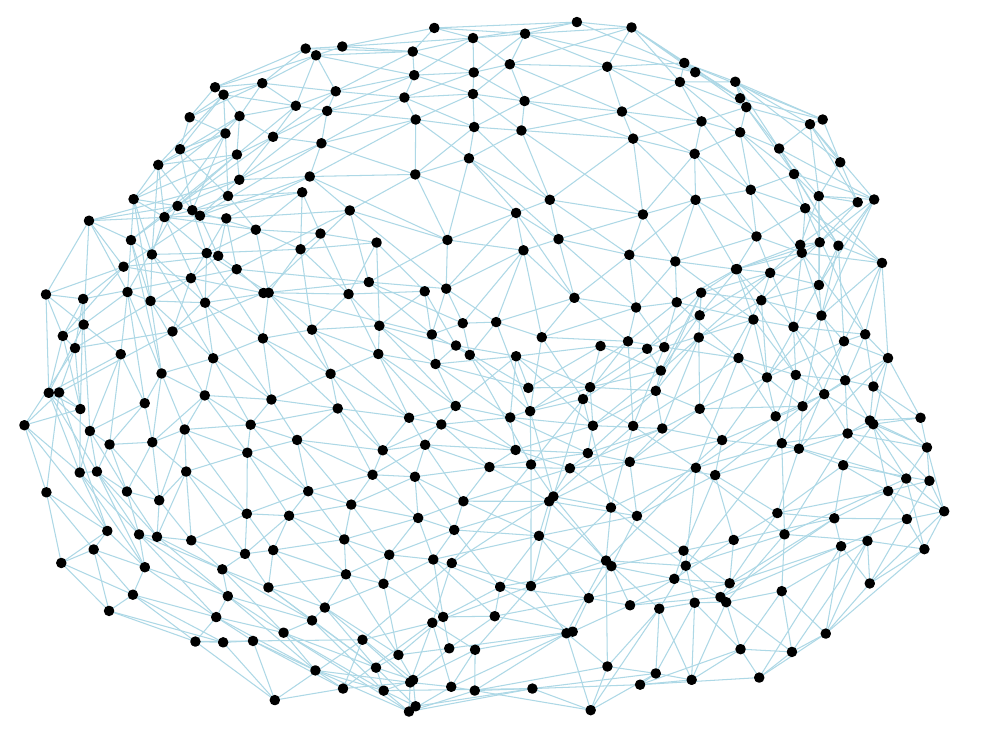} &
  \includegraphics[width=9.5mm]{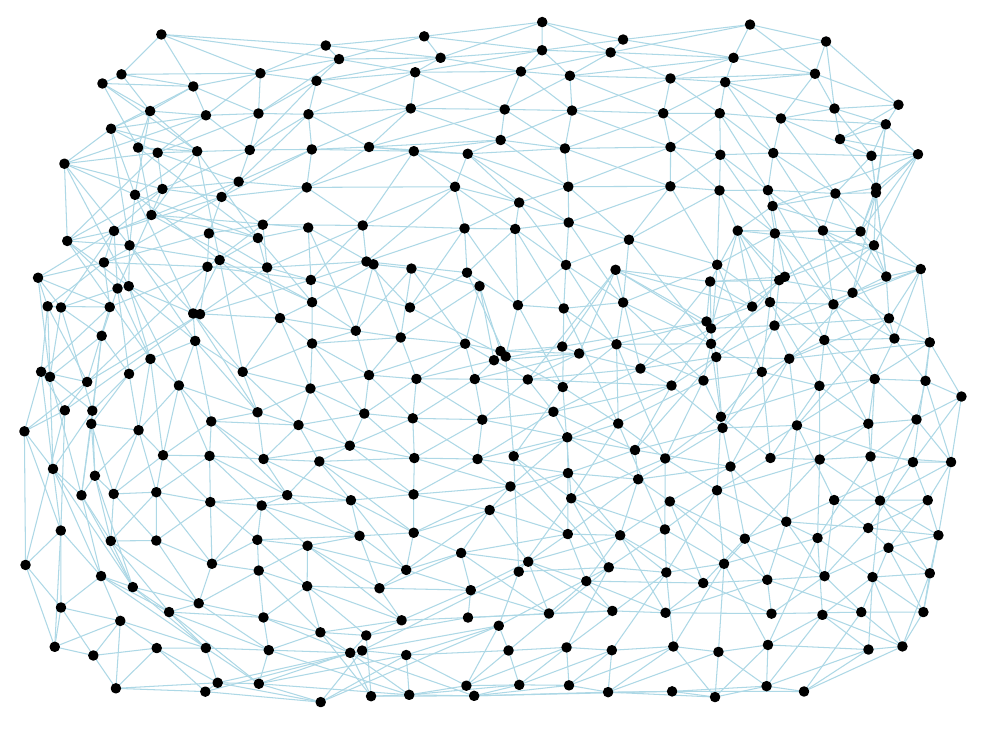}
  \\
         & \texttt{ELD-CN-AR} & & \includegraphics[width=9.5mm]{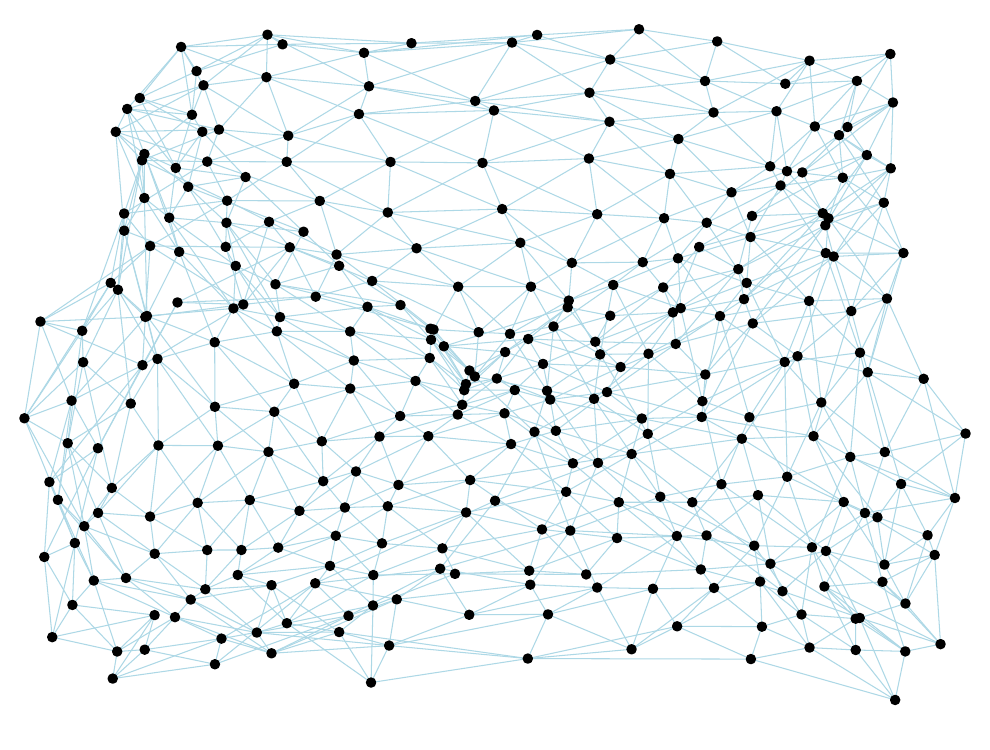} &
  \includegraphics[width=9.5mm]{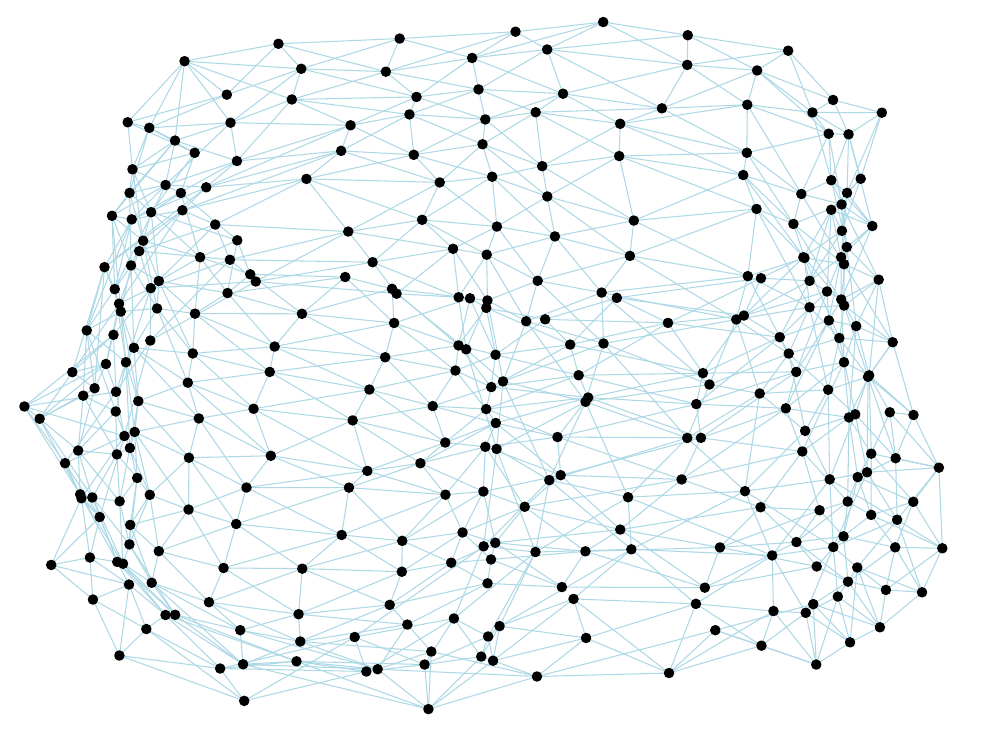} &
  \includegraphics[width=9.5mm]{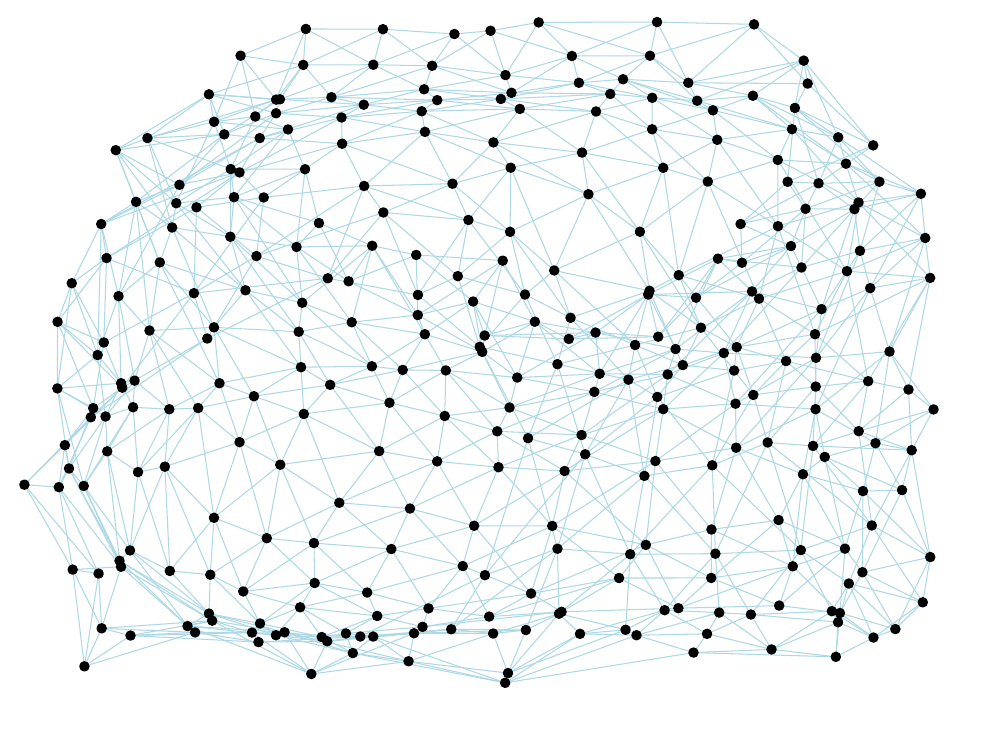} &
  \includegraphics[width=9.5mm]{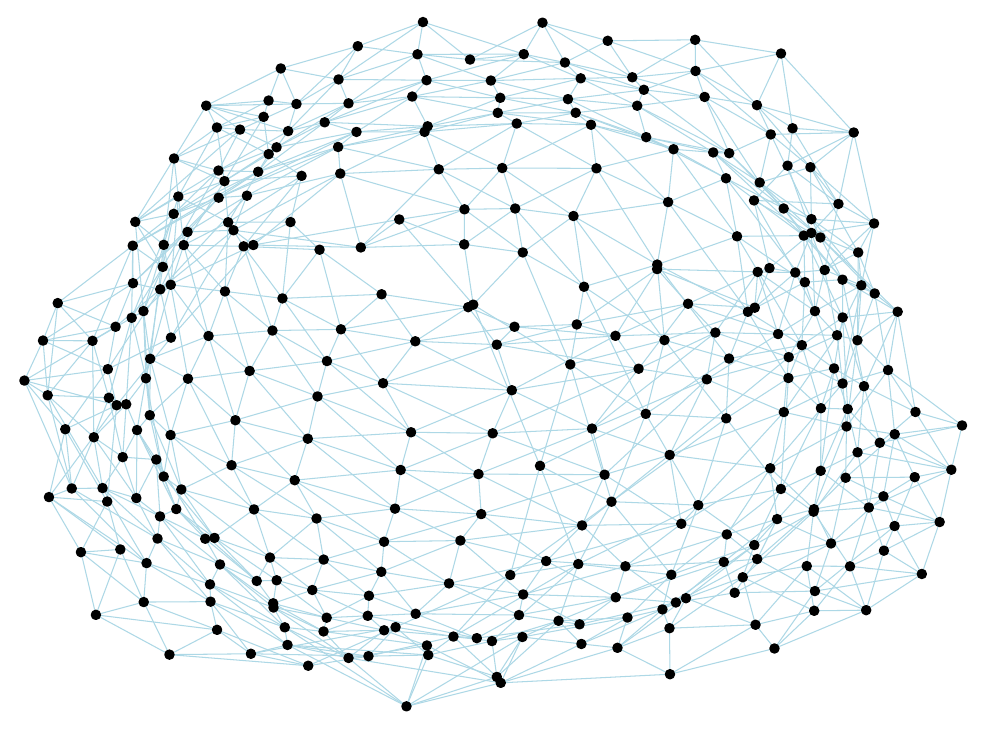} &
  \includegraphics[width=9.5mm]{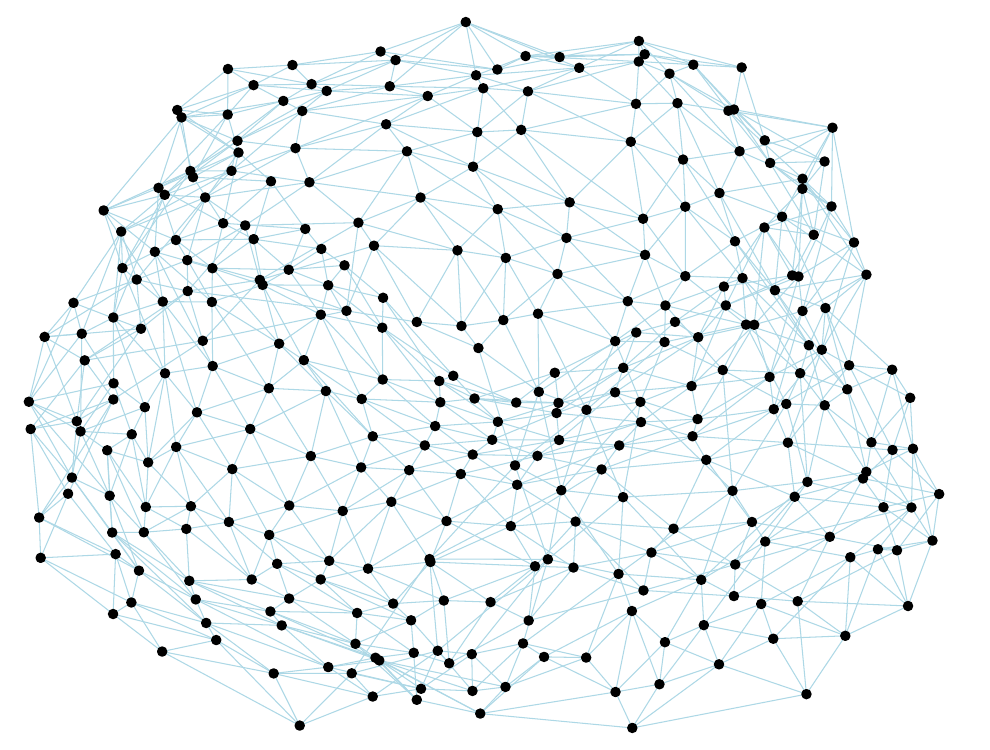} &
  \includegraphics[width=9.5mm]{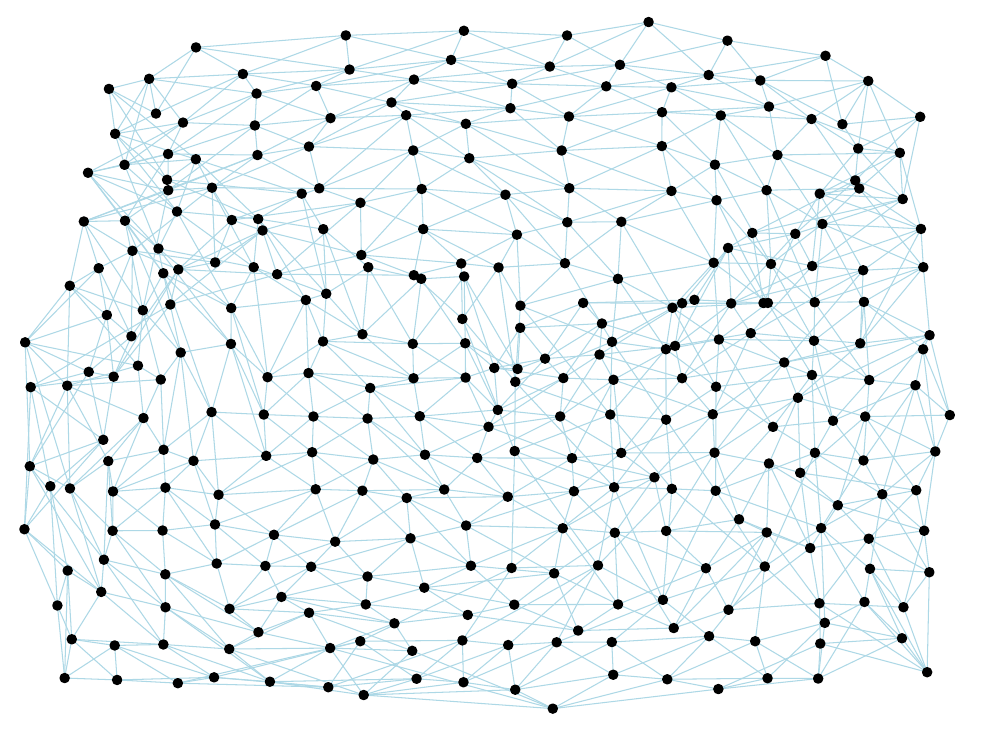}
  \\
         & \texttt{ST-ELD-CN-AR} & & \includegraphics[width=9.5mm]{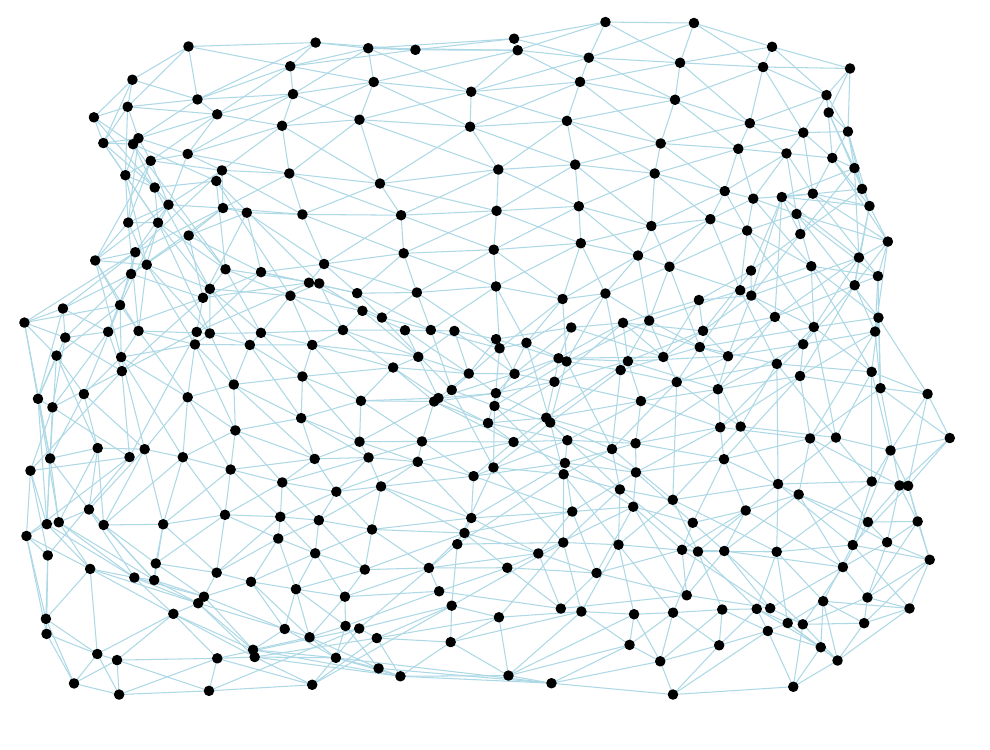} &
  \includegraphics[width=9.5mm]{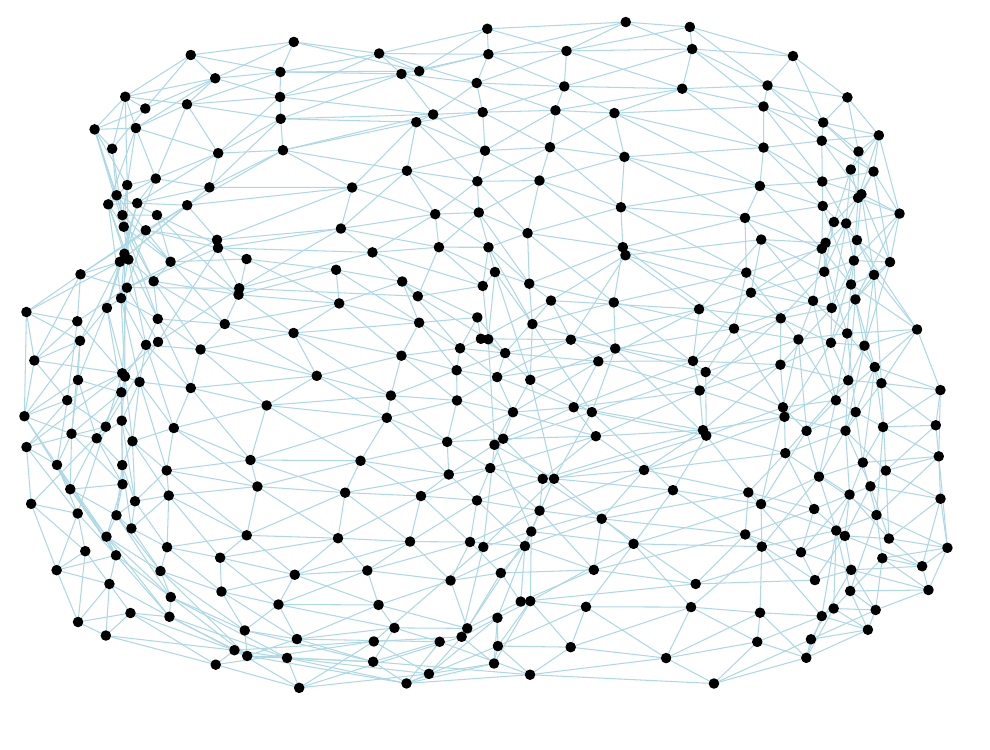} &
  \includegraphics[width=9.5mm]{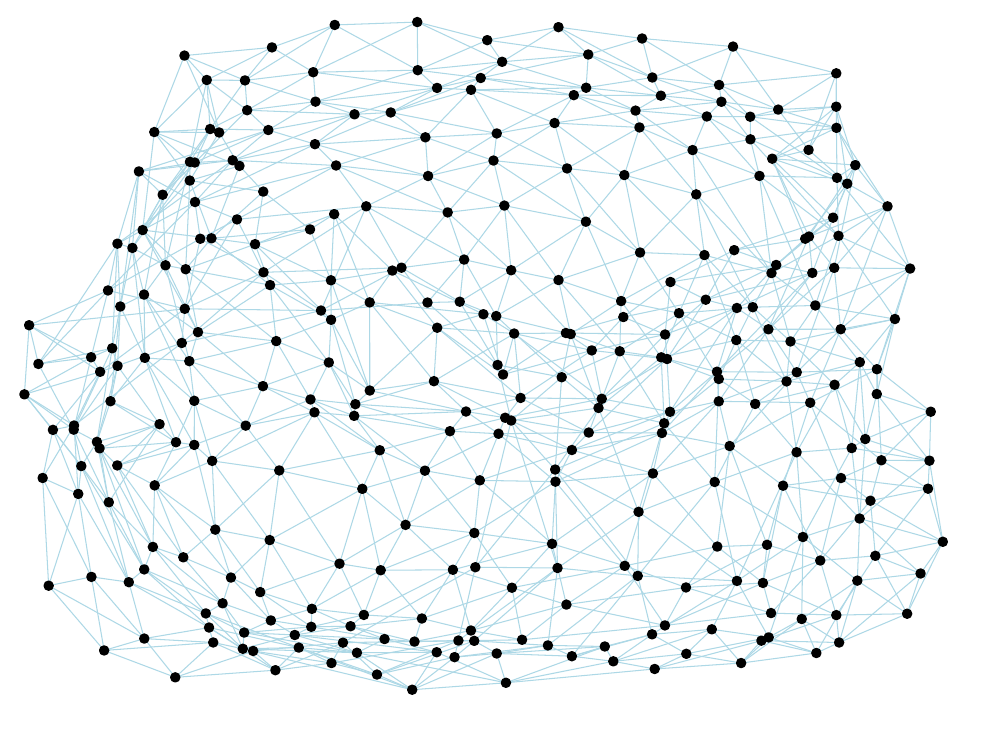} &
  \includegraphics[width=9.5mm]{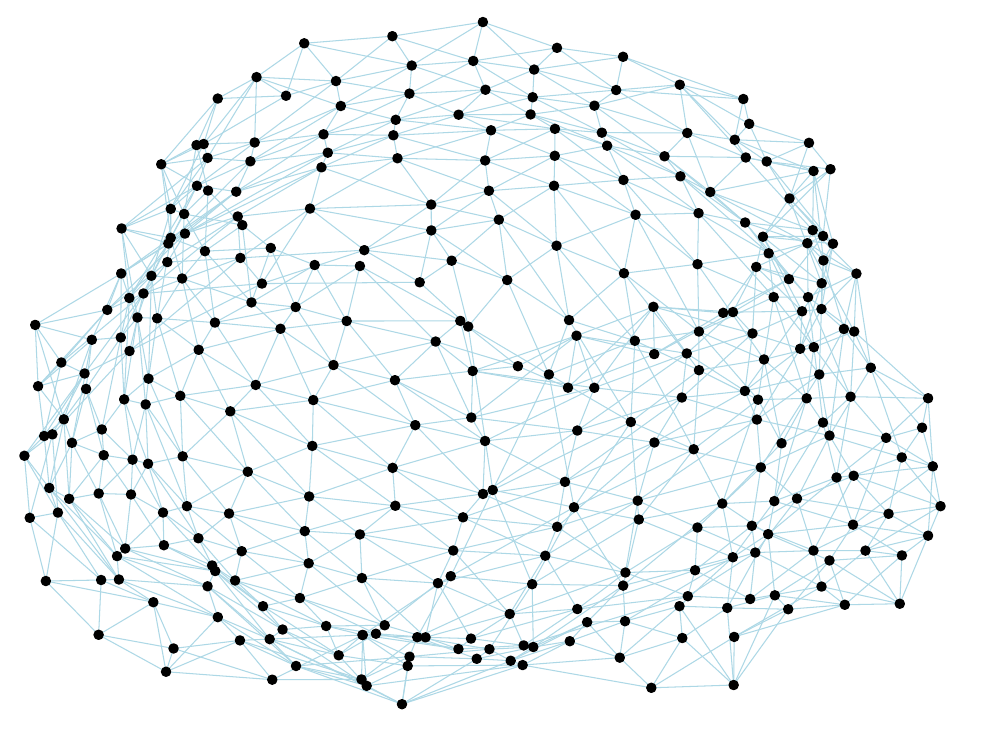} &
  \includegraphics[width=9.5mm]{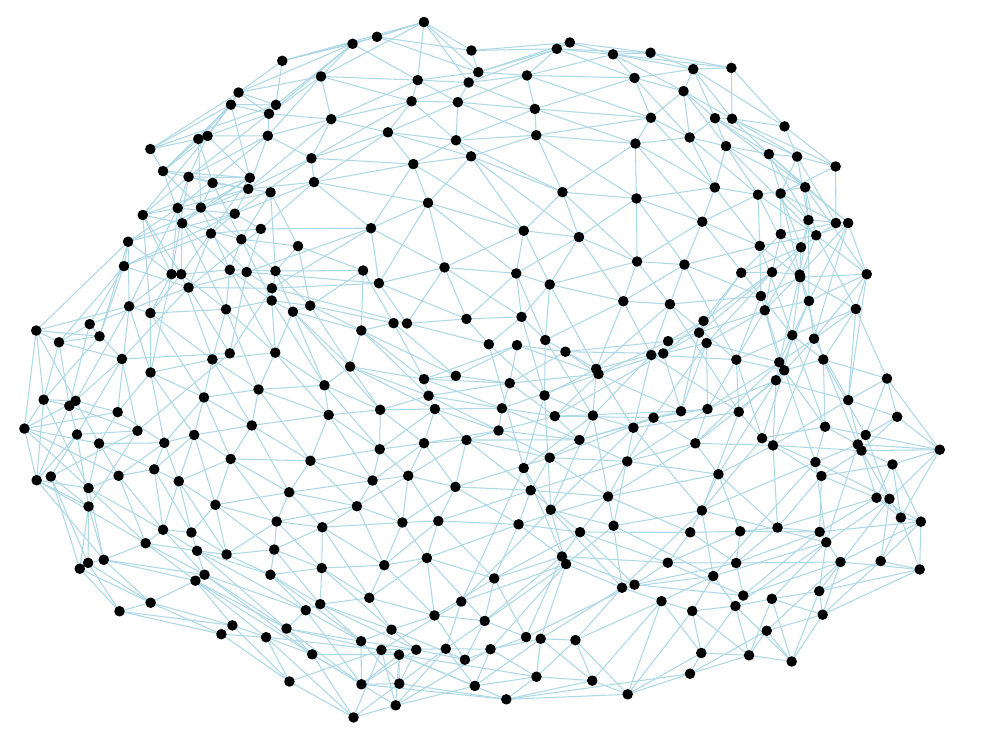} &
  \includegraphics[width=9.5mm]{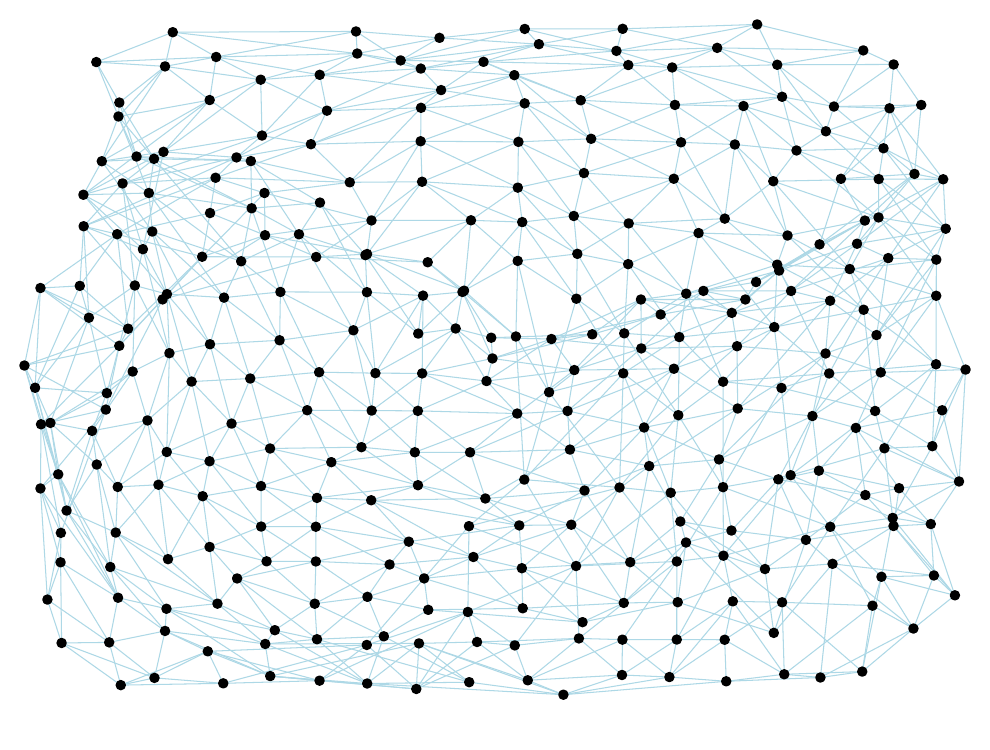}
  \\
  \hline
\end{tabular}
\caption{Collection of different graph drawings. The \texttt{START} column indicates the starting drawing of various graphs. The results of applying Alg.~\ref{alg:sim_anneal}  to \texttt{START} to six target shapes (\texttt{X}, \texttt{VERT}, \texttt{HOR}, \texttt{O}, \texttt{DINO}, \texttt{GRID}) are shown in their respective columns. The rows indicate the metrics that have $\pm\epsilon=0.0025$ for combinations of \texttt{ST,ELD,AR} and $\pm\epsilon=\texttt{CN}(\Gamma)*0.05$ for \texttt{CN}.}
\label{fig:results_combs_dwt}
\end{figure*}

\begin{figure*}[!ht]
\centering
\begin{tabular}{cc|ccccccc}
  \hline
  & & \texttt{START} & \texttt{X} & \texttt{VERT} & \texttt{HOR} & \texttt{O} & \texttt{DINO} & \texttt{GRID}\\
\hline 
    \emph{bar-albert} & \texttt{ST-ELD} & \includegraphics[width=9.5mm]{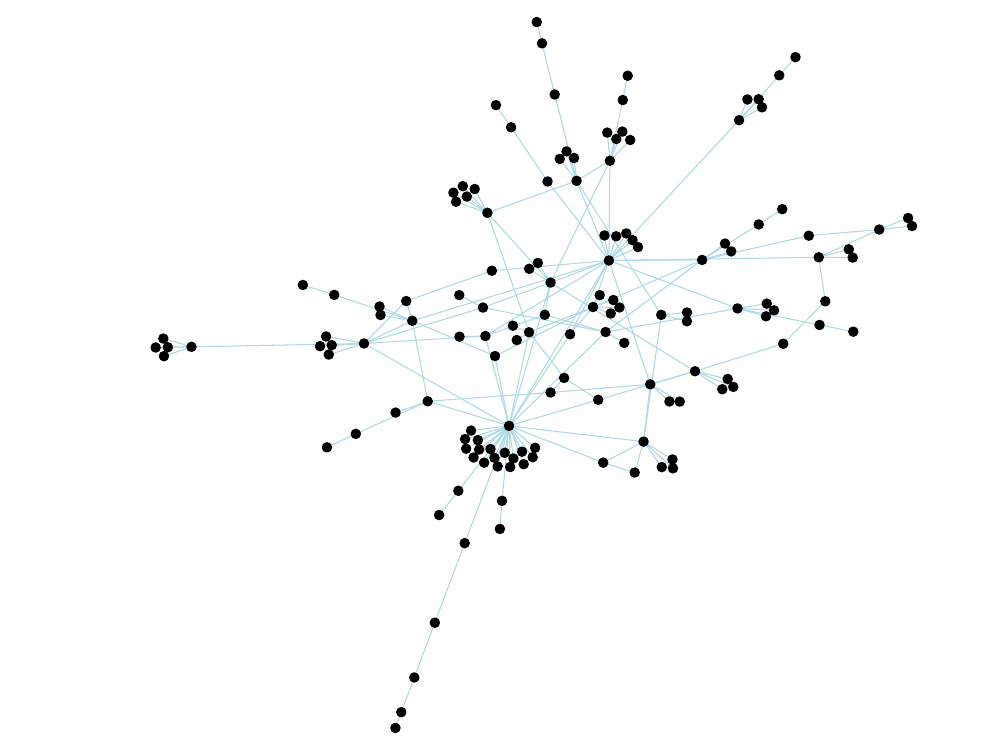} & \includegraphics[width=9.5mm]{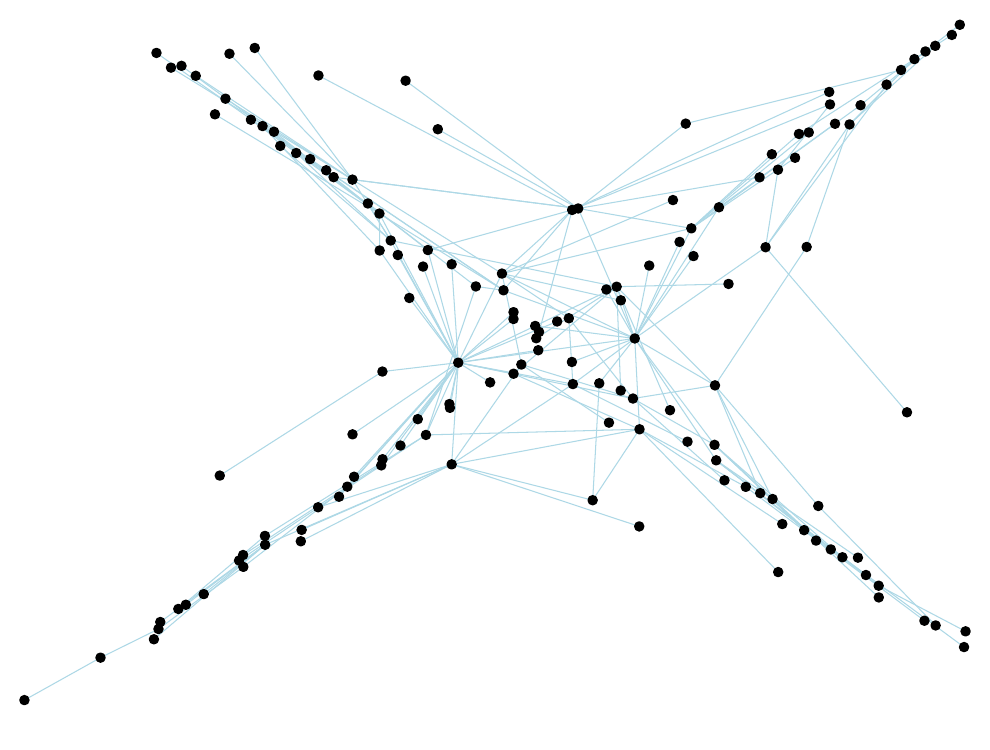} &
  \includegraphics[width=9.5mm]{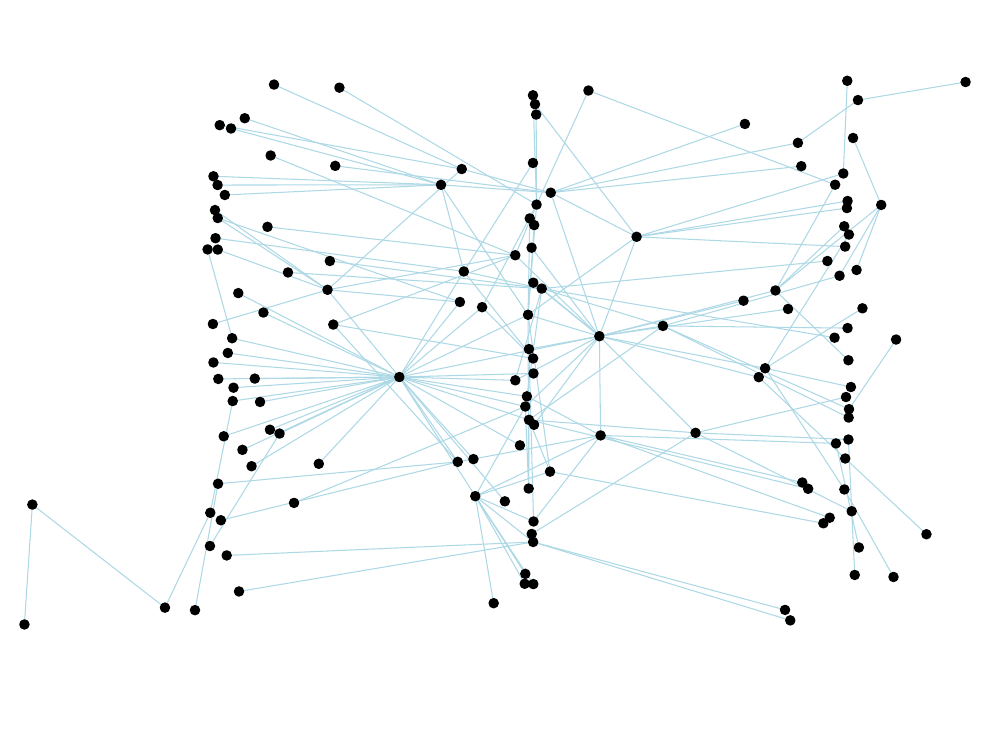} &
  \includegraphics[width=9.5mm]{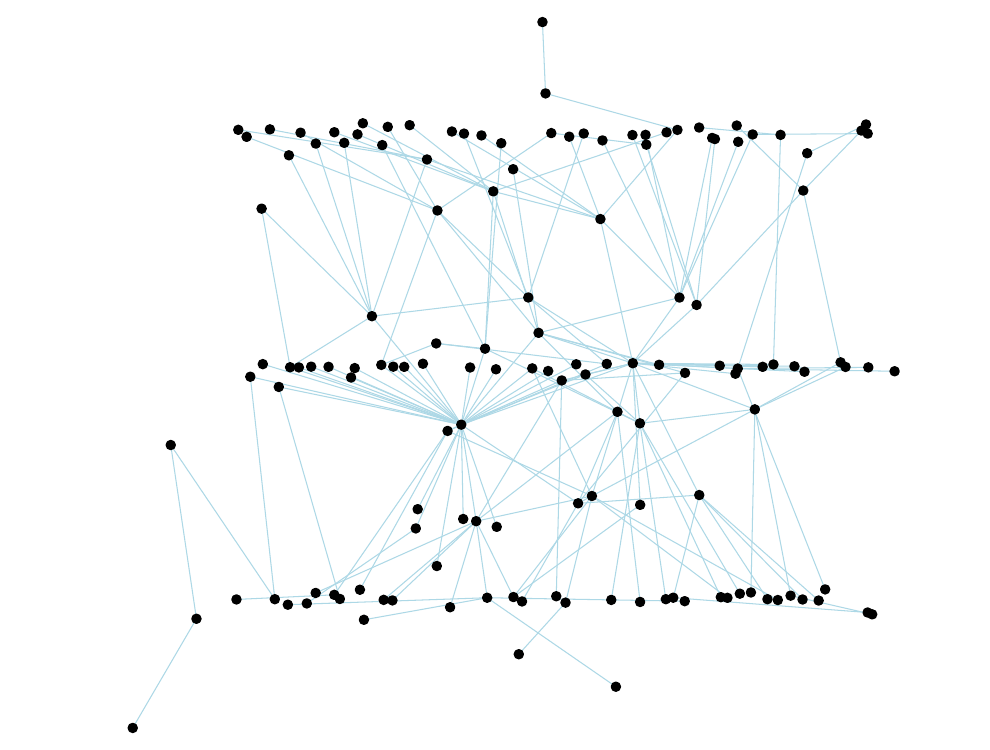} &
  \includegraphics[width=9.5mm]{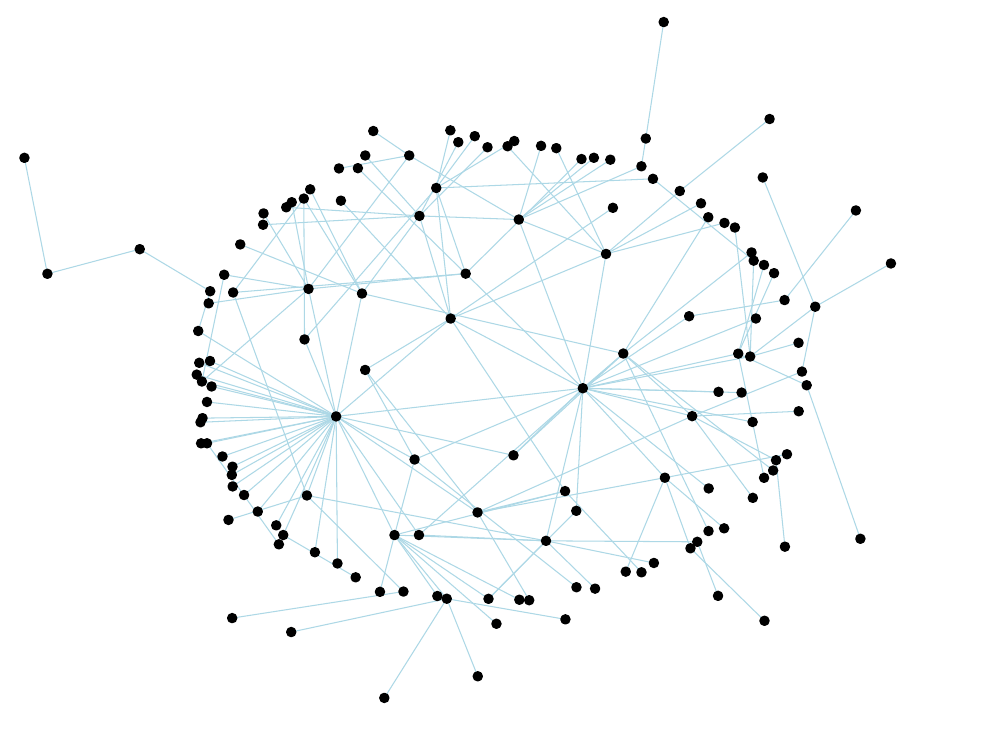} &
  \includegraphics[width=9.5mm]{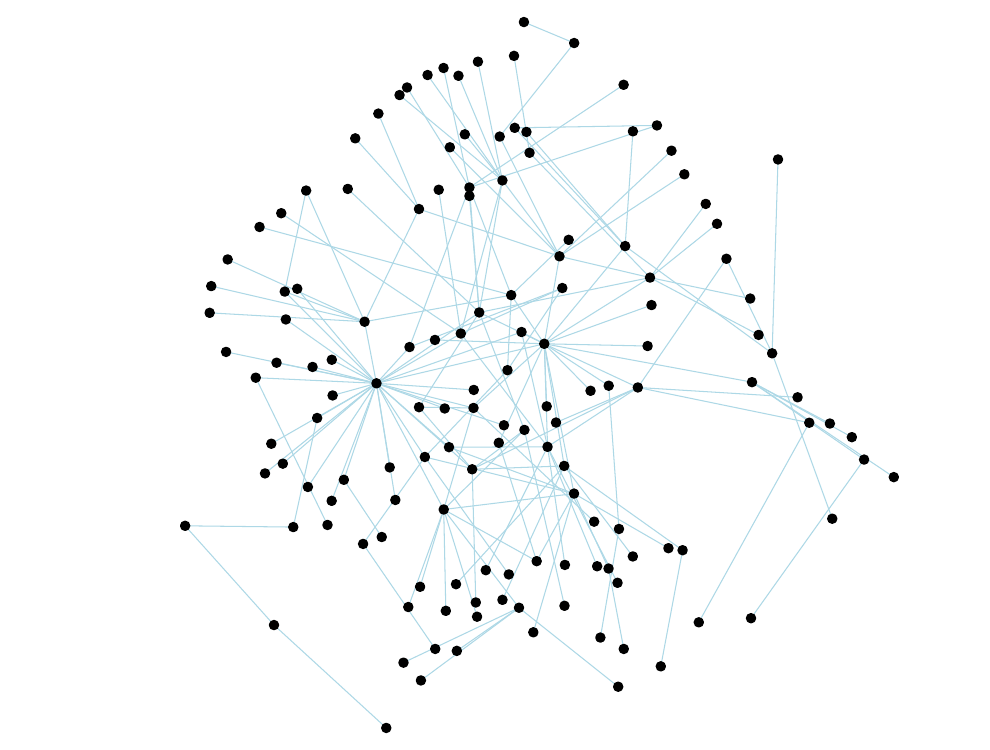} &
  \includegraphics[width=9.5mm]{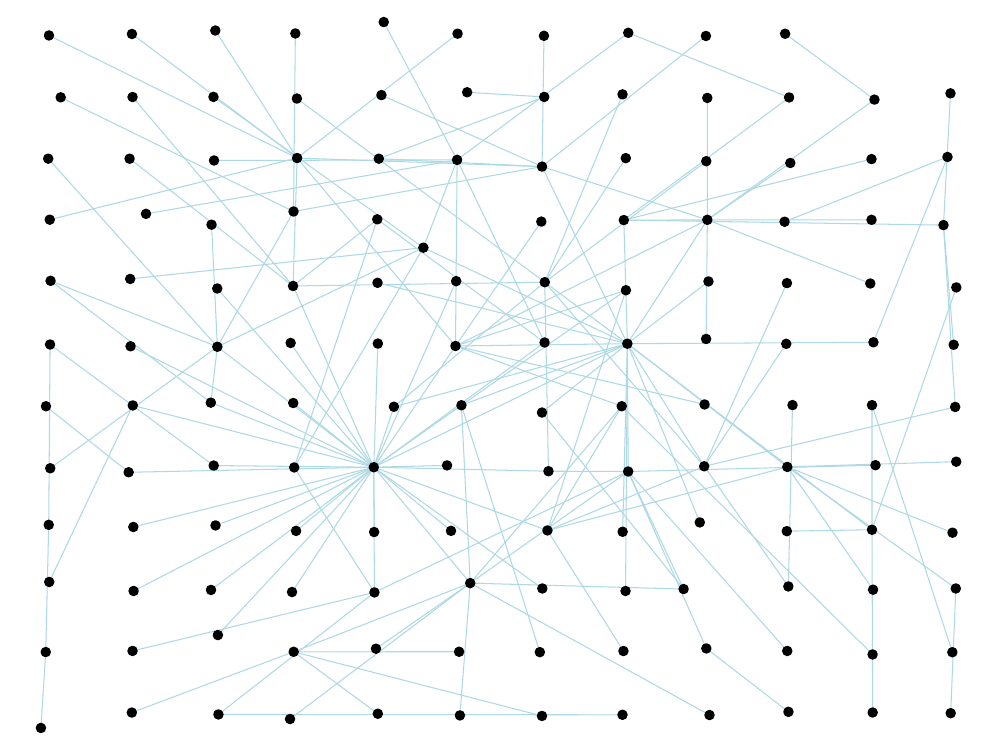}
  \\
     & \texttt{ST-CN} & & \includegraphics[width=9.5mm]{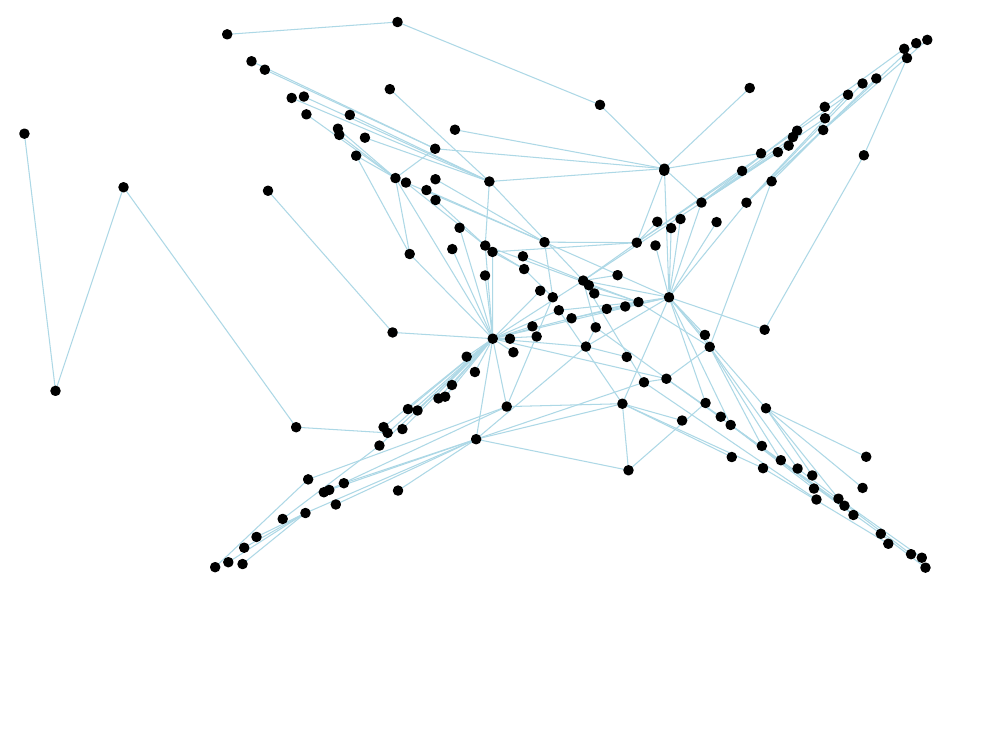} &
  \includegraphics[width=9.5mm]{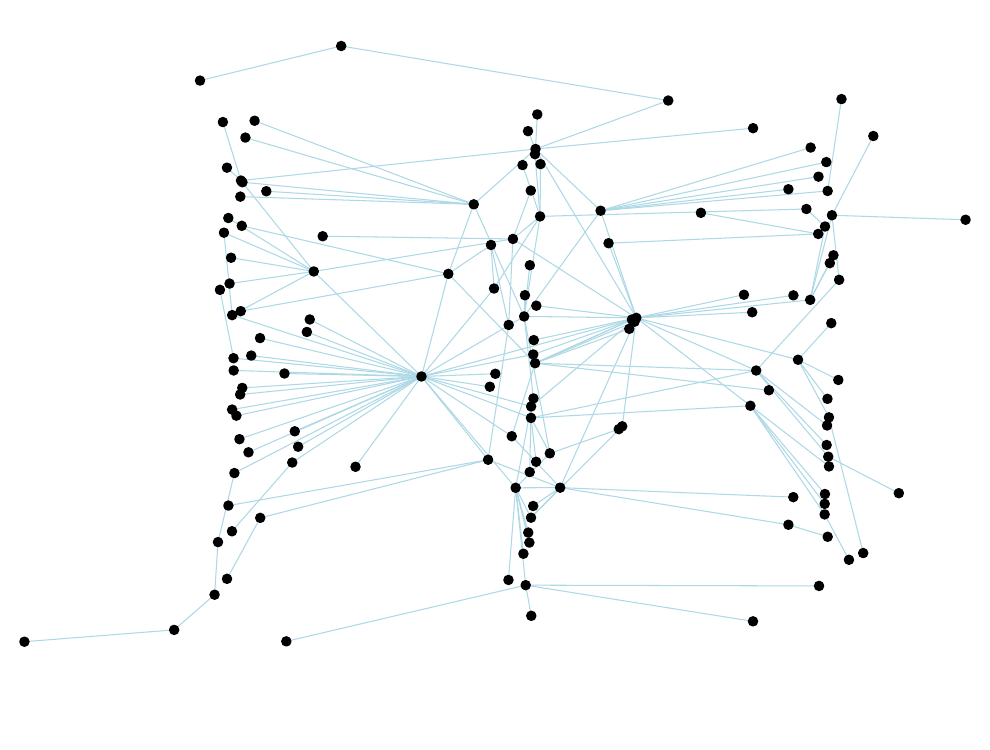} &
  \includegraphics[width=9.5mm]{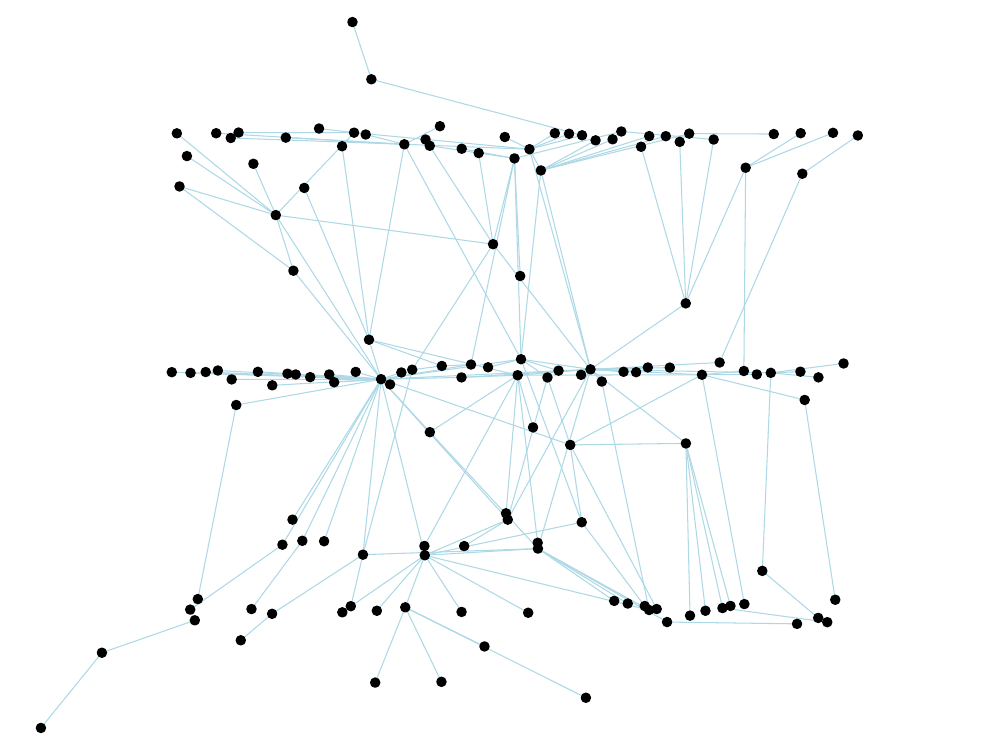} &
  \includegraphics[width=9.5mm]{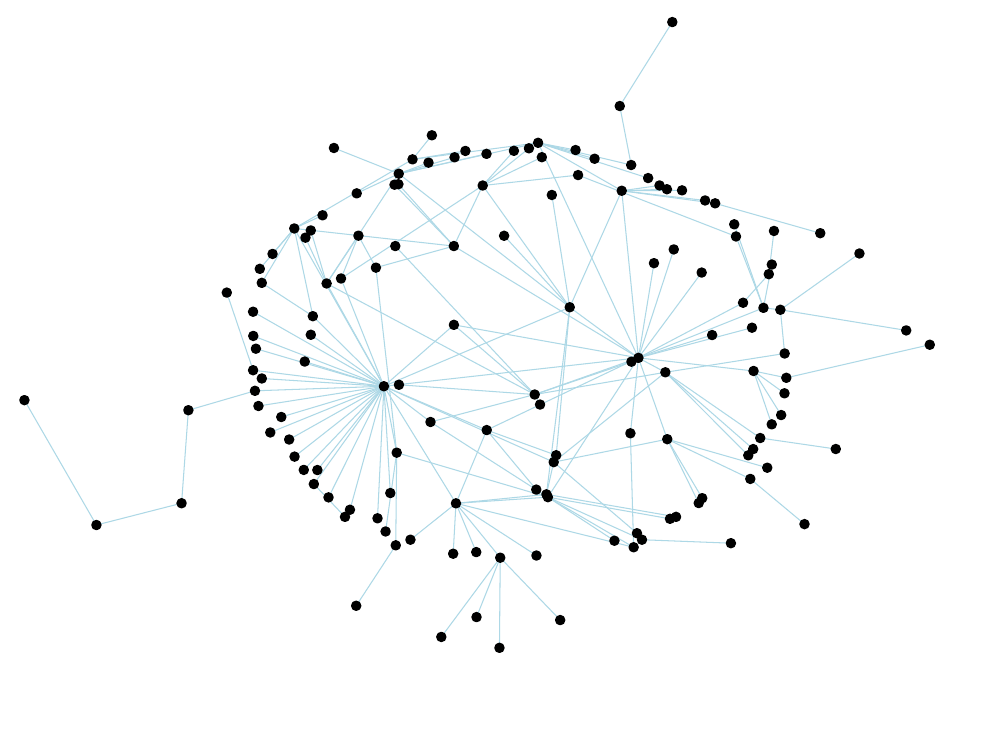} &
  \includegraphics[width=9.5mm]{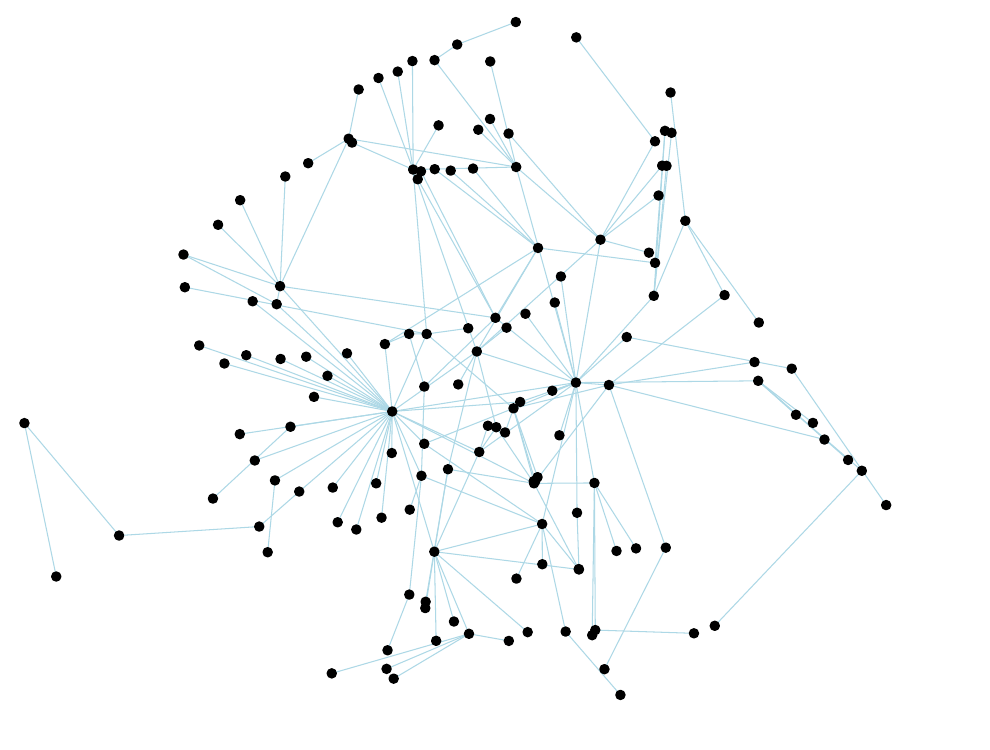} &
  \includegraphics[width=9.5mm]{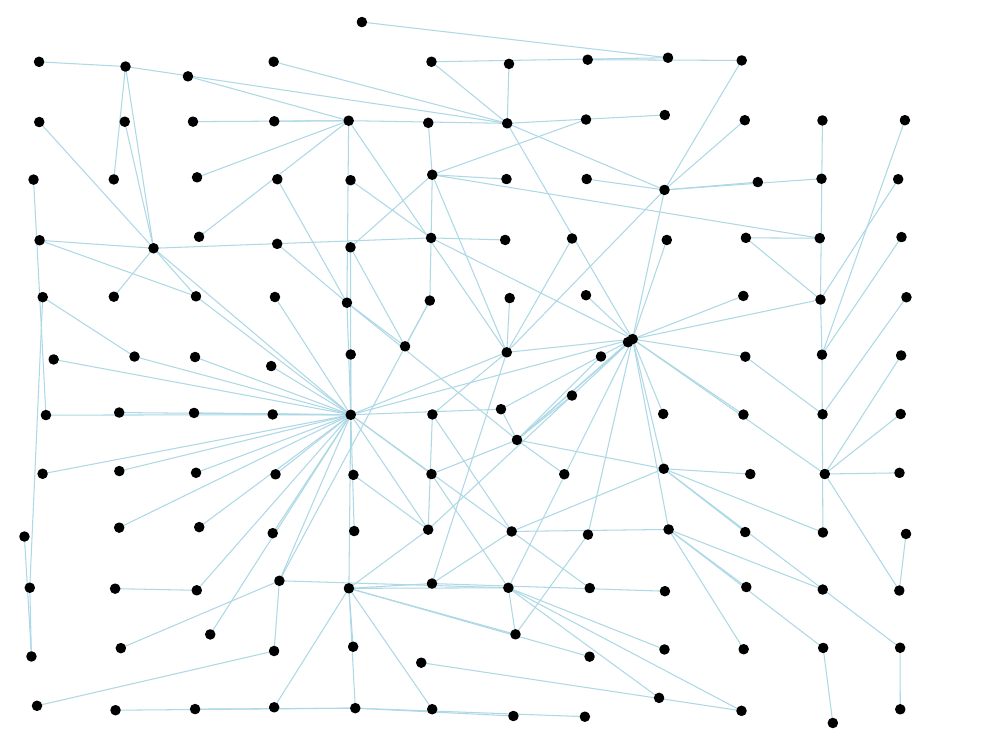}
  \\
    & \texttt{ST-AR} &
   & \includegraphics[width=9.5mm]{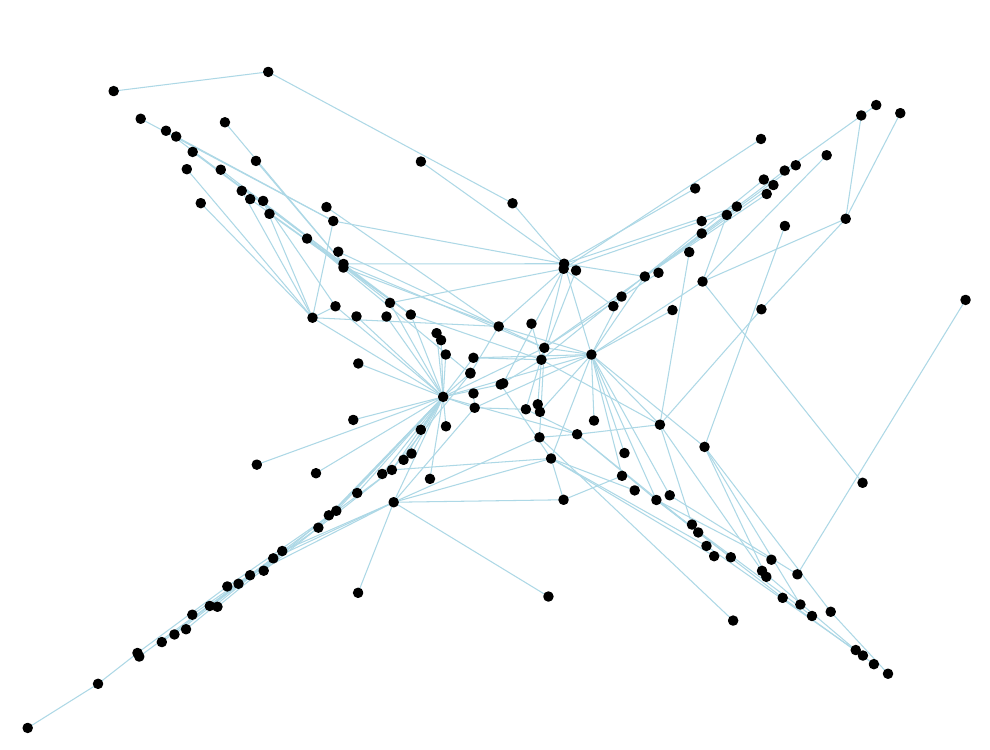} &
  \includegraphics[width=9.5mm]{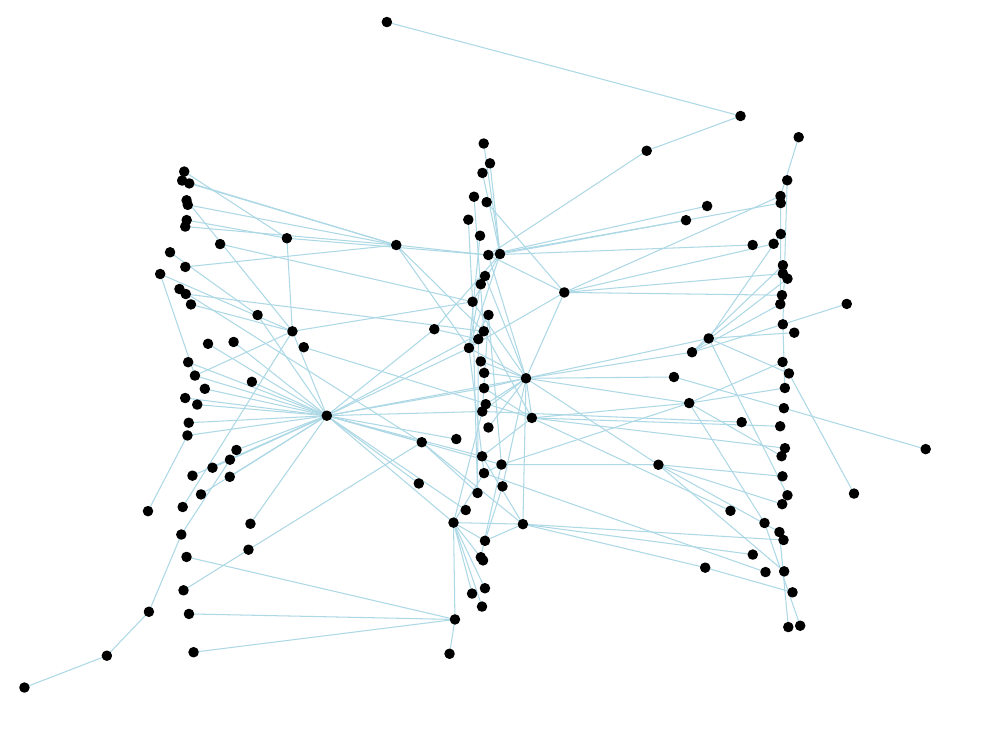} &
  \includegraphics[width=9.5mm]{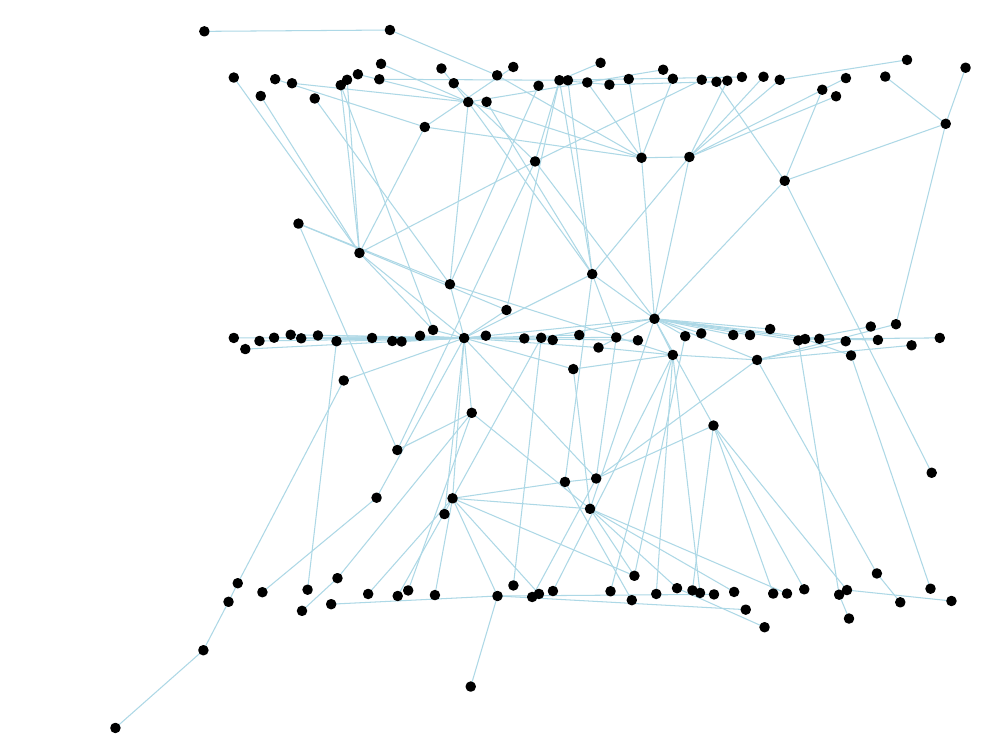} &
  \includegraphics[width=9.5mm]{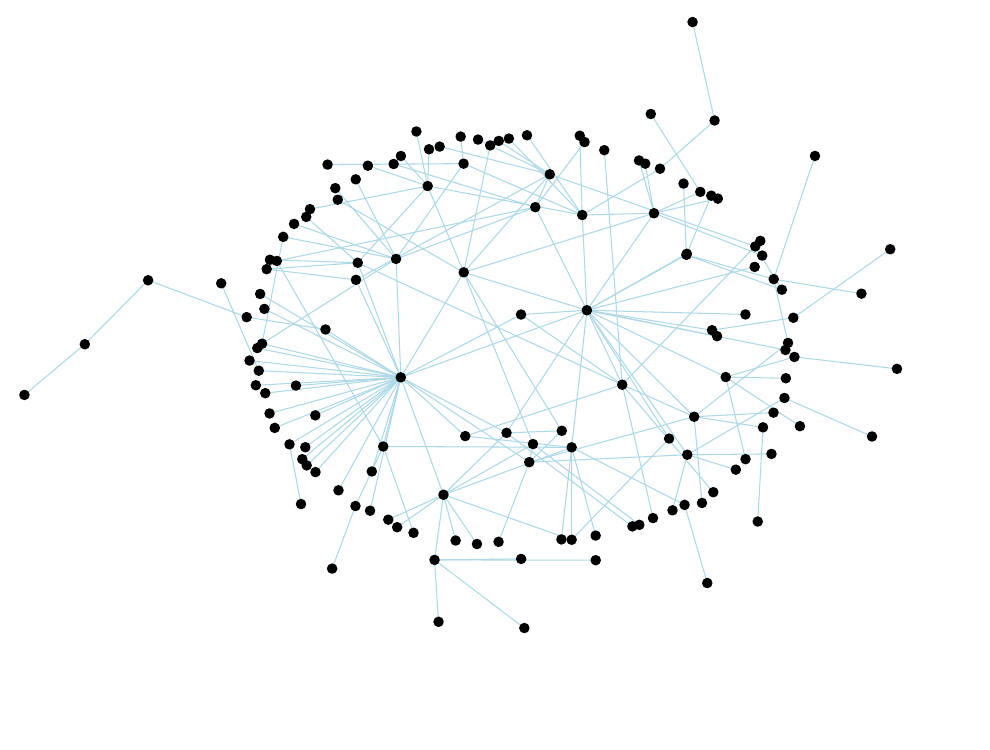} &
  \includegraphics[width=9.5mm]{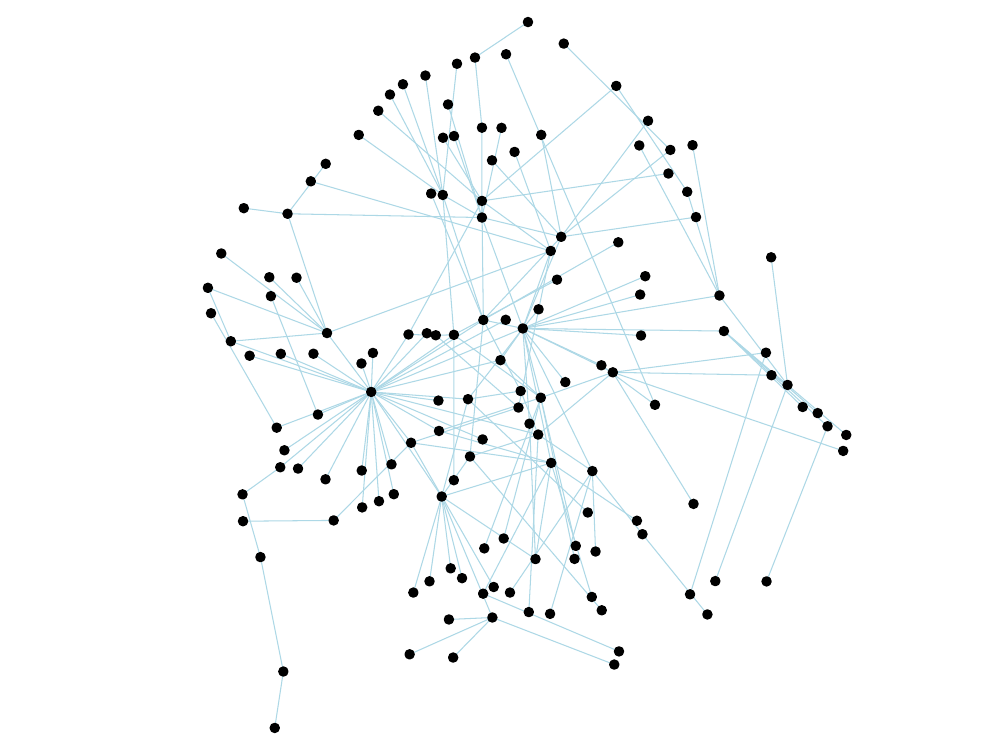} &
  \includegraphics[width=9.5mm]{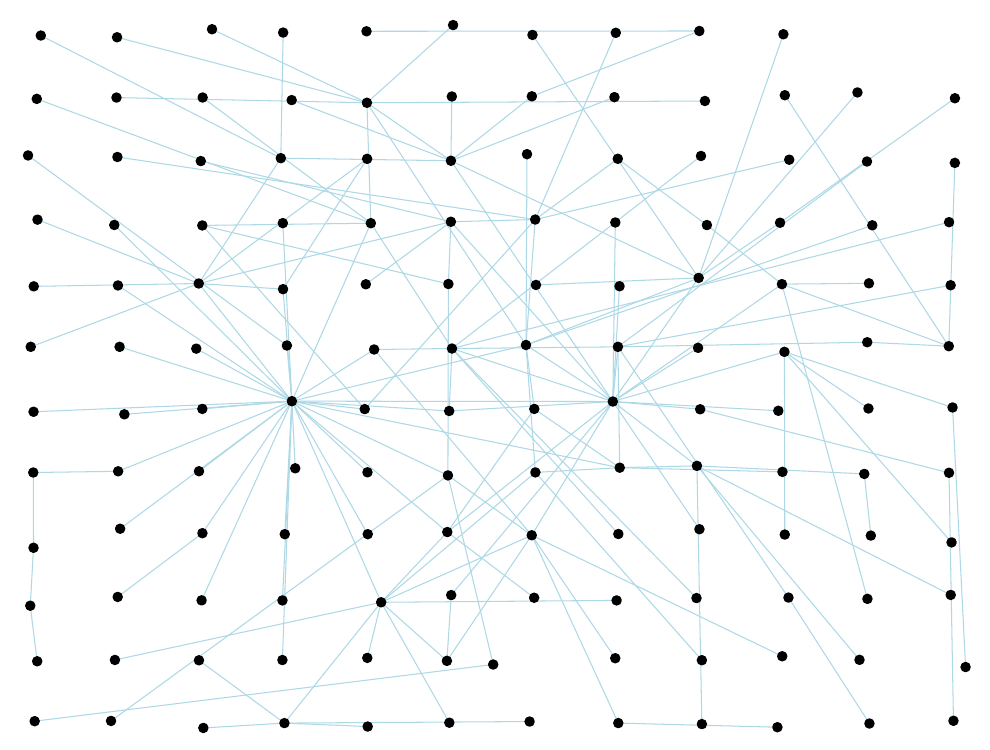}
  \\
     & \texttt{ELD-CN} & & \includegraphics[width=9.5mm]{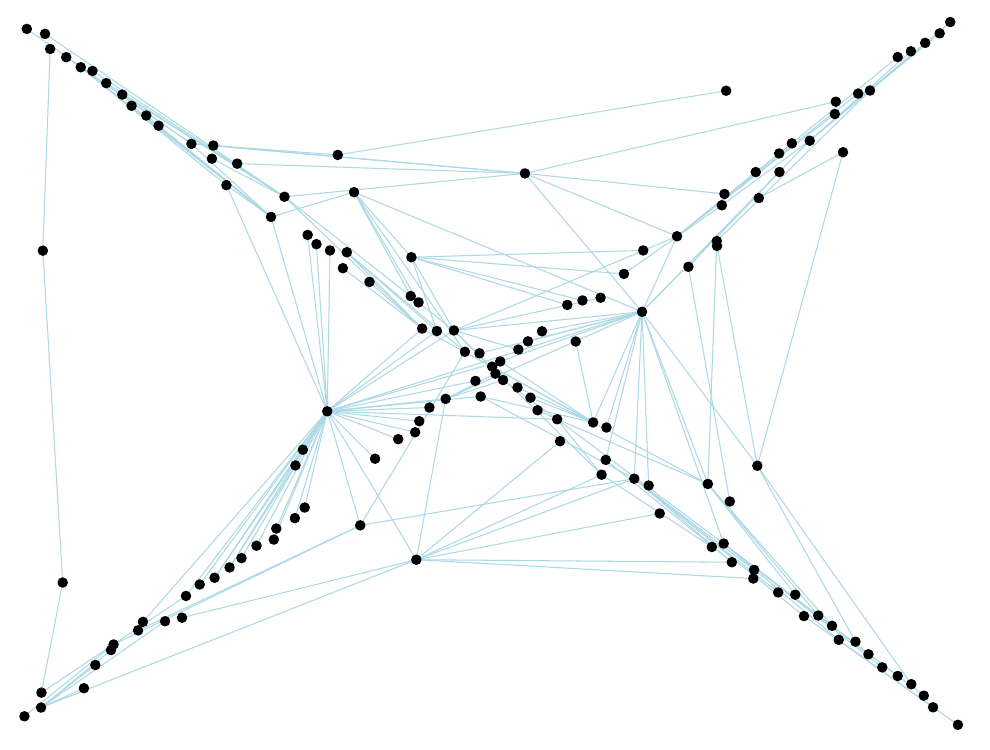} &
  \includegraphics[width=9.5mm]{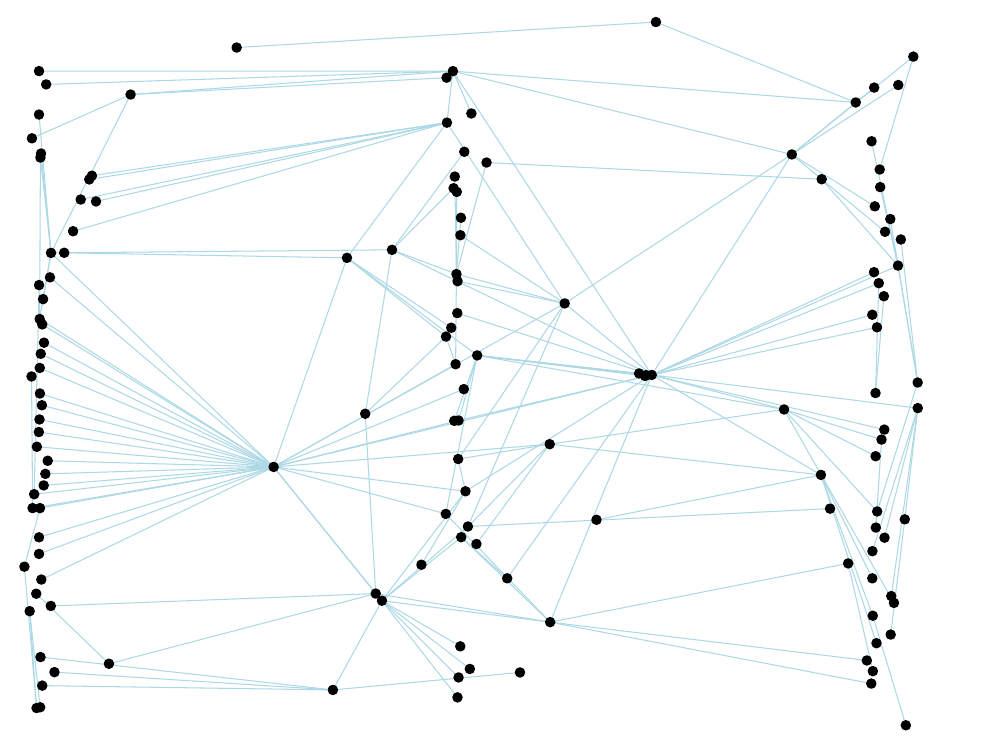} &
  \includegraphics[width=9.5mm]{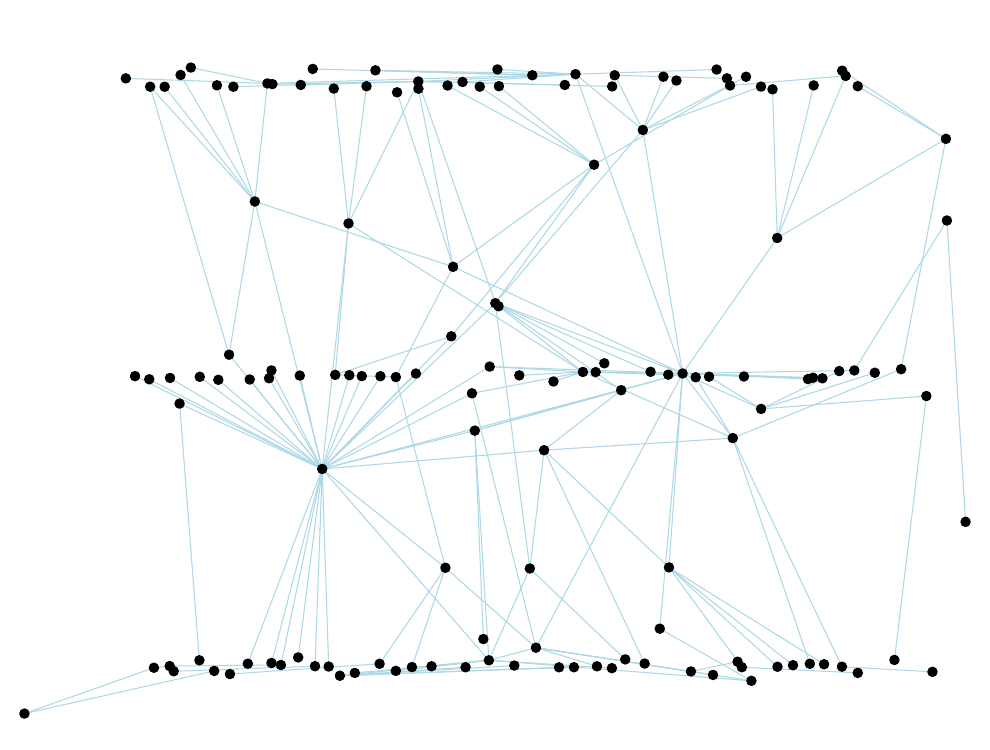} &
  \includegraphics[width=9.5mm]{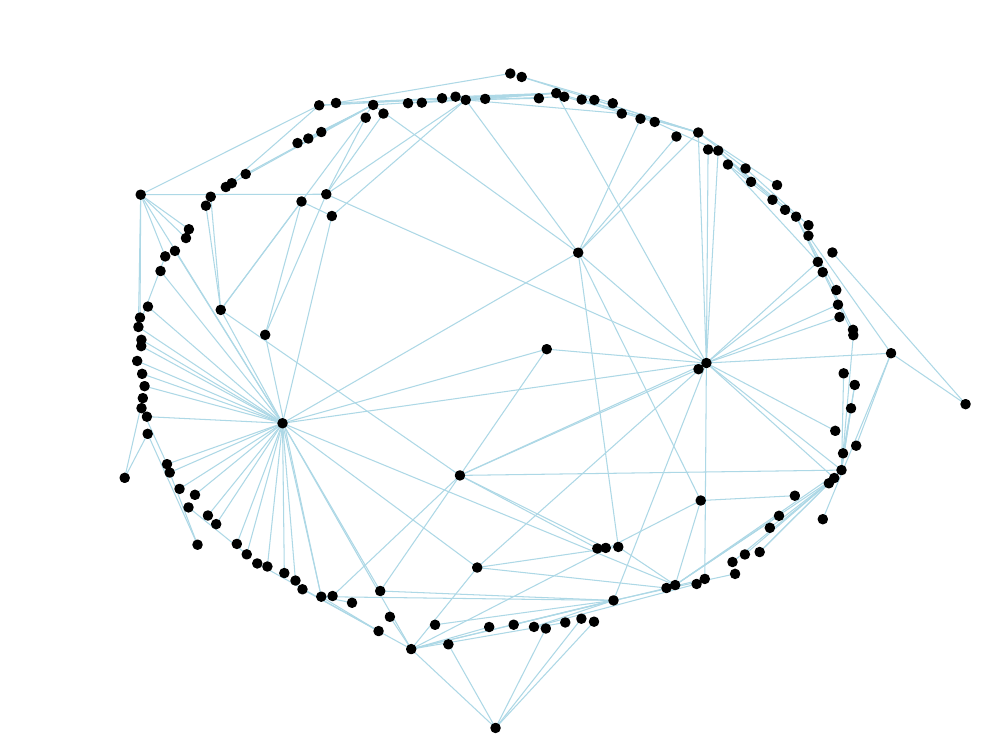} &
  \includegraphics[width=9.5mm]{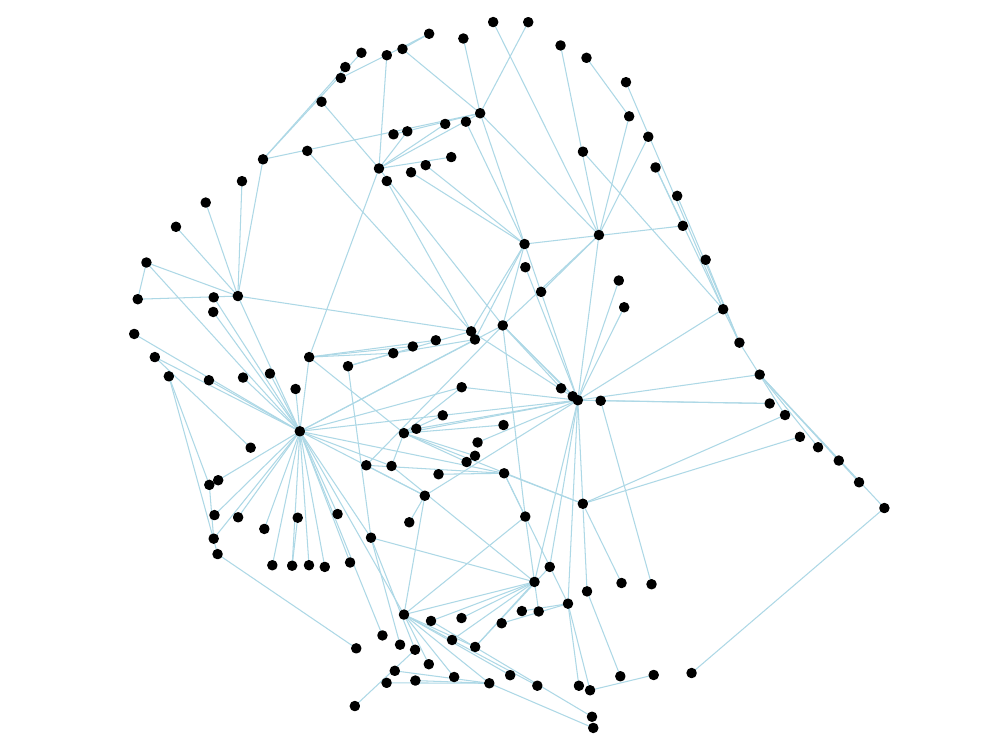} &
  \includegraphics[width=9.5mm]{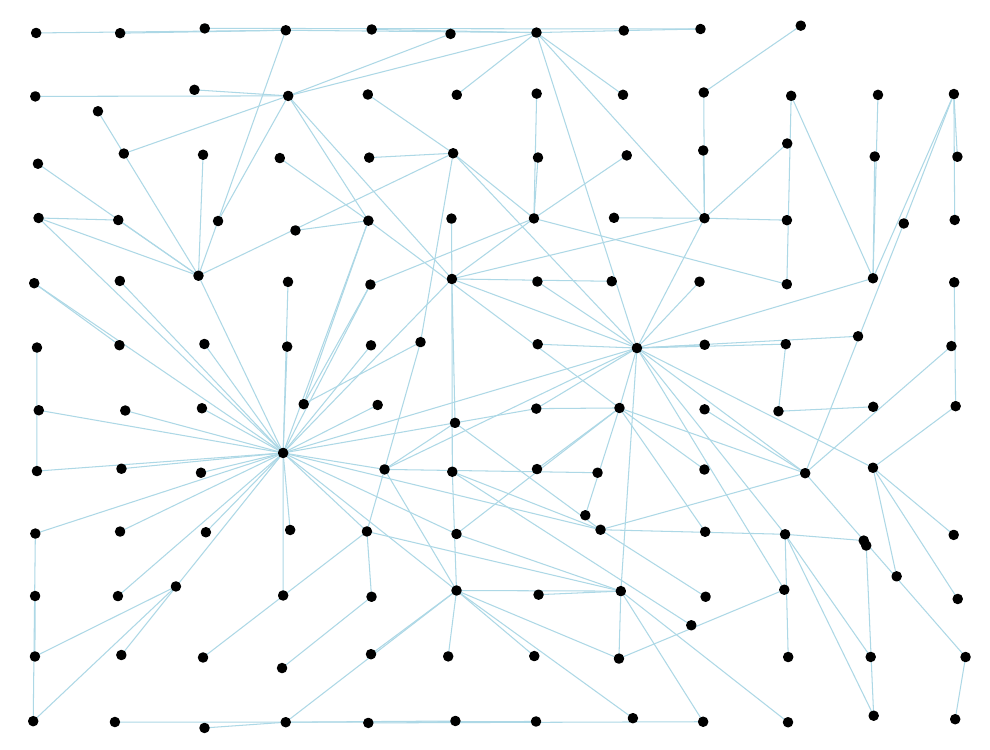}
  \\
       & \texttt{ELD-AR} & & \includegraphics[width=9.5mm]{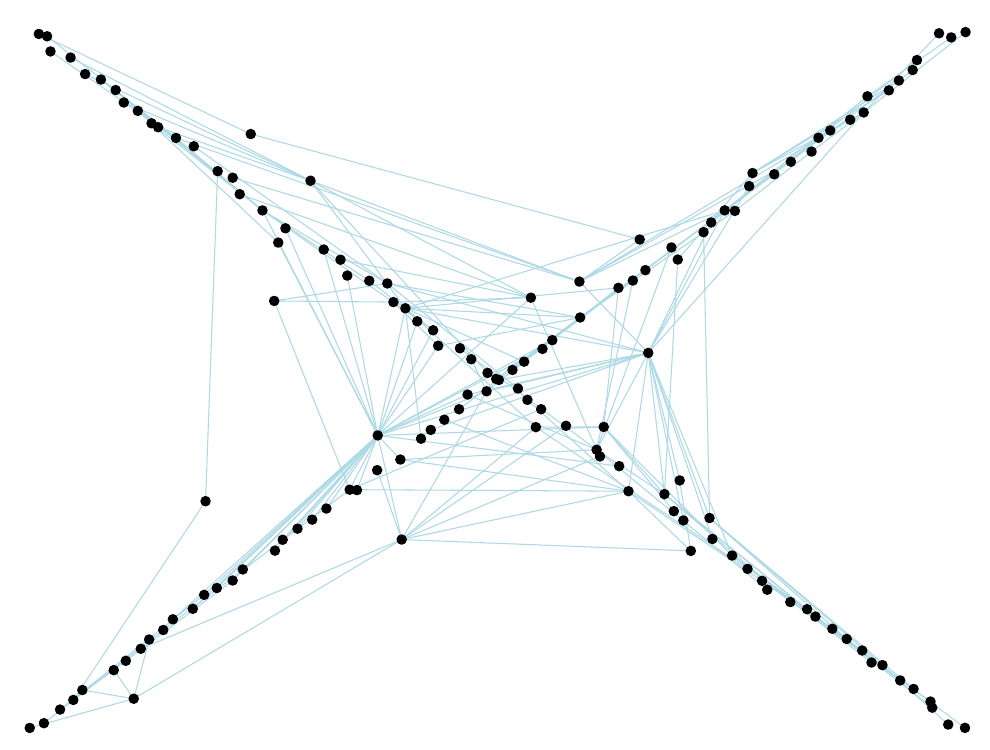} &
  \includegraphics[width=9.5mm]{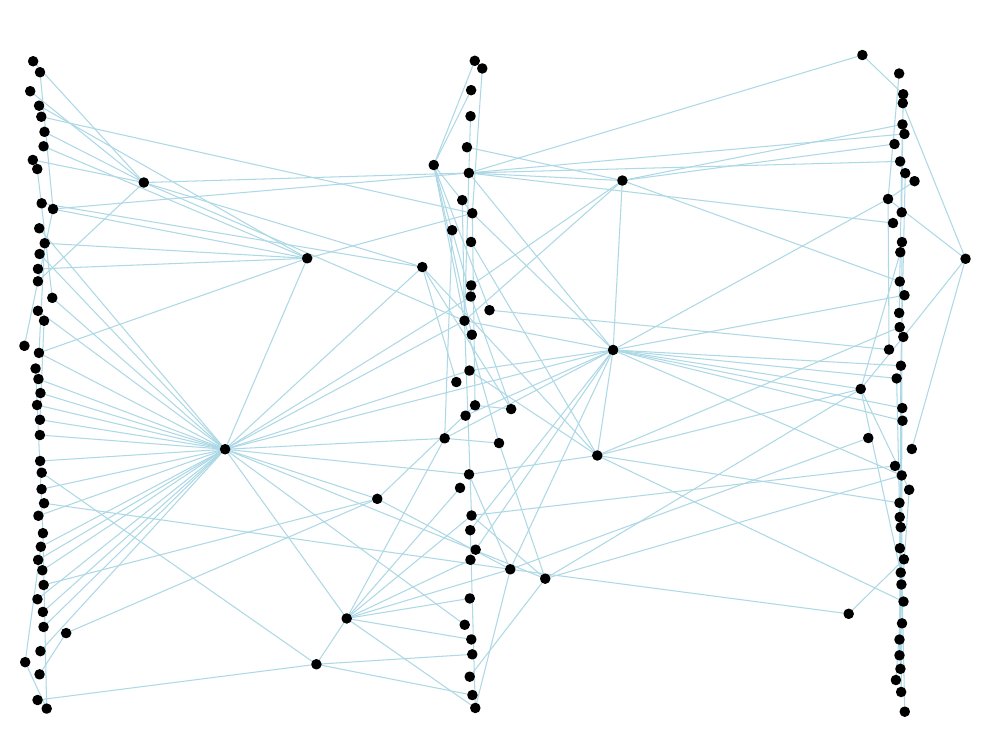} &
  \includegraphics[width=9.5mm]{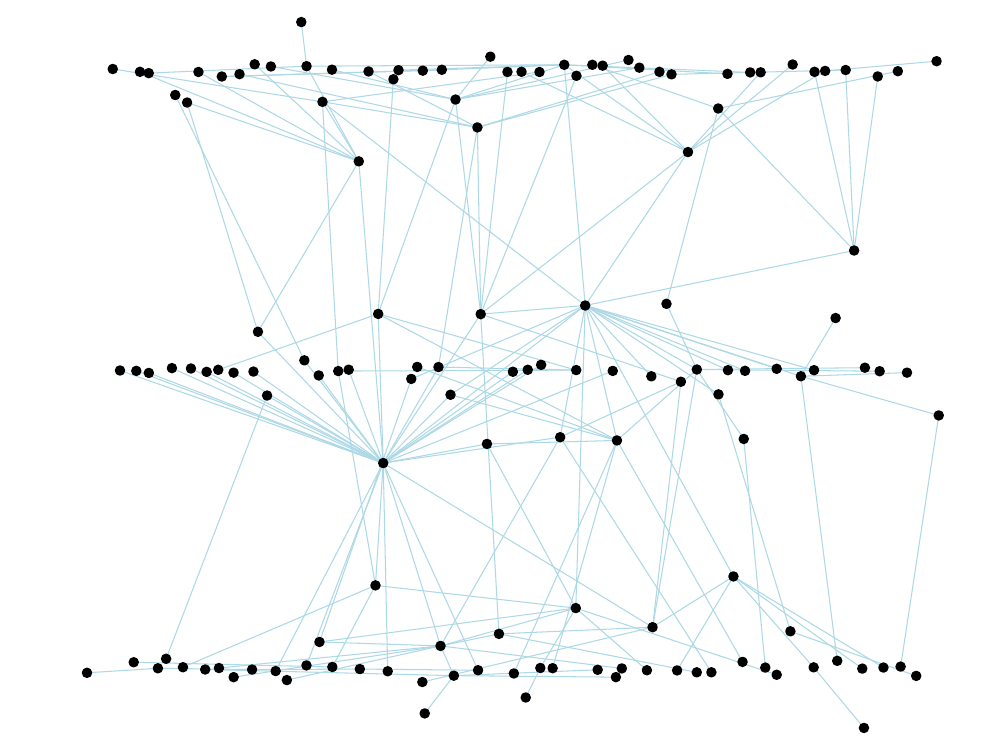} &
  \includegraphics[width=9.5mm]{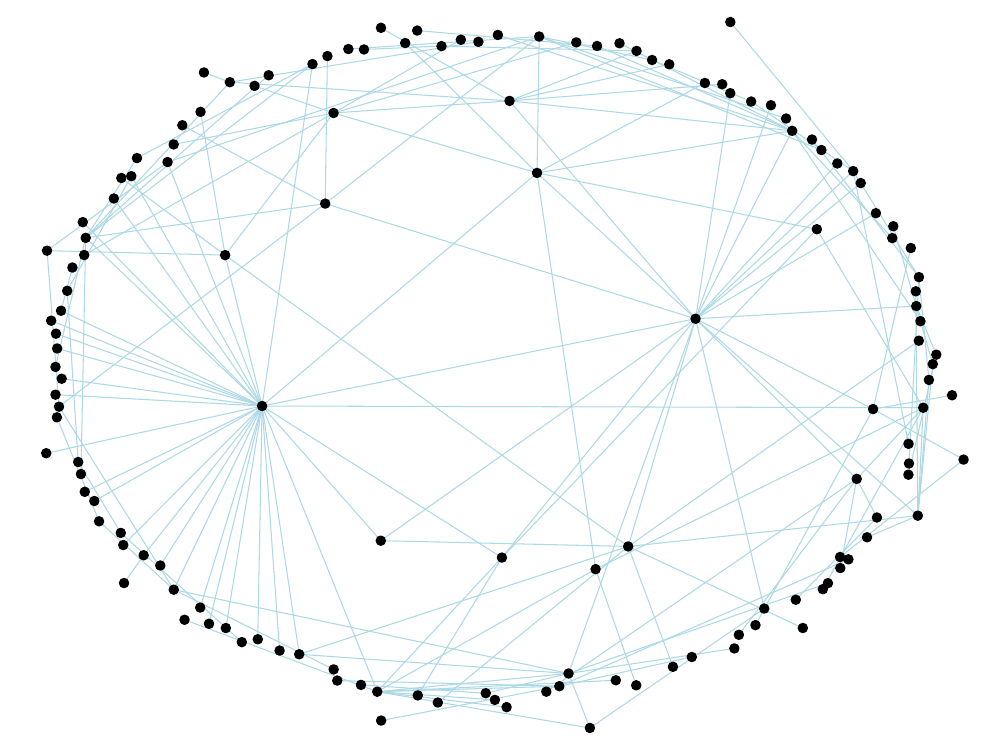} &
  \includegraphics[width=9.5mm]{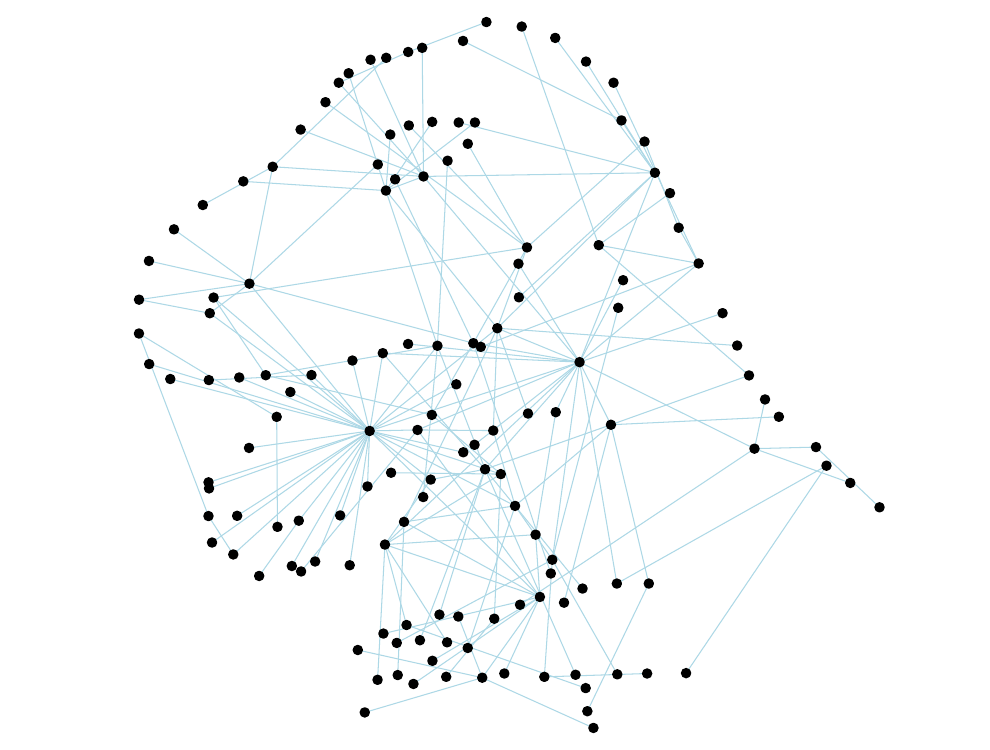} &
  \includegraphics[width=9.5mm]{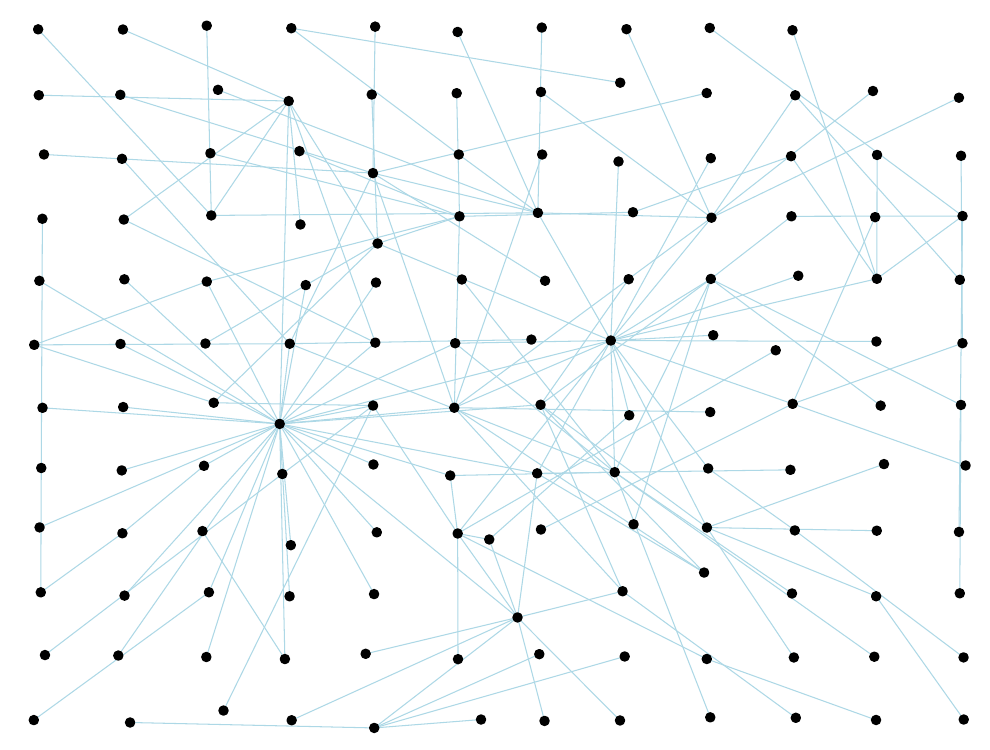}
  \\
       & \texttt{CN-AR} & & \includegraphics[width=9.5mm]{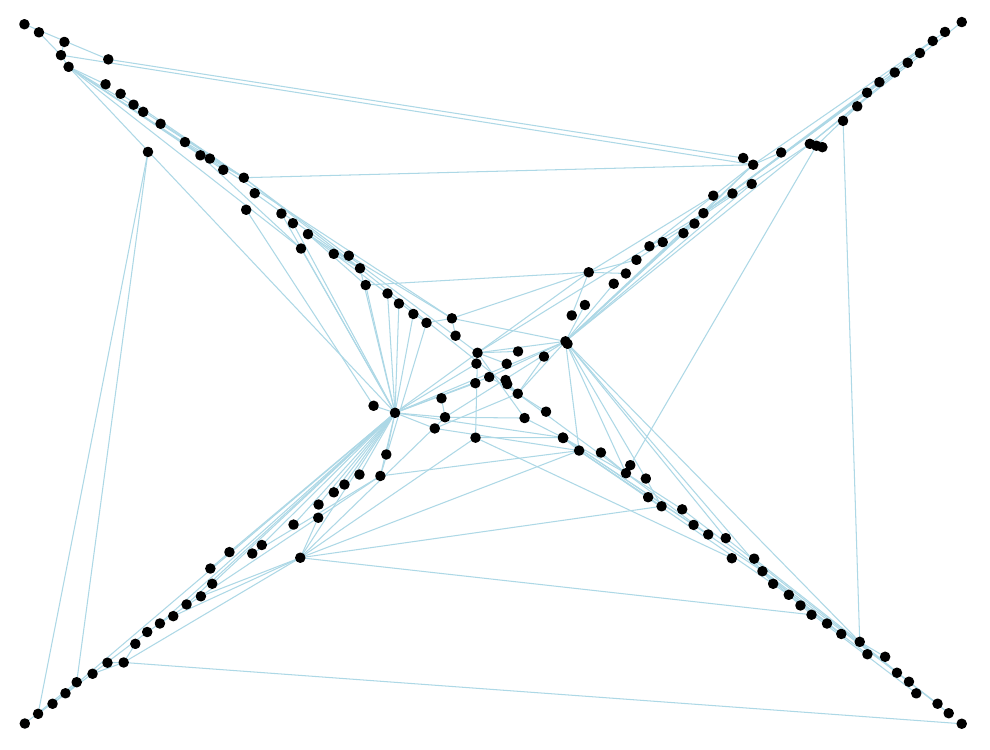} &
  \includegraphics[width=9.5mm]{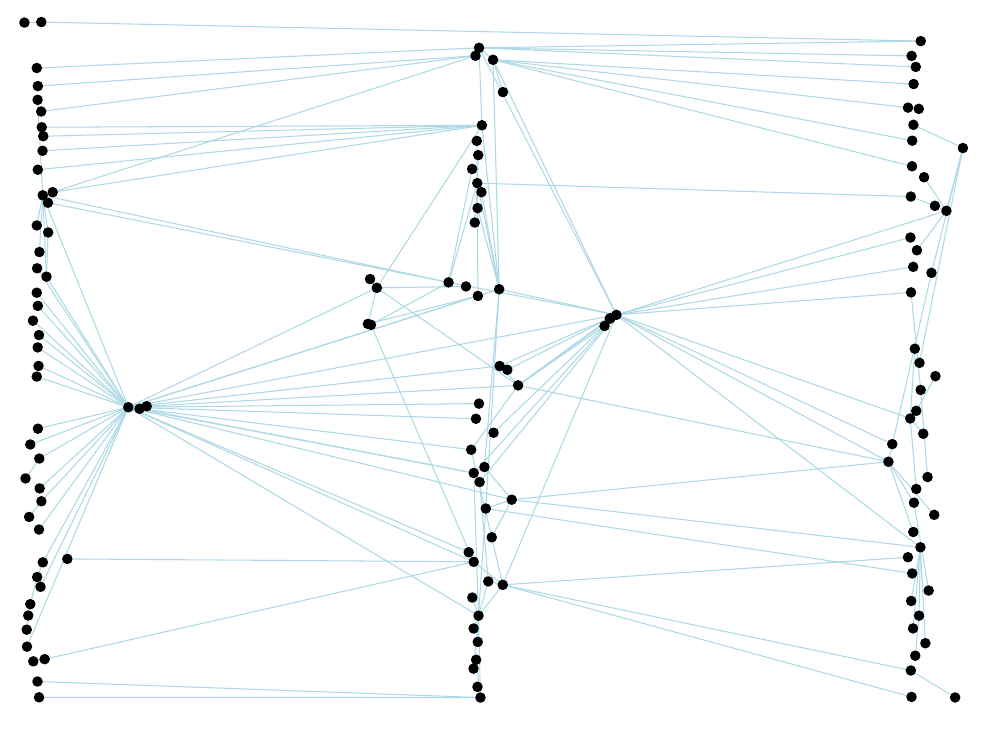} &
  \includegraphics[width=9.5mm]{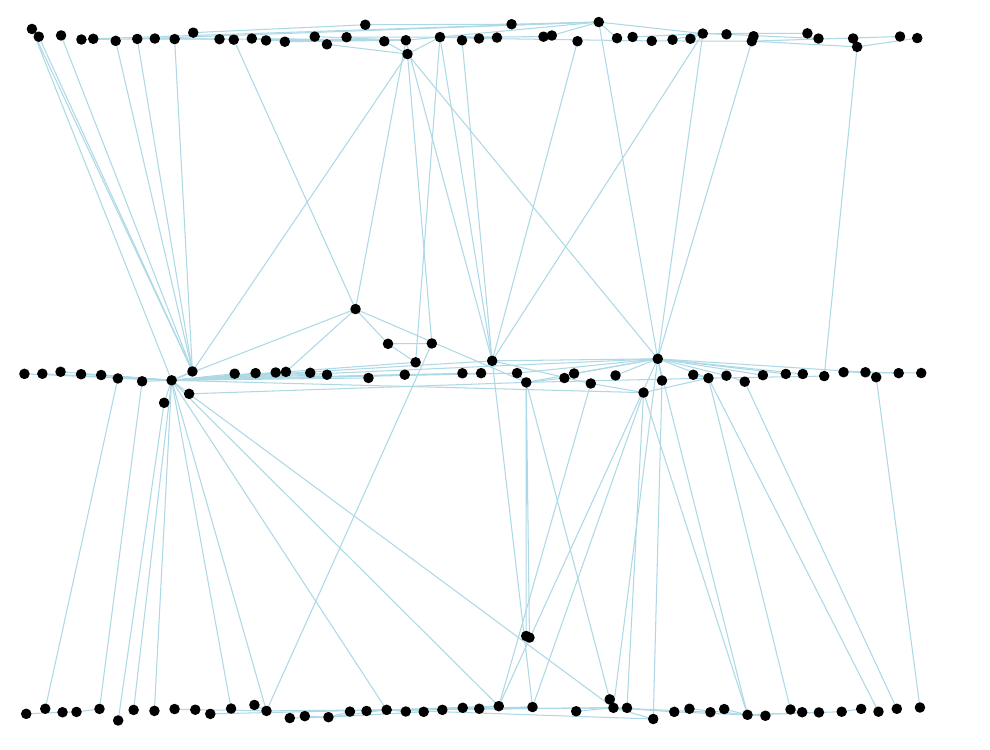} &
  \includegraphics[width=9.5mm]{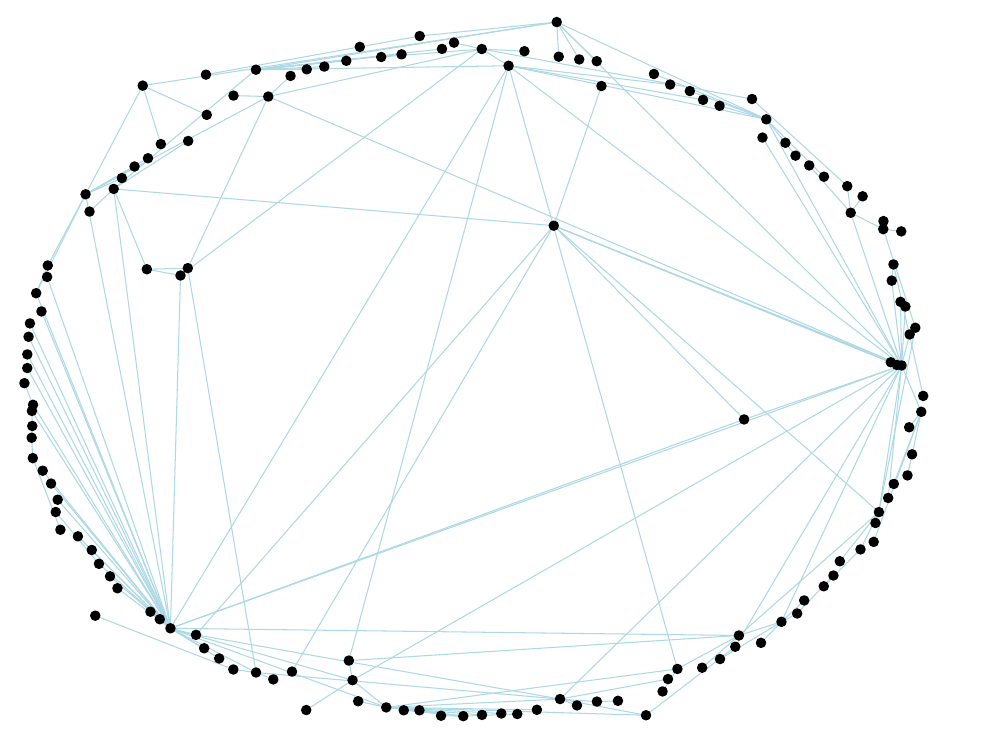} &
  \includegraphics[width=9.5mm]{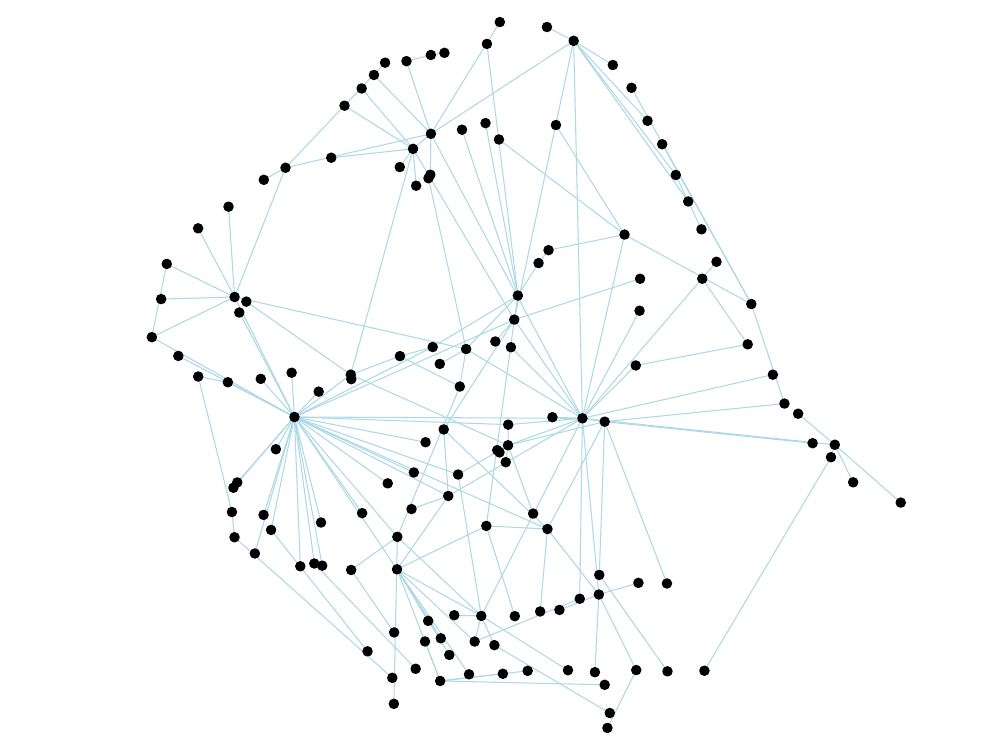} &
  \includegraphics[width=9.5mm]{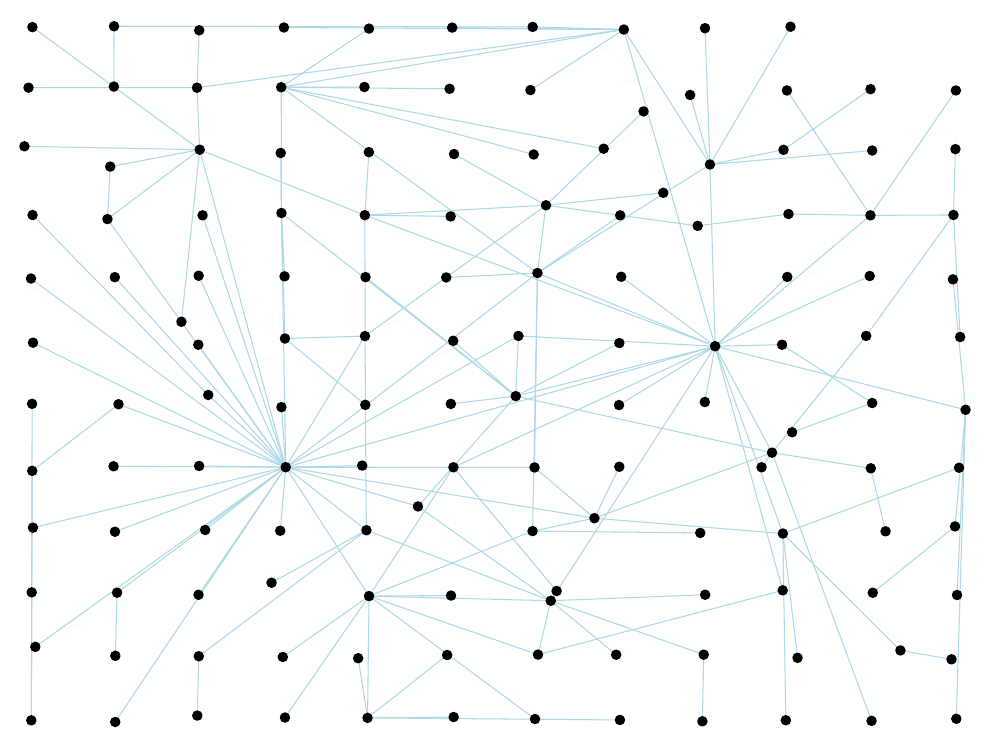}
  \\
       & \texttt{ST-ELD-CN} & &\includegraphics[width=9.5mm]{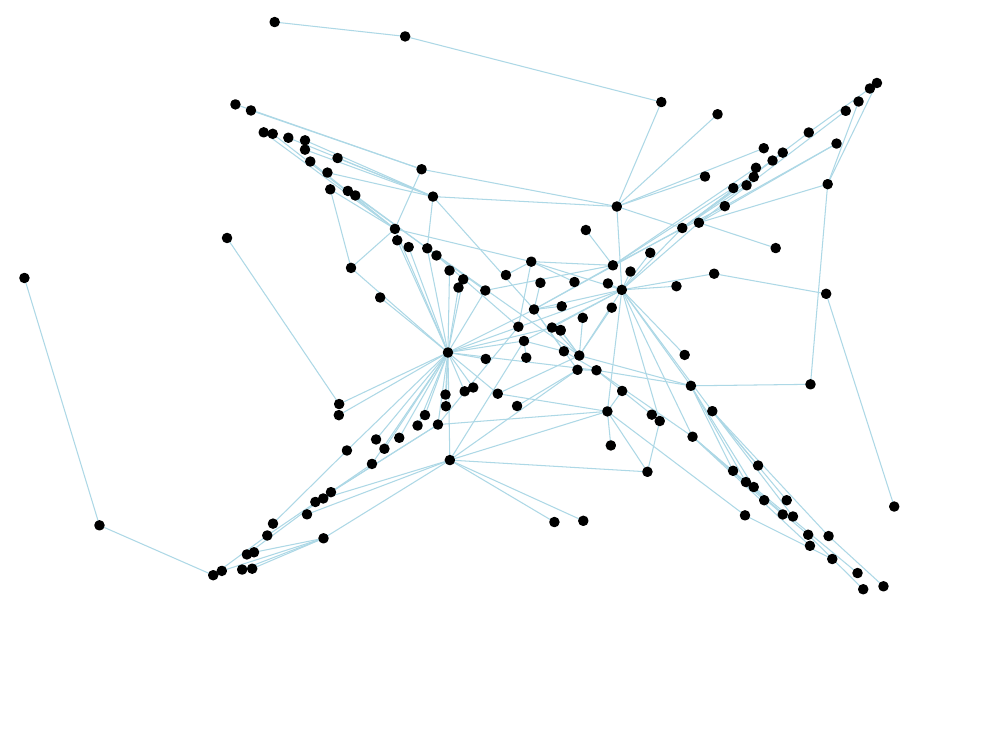} &
  \includegraphics[width=9.5mm]{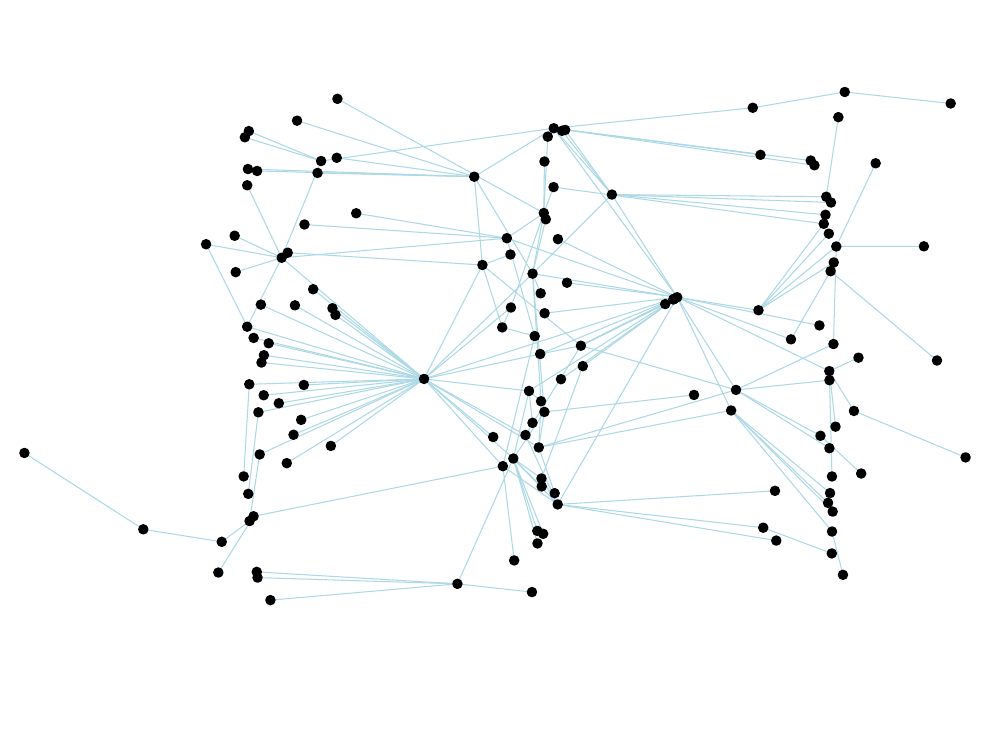} &
  \includegraphics[width=9.5mm]{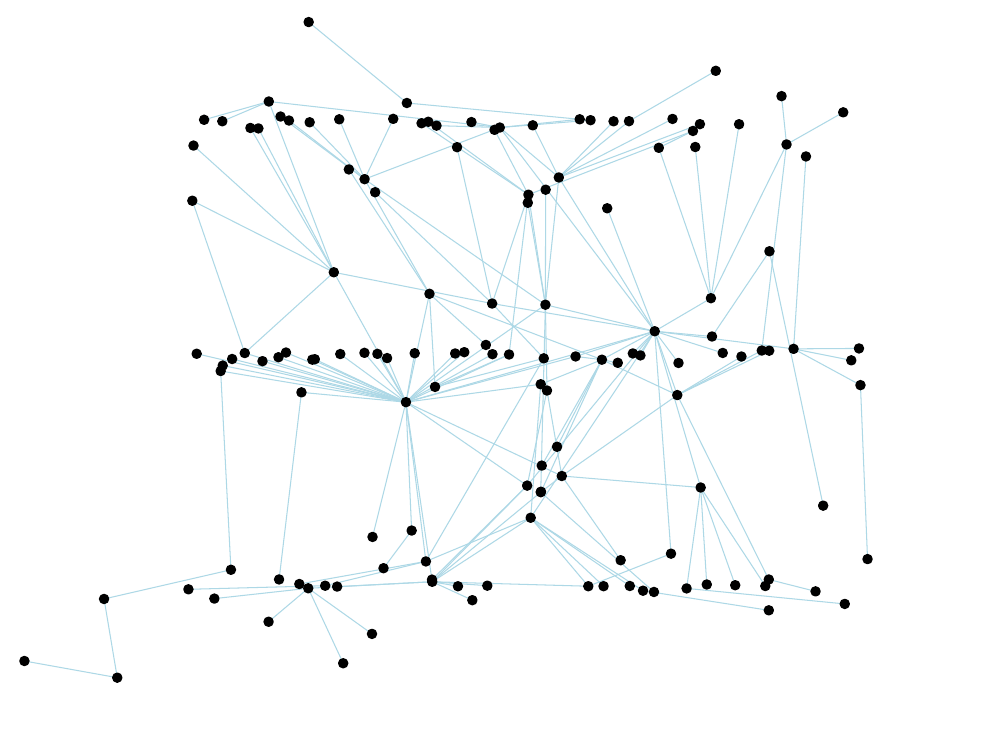} &
  \includegraphics[width=9.5mm]{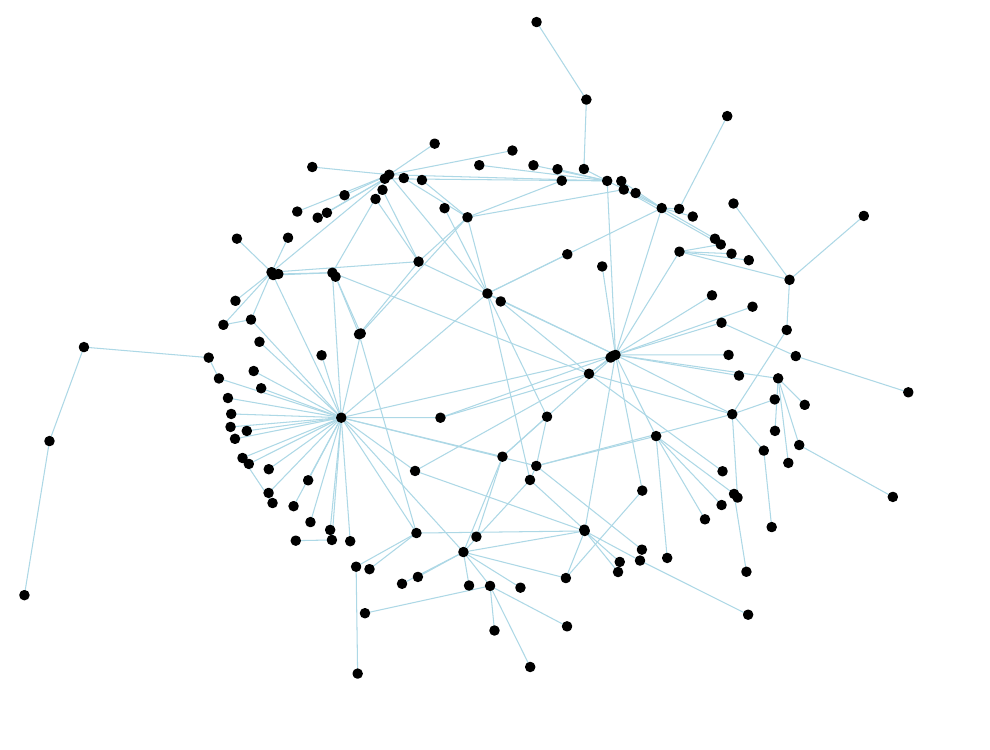} &
  \includegraphics[width=9.5mm]{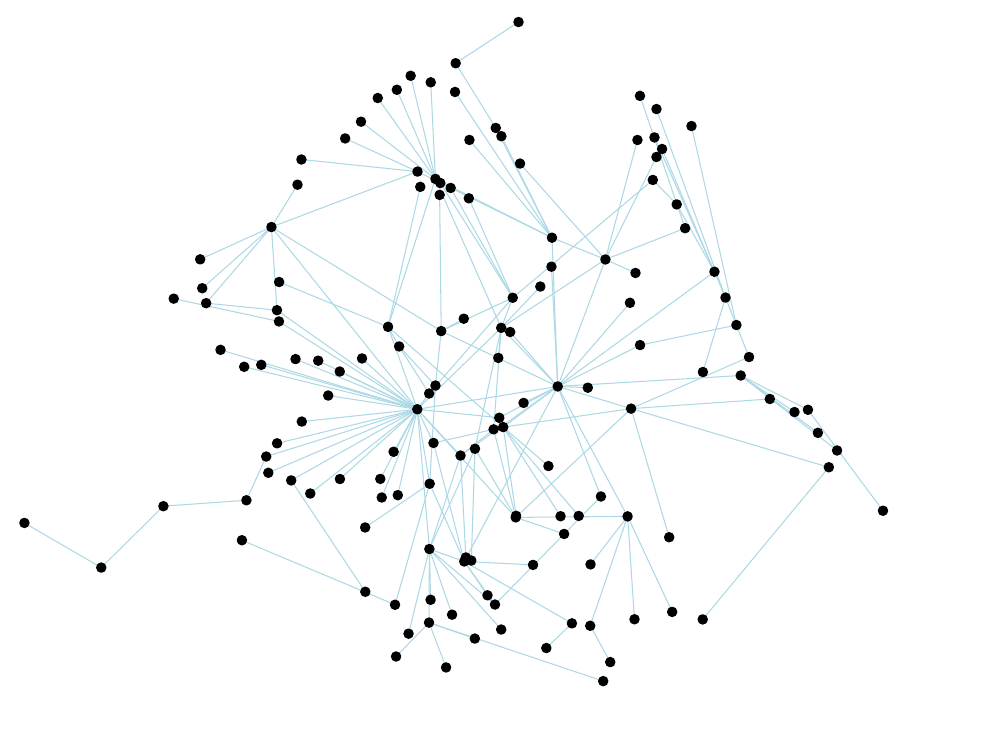} &
  \includegraphics[width=9.5mm]{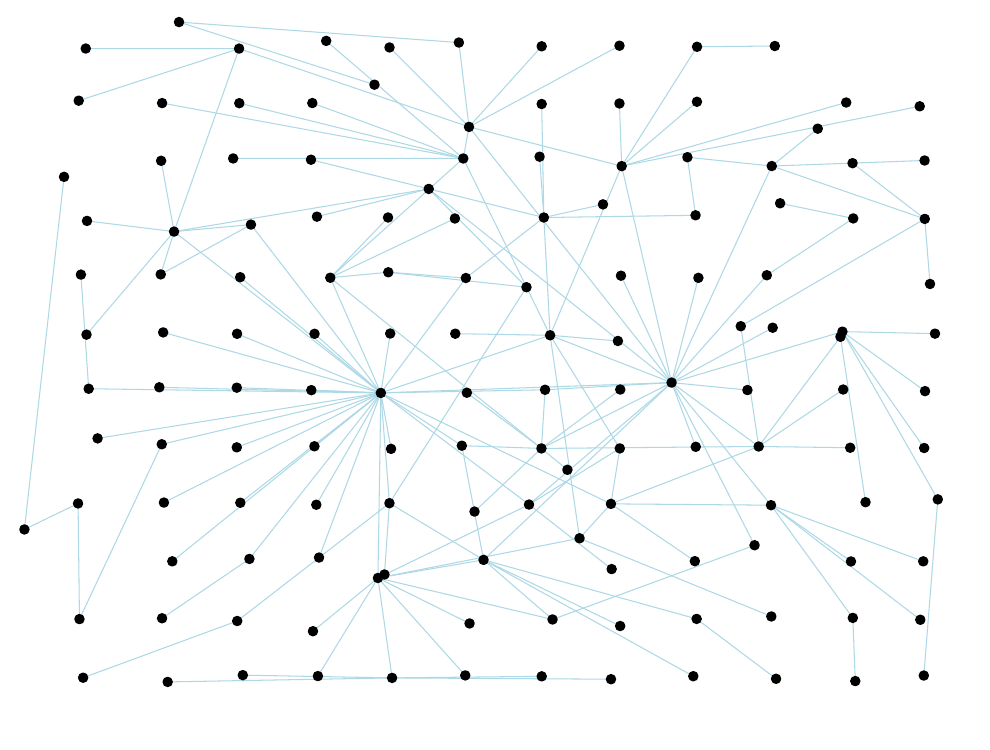}
  \\
         & \texttt{ST-ELD-AR} & &\includegraphics[width=9.5mm]{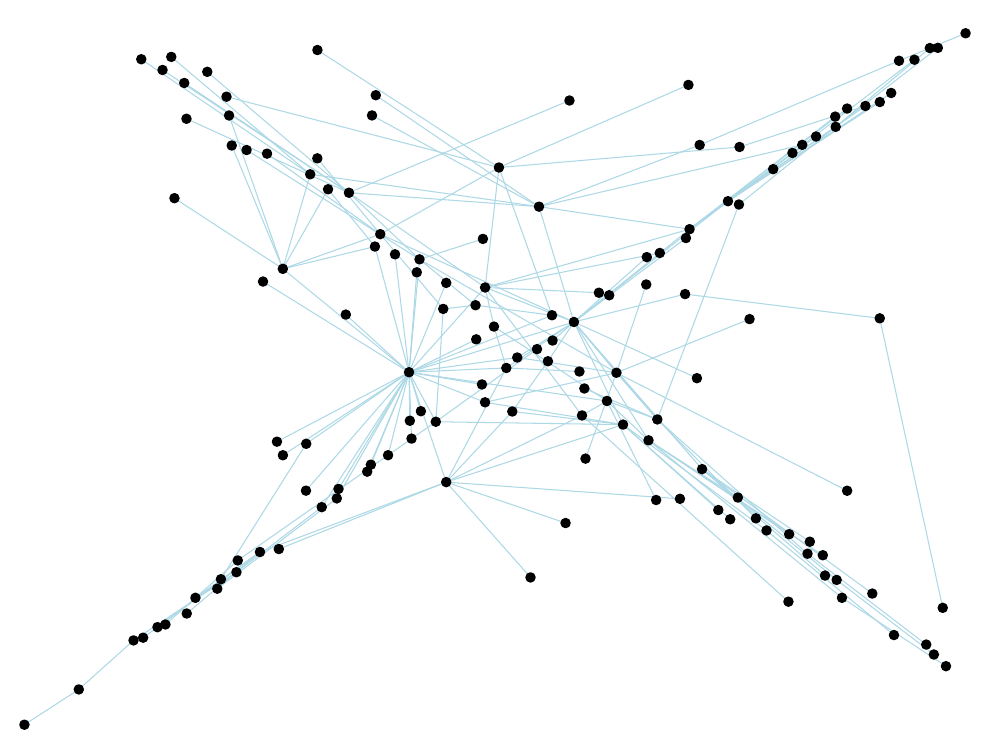} &
  \includegraphics[width=9.5mm]{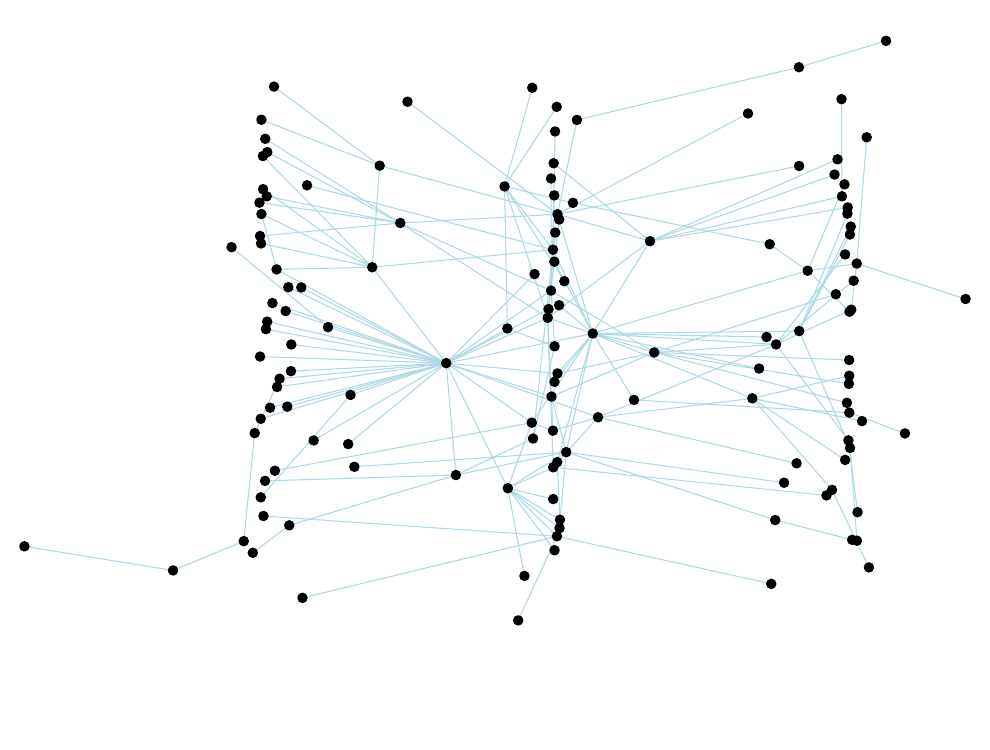} &
  \includegraphics[width=9.5mm]{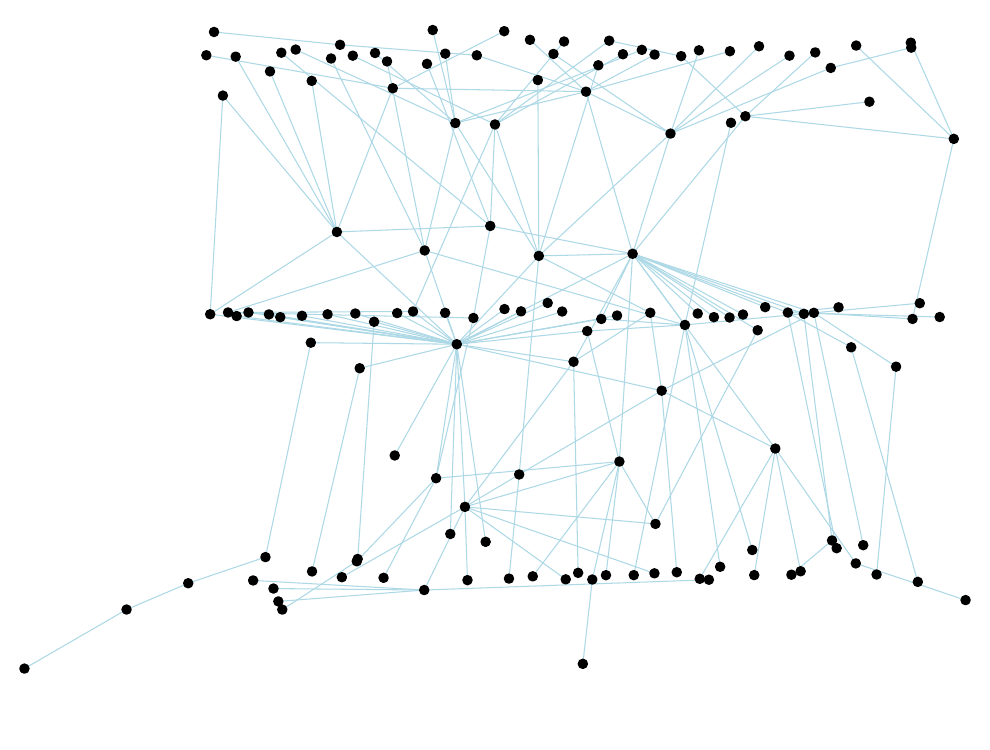} &
  \includegraphics[width=9.5mm]{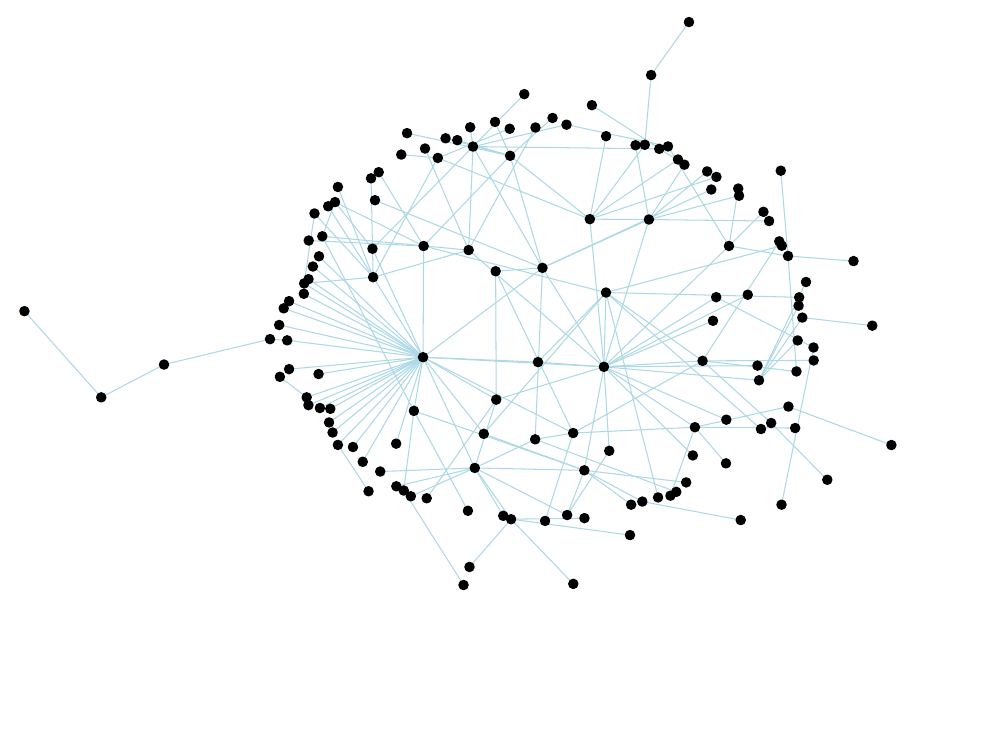} &
  \includegraphics[width=9.5mm]{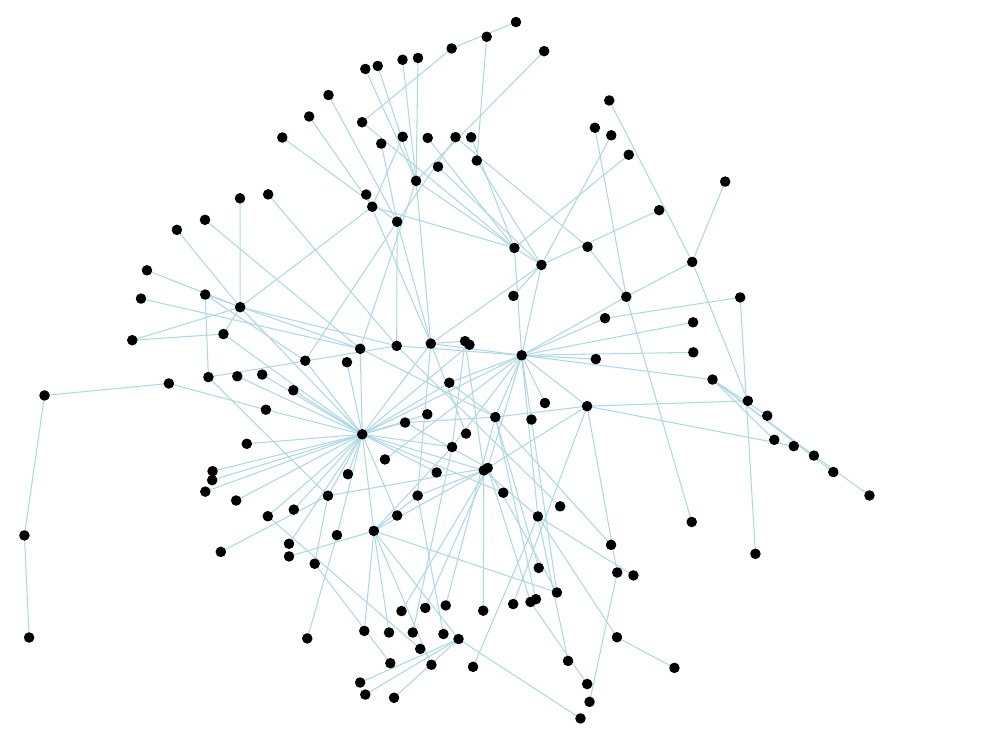} &
  \includegraphics[width=9.5mm]{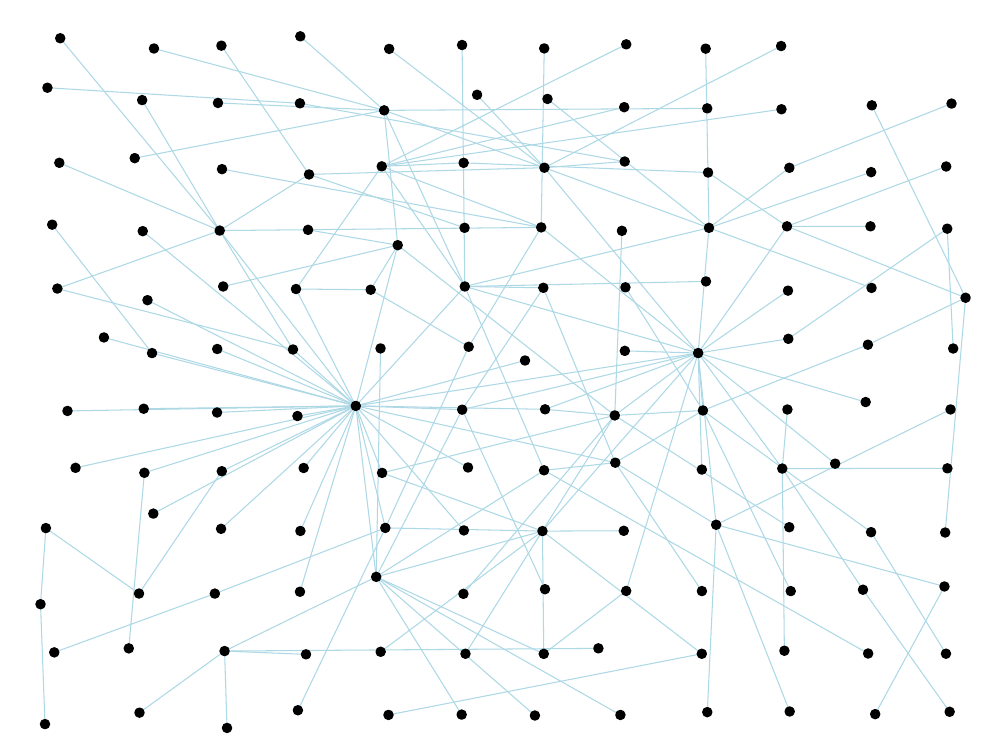}
  \\
         & \texttt{ST-CN-AR} & &\includegraphics[width=9.5mm]{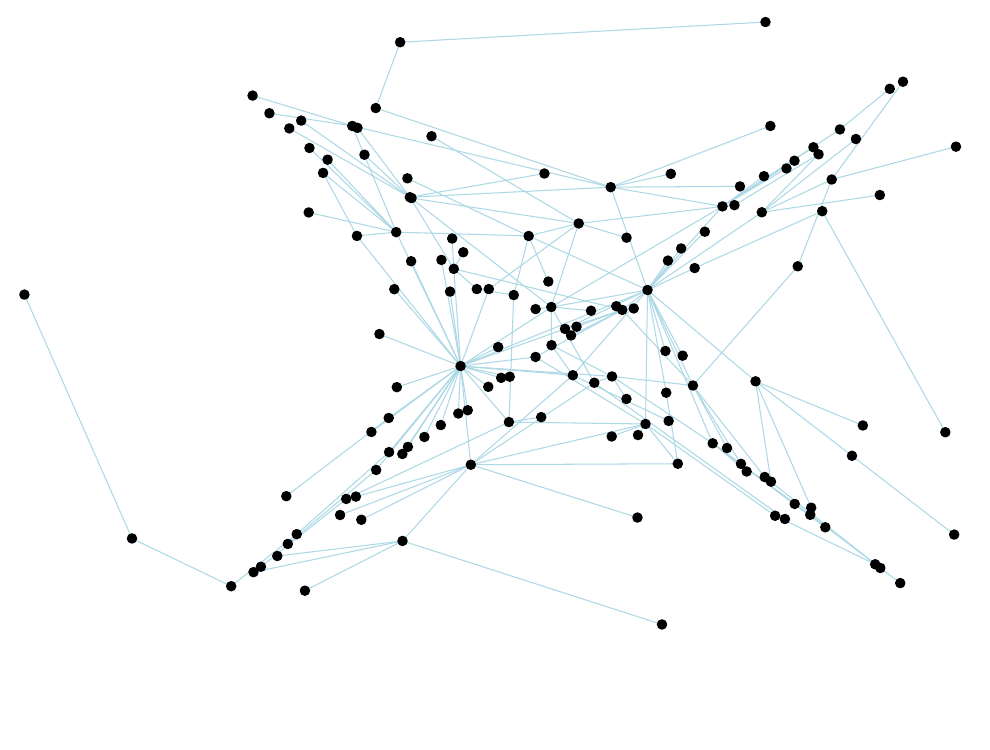} &
  \includegraphics[width=9.5mm]{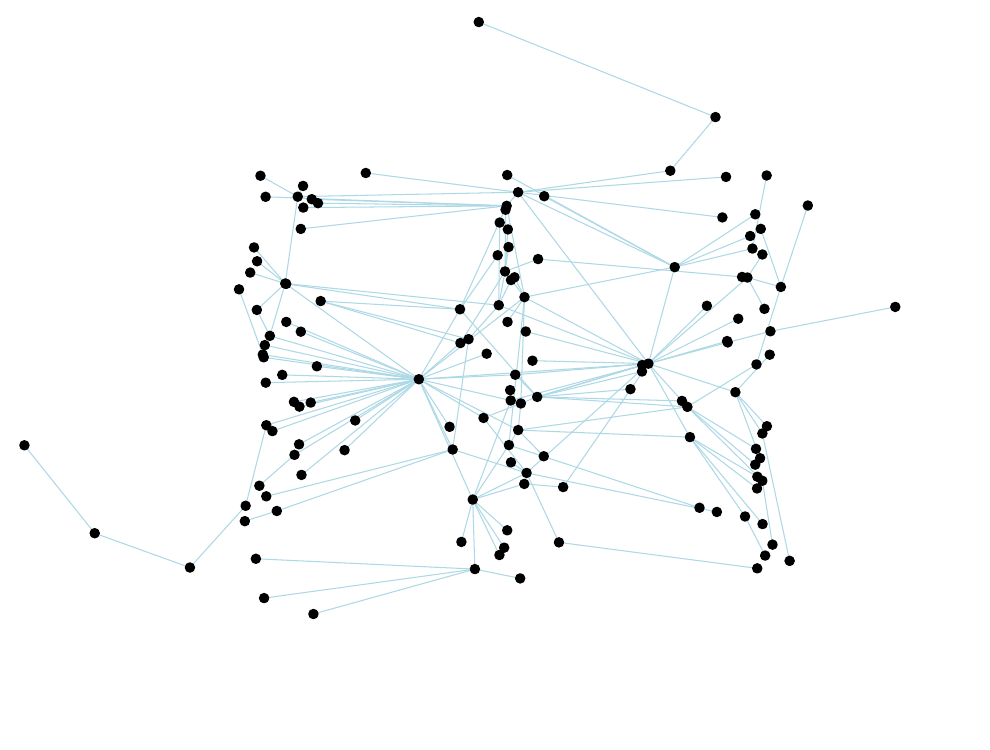} &
  \includegraphics[width=9.5mm]{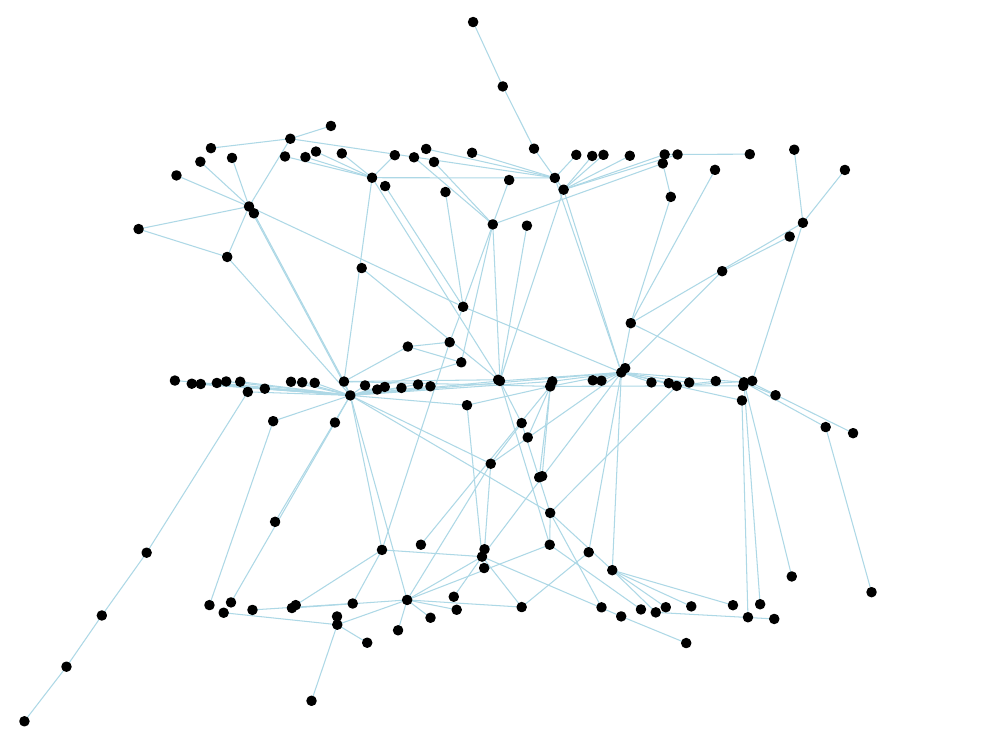} &
  \includegraphics[width=9.5mm]{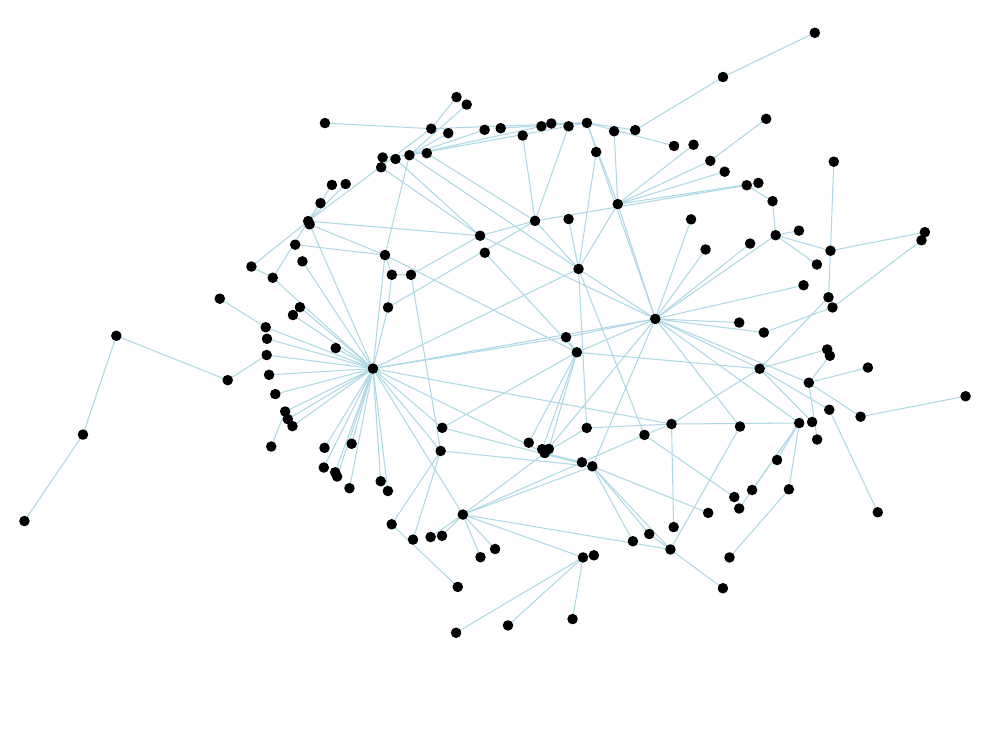} &
  \includegraphics[width=9.5mm]{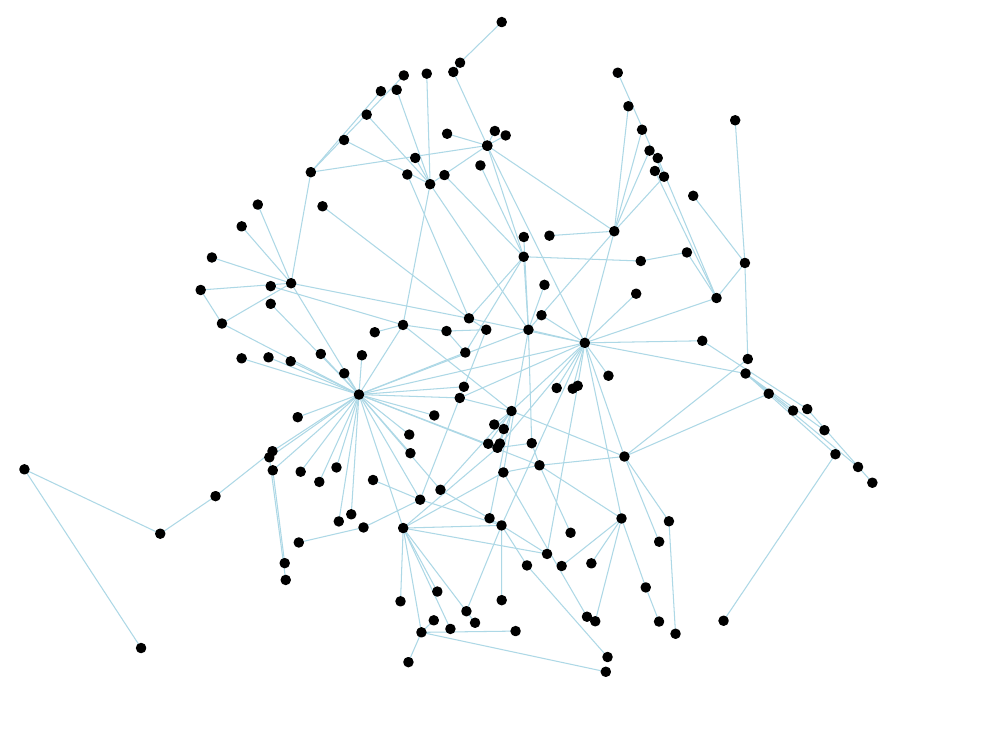} &
  \includegraphics[width=9.5mm]{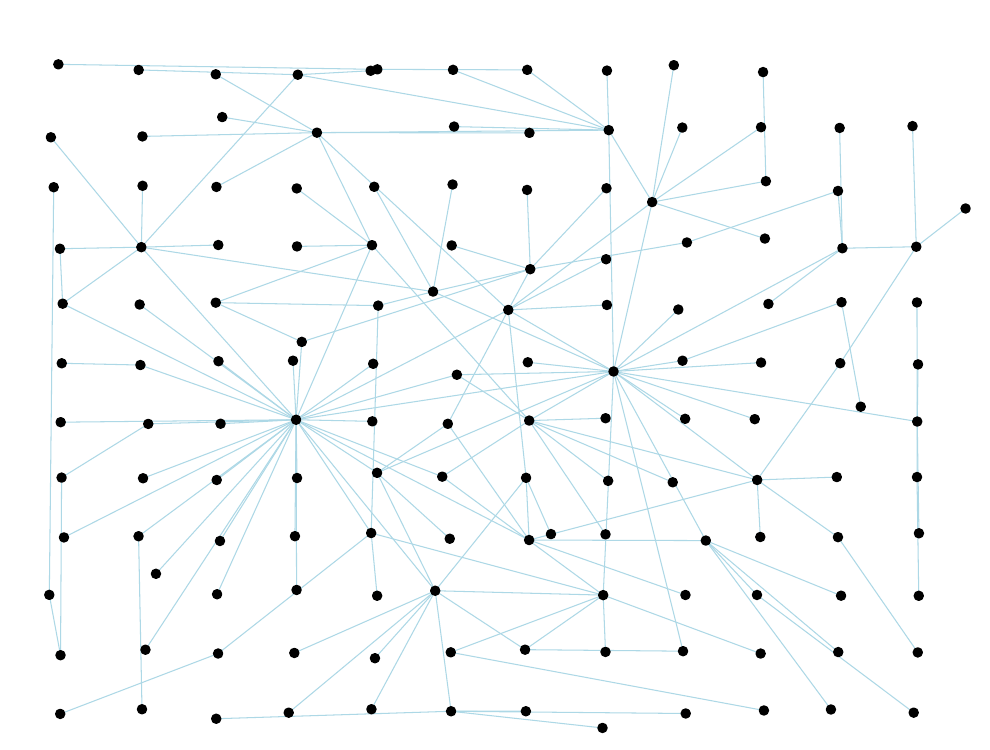}
  \\
         & \texttt{ELD-CN-AR} & & \includegraphics[width=9.5mm]{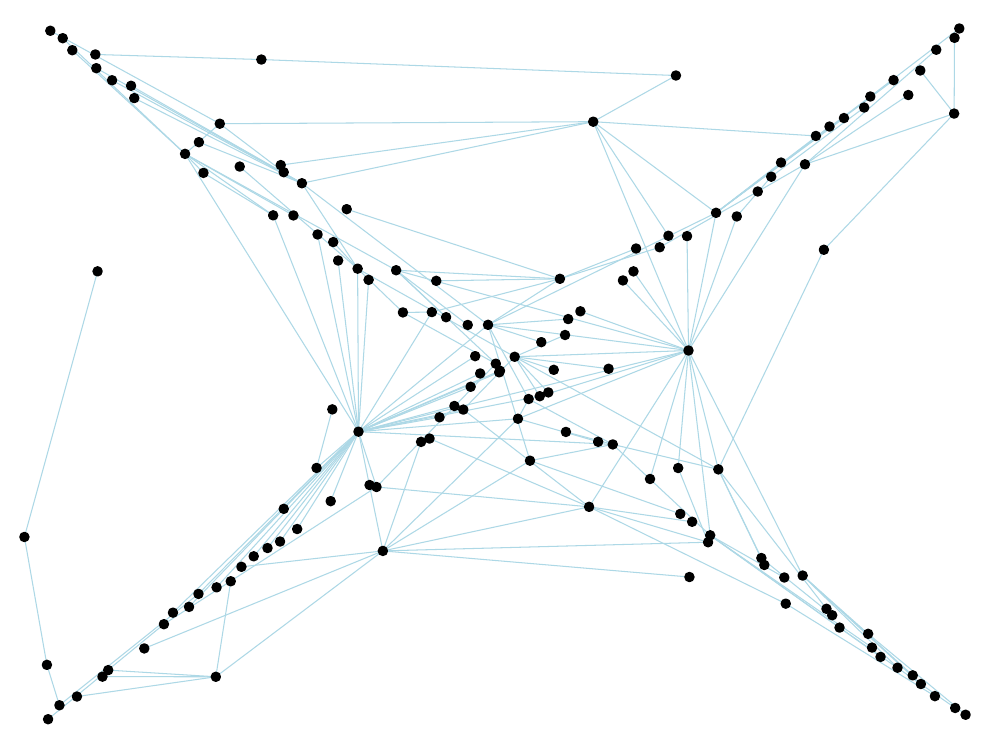} &
  \includegraphics[width=9.5mm]{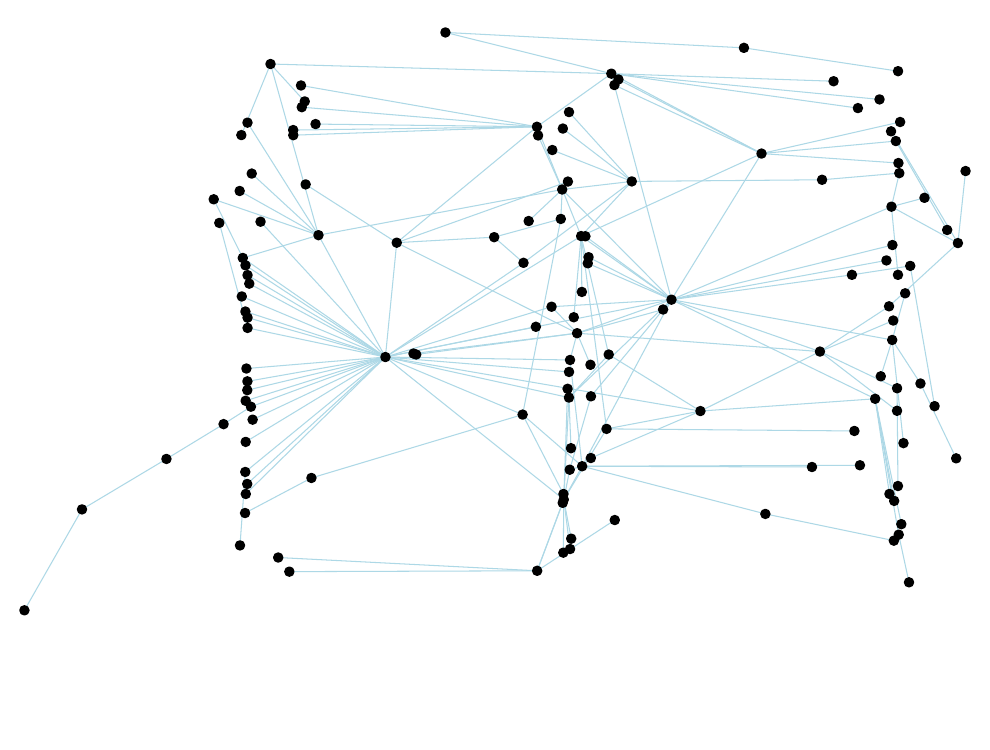} &
  \includegraphics[width=9.5mm]{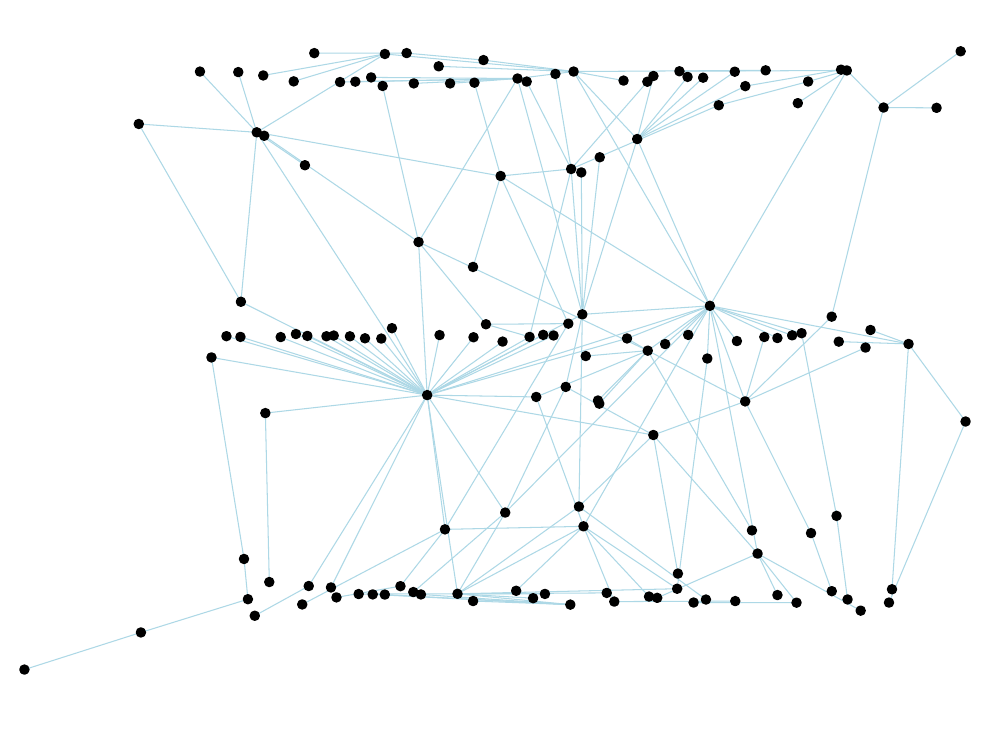} &
  \includegraphics[width=9.5mm]{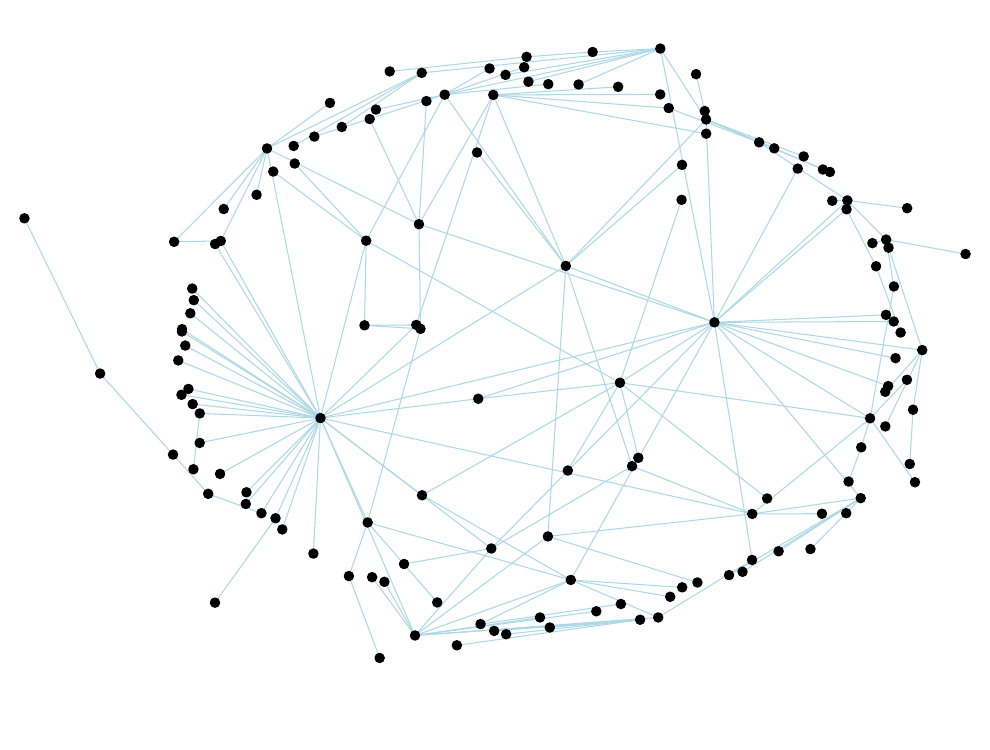} &
  \includegraphics[width=9.5mm]{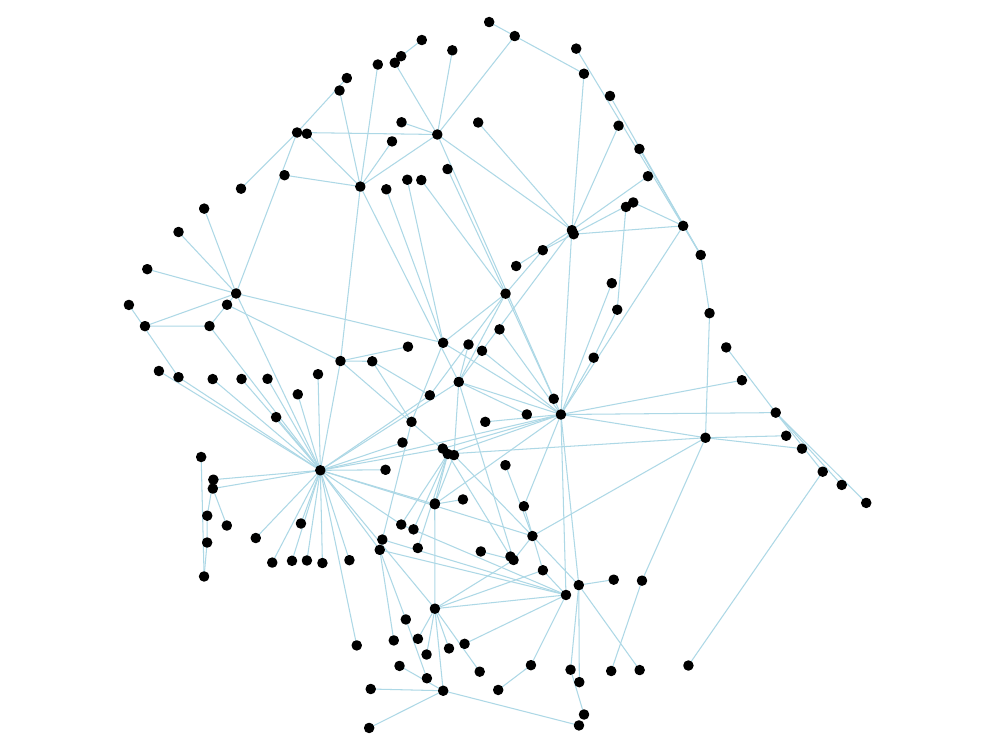} &
  \includegraphics[width=9.5mm]{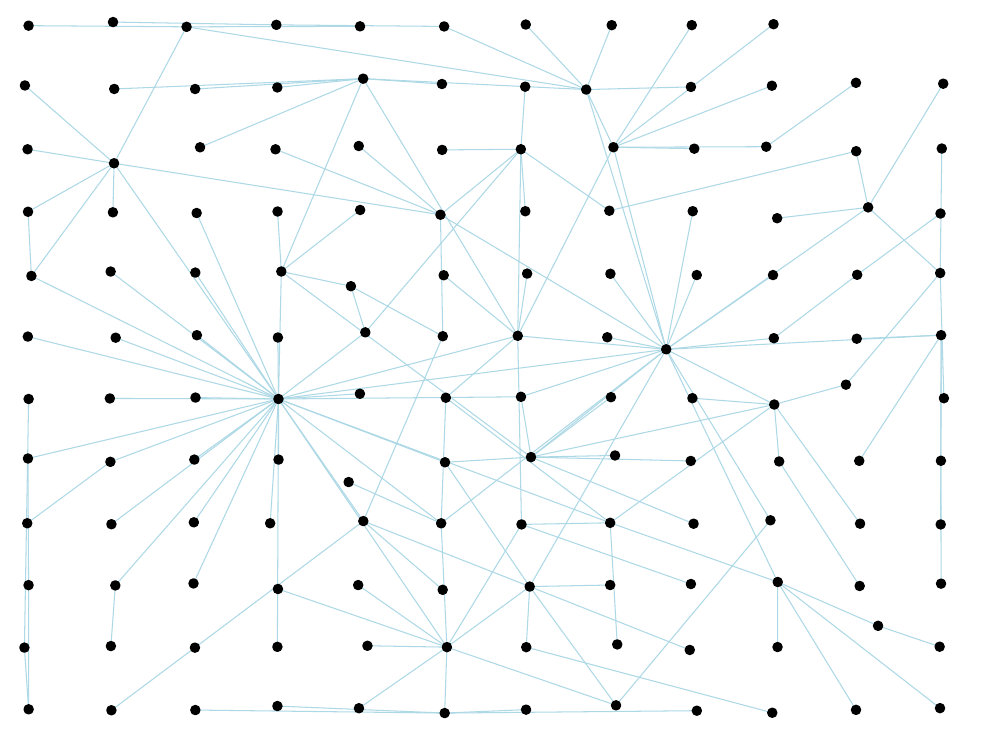}
  \\
         & \texttt{ST-ELD-CN-AR} & & \includegraphics[width=9.5mm]{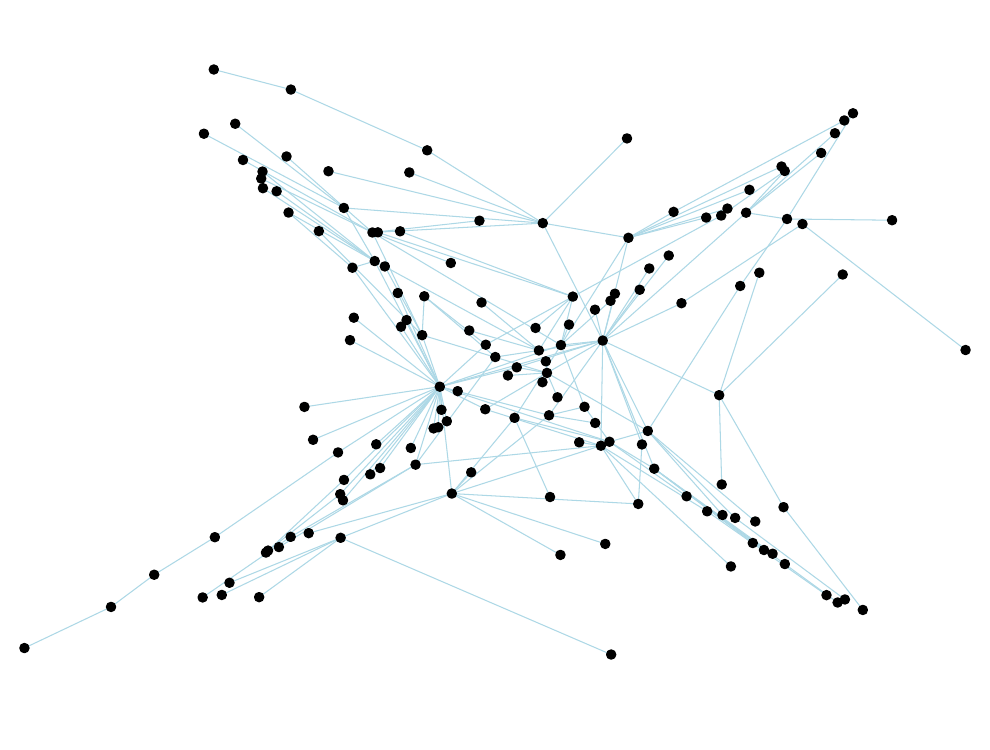} &
  \includegraphics[width=9.5mm]{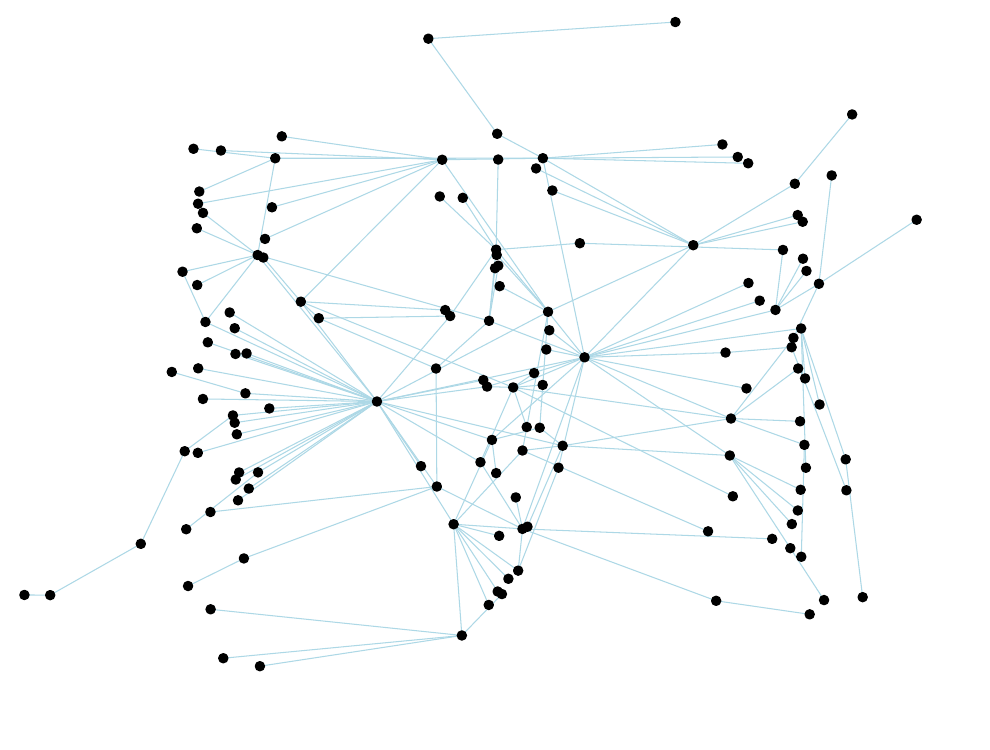} &
  \includegraphics[width=9.5mm]{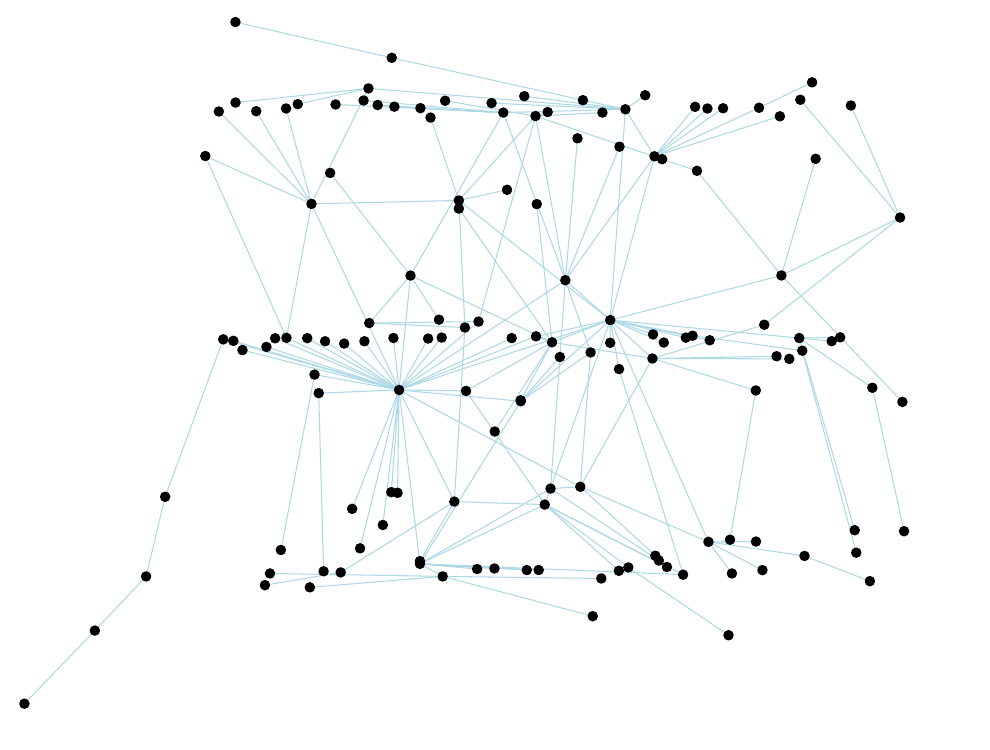} &
  \includegraphics[width=9.5mm]{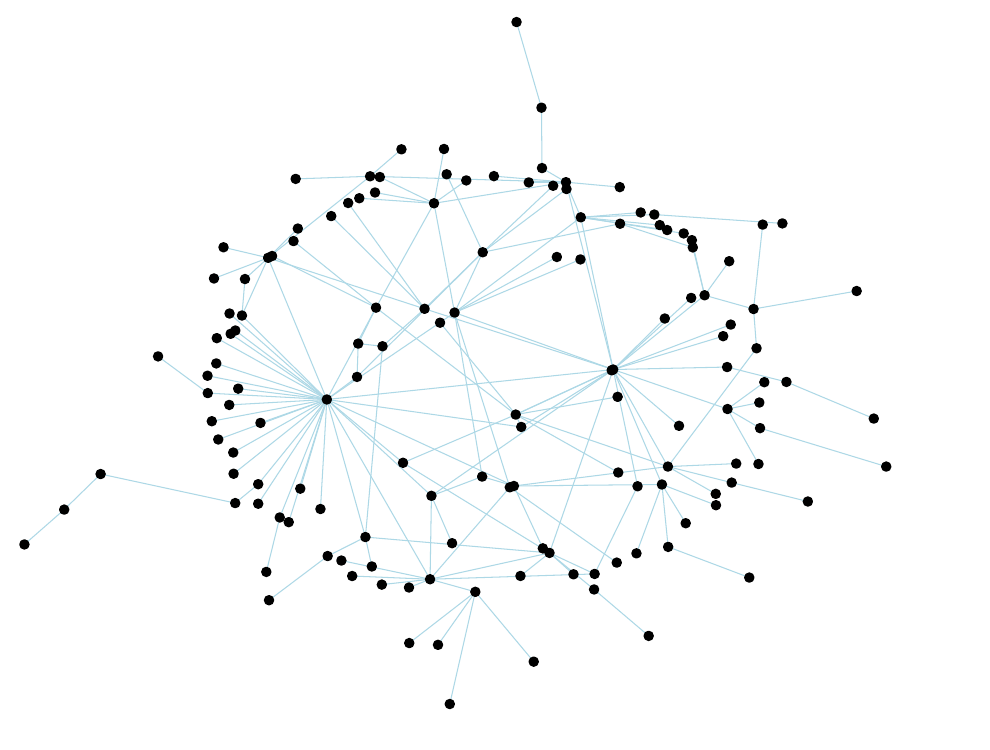} &
  \includegraphics[width=9.5mm]{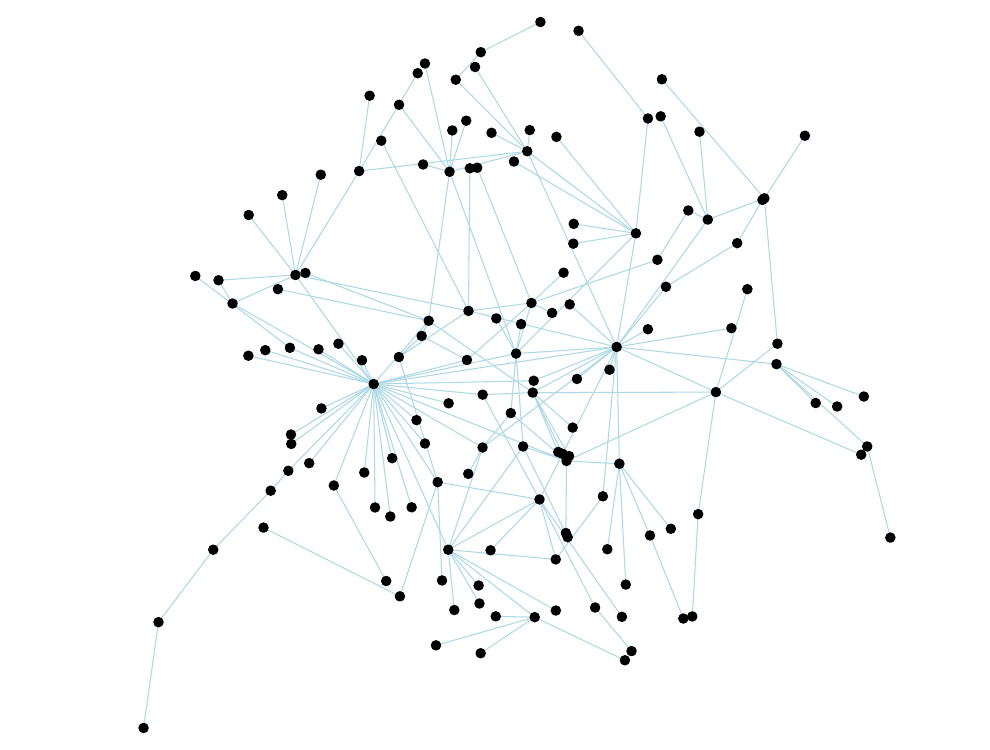} &
  \includegraphics[width=9.5mm]{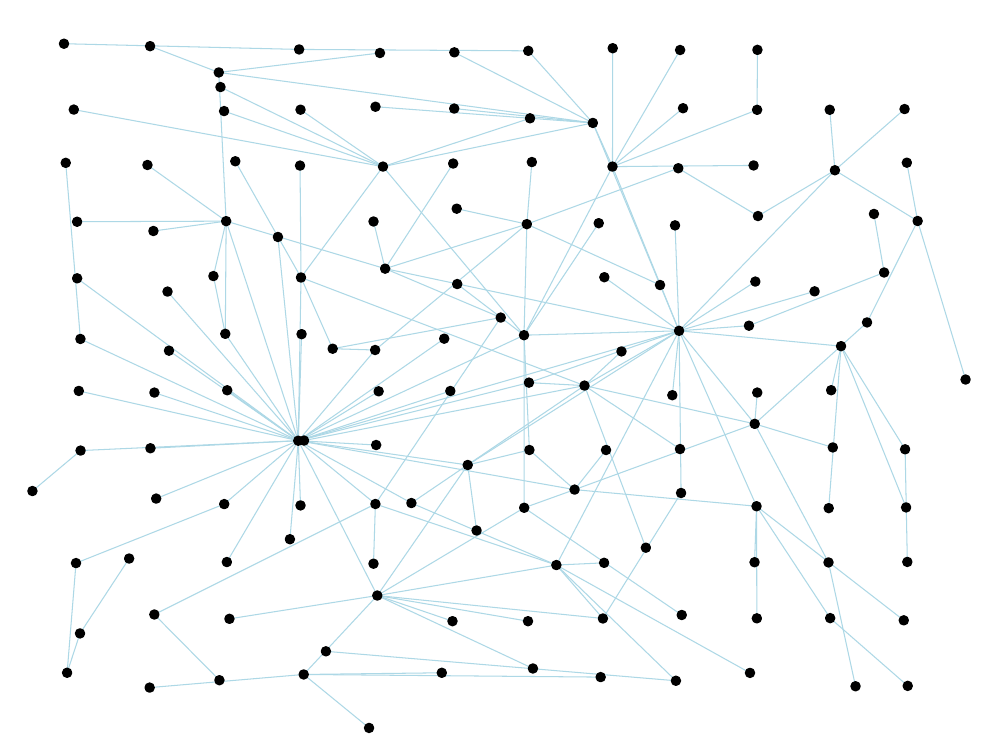}
  \\
  \hline
  \\
    \emph{polbooks} & \texttt{ST-ELD} & \includegraphics[width=9.5mm]{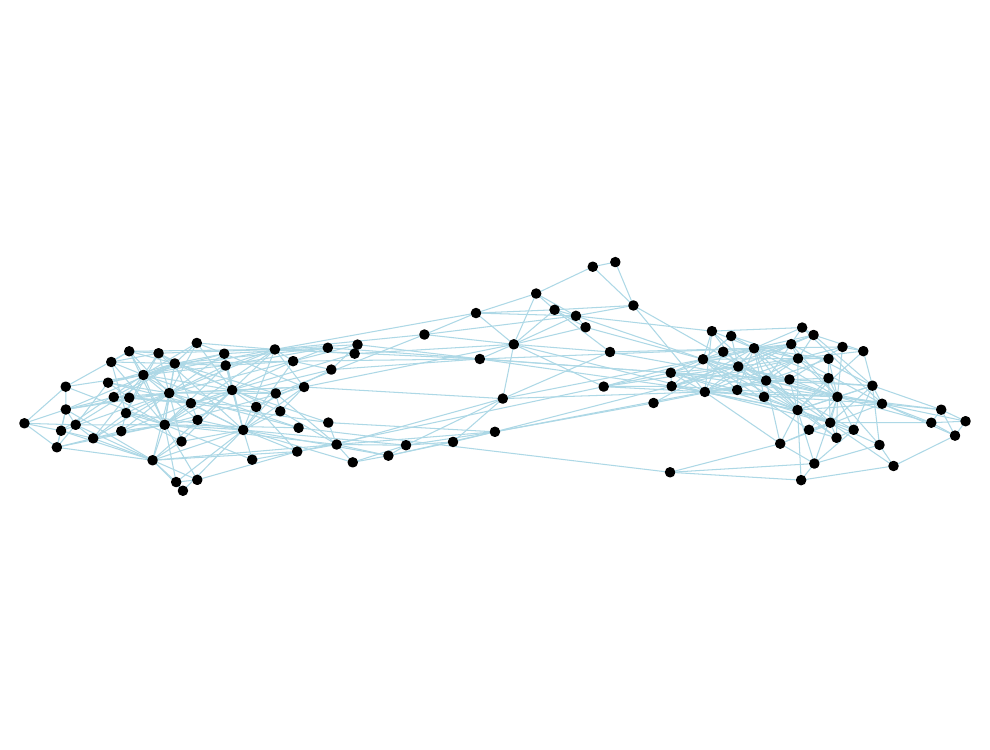} & \includegraphics[width=9.5mm]{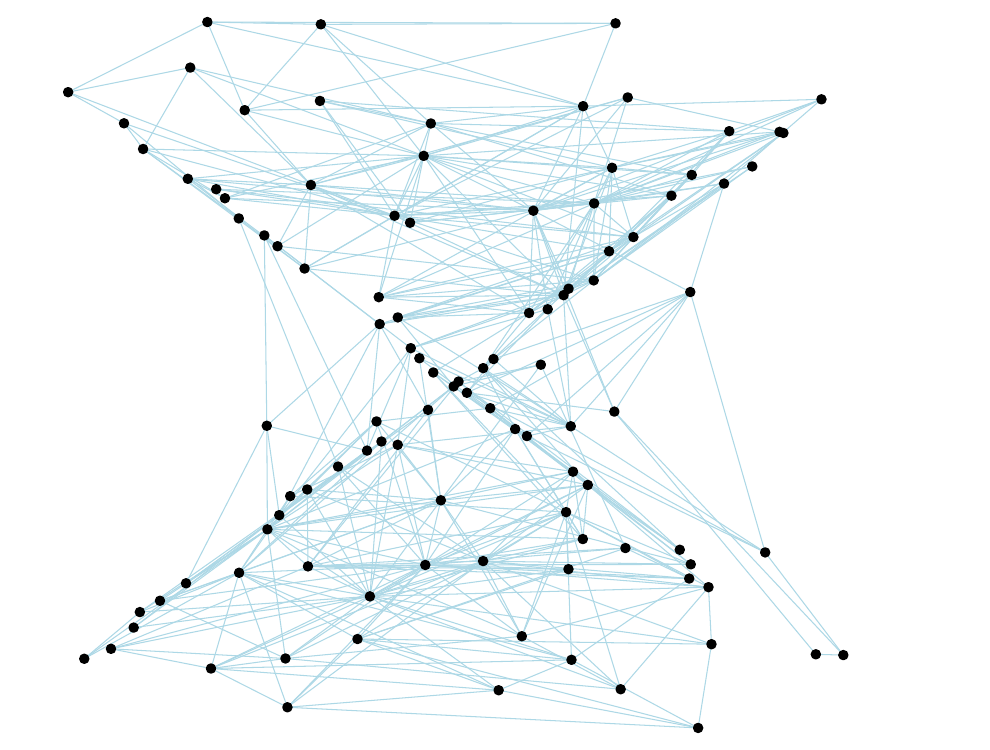} &
  \includegraphics[width=9.5mm]{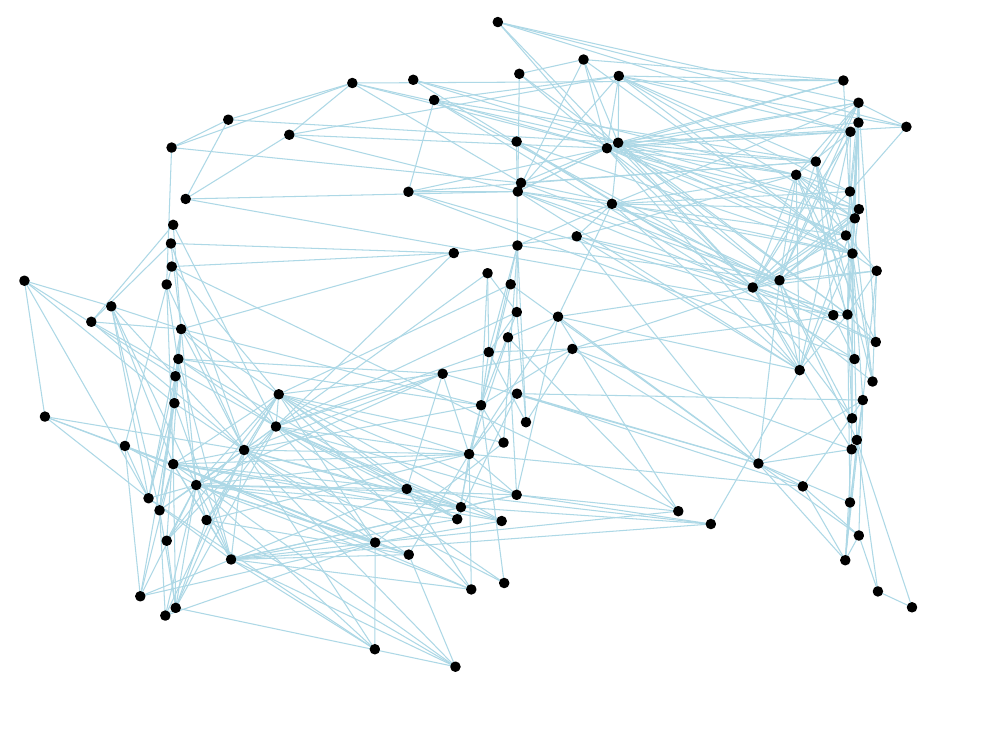} &
  \includegraphics[width=9.5mm]{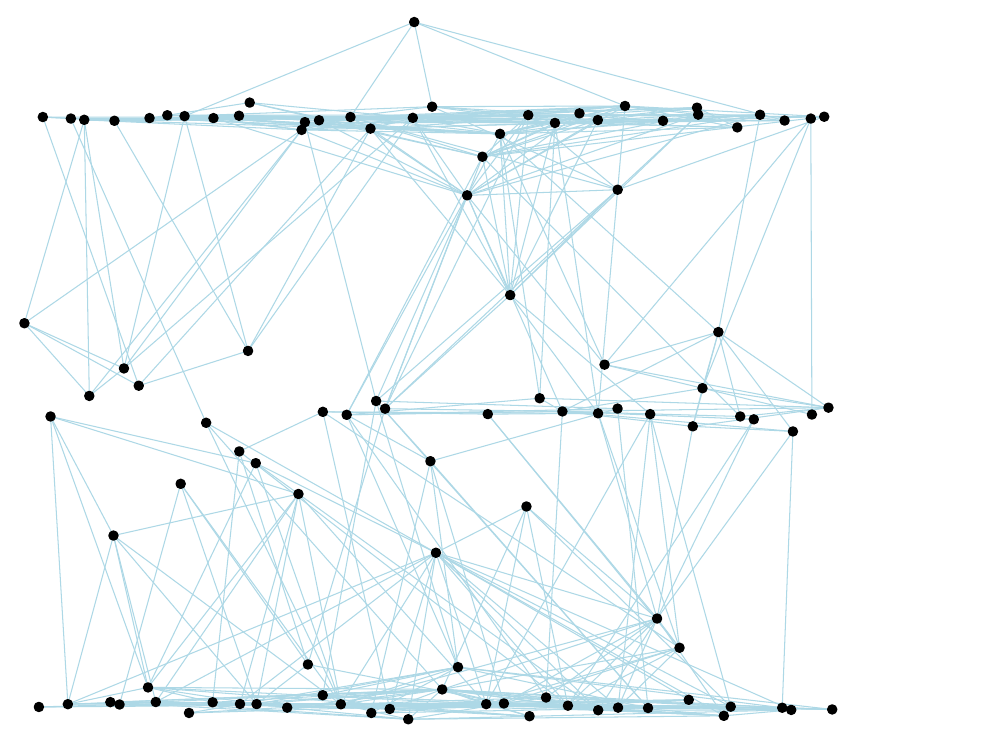} &
  \includegraphics[width=9.5mm]{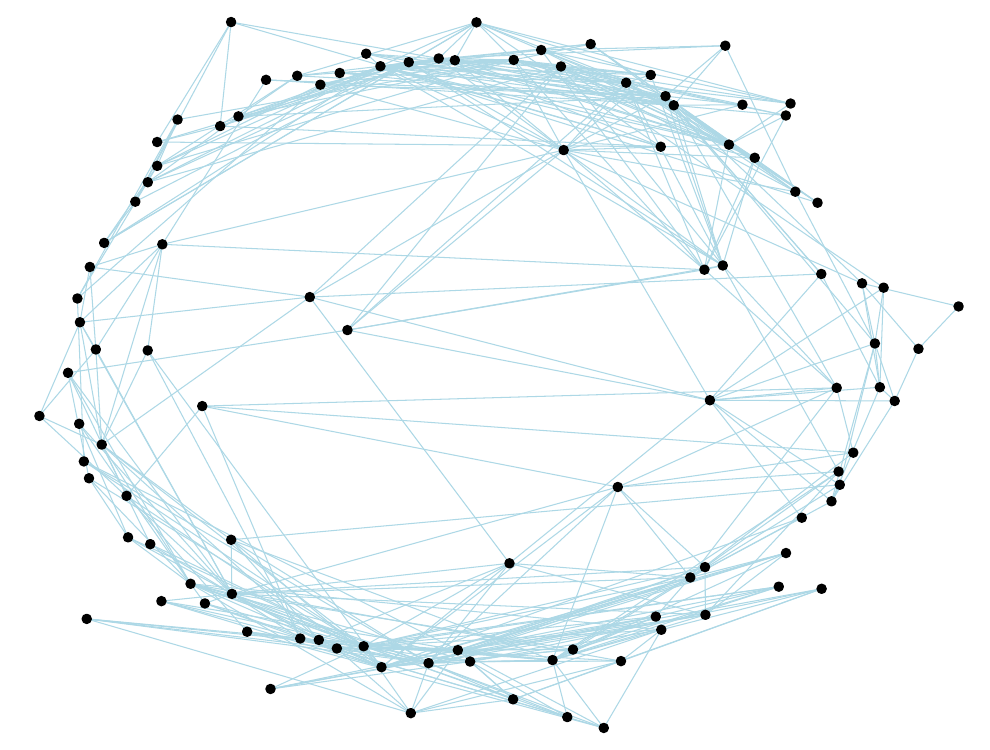} &
  \includegraphics[width=9.5mm]{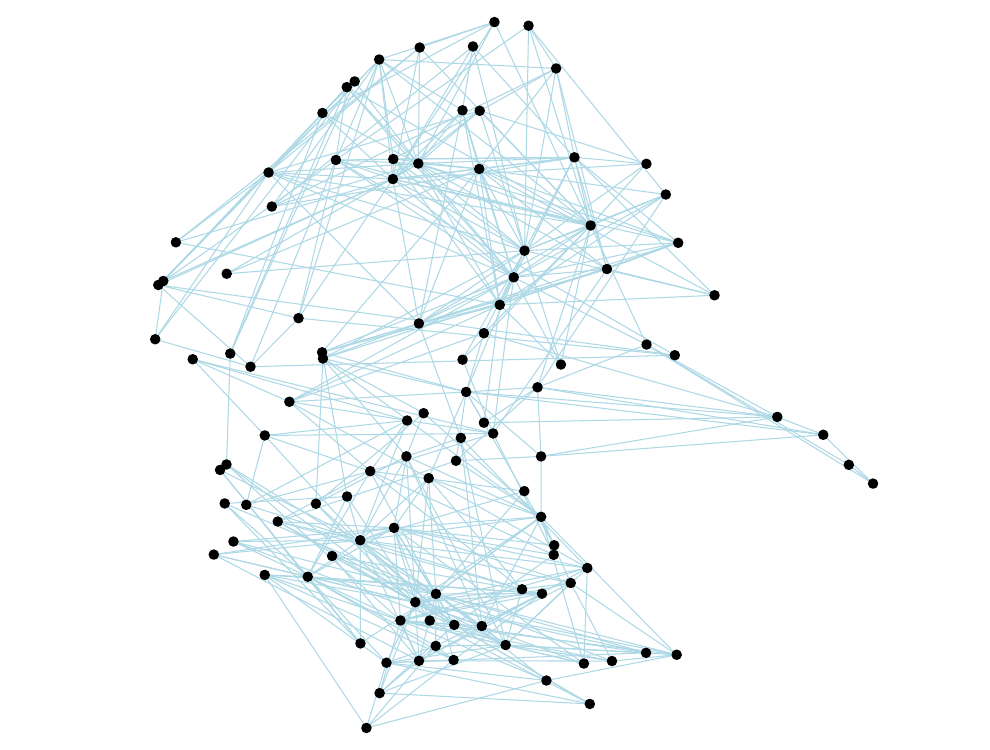} &
  \includegraphics[width=9.5mm]{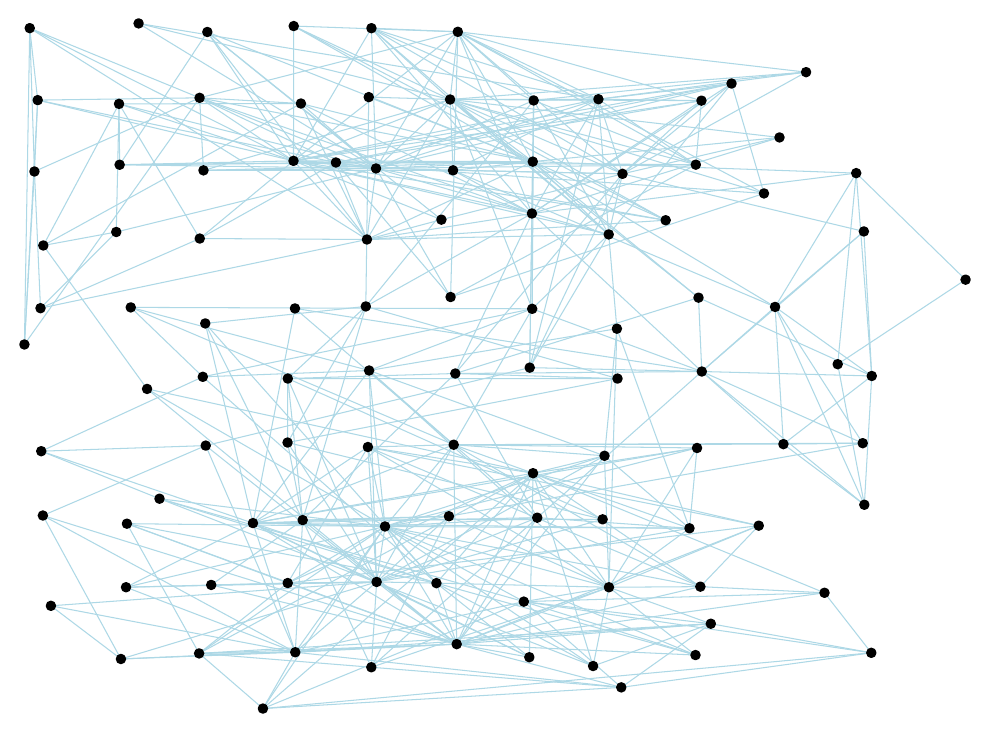}
  \\
     & \texttt{ST-CN} & & \includegraphics[width=9.5mm]{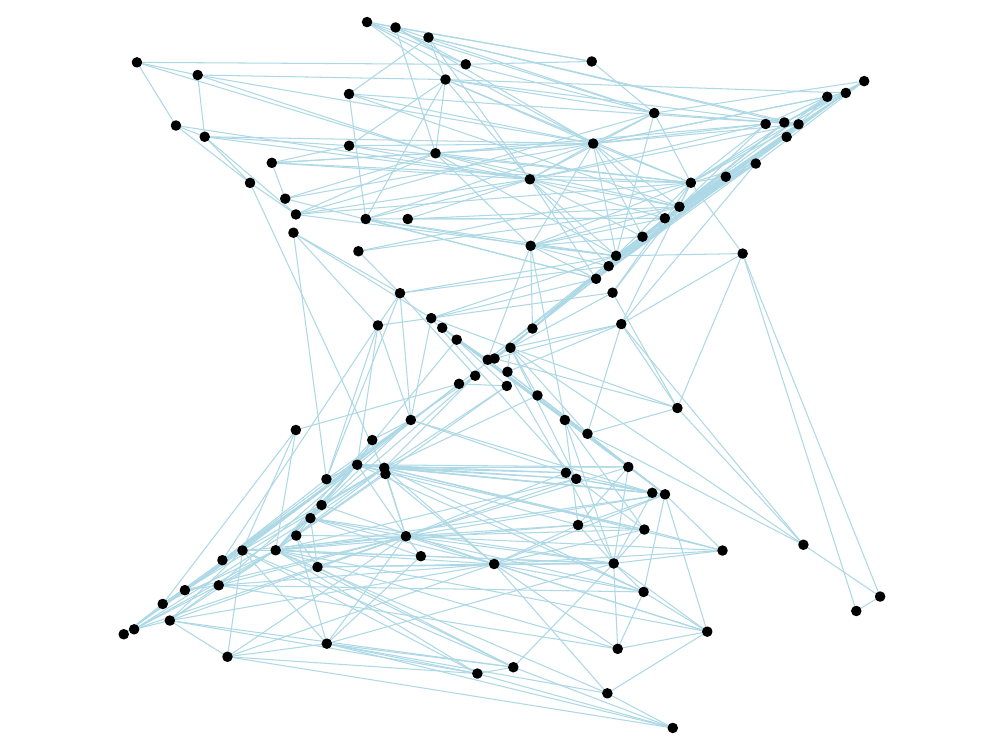} &
  \includegraphics[width=9.5mm]{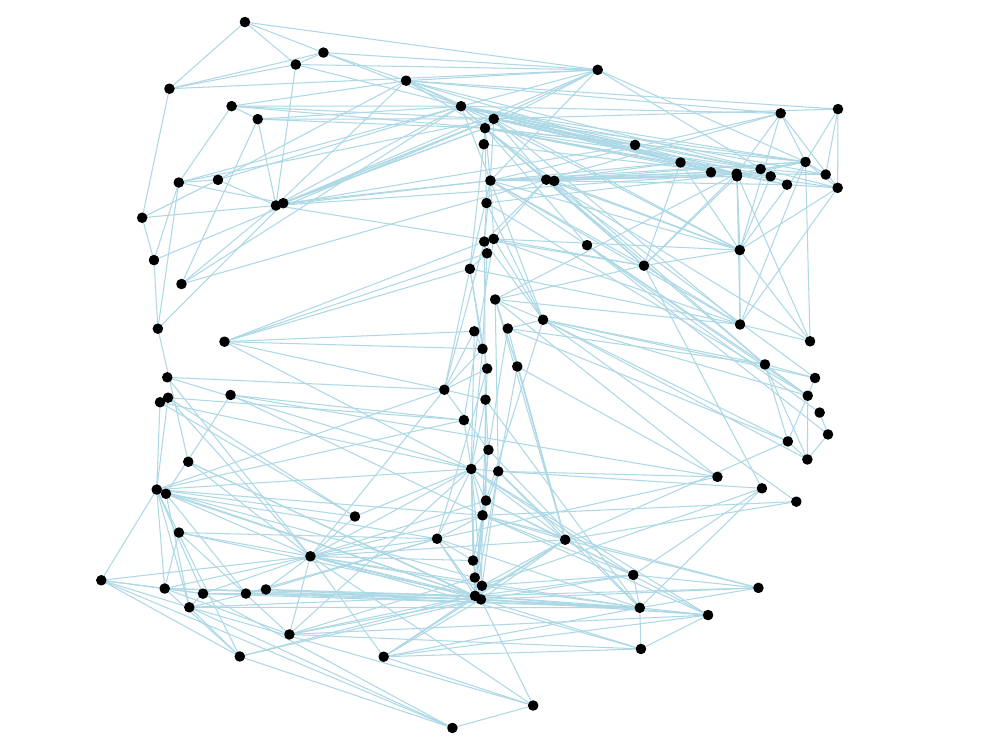} &
  \includegraphics[width=9.5mm]{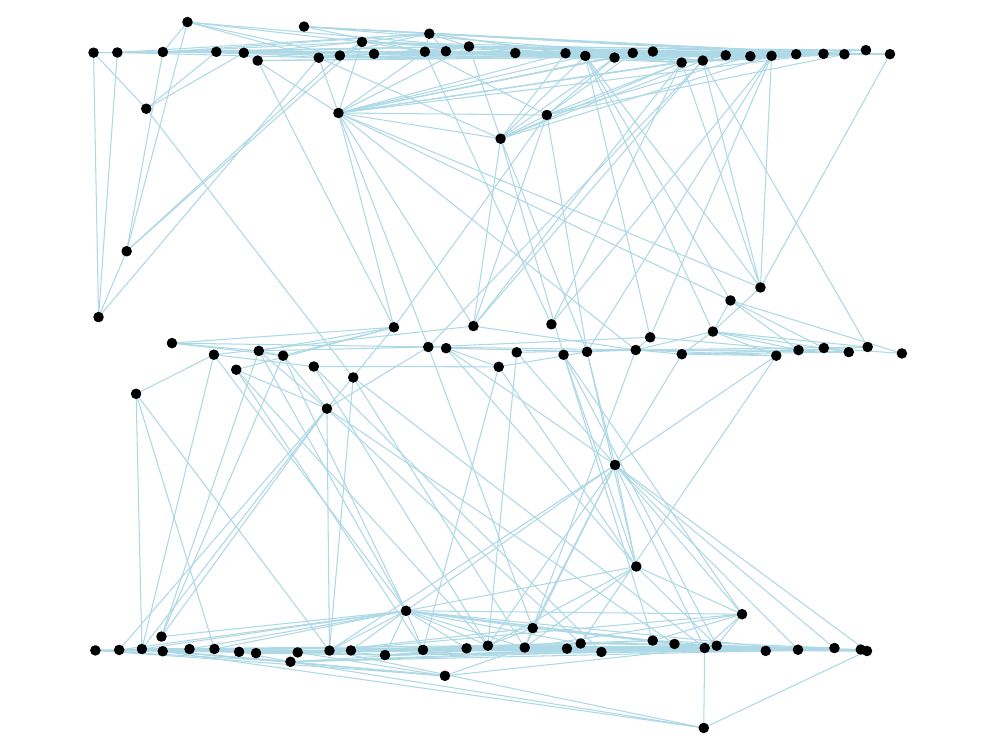} &
  \includegraphics[width=9.5mm]{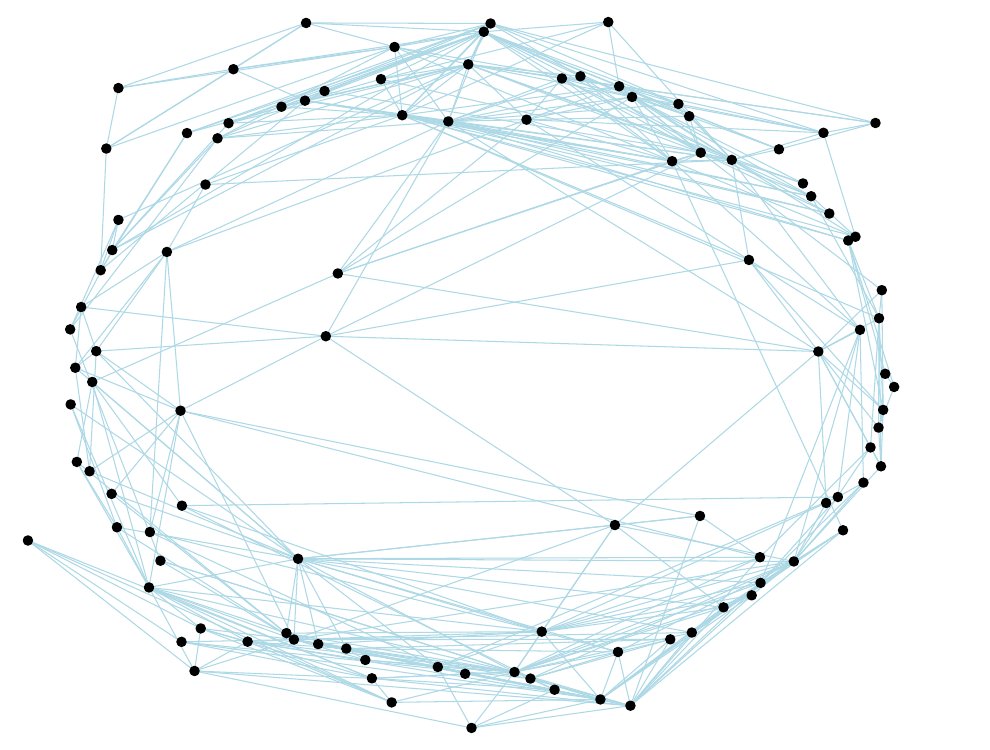} &
  \includegraphics[width=9.5mm]{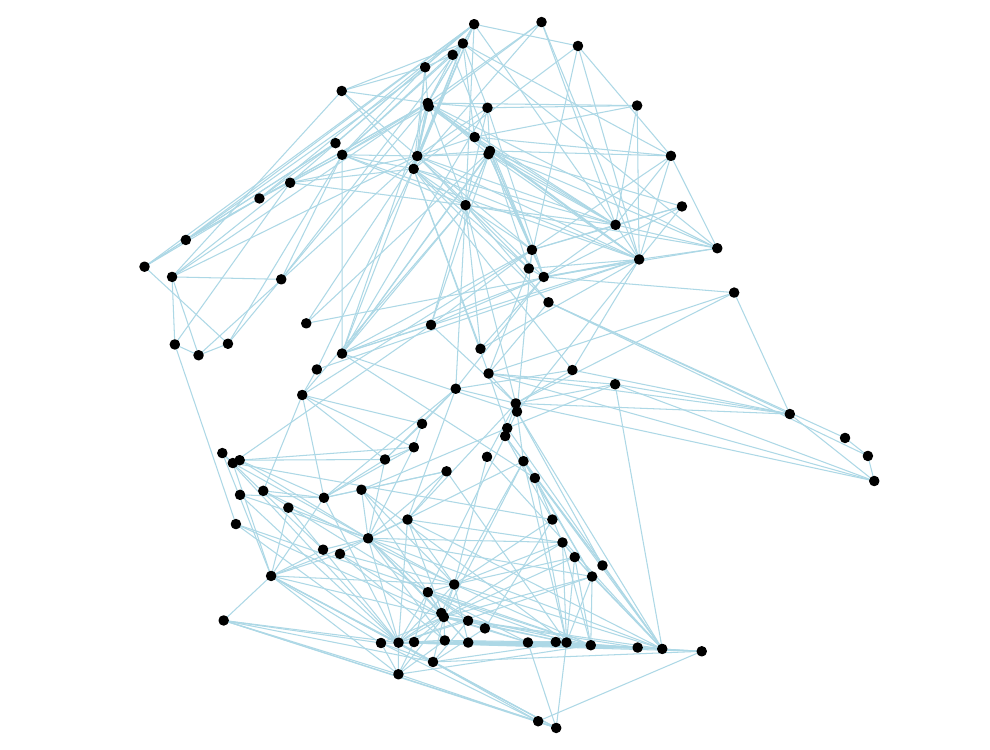} &
  \includegraphics[width=9.5mm]{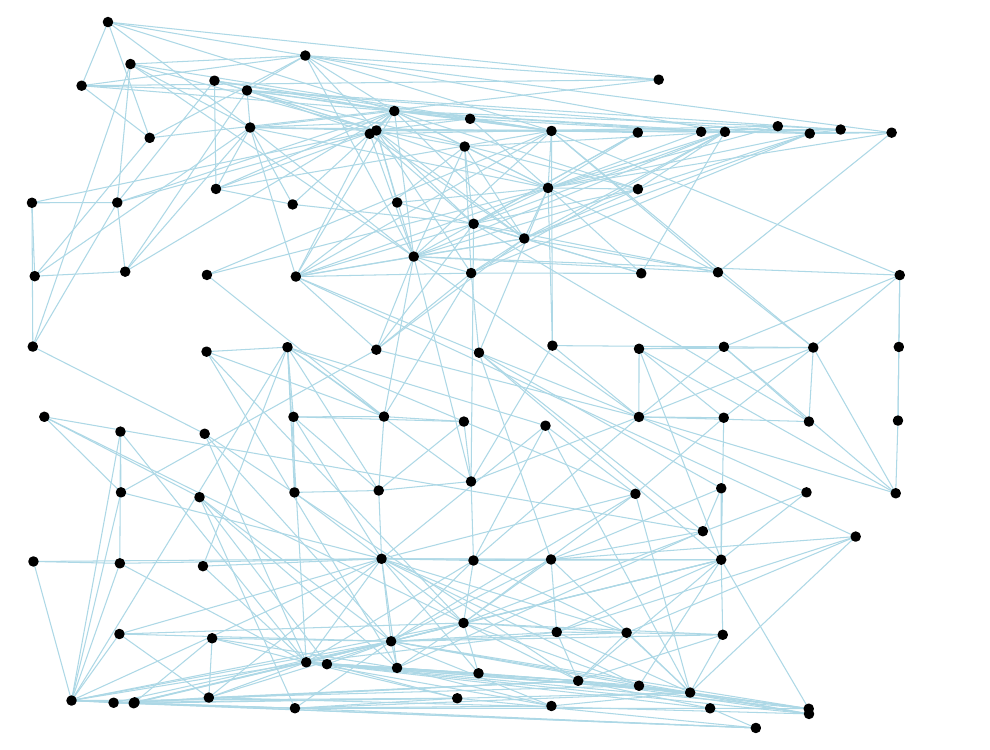}
  \\
       & \texttt{ST-AR} &  & \includegraphics[width=9.5mm]{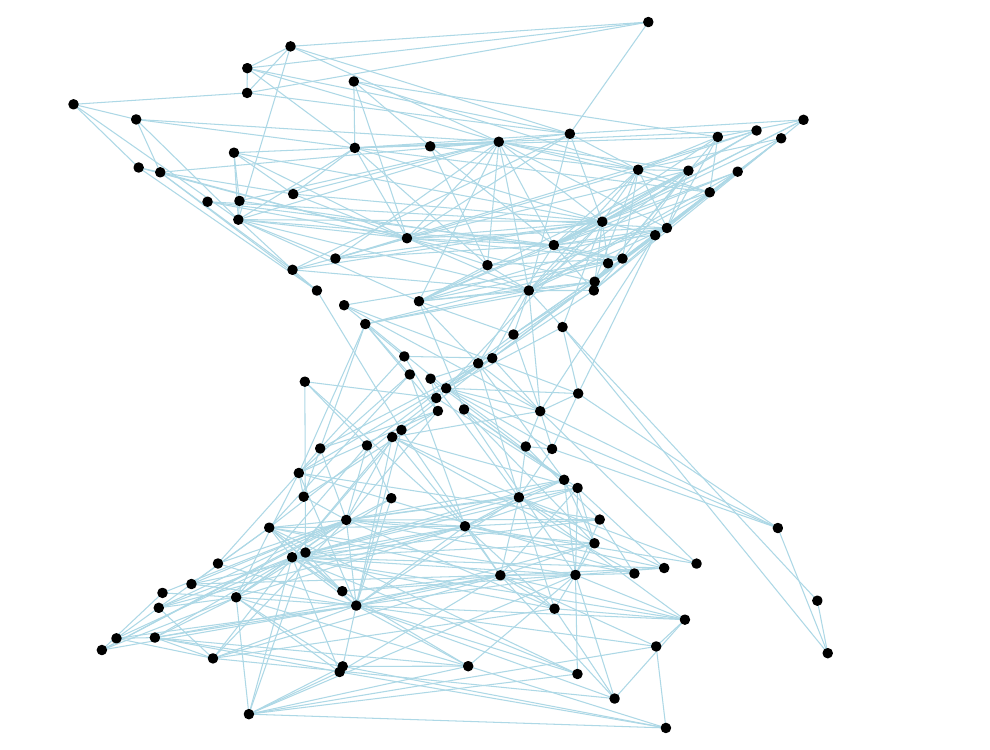} &
  \includegraphics[width=9.5mm]{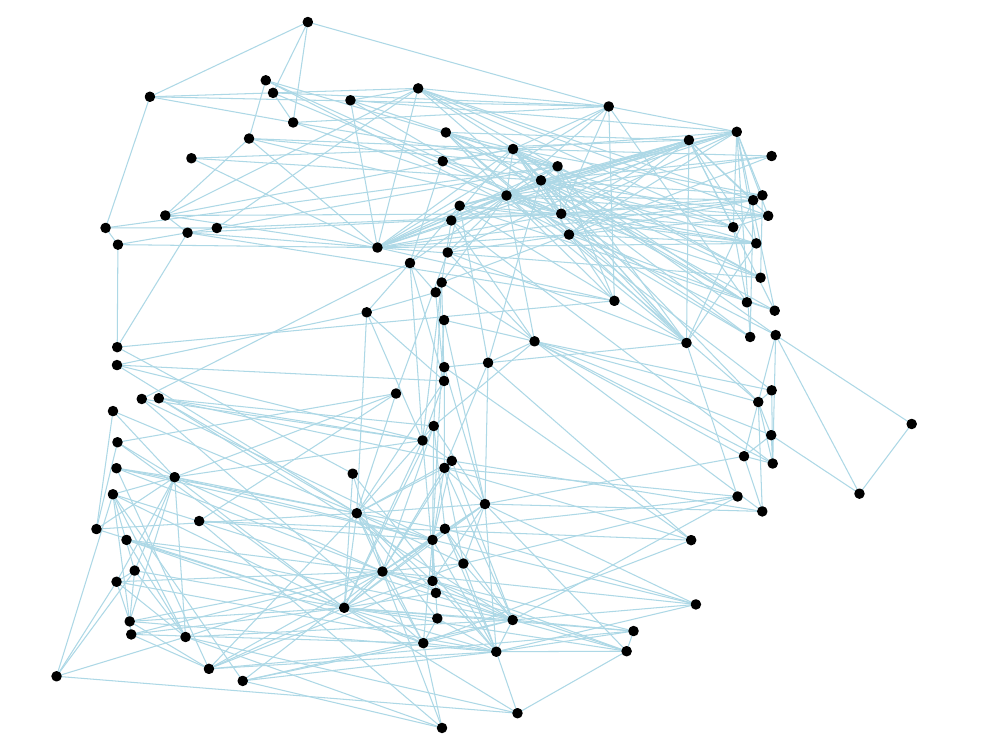} &
  \includegraphics[width=9.5mm]{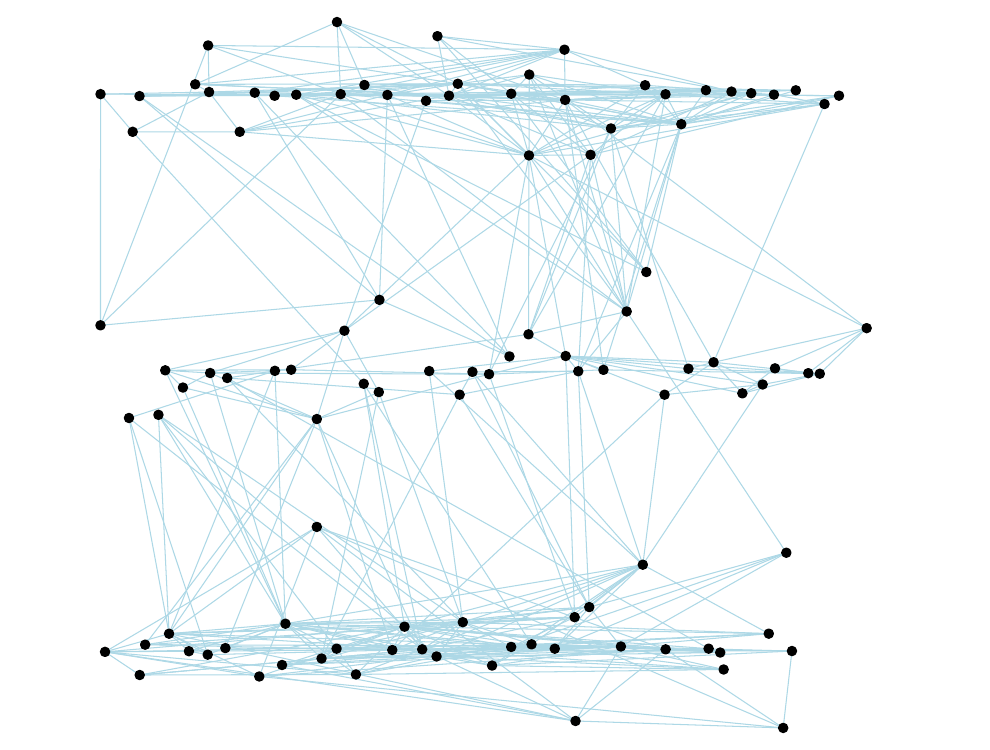} &
  \includegraphics[width=9.5mm]{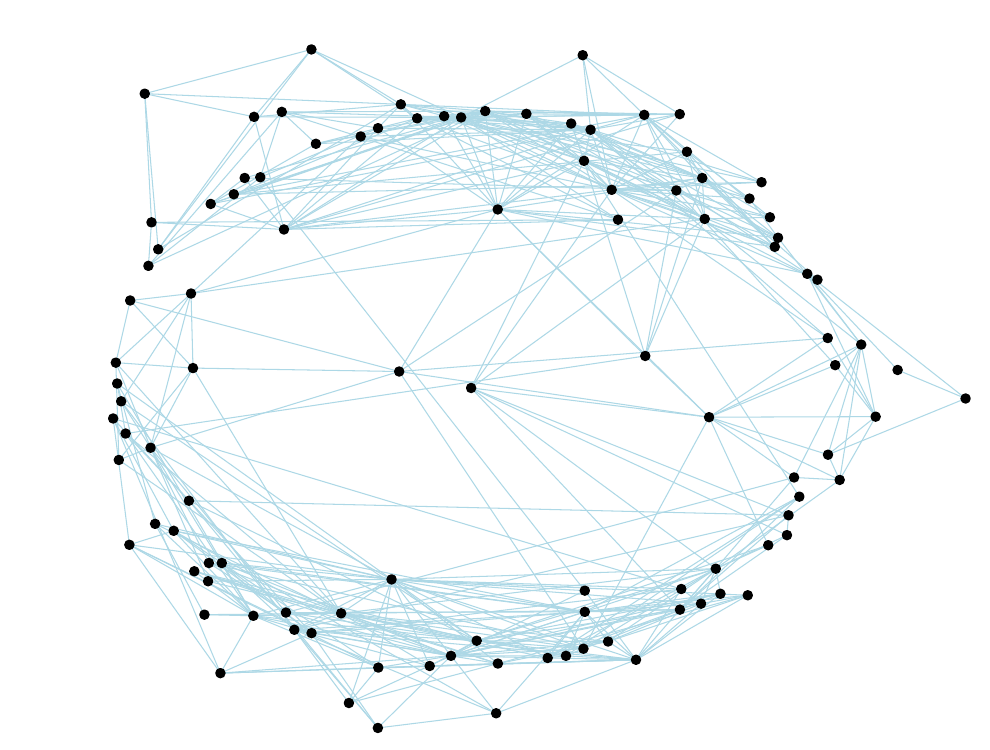} &
  \includegraphics[width=9.5mm]{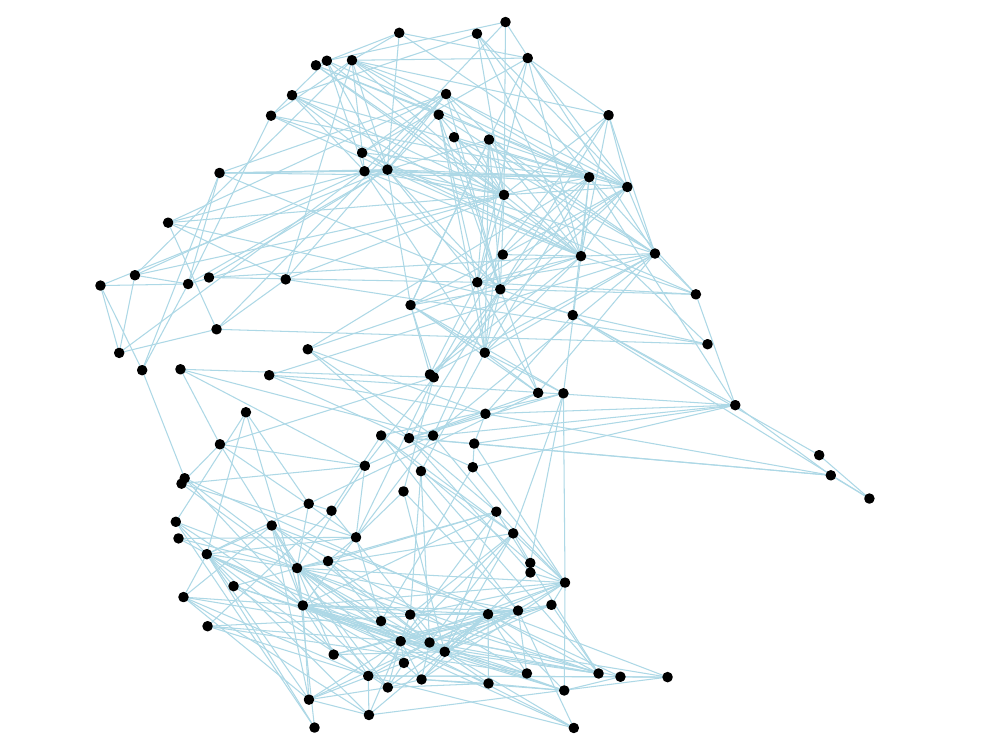} &
  \includegraphics[width=9.5mm]{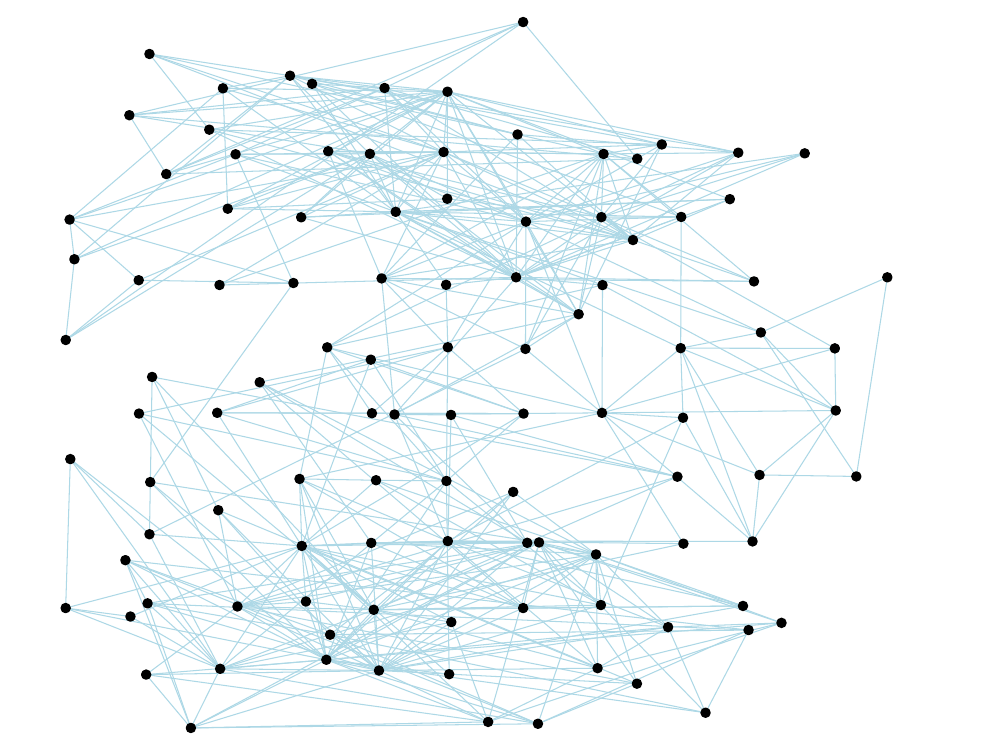}
  \\
     & \texttt{ELD-CN} & & \includegraphics[width=9.5mm]{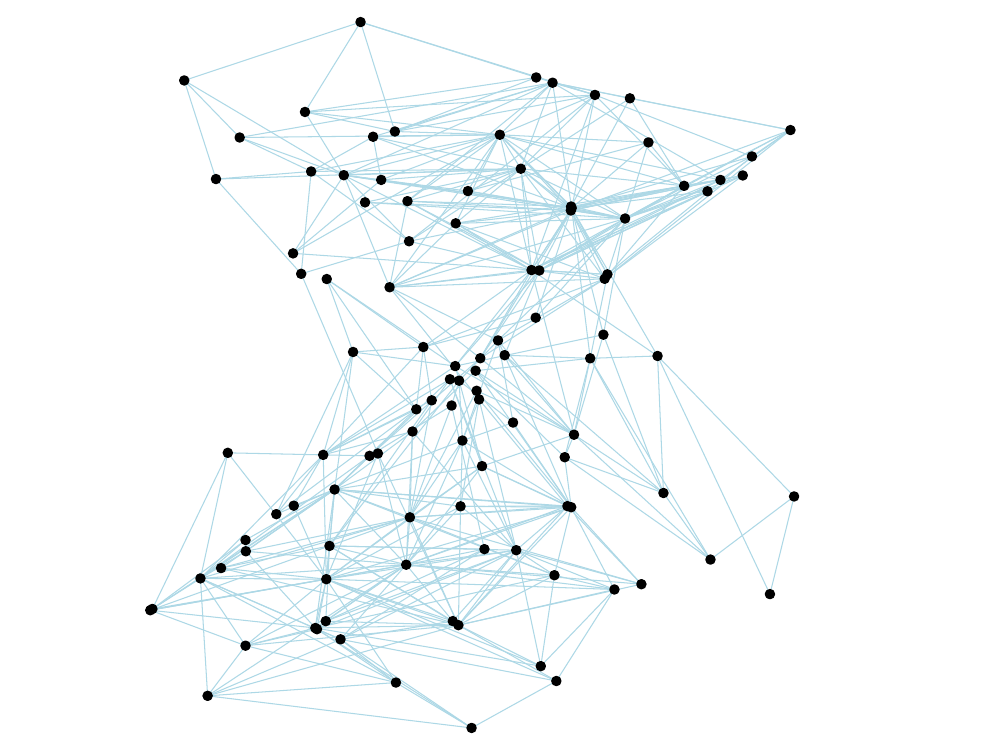} &
  \includegraphics[width=9.5mm]{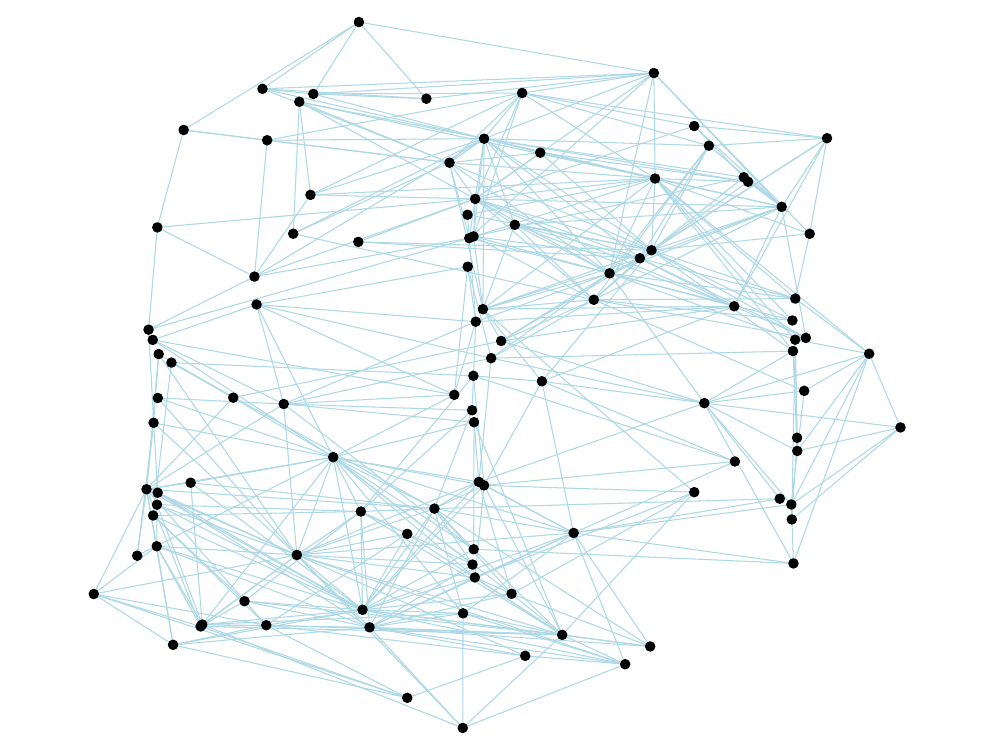} &
  \includegraphics[width=9.5mm]{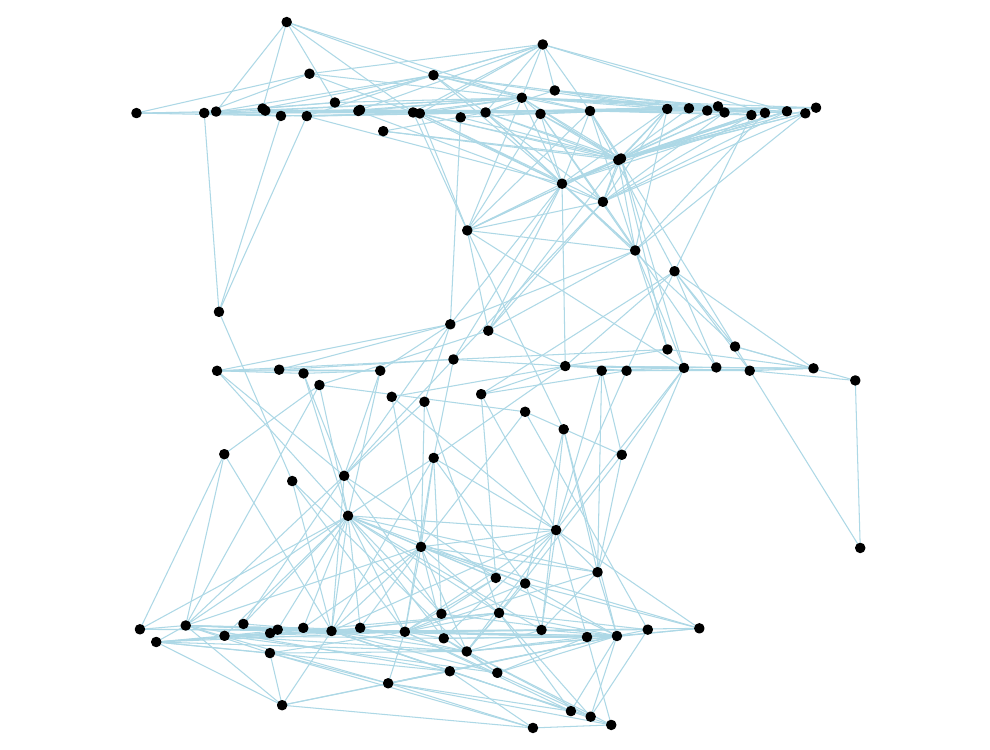} &
  \includegraphics[width=9.5mm]{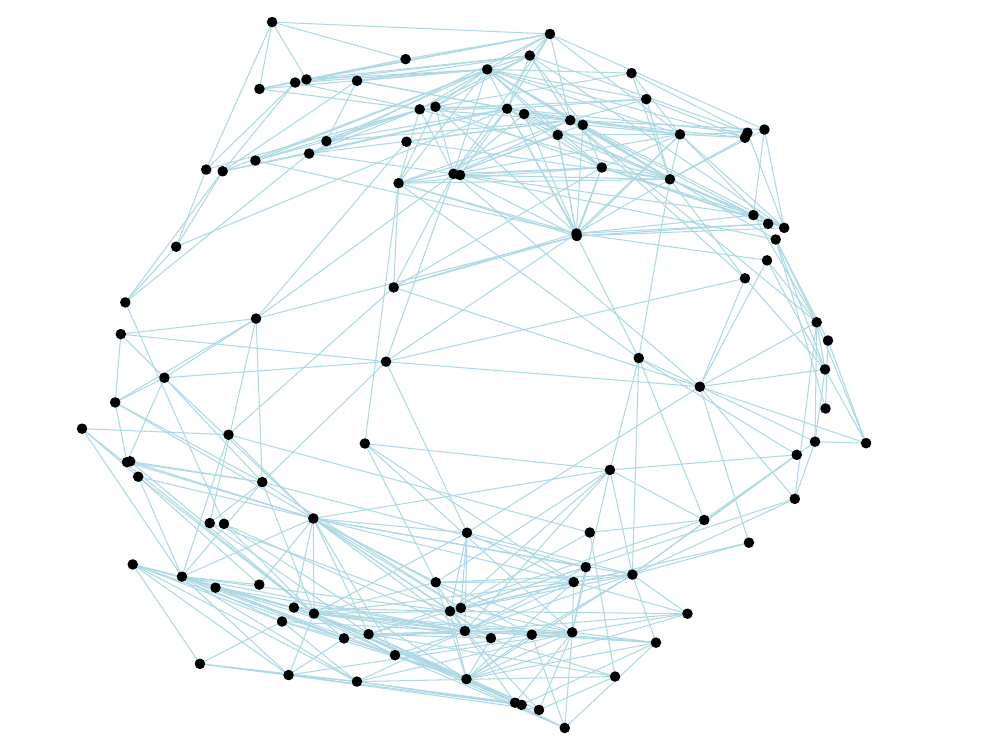} &
  \includegraphics[width=9.5mm]{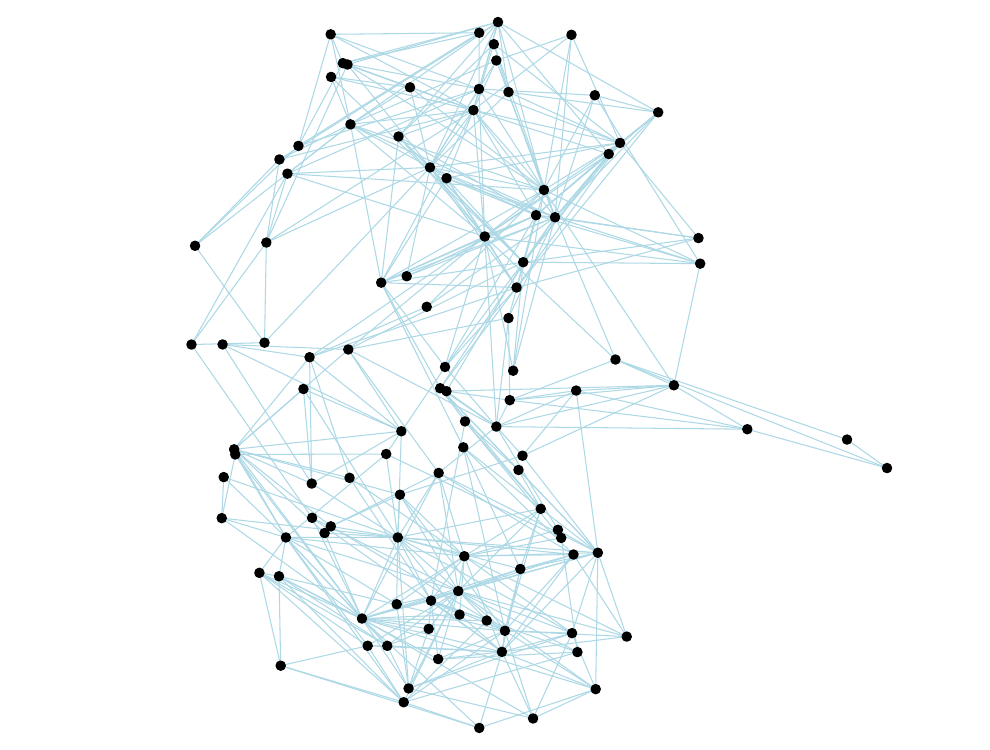} &
  \includegraphics[width=9.5mm]{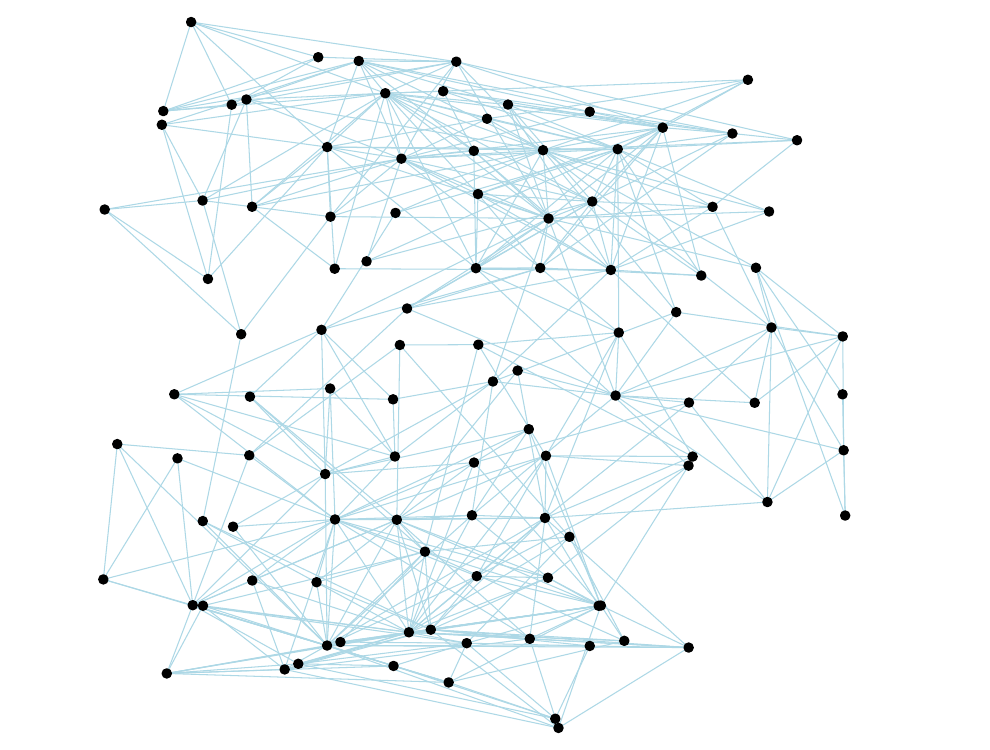}
  \\
       & \texttt{ELD-AR} & & \includegraphics[width=9.5mm]{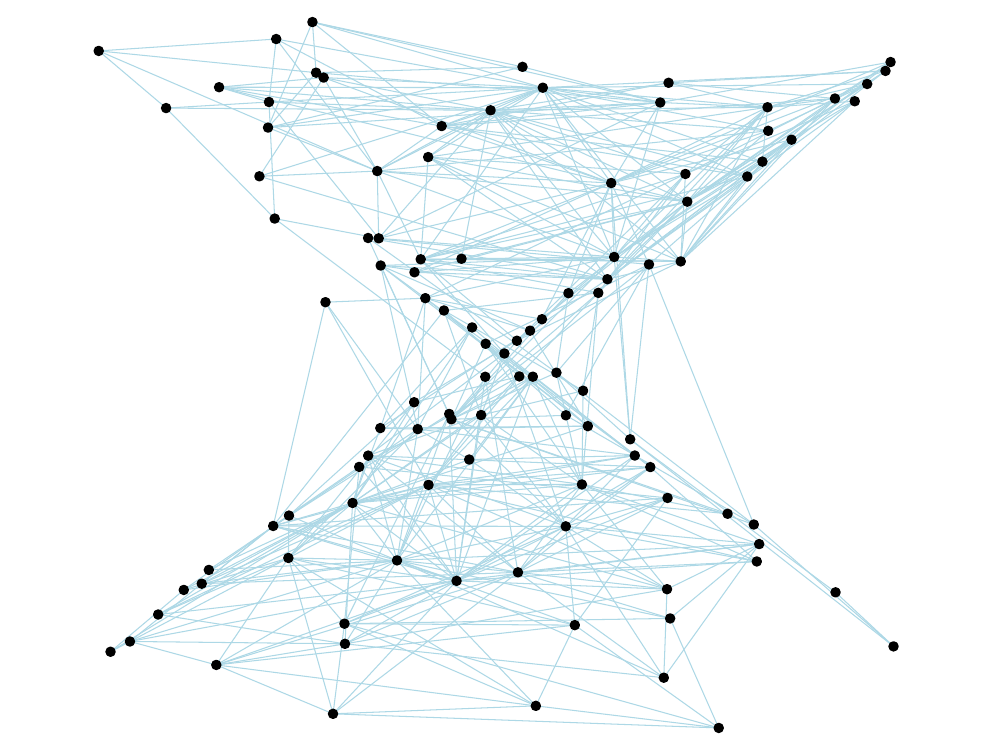} &
  \includegraphics[width=9.5mm]{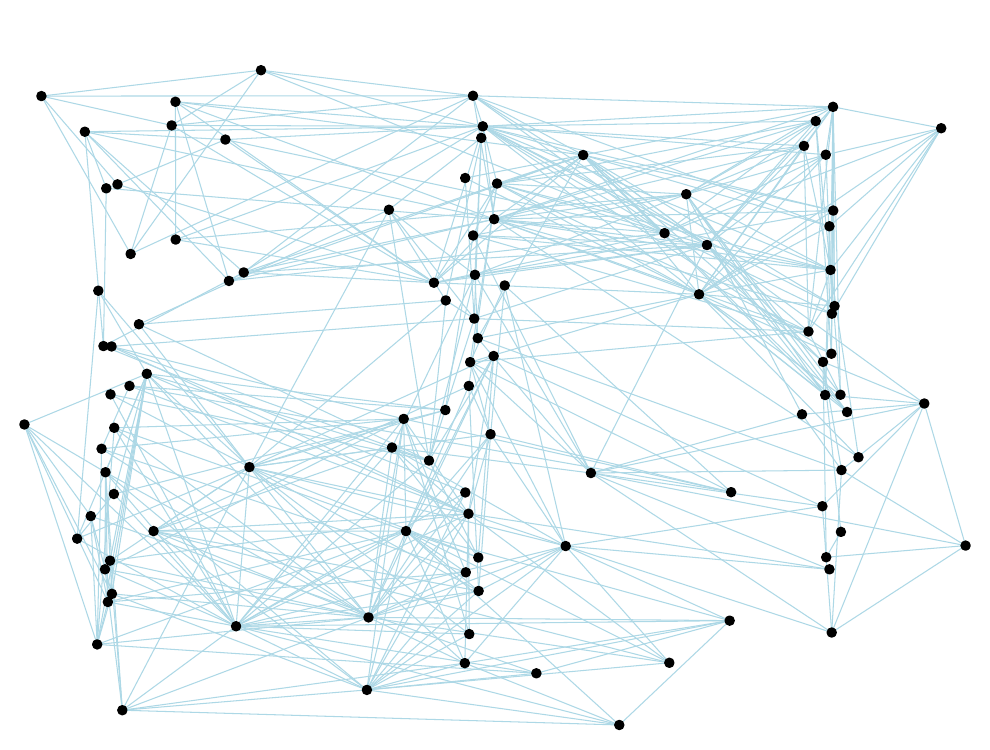} &
  \includegraphics[width=9.5mm]{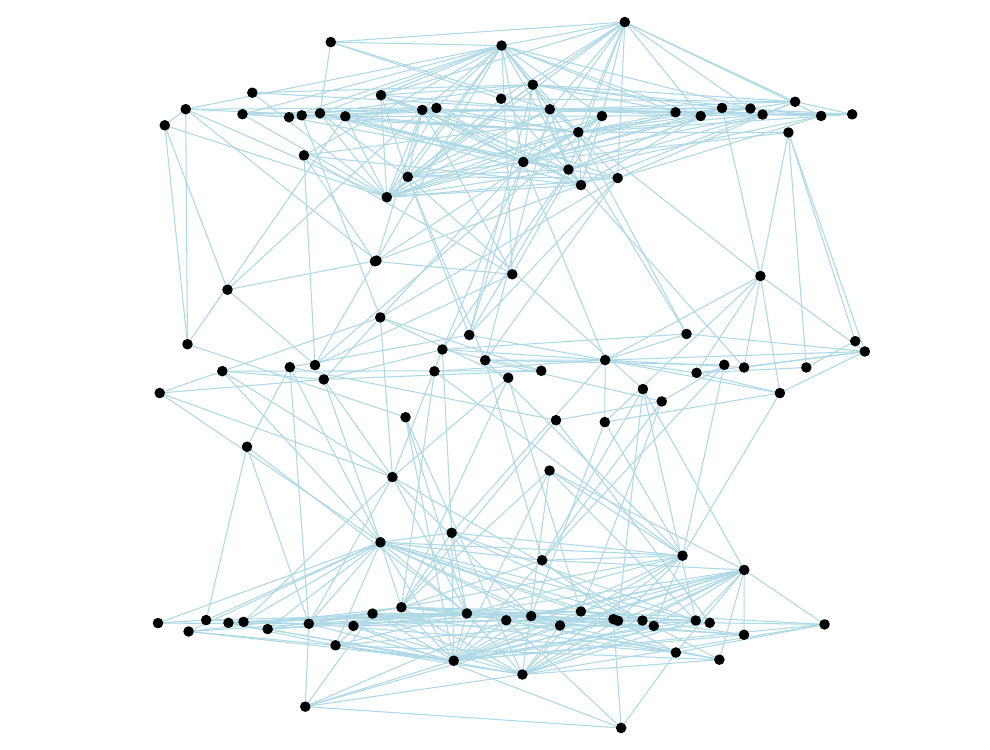} &
  \includegraphics[width=9.5mm]{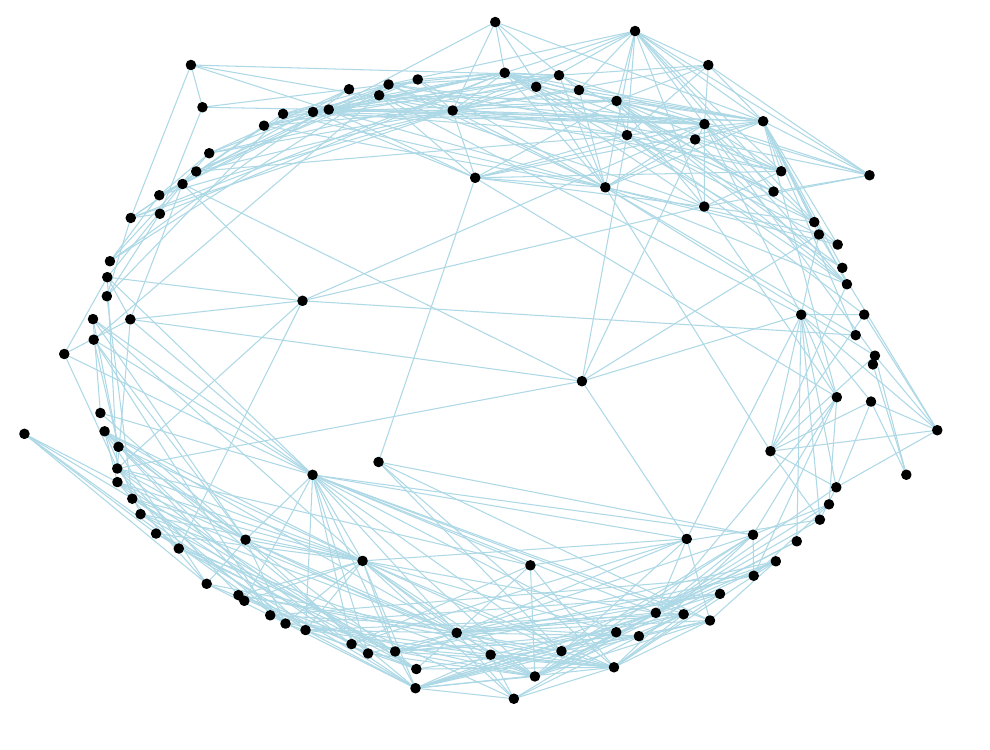} &
  \includegraphics[width=9.5mm]{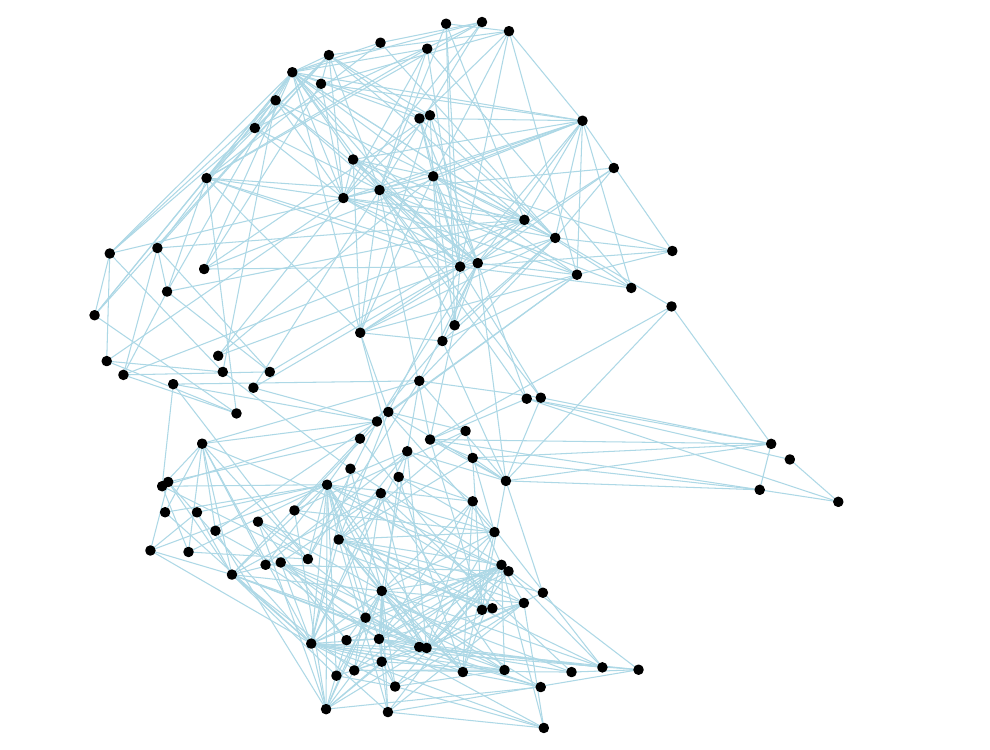} &
  \includegraphics[width=9.5mm]{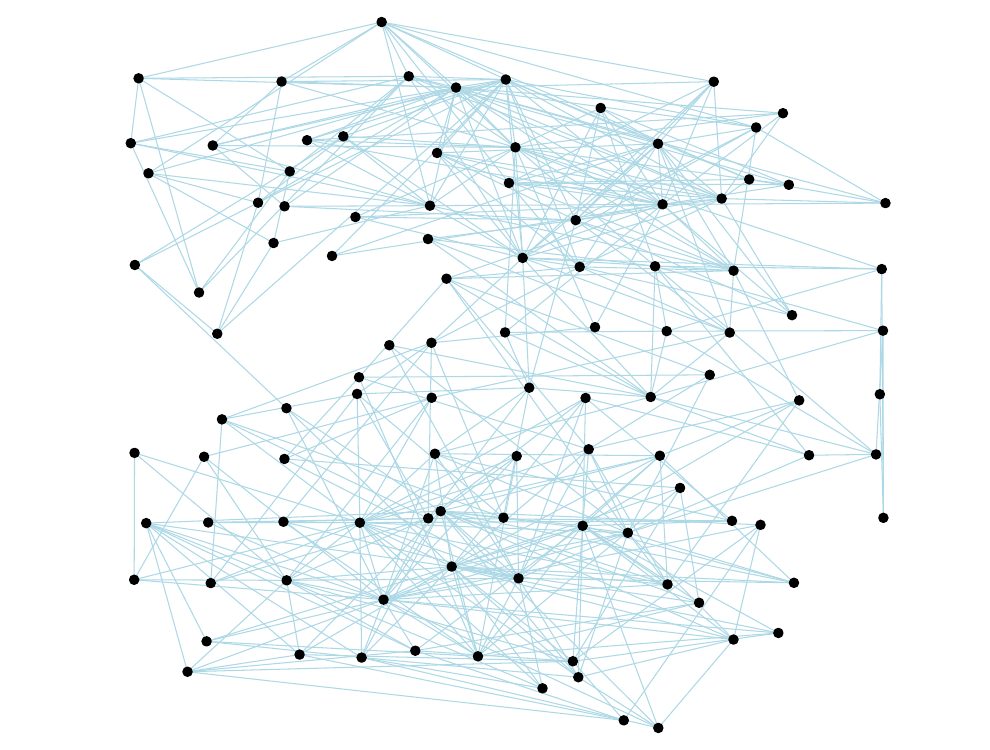}
  \\
       & \texttt{CN-AR} & & \includegraphics[width=9.5mm]{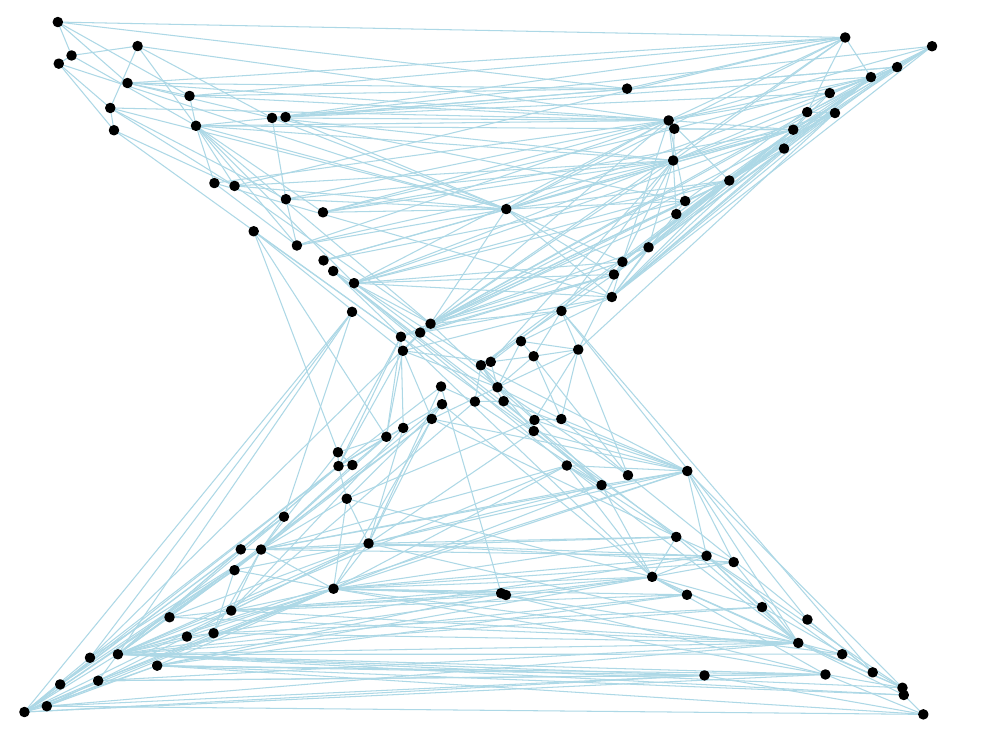} &
  \includegraphics[width=9.5mm]{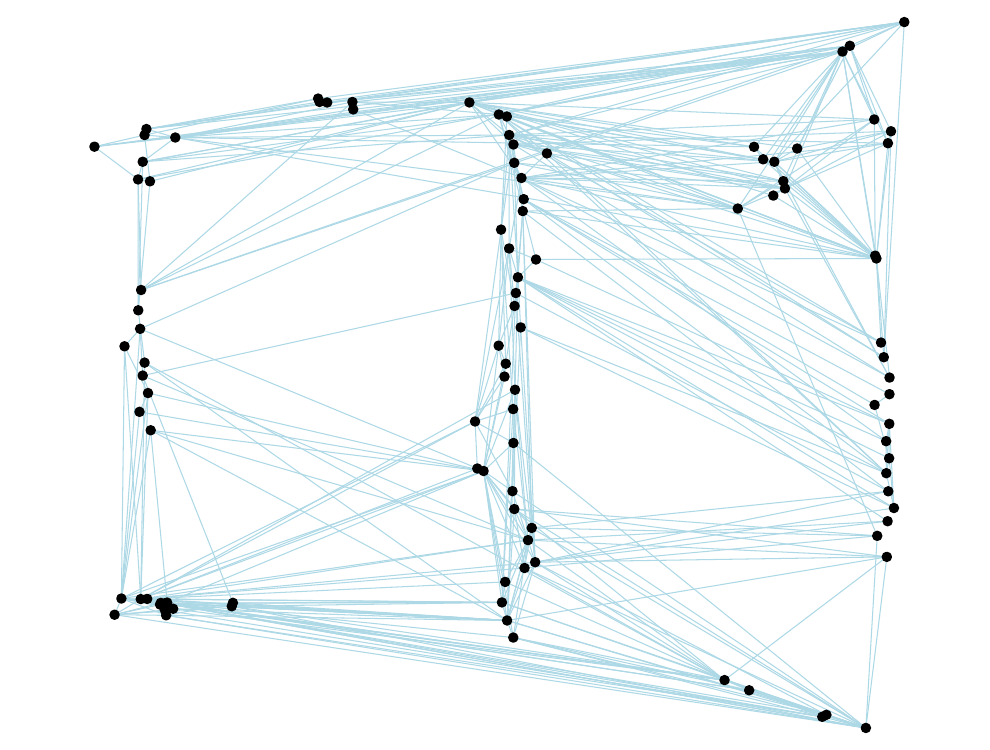} &
  \includegraphics[width=9.5mm]{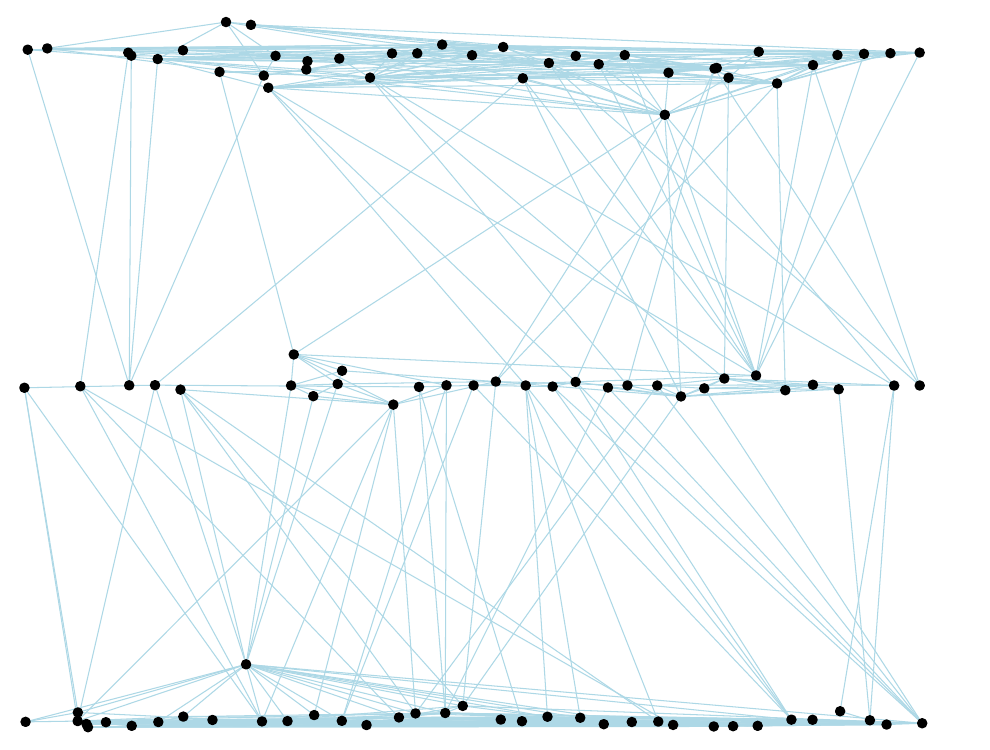} &
  \includegraphics[width=9.5mm]{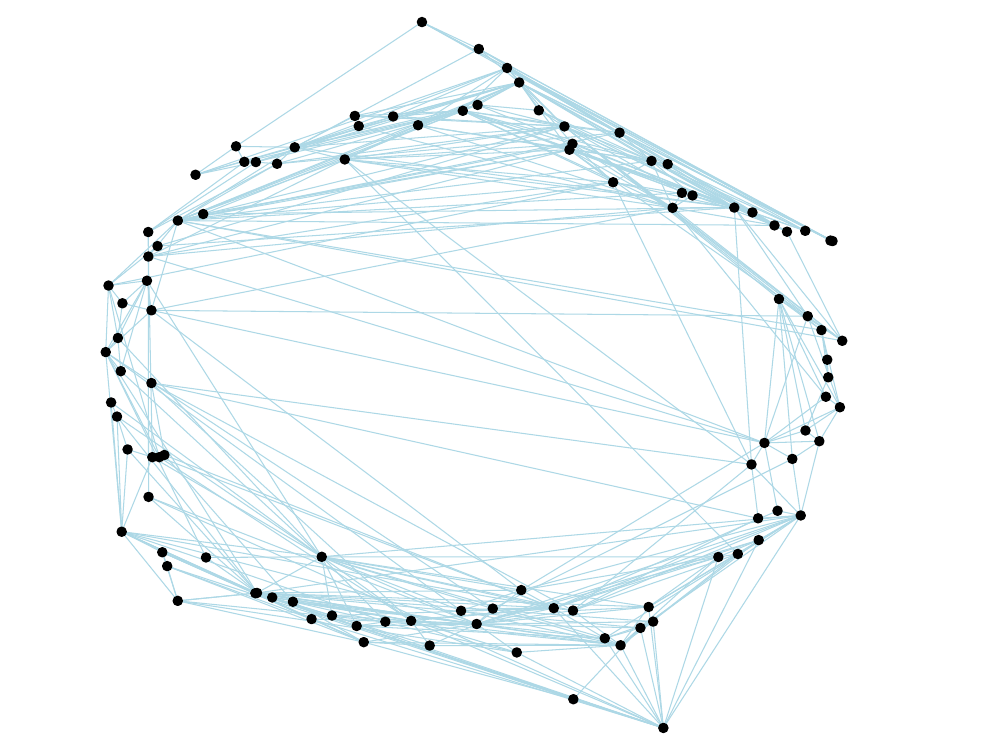} &
  \includegraphics[width=9.5mm]{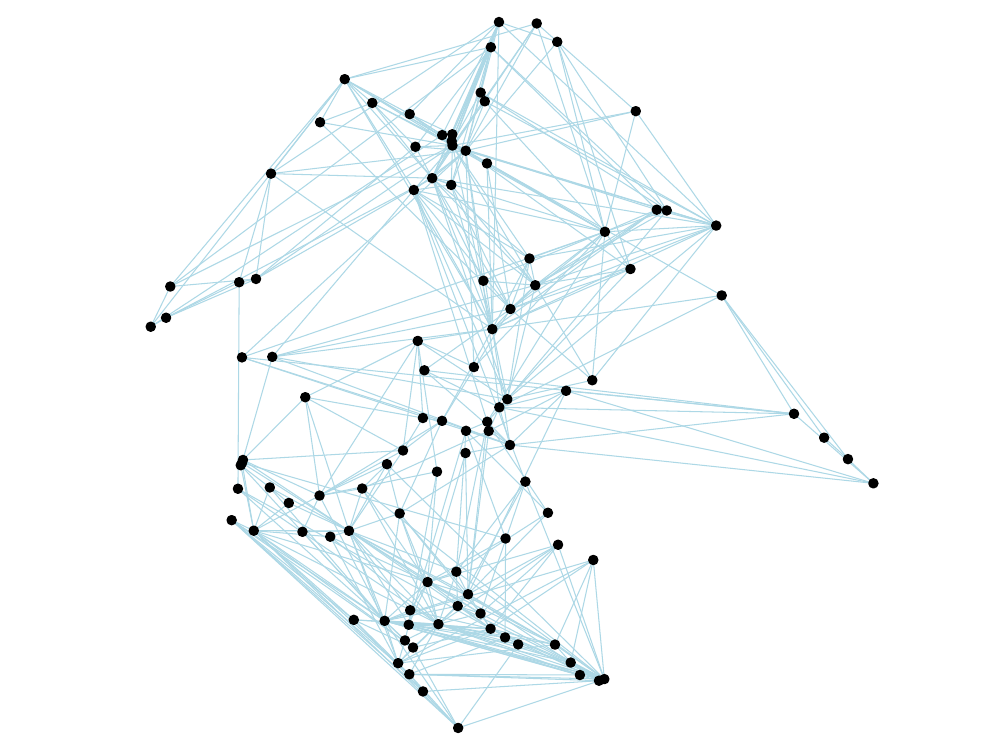} &
  \includegraphics[width=9.5mm]{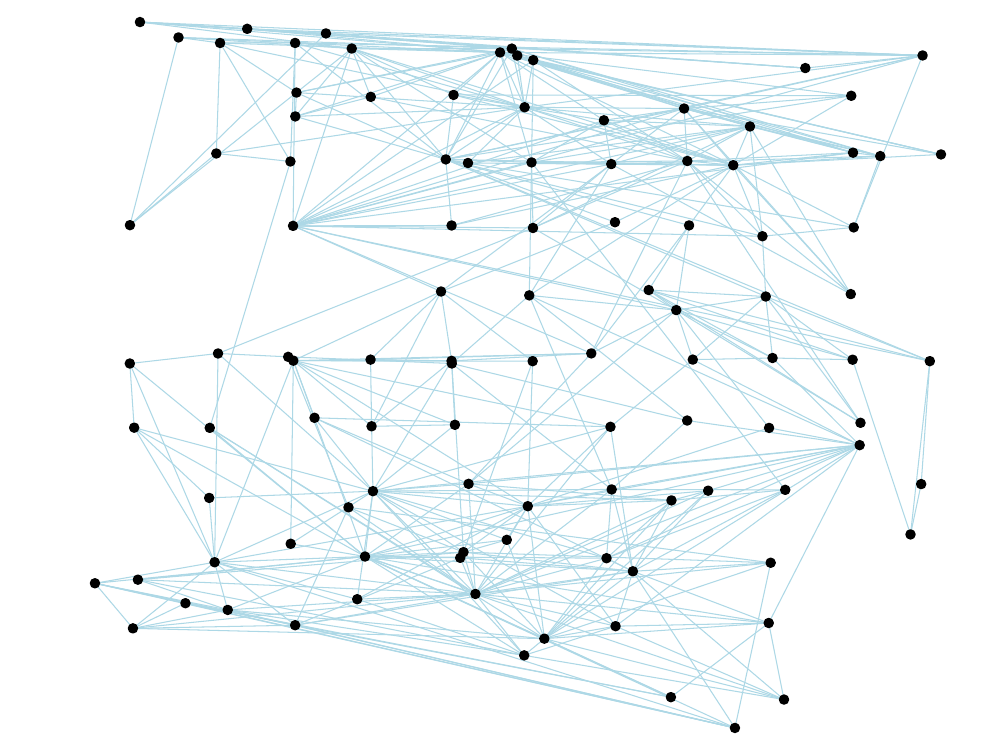}
  \\
       & \texttt{ST-ELD-CN} & &\includegraphics[width=9.5mm]{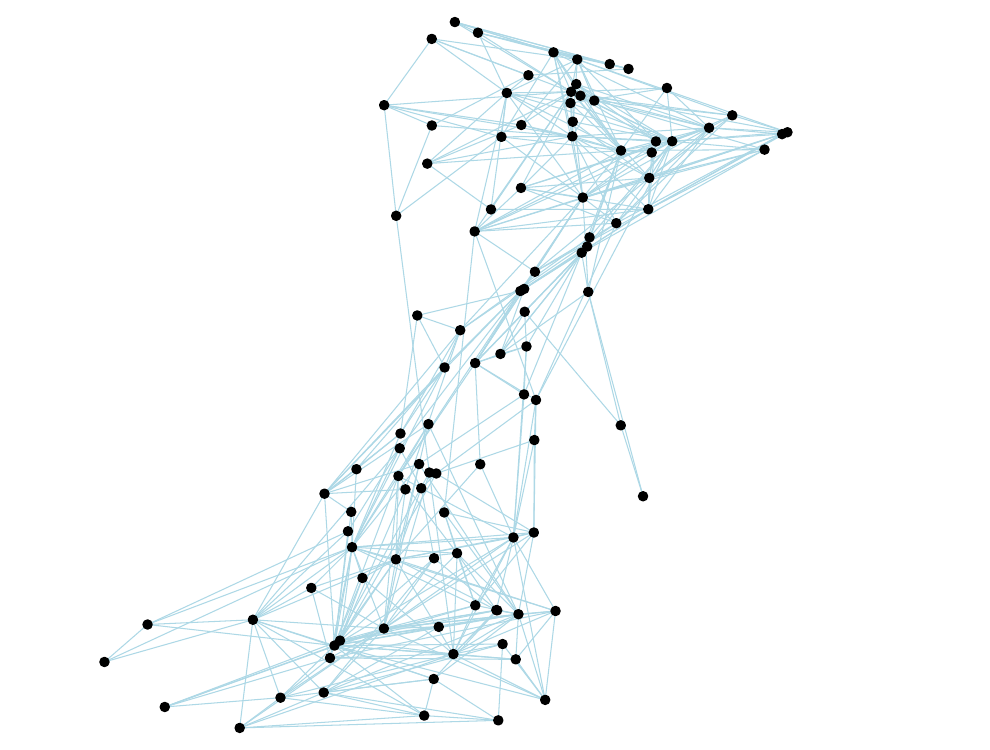} &
  \includegraphics[width=9.5mm]{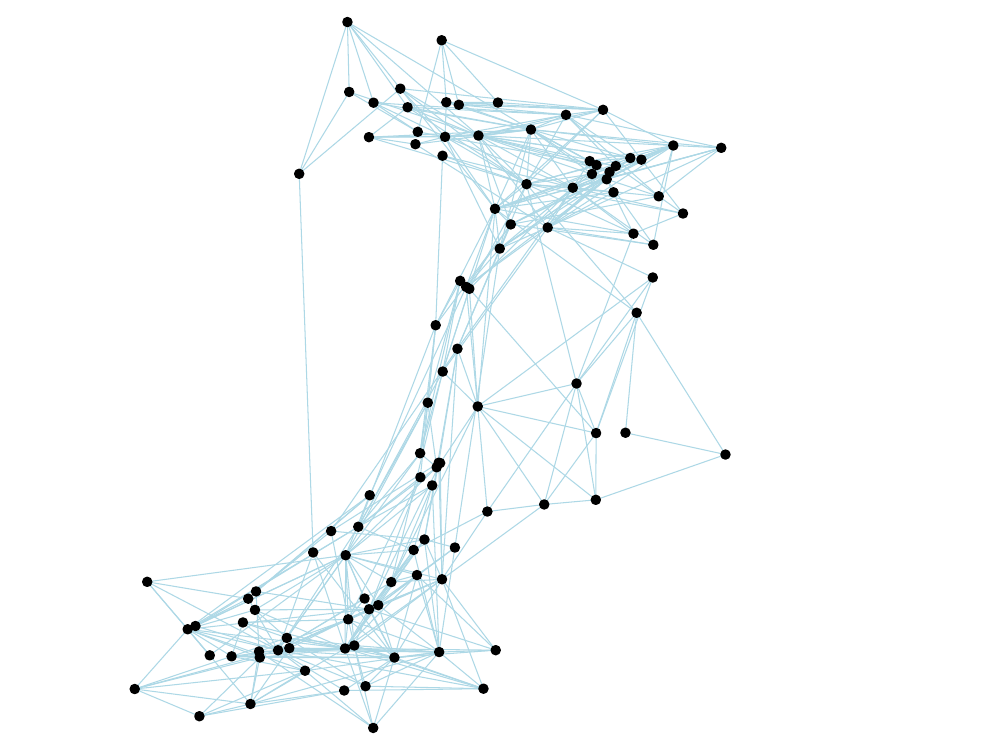} &
  \includegraphics[width=9.5mm]{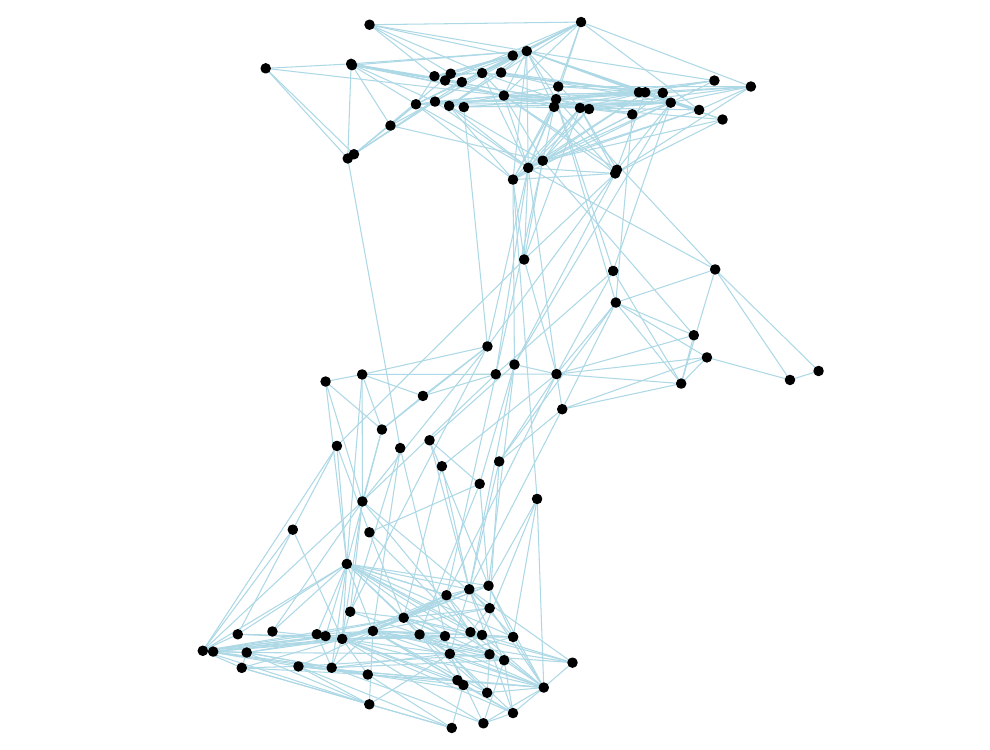} &
  \includegraphics[width=9.5mm]{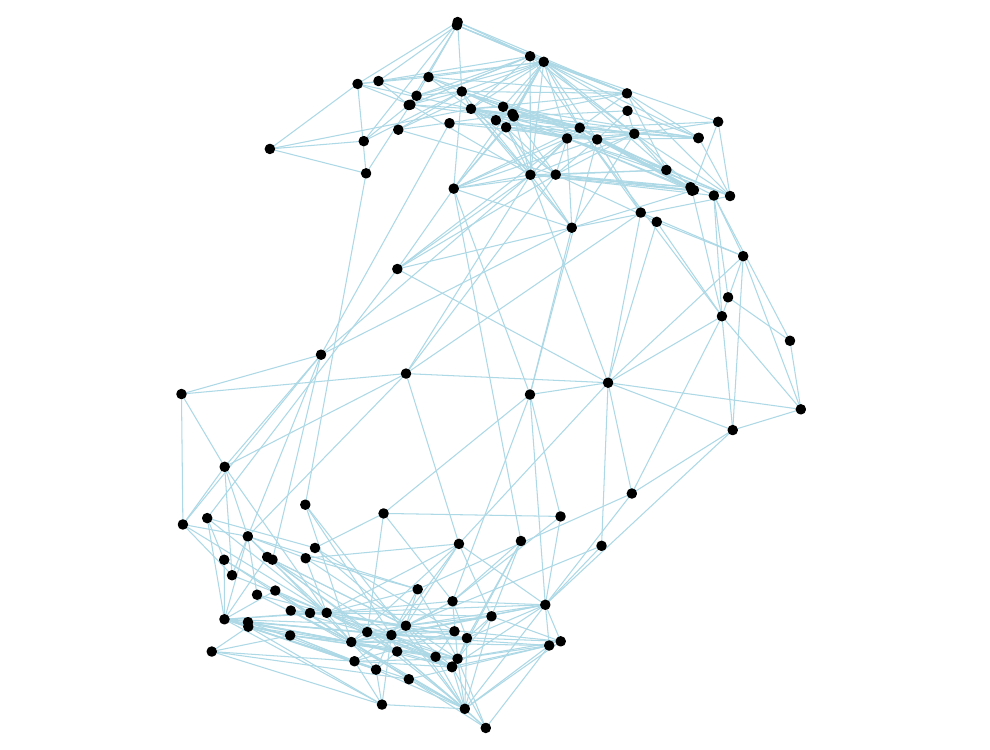} &
  \includegraphics[width=9.5mm]{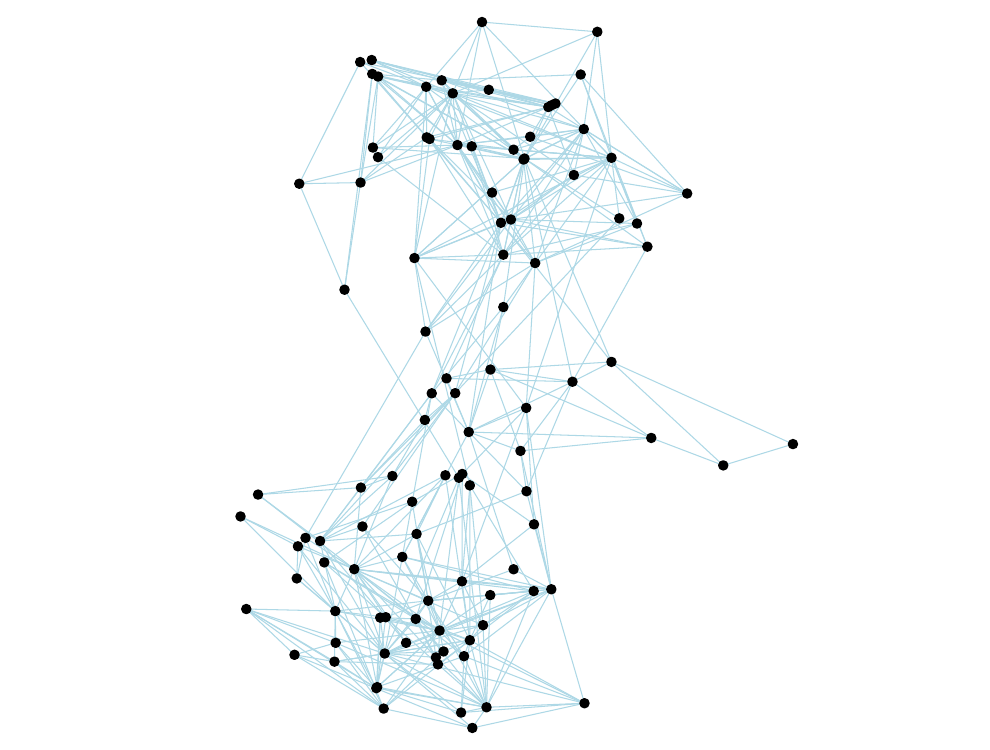} &
  \includegraphics[width=9.5mm]{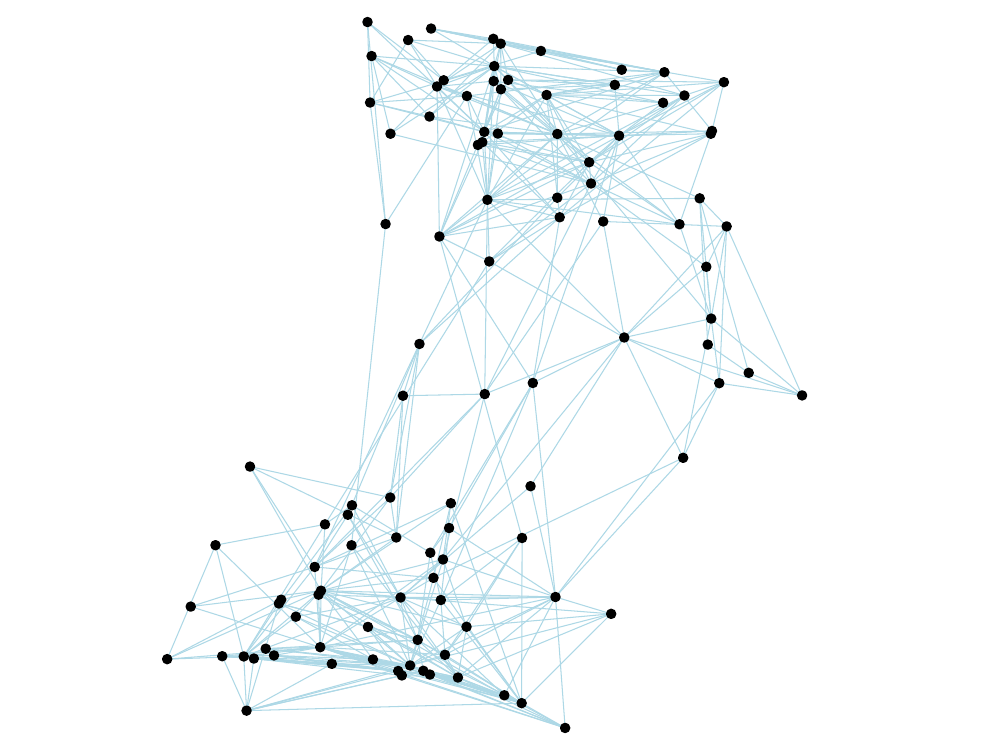}
  \\
         & \texttt{ST-ELD-AR} & &\includegraphics[width=9.5mm]{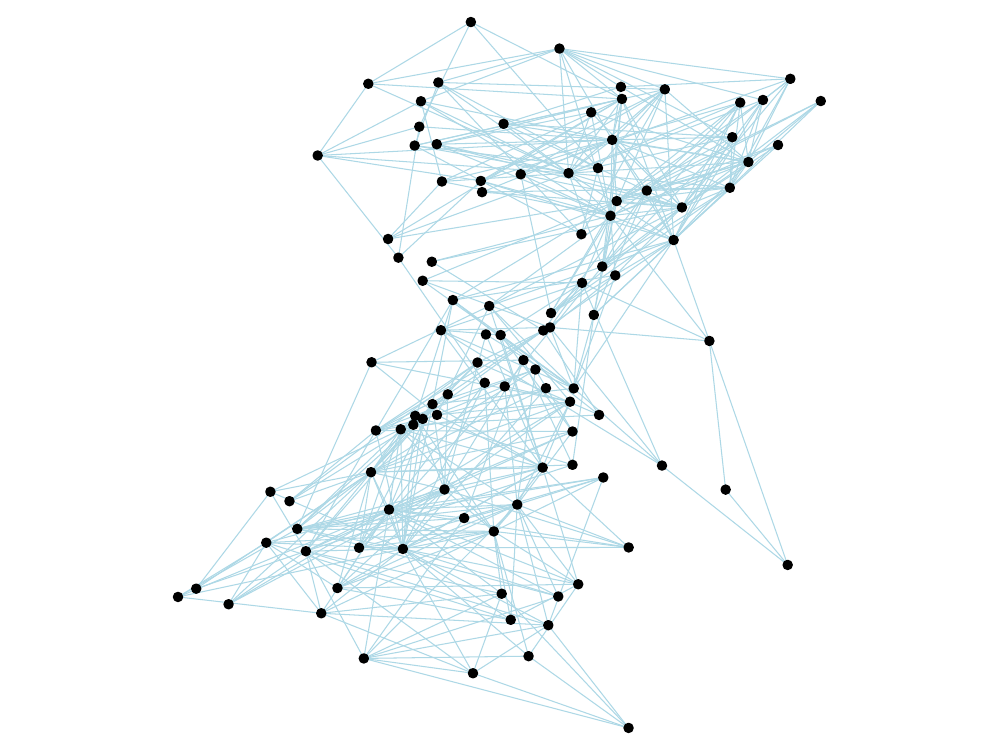} &
  \includegraphics[width=9.5mm]{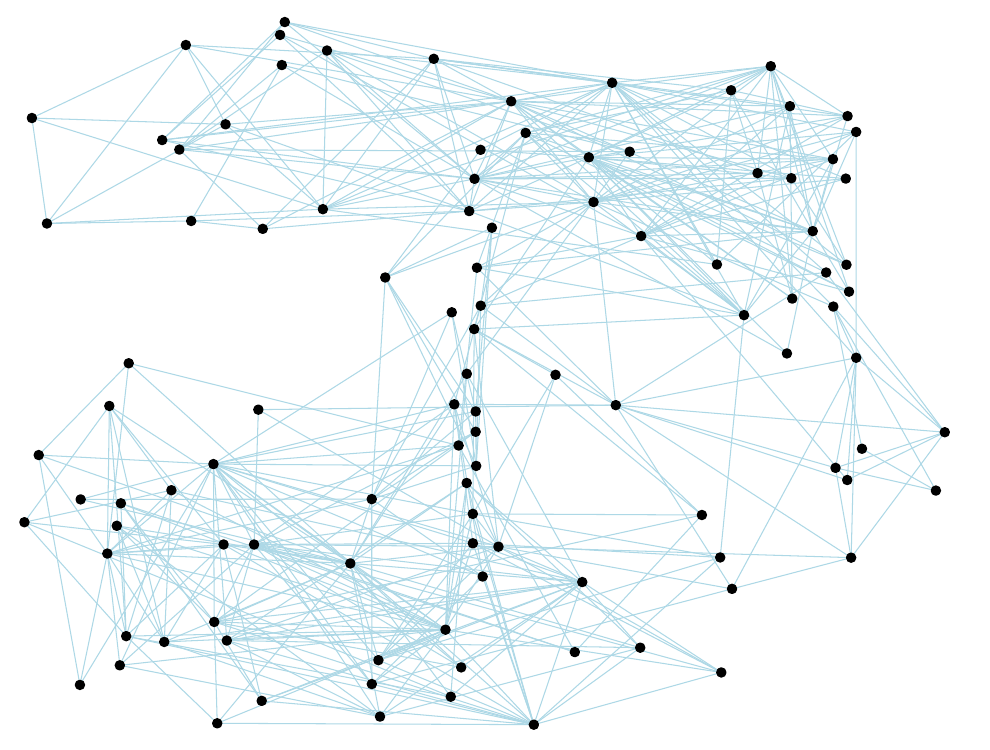} &
  \includegraphics[width=9.5mm]{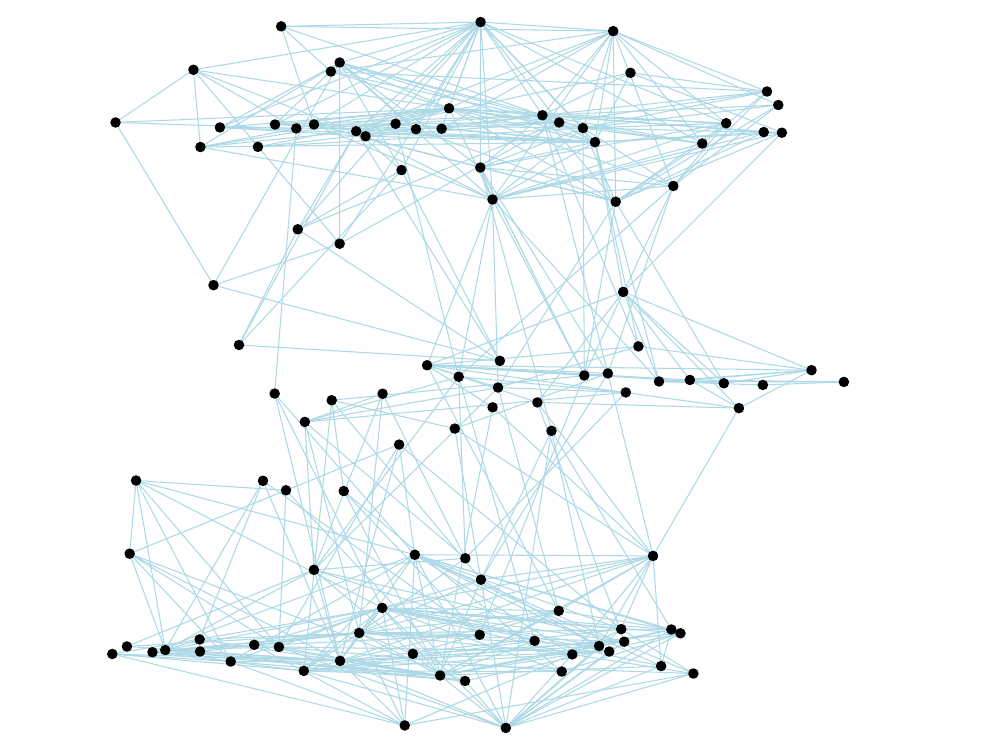} &
  \includegraphics[width=9.5mm]{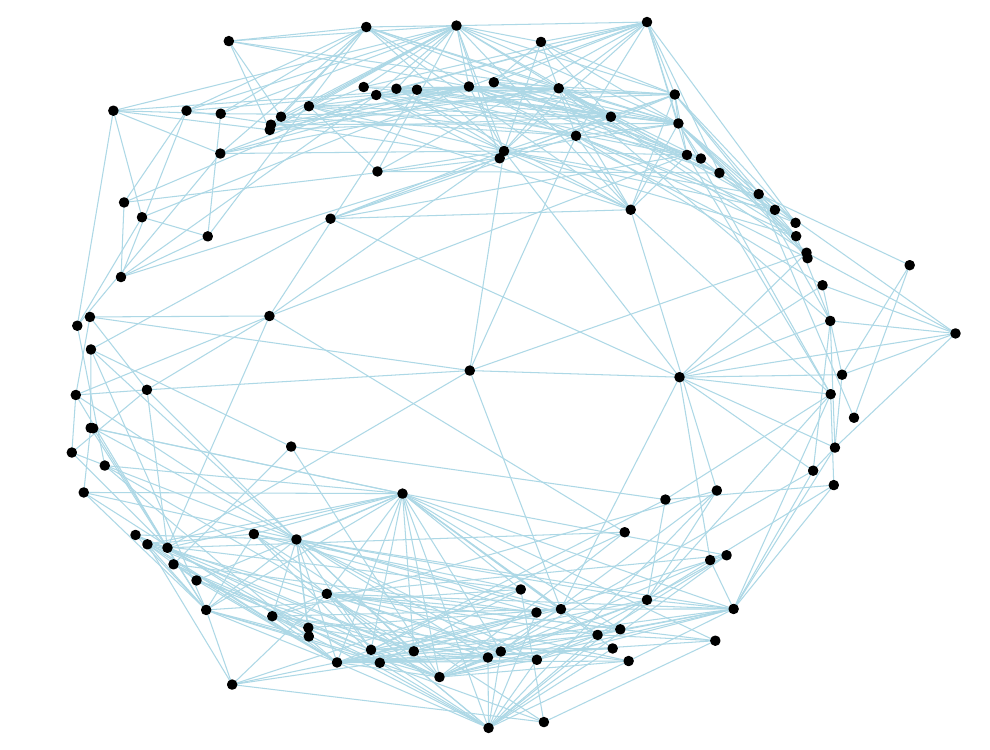} &
  \includegraphics[width=9.5mm]{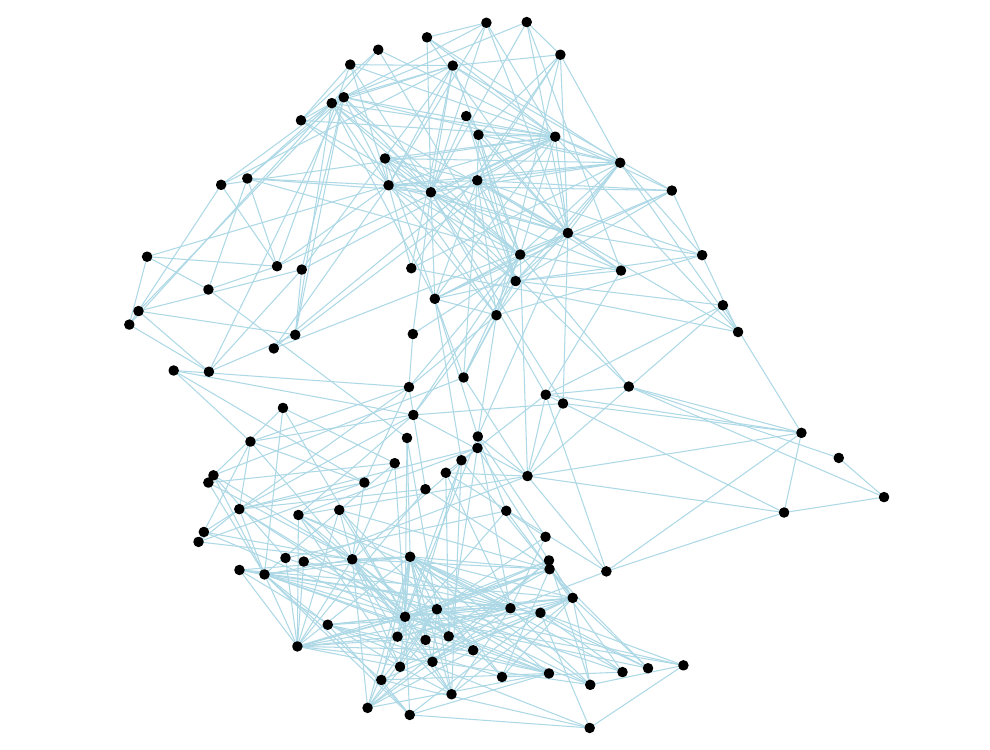} &
  \includegraphics[width=9.5mm]{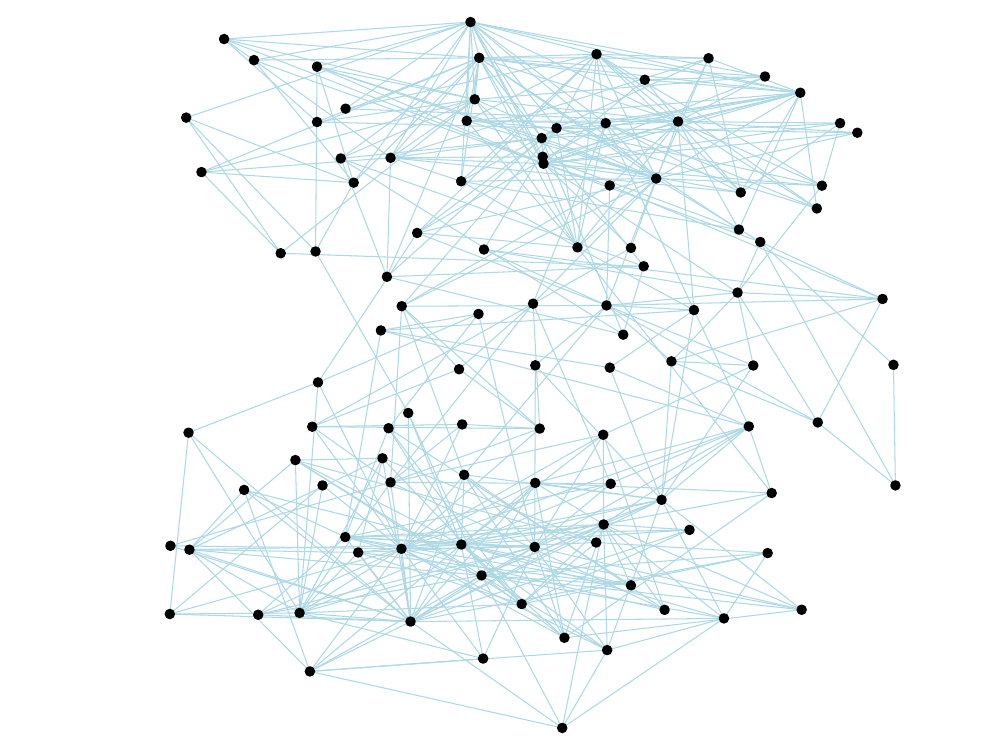}
  \\
         & \texttt{ST-CN-AR} & &\includegraphics[width=9.5mm]{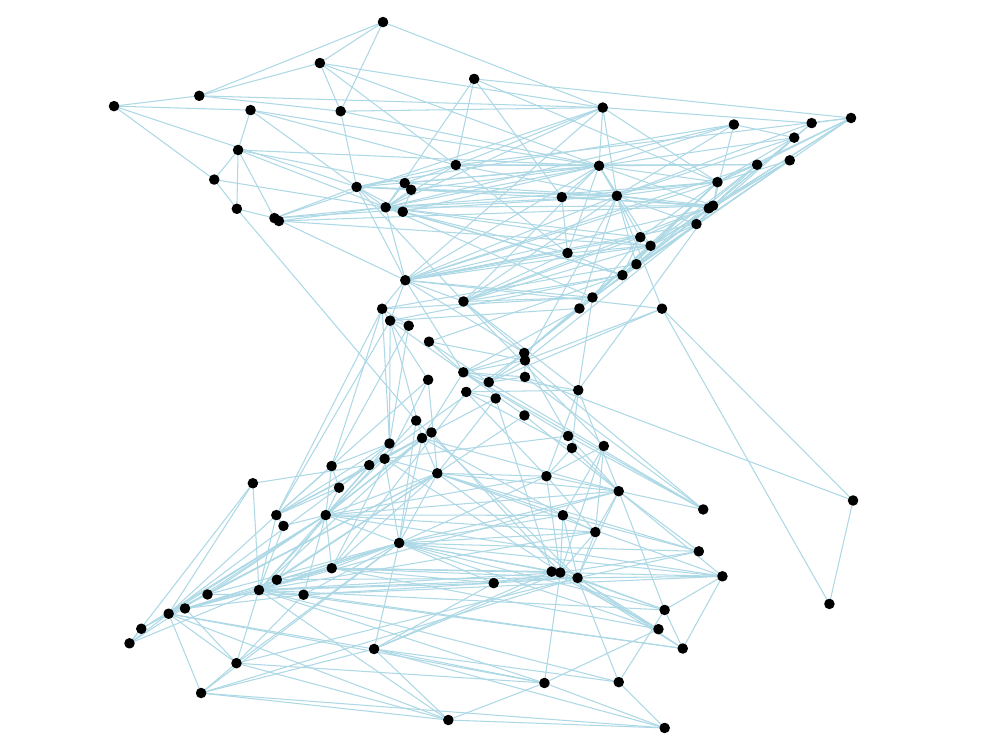} &
  \includegraphics[width=9.5mm]{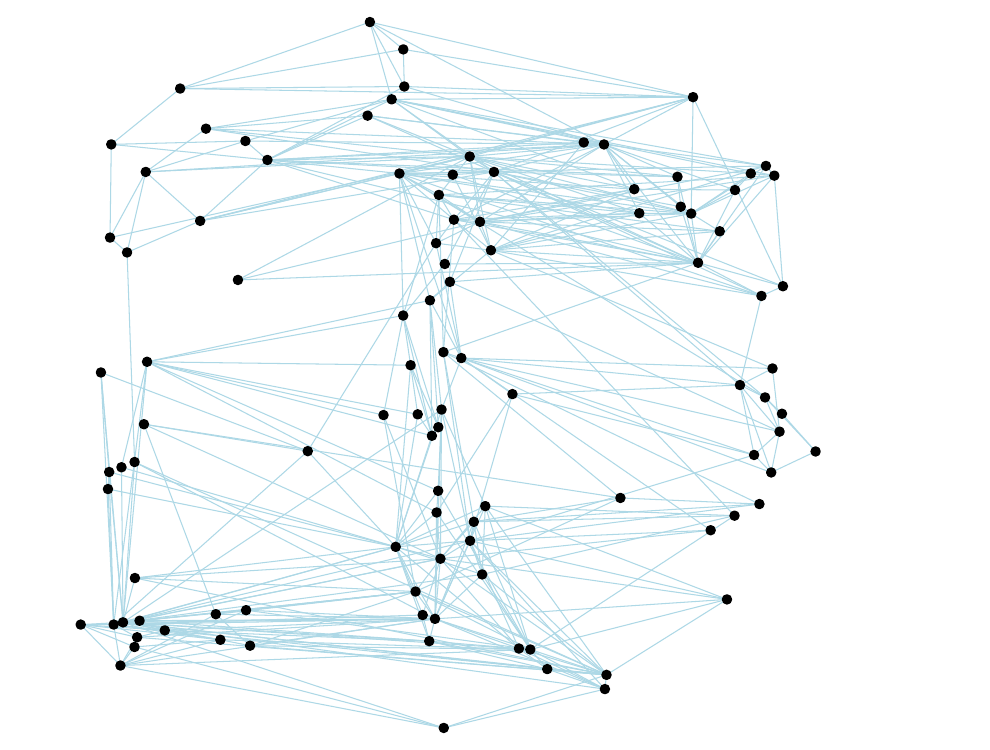} &
  \includegraphics[width=9.5mm]{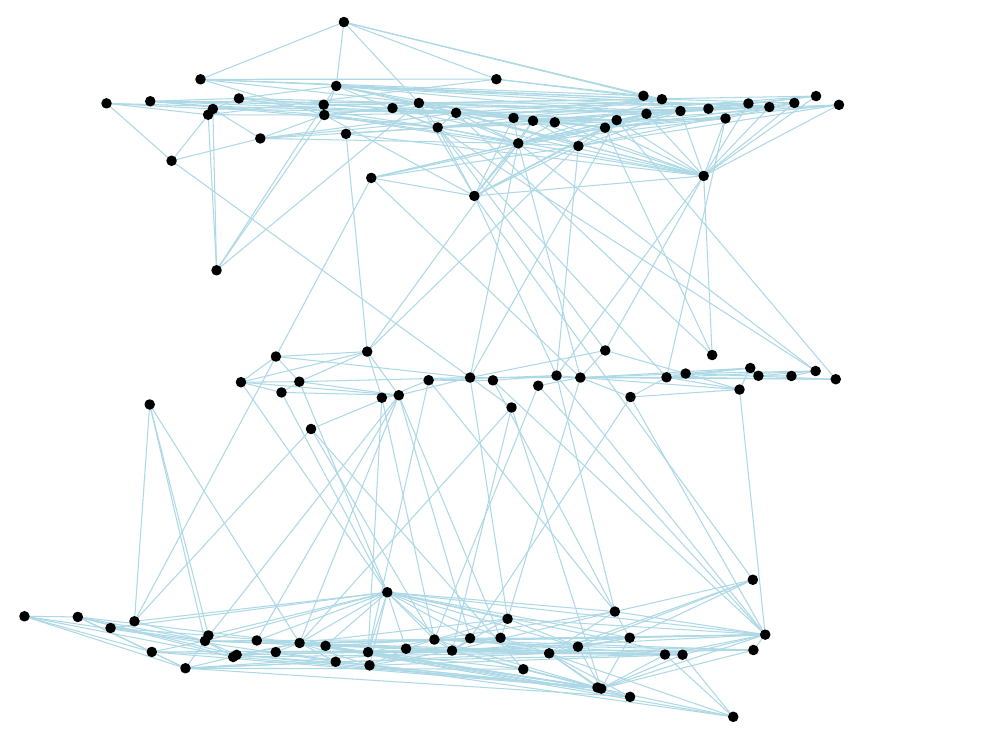} &
  \includegraphics[width=9.5mm]{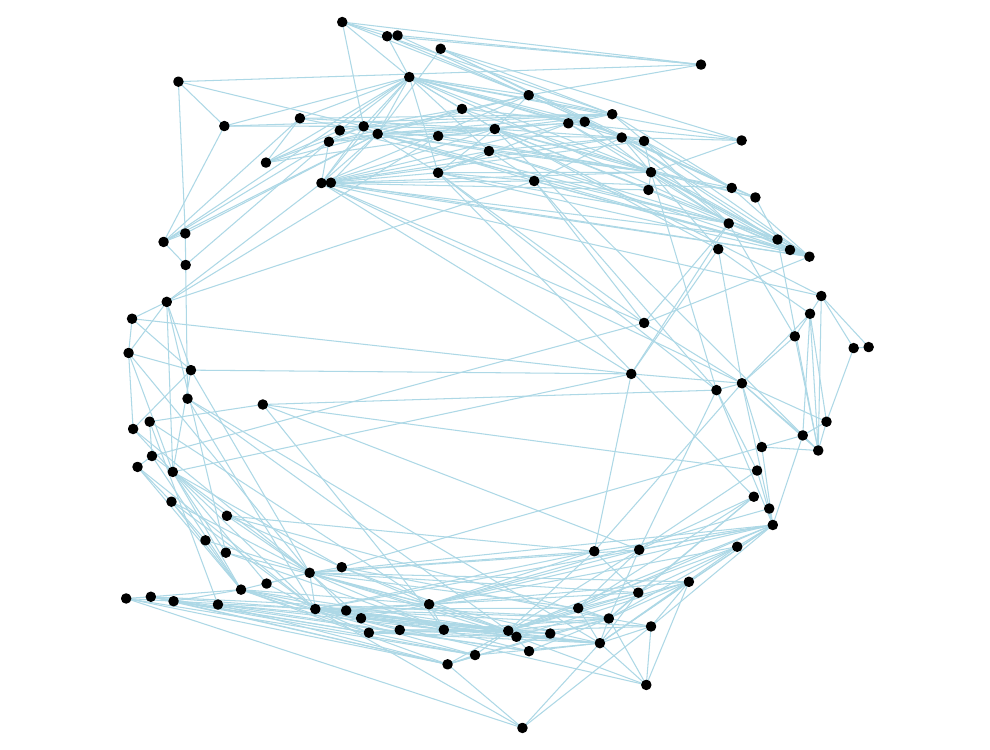} &
  \includegraphics[width=9.5mm]{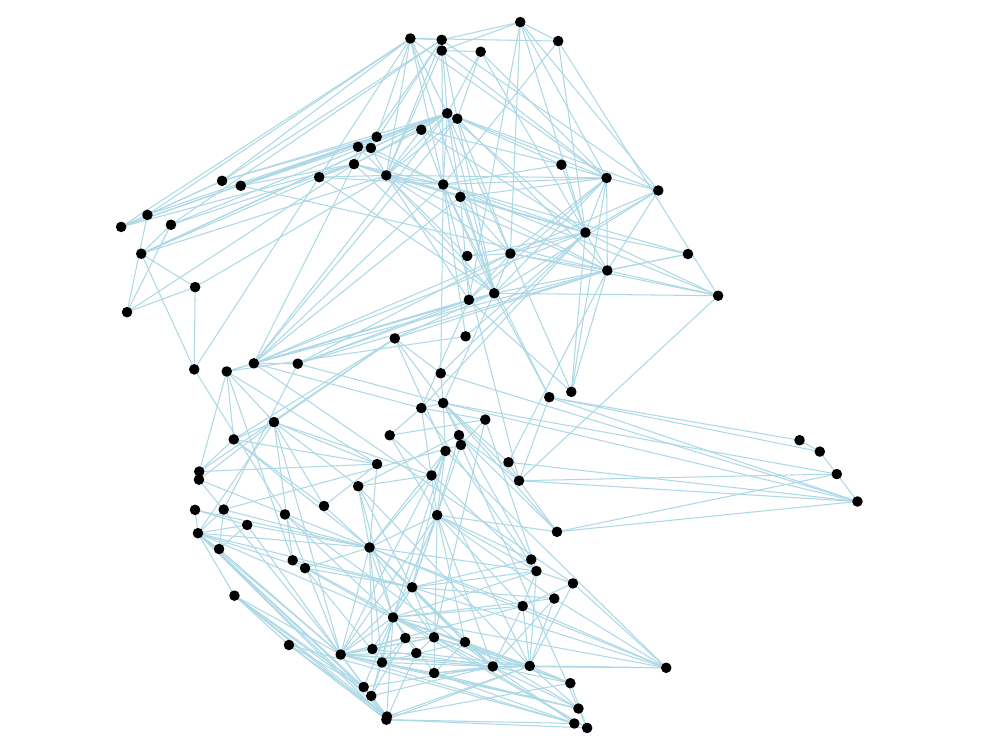} &
  \includegraphics[width=9.5mm]{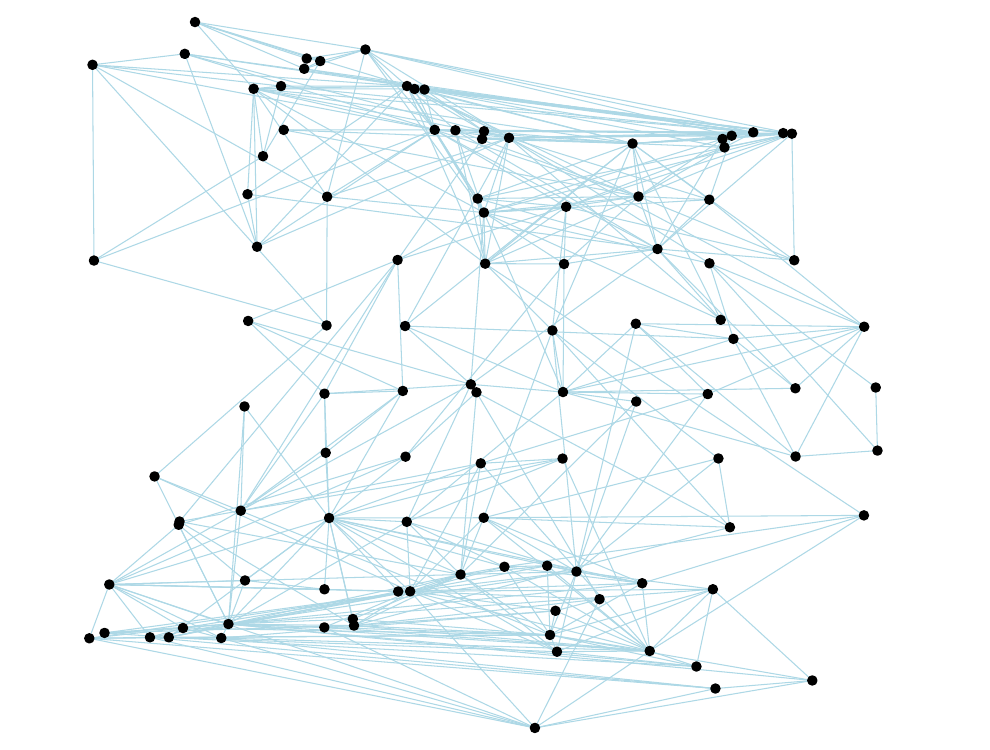}
  \\
         & \texttt{ELD-CN-AR} & & \includegraphics[width=9.5mm]{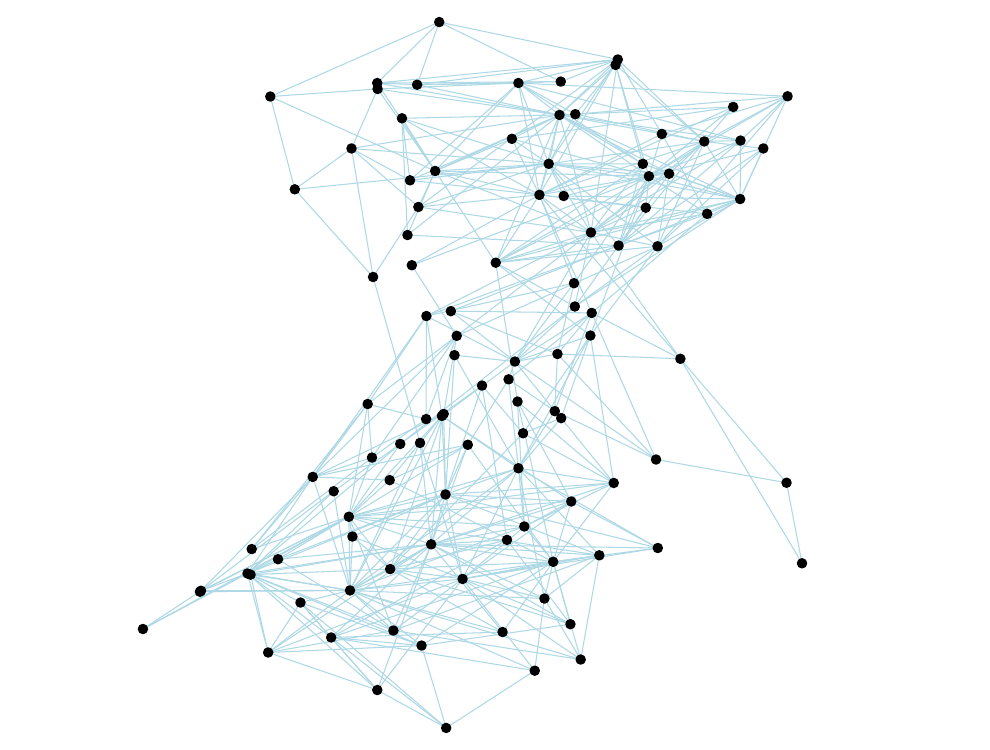} &
  \includegraphics[width=9.5mm]{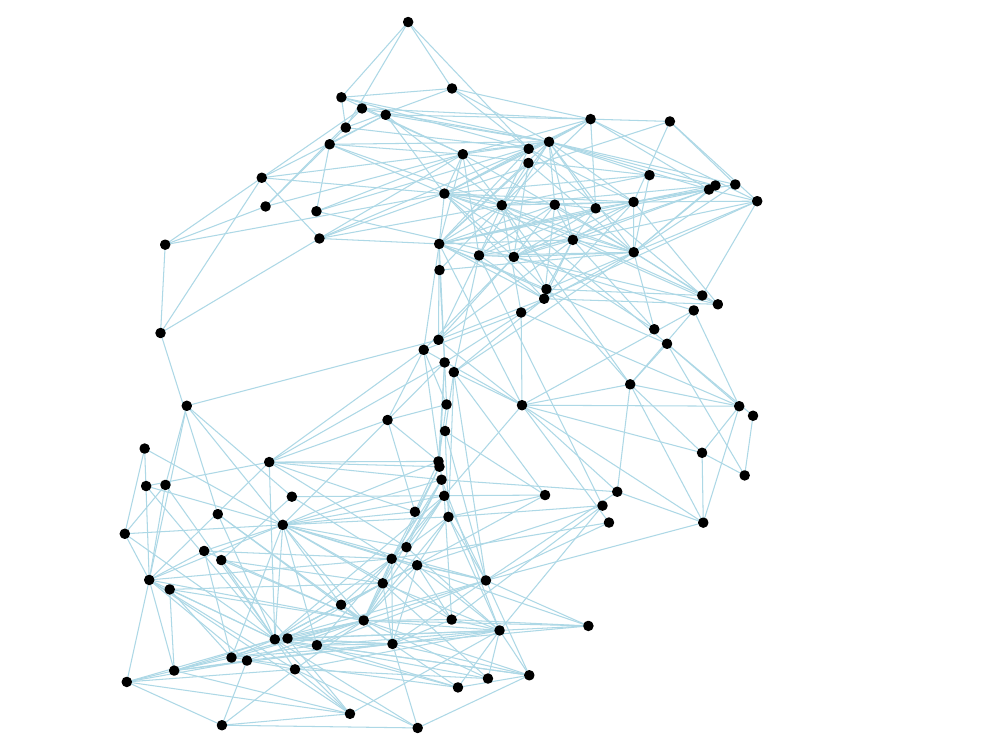} &
  \includegraphics[width=9.5mm]{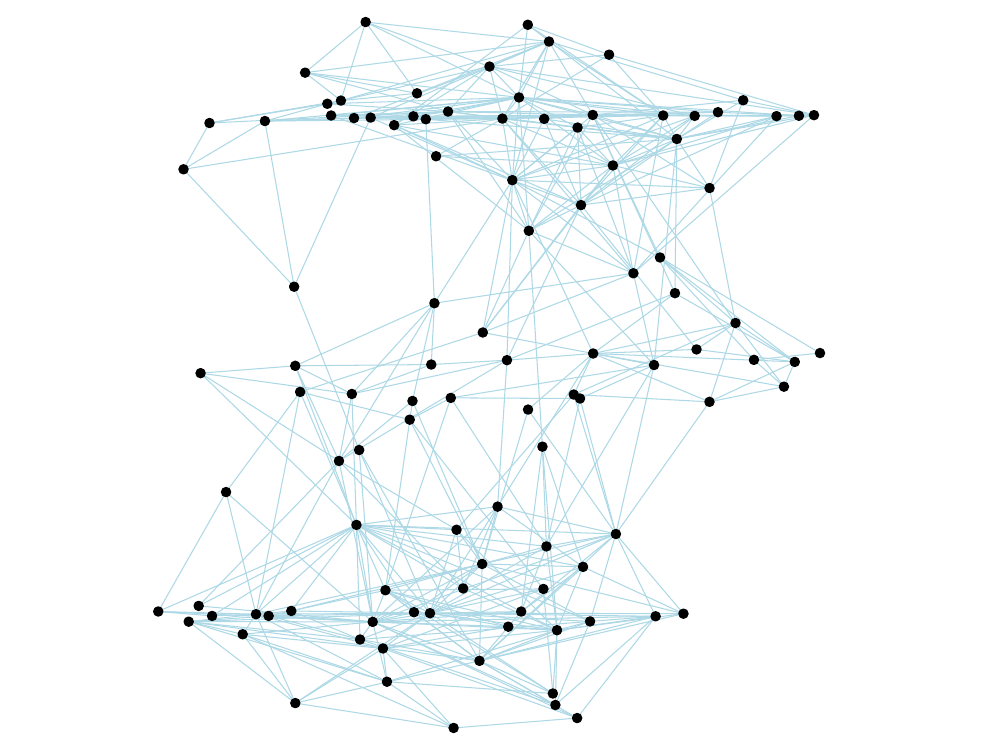} &
  \includegraphics[width=9.5mm]{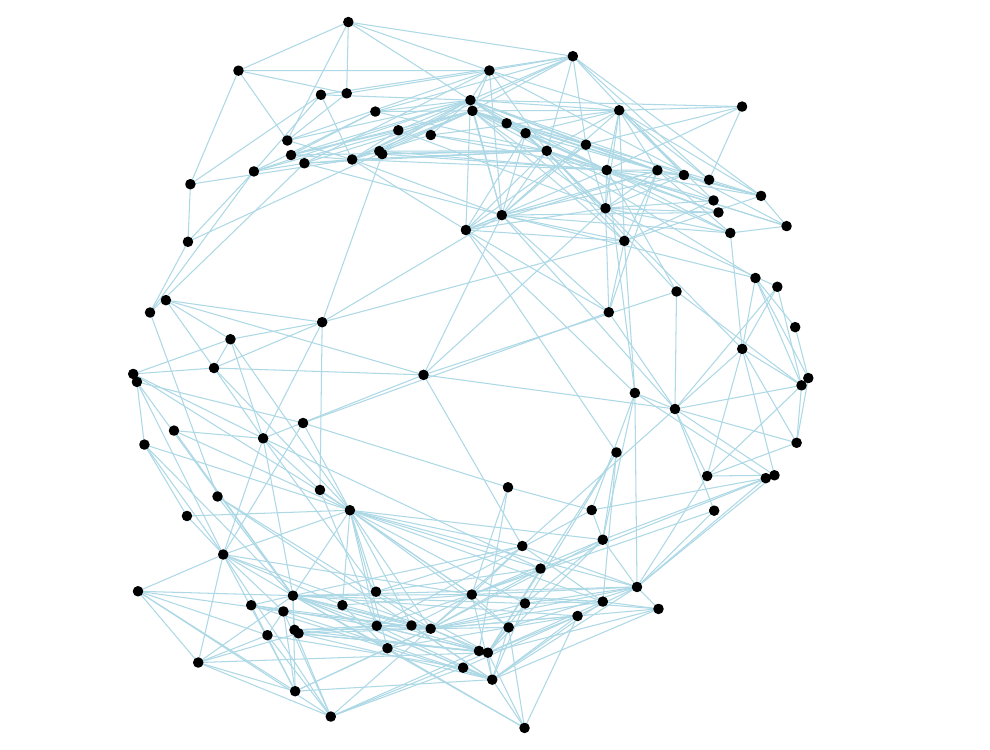} &
  \includegraphics[width=9.5mm]{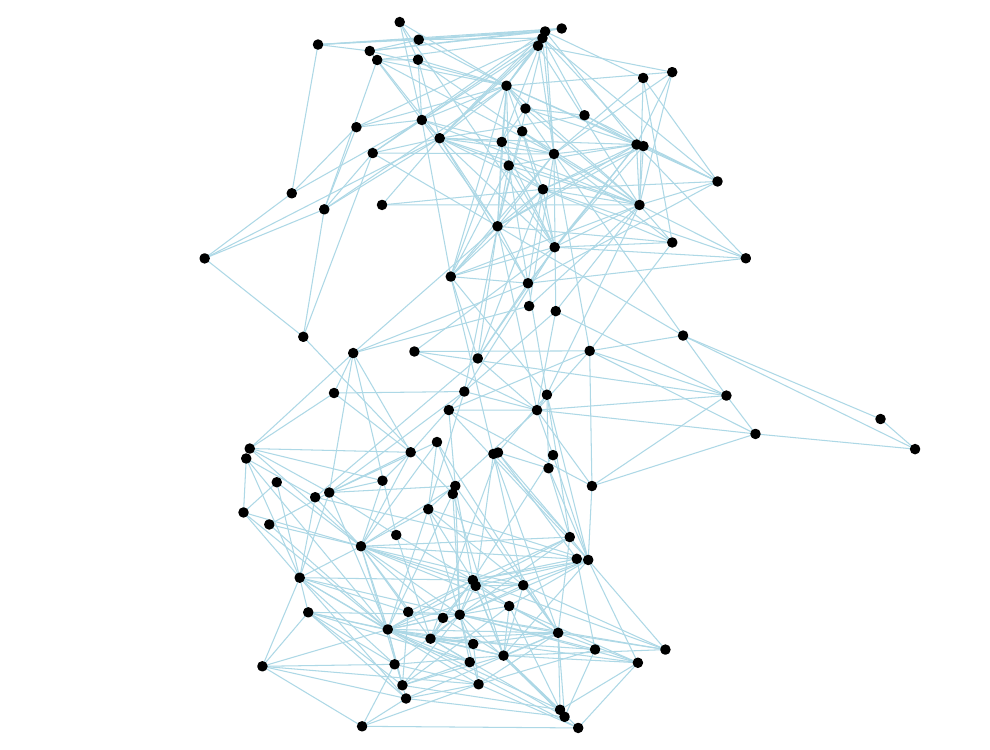} &
  \includegraphics[width=9.5mm]{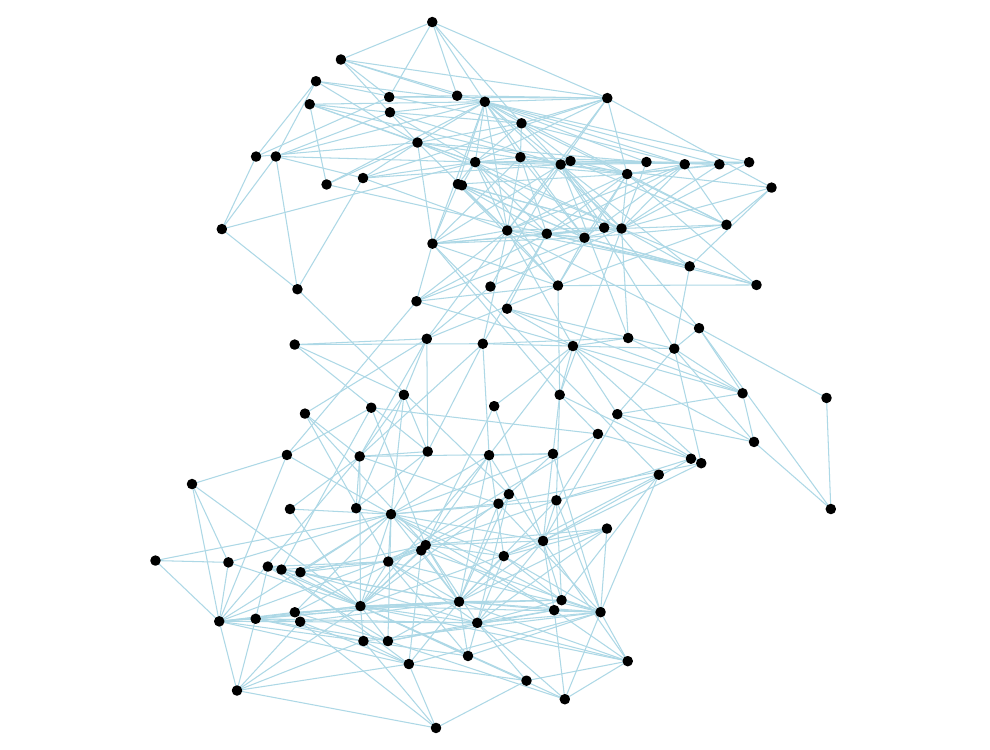}
  \\
         & \texttt{ST-ELD-CN-AR} & & \includegraphics[width=9.5mm]{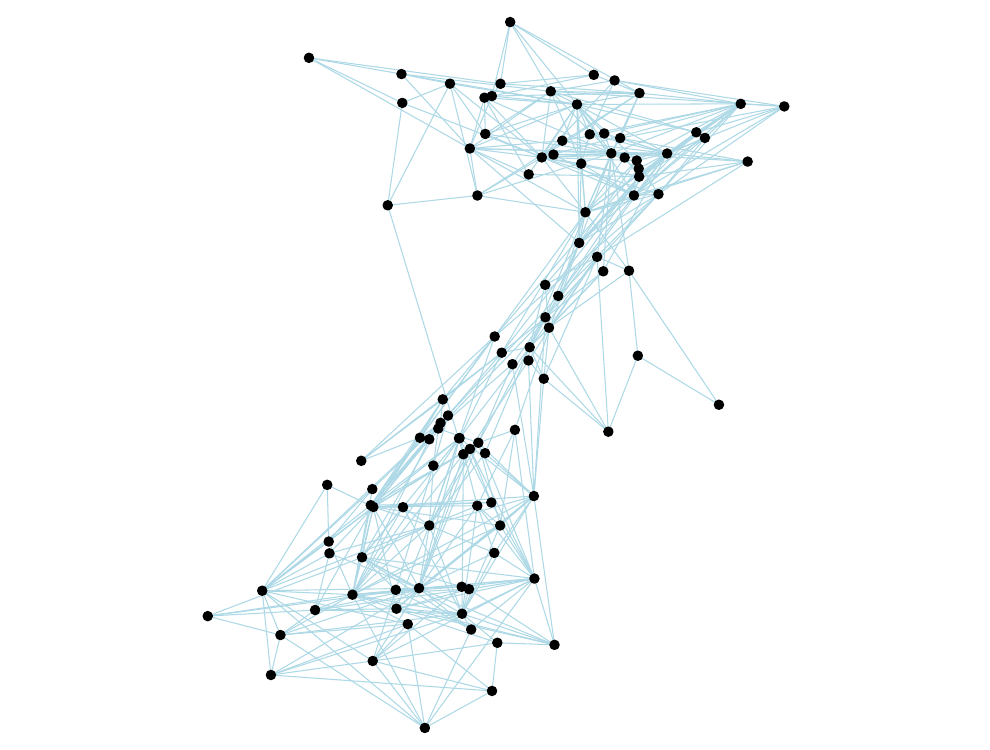} &
  \includegraphics[width=9.5mm]{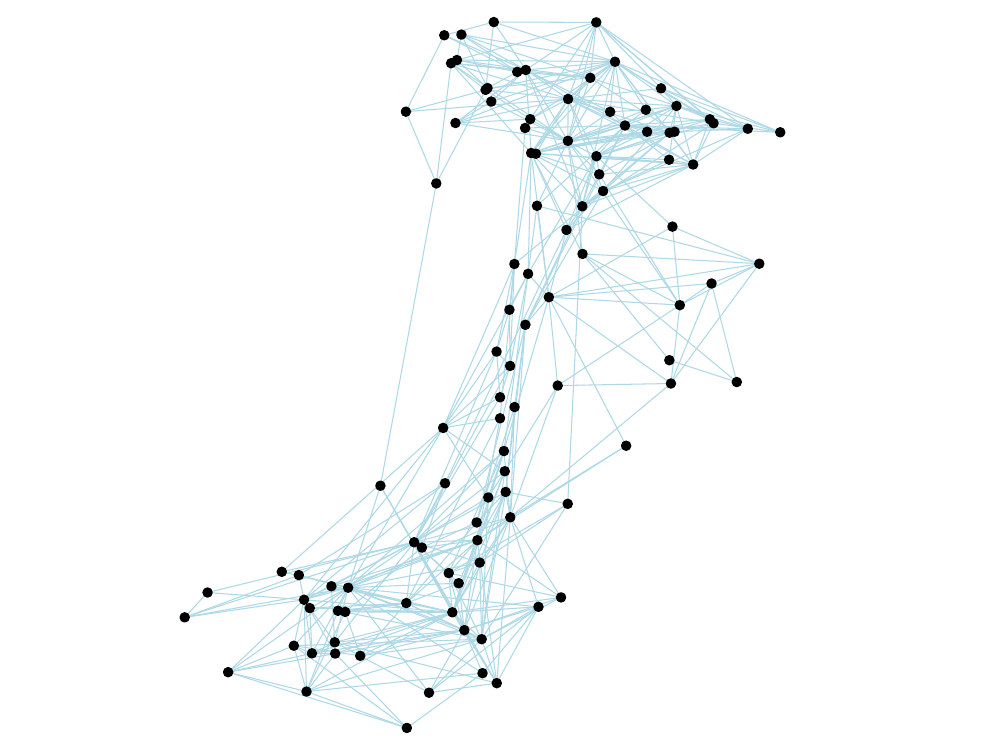} &
  \includegraphics[width=9.5mm]{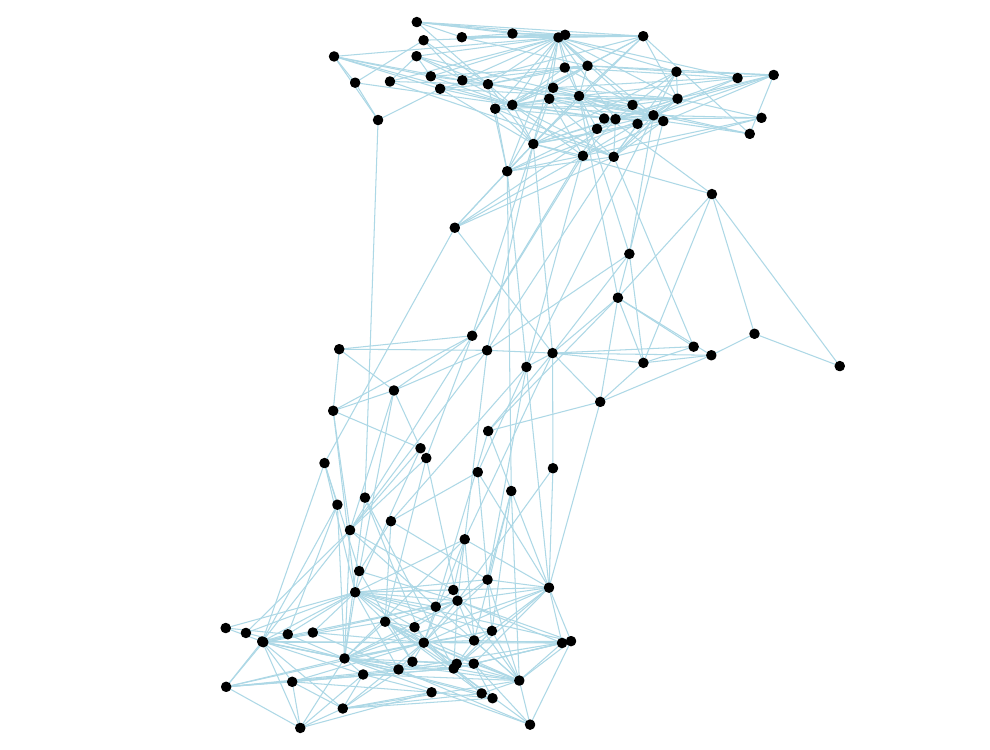} &
  \includegraphics[width=9.5mm]{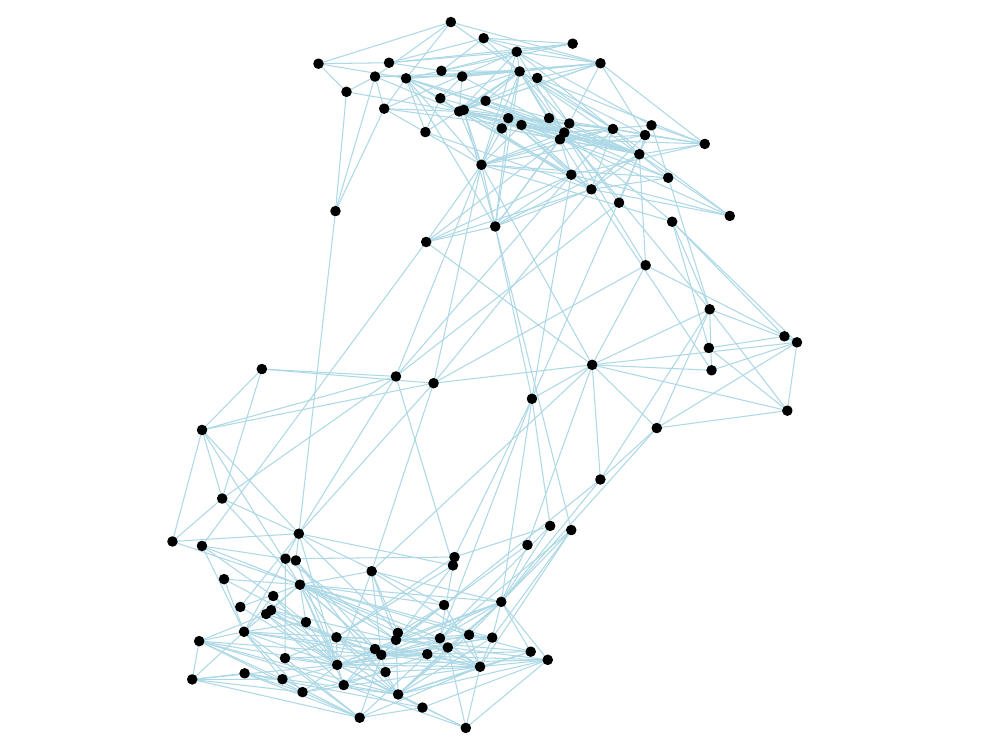} &
  \includegraphics[width=9.5mm]{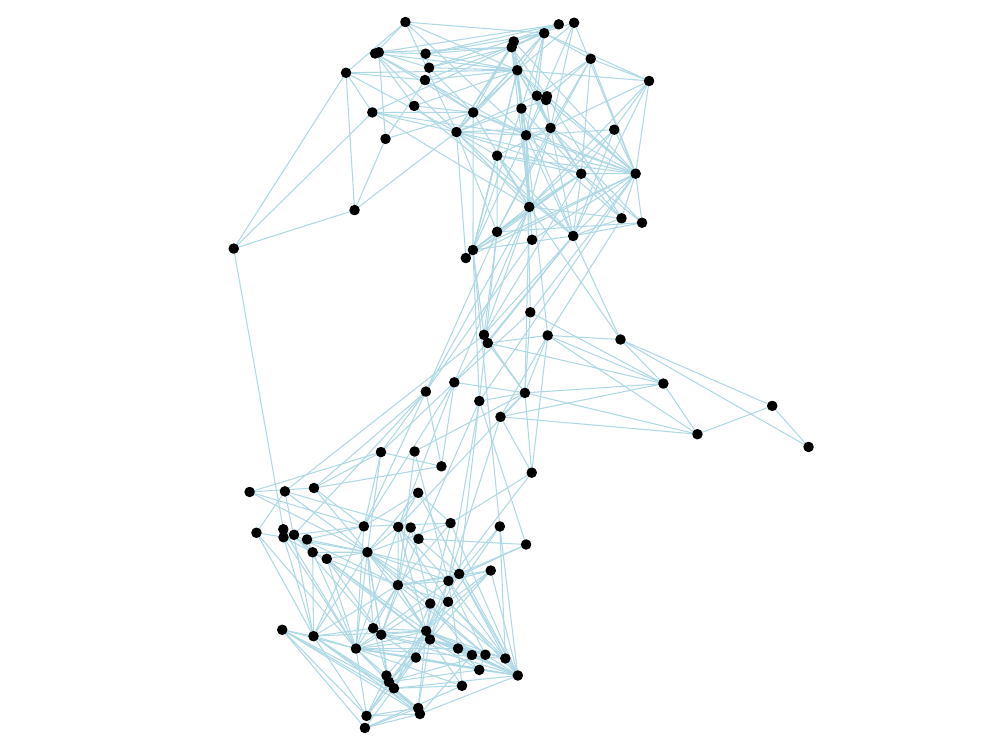} &
  \includegraphics[width=9.5mm]{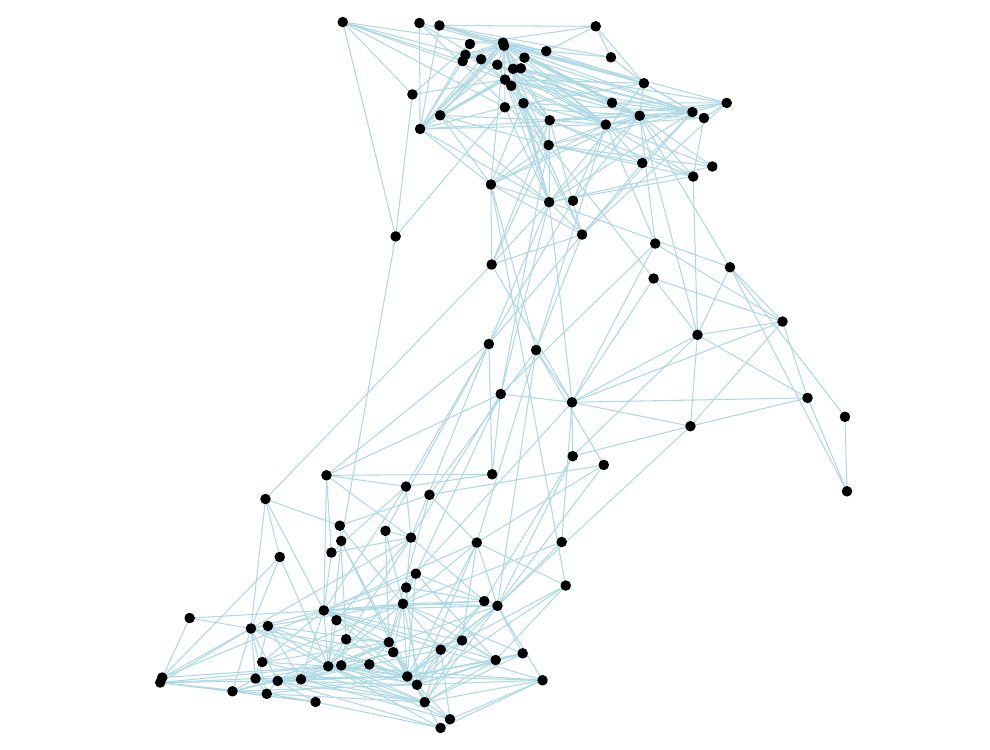}
  \\
  \hline
\end{tabular}
\caption{Collection of different graph drawings. The \texttt{START} column indicates the starting drawing of various graphs. The results of applying Alg.~\ref{alg:sim_anneal}  to \texttt{START} to six target shapes (\texttt{X}, \texttt{VERT}, \texttt{HOR}, \texttt{O}, \texttt{DINO}, \texttt{GRID}) are shown in their respective columns. The rows indicate the metrics that have $\pm\epsilon=0.0025$ for combinations of \texttt{ST,ELD,AR} and $\pm\epsilon=\texttt{CN}(\Gamma)*0.05$ for \texttt{CN}.}
\label{fig:results_combs_bar_polbooks}
\end{figure*}

\subparagraph*{Combination of metrics:} We observe that our technique is almost never able to fool all four metrics at once, except for the \emph{bar-albert} graph. Looking at the drawings in Figs.~\ref{fig:results_combs_dwt}-\ref{fig:results_combs_bar_polbooks}$\star$ and the similarity values in Fig.~\ref{fig:jitter_all}$\star$, we see that fooling three metrics in the same time is also hard. In contrast, fooling combinations of two metrics is sometimes possible but proves to be hard for \emph{dwt\_307} and \emph{gam10am}. We  observe that the combination \texttt{ST-CN} is rather resistant to fooling; every other metric pair is much easier to fool than \texttt{ST-CN} except for \texttt{ELD-CN}.

\subparagraph*{Graph type:} The spread of the similarity values in Figure~\ref{fig:jitter_all}$\star$, and in particular the low values for \emph{dwt\_307} and \emph{bar-albert}, support our hypothesis that the graph \emph{type} can influence how difficult it is to fool its quality metrics. For \emph{dwt\_307}, we see that the target shapes (except \texttt{GRID}) are much less visible for most metrics compared to results of other graphs. In Fig.~\ref{fig:sig_tests_graphs}$\star$, we see that the metrics and their combinations for this graph are significantly harder to fool than other graphs, most likely due to its strong mesh-like structure. The metrics and their combinations for the \emph{bar-albert} graph, whose tree-like structure allows for a lot of layout flexibility, are significantly easier to fool compared to all other graphs in our dataset.

\subparagraph*{Target shape:} Lastly, we observe that some target \emph{shapes} are harder to morph to. For instance, \texttt{DINO} is much more complex than \texttt{O} or \texttt{X}. This leads to messy drawings for some graphs -- \emph{dwt\_307} is unable to even come close. When aggregating the results over all graphs and metric combinations (see  Fig.~\ref{fig:sig_tests_target}$\star$), we see that only the \texttt{GRID} shape is significantly easier to morph to than all other shapes.

\section{Conclusion}
\label{sec:conclusion}

In this work we provide evidence against the assumptions that (1) `high' quality metric values indicate `good' graph drawings; and that (2) drawings of the same graph with similar metric values are similar in their quality.
We use simulated annealing to morph existing graph drawings into six different, arbitrary, target shapes without substantially altering one or more quality metric values. To our knowledge, this is the first time that a systematic fooling of graph drawing quality metrics has been explored. Depending on the graph and metric, most graph drawings can easily be morphed towards such targets. We observe that the structure of the graph has an influence on how well a graph drawing can fooled. Furthermore, we find evidence that some metrics (\emph{stress}) are more difficult to fool than others, depending on the target shape; and that some combinations of metrics are easier to fool. All in all, our results suggest that current quality metrics are not enough to capture the essence of a `good' graph drawing on their own. Similar to Huang et al.\,\cite{Huang_Eades_Hong_Lin_2013}, we emphasize the importance of combinations of two or more quality metrics that capture both the graph structure (faithfulness) and readability of the drawing. Future work can consider advancing our fooling method by using more sophisticated techniques such as gradient descent or deep learning approaches. Furthermore, more experiments can be conducted with more graphs, target drawings, and quality metrics, as well as linear combinations of multiple quality metrics. Lastly, a perceptual user-study could explore how readable or misleading these morphed drawings are.

\newpage
\bibliography{references}

\begin{thebibliography}{10}

\bibitem{Ahmed_Luca_Devkota_Kobourov_Li_2022_sgd}
Reyan Ahmed, Felice~De Luca, Sabin Devkota, Stephen Kobourov, and Mingwei Li.
\newblock {Multicriteria Scalable Graph Drawing via Stochastic Gradient Descent, {$(SGD)^{2}$}}.
\newblock {\em IEEE TVCG}, 28(06):2388--2399, 2022.
\newblock \href {https://doi.org/10.1109/TVCG.2022.3155564} {\path{doi:10.1109/TVCG.2022.3155564}}.

\bibitem{Anscombe_1973_stats}
Francis~J. Anscombe.
\newblock {Graphs in Statistical Analysis}.
\newblock {\em The American Statistician}, 27(1):17--21, 1973.
\newblock \href {https://doi.org/10.1080/00031305.1973.10478966} {\path{doi:10.1080/00031305.1973.10478966}}.

\bibitem{Boger_et_al_2021_stats}
Tal Boger, Steven~B. Most, and Steven~L. Franconeri.
\newblock {Jurassic Mark: Inattentional Blindness for a Datasaurus Reveals that Visualizations are Explored, not Seen}.
\newblock In {\em IEEE Vis}, pages 71--75, 2021.
\newblock \href {https://doi.org/10.1109/VIS49827.2021.9623273} {\path{doi:10.1109/VIS49827.2021.9623273}}.

\bibitem{Burch_2021_usereval}
Michael Burch, Weidong Huang, Mathew Wakefield, Helen~C. Purchase, Daniel Weiskopf, and Jie Hua.
\newblock {The State of the Art in Empirical User Evaluation of Graph Visualizations}.
\newblock {\em IEEE Access}, 9:4173--4198, 2021.
\newblock \href {https://doi.org/10.1109/ACCESS.2020.3047616} {\path{doi:10.1109/ACCESS.2020.3047616}}.

\bibitem{Cairo_2016_Datausaurus}
Albert Cairo.
\newblock {Download the Datasaurus: Never trust summary statistics alone; always visualize your data}, 2016.
\newblock URL: \url{https://thefunctionalart.blogspot.com/2016/08/download-datasaurus-never-trust-summary.html}.

\bibitem{Chari_Pachter_2023_elephant}
Tara Chari and Lior Pachter.
\newblock {The specious art of single-cell genomics}.
\newblock {\em Computational Biology}, 19(8):1--20, 2023.
\newblock \href {https://doi.org/10.1371/journal.pcbi.1011288} {\path{doi:10.1371/journal.pcbi.1011288}}.

\bibitem{Chatterjee_Aykut_2007_stats}
Sangit Chatterjee and Aykut Firat.
\newblock {Generating Data with Identical Statistics but Dissimilar Graphics}.
\newblock {\em The American Statistician}, 61(3):248--254, 2007.
\newblock \href {https://doi.org/10.1198/000313007X220057} {\path{doi:10.1198/000313007X220057}}.

\bibitem{Chen_2021_graphstats}
Hang Chen, Utkarsh Soni, Yafeng Lu, Vahan Huroyan, Ross Maciejewski, and Stephen~G. Kobourov.
\newblock { Same Stats, Different Graphs: Exploring the Space of Graphs in Terms of Graph Properties }.
\newblock {\em IEEE TVCG}, 27(03):2056--2072, 2021.
\newblock \href {https://doi.org/10.1109/TVCG.2019.2946558} {\path{doi:10.1109/TVCG.2019.2946558}}.

\bibitem{Chimani_etal2014_stressfavor}
Markus Chimani, Patrick Eades, Peter Eades, Seok-Hee Hong, Weidong Huang, Karsten Klein, Michael~R. Marner, Ross~T. Smith, and Bruce~H. Thomas.
\newblock People prefer less stress and fewer crossings.
\newblock In {\em Graph Drawing}, volume 8871 of {\em LNCS}, pages 523--524. Springer Berlin Heidelberg, 2014.

\bibitem{Cuturi_2013_sinkhorn}
Marco Cuturi.
\newblock {Sinkhorn distances: lightspeed computation of optimal transport}.
\newblock In {\em NIPS}, volume~2, page 2292–2300. Curran Associates Inc., 2013.
\newblock \href {https://doi.org/10.48550/arXiv.1306.0895} {\path{doi:10.48550/arXiv.1306.0895}}.

\bibitem{stressplusx}
Sabin Devkota, Reyan Ahmed, Felice De~Luca, Katherine~E. Isaacs, and Stephen~G. Kobourov.
\newblock {Stress-Plus-X (SPX) Graph Layout}.
\newblock In {\em Graph Drawing}, volume 11904 of {\em LNCS}, pages 291--304. Springer Berlin Heidelberg, 2019.
\newblock \href {https://doi.org/10.1007/978-3-030-35802-0_23} {\path{doi:10.1007/978-3-030-35802-0_23}}.

\bibitem{gdnvcontest}
Sara {Di Bartolomeo}, Fabian Klute, Debajyoti Mondal, and Jules Wulms.
\newblock Graph drawing contest report.
\newblock In {\em Graph Drawing}, LIPIcs, pages 41:1--41:13. Schloss Dagstuhl, October 2024.
\newblock \href {https://doi.org/10.4230/LIPICS.GD.2024.41} {\path{doi:10.4230/LIPICS.GD.2024.41}}.

\bibitem{bartolomeo_Lang_Dunne_2022_worst}
Sara Di~Bartolomeo, Mat{\v{e}}j Lang, and Cody Dunne.
\newblock {The worst graph layout algorithm ever}.
\newblock In {\em Proc. alt. VIS workshop at IEEE VIS}, 2022.
\newblock \href {https://doi.org/10.31219/osf.io/4hfy9} {\path{doi:10.31219/osf.io/4hfy9}}.

\bibitem{espadoto19}
Mateus Espadoto, Rafael~M. Martins, Andreas Kerren, Nina S.~T. Hirata, and Alexandru~C. Telea.
\newblock {Toward a Quantitative Survey of Dimension Reduction Techniques}.
\newblock {\em IEEE TVCG}, 27(3):2153--2173, 2919.
\newblock \href {https://doi.org/10.1109/TVCG.2019.2944182} {\path{doi:10.1109/TVCG.2019.2944182}}.

\bibitem{Fruchterman_Reingold_fd_1991}
Thomas M.~J. Fruchterman and Edward~M. Reingold.
\newblock {Graph drawing by force-directed placement}.
\newblock {\em Software: Practice and Experience}, 21(11):1129--1164, 1991.
\newblock \href {https://doi.org/10.1002/spe.4380211102} {\path{doi:10.1002/spe.4380211102}}.

\bibitem{Gansner_Koren_North_SM_2005}
Emden~R. Gansner, Yehuda Koren, and Stephen~C. North.
\newblock {Graph Drawing by Stress Majorization}.
\newblock In {\em Graph Drawing}, volume 3383 of {\em LNCS}, page 239–250. Springer Berlin Heidelberg, 2005.
\newblock \href {https://doi.org/10.1007/978-3-540-31843-9_25} {\path{doi:10.1007/978-3-540-31843-9_25}}.

\bibitem{groetschla2024coregd}
Florian Gr{\"o}tschla, Jo{\"e}l Mathys, Robert Veres, and Roger Wattenhofer.
\newblock Core-{GD}: A hierarchical framework for scalable graph visualization with {GNN}s.
\newblock In {\em The Twelfth International Conference on Learning Representations}, 2024.
\newblock \href {https://doi.org/10.48550/arXiv.2402.06706} {\path{doi:10.48550/arXiv.2402.06706}}.

\bibitem{Huang07}
Weidong Huang.
\newblock {Using eye tracking to investigate graph layout effects}.
\newblock In {\em {APVIS}}, pages 97--100. {IEEE} Computer Society, 2007.
\newblock \href {https://doi.org/10.1109/APVIS.2007.329282} {\path{doi:10.1109/APVIS.2007.329282}}.

\bibitem{Huang_Eades_Hong_Lin_2013}
Weidong Huang, Peter Eades, Seok-Hee Hong, and Chun-Cheng Lin.
\newblock {Improving multiple aesthetics produces better graph drawings}.
\newblock {\em Journal of Visual Languages and Computing}, 24(4):262–272, 2013.
\newblock \href {https://doi.org/10.1016/j.jvlc.2011.12.002} {\path{doi:10.1016/j.jvlc.2011.12.002}}.

\bibitem{Jacomy_Venturini_Heymann_Bastian__FA2_2014}
Mathieu Jacomy, Tommaso Venturini, Sebastien Heymann, and Mathieu Bastian.
\newblock {ForceAtlas2, a Continuous Graph Layout Algorithm for Handy Network Visualization Designed for the Gephi Software}.
\newblock {\em PLOS ONE}, 9(6):e98679, 2014.
\newblock \href {https://doi.org/10.1371/journal.pone.0098679} {\path{doi:10.1371/journal.pone.0098679}}.

\bibitem{KamadaKawai1989}
Tomihisa Kamada and Satoru Kawai.
\newblock {An algorithm for drawing general undirected graphs}.
\newblock {\em Information Processing Letters}, 31(1):7--15, 1989.
\newblock \href {https://doi.org/10.1016/0020-0190(89)90102-6} {\path{doi:10.1016/0020-0190(89)90102-6}}.

\bibitem{machado2025_metrics}
Alister Machado, Michael Behrisch, and Alexandru~C. Telea.
\newblock Necessary but not sufficient: Limitations of projection quality metrics.
\newblock {\em Computer Graphics Forum}, page e70101, 2015.
\newblock \href {https://doi.org/10.1111/cgf.70101} {\path{doi:10.1111/cgf.70101}}.

\bibitem{sharp}
Alister Machado, Alexandru~C. Telea, and Michael Behrisch.
\newblock {ShaRP: Shape-Regularized Multidimensional Projections}.
\newblock In {\em EuroVis Workshop on Visual Analytics (EuroVA)}. The Eurographics Association, 2023.
\newblock \href {https://doi.org/10.2312/eurova.20231088} {\path{doi:10.2312/eurova.20231088}}.

\bibitem{Matejka_Fitzmaurice_2017_stats_sim_anneal}
Justin Matejka and George Fitzmaurice.
\newblock {Same Stats, Different Graphs: Generating Datasets with Varied Appearance and Identical Statistics through Simulated Annealing}.
\newblock In {\em CHI}, page 1290–1294. Association for Computing Machinery, 2017.
\newblock \href {https://doi.org/10.1145/3025453.3025912} {\path{doi:10.1145/3025453.3025912}}.

\bibitem{Mooney_Purchase_Wybrow_Kobourov_2024_metriclandscape}
Gavin~J. Mooney, Helen~C. Purchase, Michael Wybrow, and Stephen~G. Kobourov.
\newblock {The Multi-Dimensional Landscape of Graph Drawing Metrics}.
\newblock In {\em IEEE PacificVis}, pages 122--131, 2024.
\newblock \href {https://doi.org/10.1109/PacificVis60374.2024.00022} {\path{doi:10.1109/PacificVis60374.2024.00022}}.

\bibitem{Mooney_Purchase_Wybrow_Kobourov_Miller_2024_stressperception}
Gavin~J. Mooney, Helen~C. Purchase, Michael Wybrow, Stephen~G. Kobourov, and Jacob Miller.
\newblock {The Perception of Stress in Graph Drawings}.
\newblock In {\em Graph Drawing}, volume 320 of {\em LIPIcs}, pages 21:1--21:17. Schloss Dagstuhl, 2024.
\newblock \href {https://doi.org/10.4230/LIPIcs.GD.2024.21} {\path{doi:10.4230/LIPIcs.GD.2024.21}}.

\bibitem{Moshiri_2018_dual_barabasi}
Niema Moshiri.
\newblock {The dual-Barab\'asi-Albert model}.
\newblock {\em arXiv preprint arXiv:1810.10538}, 2018.

\bibitem{Nguyen_Eades_Hong_2013}
Quan Nguyen, Peter Eades, and Seok-Hee Hong.
\newblock On the faithfulness of graph visualizations.
\newblock In {\em 2013 IEEE PacificVis}, pages 209--216, 2013.
\newblock \href {https://doi.org/10.1109/PacificVis.2013.6596147} {\path{doi:10.1109/PacificVis.2013.6596147}}.

\bibitem{Purchase_Performance_1997}
Helen~C. Purchase.
\newblock Which aesthetic has the greatest effect on human understanding?
\newblock In {\em Graph Drawing}, volume 1353 of {\em LNCS}, page 248–261. Springer Berlin Heidelberg, 1997.
\newblock \href {https://doi.org/10.1007/3-540-63938-1_67} {\path{doi:10.1007/3-540-63938-1_67}}.

\bibitem{Purchase_2002_metrics}
Helen~C. Purchase.
\newblock {Metrics for Graph Drawing Aesthetics}.
\newblock {\em Journal of Visual Languages and Computing}, 13(5):501--516, 2002.
\newblock \href {https://doi.org/10.1006/jvlc.2002.0232} {\path{doi:10.1006/jvlc.2002.0232}}.

\bibitem{Purchase1996_aesthetics}
{Helen C.} Purchase, {Robert F.} Cohen, and Murray James.
\newblock {Validating graph drawing aesthetics}.
\newblock In {\em Graph Drawing}, volume 1027 of {\em LNCS}, pages 435--446. Springer Berlin Heidelberg, 1996.
\newblock \href {https://doi.org/10.1007/bfb0021827} {\path{doi:10.1007/bfb0021827}}.

\bibitem{networkrepository}
Ryan~A. Rossi and Nesreen~K. Ahmed.
\newblock {The Network Data Repository with Interactive Graph Analytics and Visualization}.
\newblock In {\em AAAI}, 2015.
\newblock URL: \url{https://networkrepository.com}.

\bibitem{vanWageningen_Mchedlidze_Telea_2024}
Simon van Wageningen, Tamara Mchedlidze, and Alexandru~C. Telea.
\newblock {An Experimental Evaluation of Viewpoint‐Based 3D Graph Drawing}.
\newblock {\em Computer Graphics Forum}, 43(3):e15077, 2024.
\newblock \href {https://doi.org/10.1111/cgf.15077} {\path{doi:10.1111/cgf.15077}}.

\bibitem{githubpage}
Simon van Wageningen, Tamara Mchedlidze, and Alexandru~C. Telea.
\newblock {Same Quality Metrics} source code and videos, 2025.
\newblock URL: \url{https://github.com/simonvw95/Same-Quality-Metric--Different-Graph-Drawings}.

\bibitem{Villani2009_wasserstein}
C{\'e}dric Villani.
\newblock {\em {The Wasserstein distances}}, pages 93--111.
\newblock Springer, 2009.
\newblock \href {https://doi.org/10.1007/978-3-540-71050-9_6} {\path{doi:10.1007/978-3-540-71050-9_6}}.

\bibitem{deepgd}
Xiaoqi Wang, Kevin Yen, Yifan Hu, and Han-Wei Shen.
\newblock {DeepGD}: A deep learning framework for graph drawing using {GNN}.
\newblock {\em IEEE CGA}, 41(05):32--44, 2021.
\newblock \href {https://doi.org/10.1109/MCG.2021.3093908} {\path{doi:10.1109/MCG.2021.3093908}}.

\bibitem{deepdrawing}
Yong Wang, Zhihua Jin, Qianwen Wang, Weiwei Cui, Tengfei Ma, and Huamin Qu.
\newblock {DeepDrawing: A Deep Learning Approach to Graph Drawing}.
\newblock {\em IEEE TVCG}, pages 1--1, 2019.
\newblock \href {https://doi.org/10.1109/TVCG.2019.2934798} {\path{doi:10.1109/TVCG.2019.2934798}}.

\end{thebibliography}



\clearpage
\section{Appendix}
\label{sec:appendix}

\subsection{Video results}
In this work we attempt to morph drawings into a few specific shapes while keeping one or more quality metric values nearly the same. In this section we show that we can do the same process but not for a single target shapes but for a whole video.

\begin{figure}[!ht]
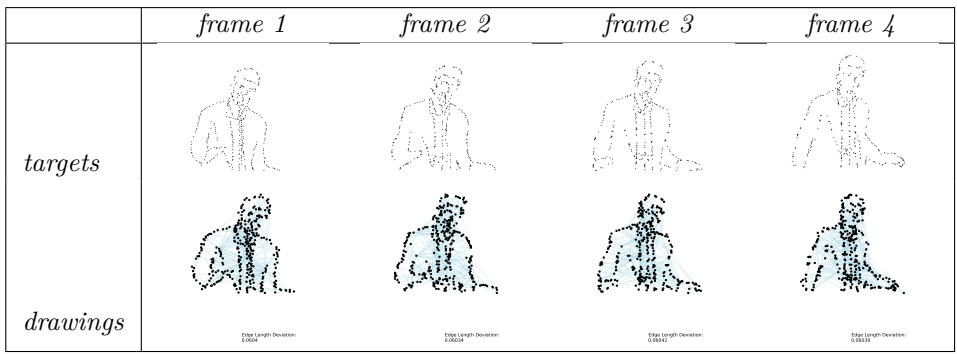

\centering
\begin{tabular}{|l|cccc|}
  \hline
  & \emph{frame 1} & \emph{frame 2} & \emph{frame 3} & \emph{frame 4}\\
\hline
  \emph{targets} &
  \includegraphics[width=22.5mm]{results/gif_0.png} & \includegraphics[width=22.5mm]{results/gif_1.png} & \includegraphics[width=22.5mm]{results/gif_2.png} & \includegraphics[width=22.5mm]{results/gif_3.png}\\
  
  \emph{drawings} &
  \includegraphics[width=22.5mm, angle=270, origin=c]{results/ba_rr-gif_0_coords.csvELD.png} & \includegraphics[width=22.5mm, angle=270, origin=c]{results/ba_rr-gif_1_coords.csvELD.png} & \includegraphics[width=22.5mm, angle=270, origin=c]{results/ba_rr-gif_2_coords.csvELD.png} & \includegraphics[width=22.5mm, angle=270, origin=c]{results/ba_rr-gif_3_coords.csvELD.png}\\
  \hline
\end{tabular}
\caption{Four sequential target frames from the video and the resulting sequential drawings from attempting to fool a quality metric (\texttt{ELD}) for the target frames.}

\label{fig:rickroll2}
\end{figure}

We perform the following steps to fool quality metrics in a video setting: We extract individual frames as images from a short video and turn the contours found in each frame into a fixed number of points in the image. This gives us 24 sequenced images as can be see in the first row of Figure~\ref{fig:rickroll2}$\star$. We then apply Alg.~\ref{alg:sim_anneal} to each image to produce 24 sequenced graph drawings which we can then splice together to create a video, as can be seen in \,\cite{githubpage}.

\clearpage
\subsection{Combinations of Metrics}


\begin{figure*}[!ht]
\centering
\begin{tabular}{cc|ccccccc}
  \hline
  & & \texttt{START} & \texttt{X} & \texttt{VERT} & \texttt{HOR} & \texttt{O} & \texttt{DINO} & \texttt{GRID}\\
\hline

    \emph{lnsp\_131} & \texttt{ST-ELD} & \includegraphics[width=9.5mm]{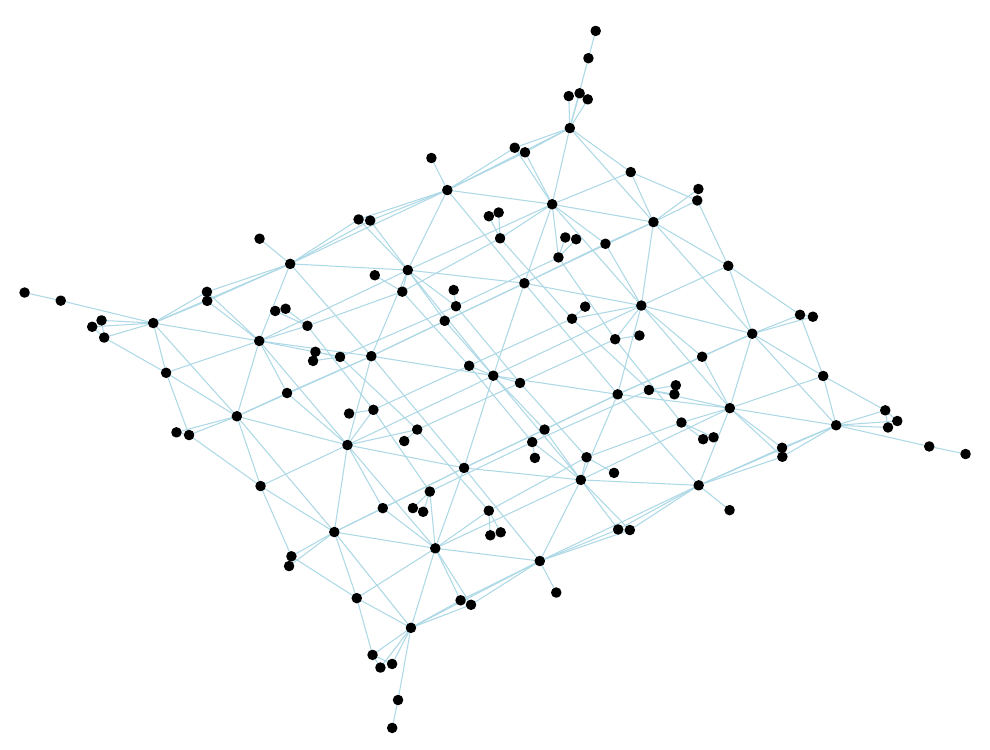} & \includegraphics[width=9.5mm]{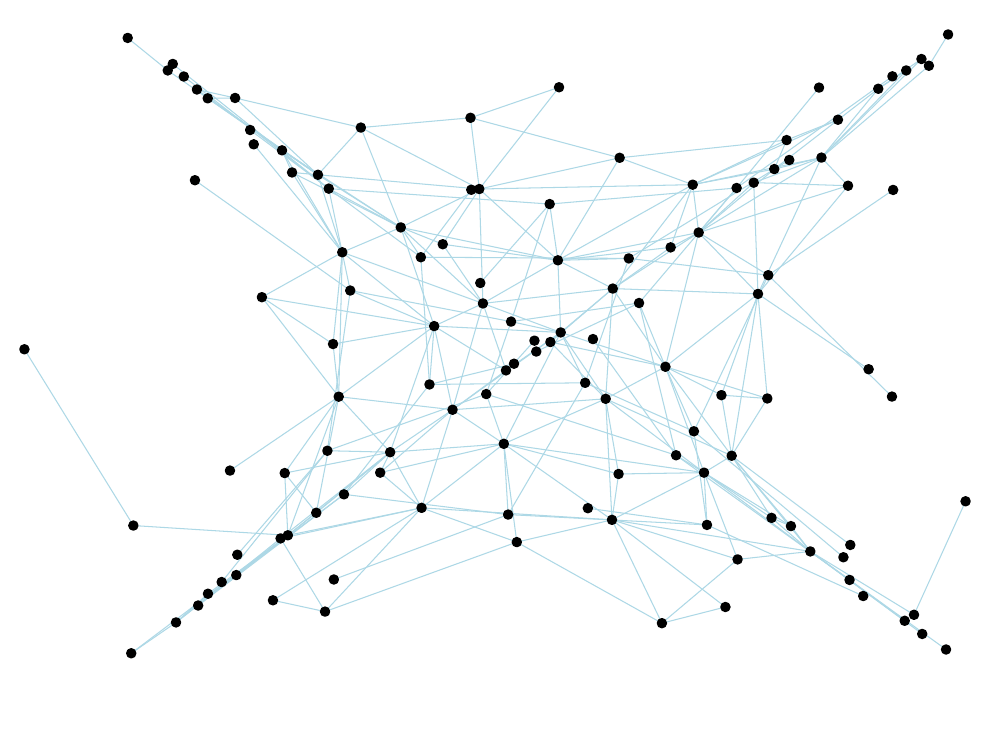} &
  \includegraphics[width=9.5mm]{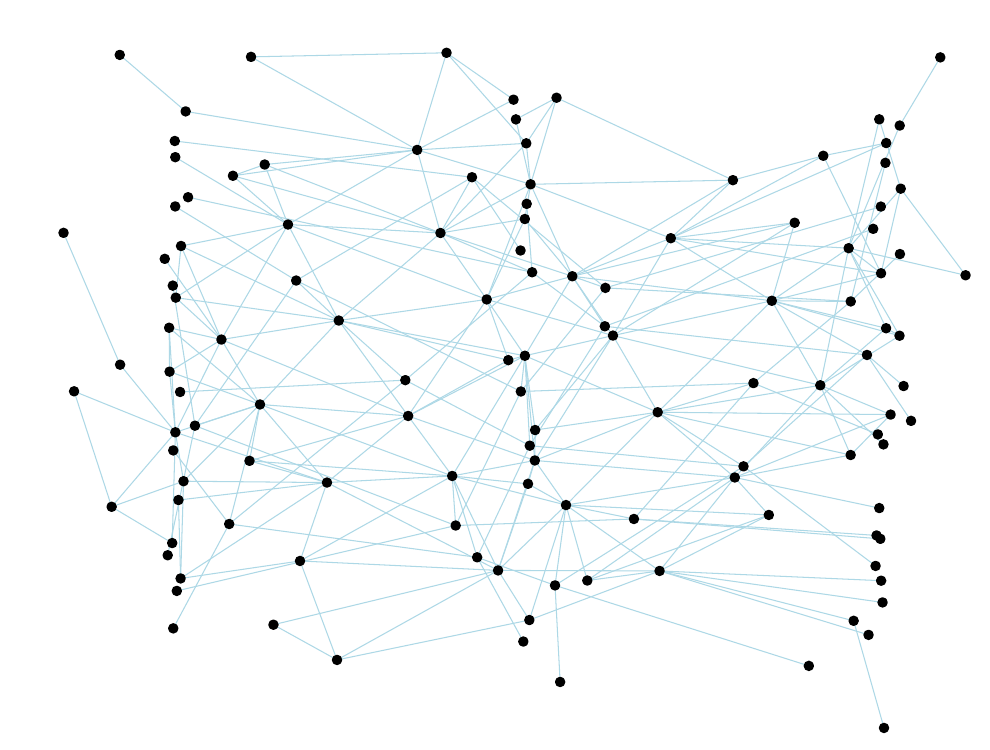} &
  \includegraphics[width=9.5mm]{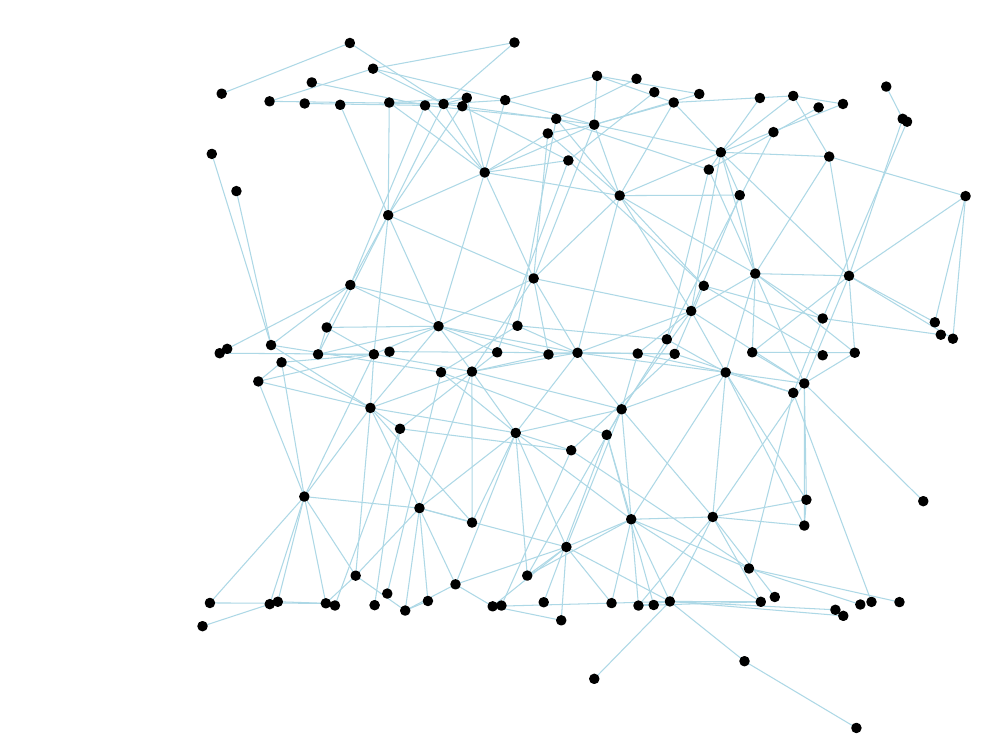} &
  \includegraphics[width=9.5mm]{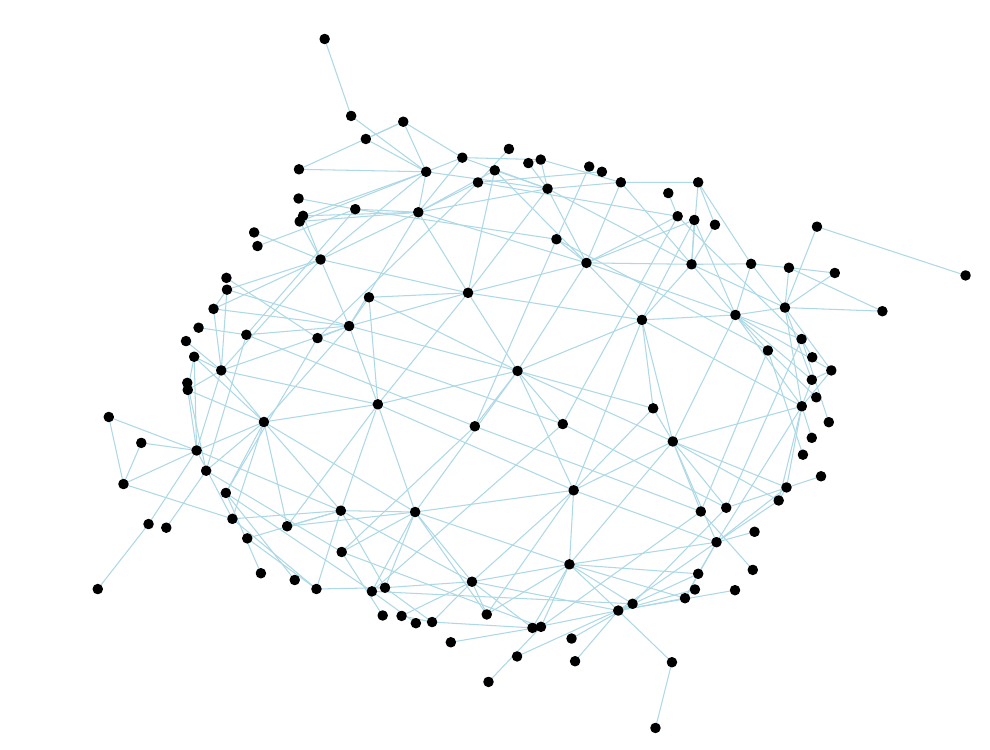} &
  \includegraphics[width=9.5mm]{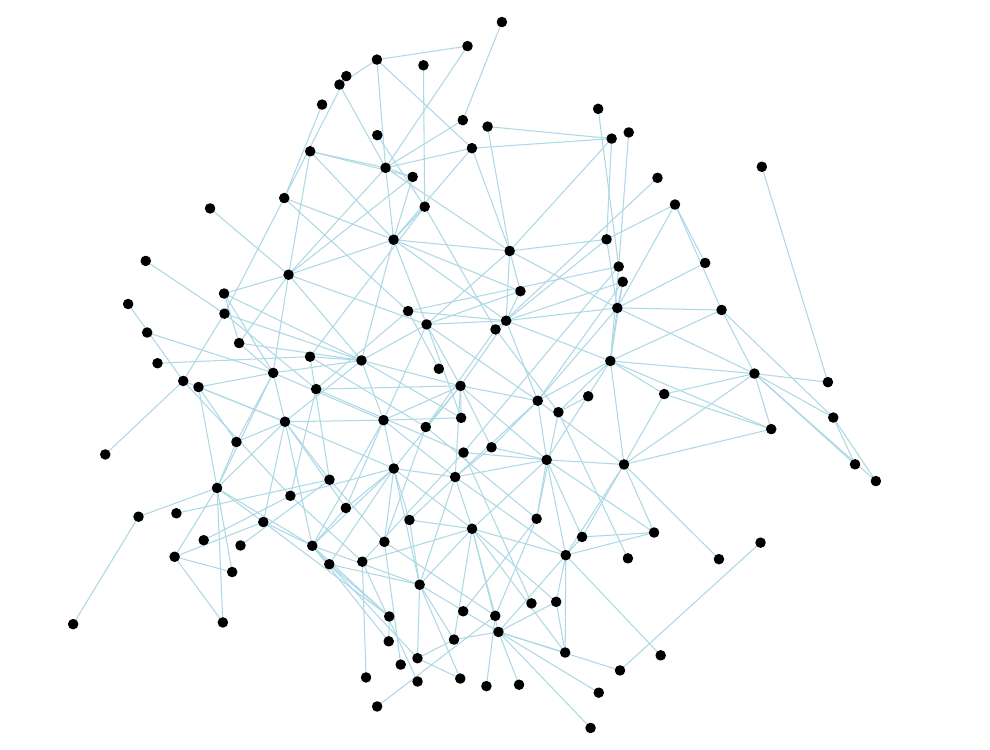} &
  \includegraphics[width=9.5mm]{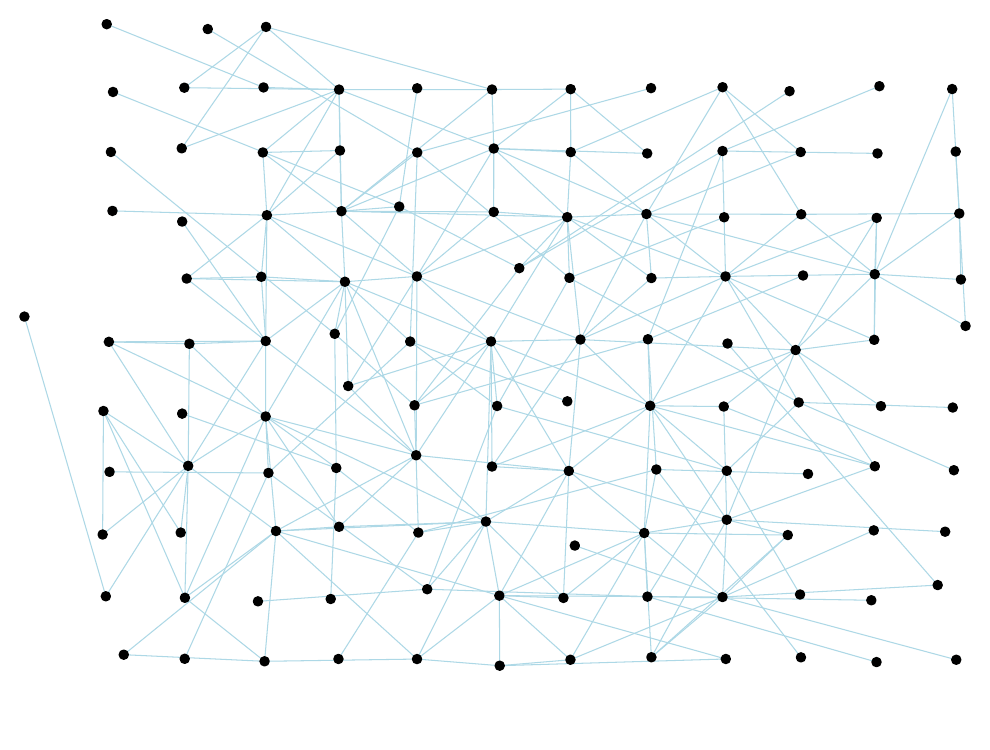}
  \\
     & \texttt{ST-CN} & & \includegraphics[width=9.5mm]{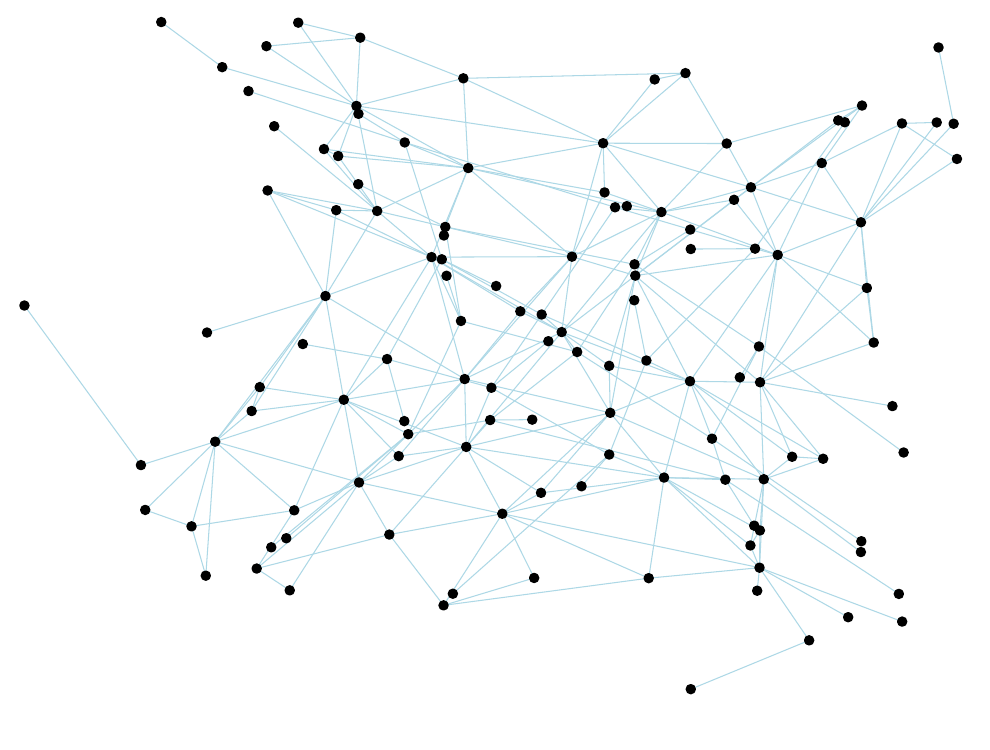} &
  \includegraphics[width=9.5mm]{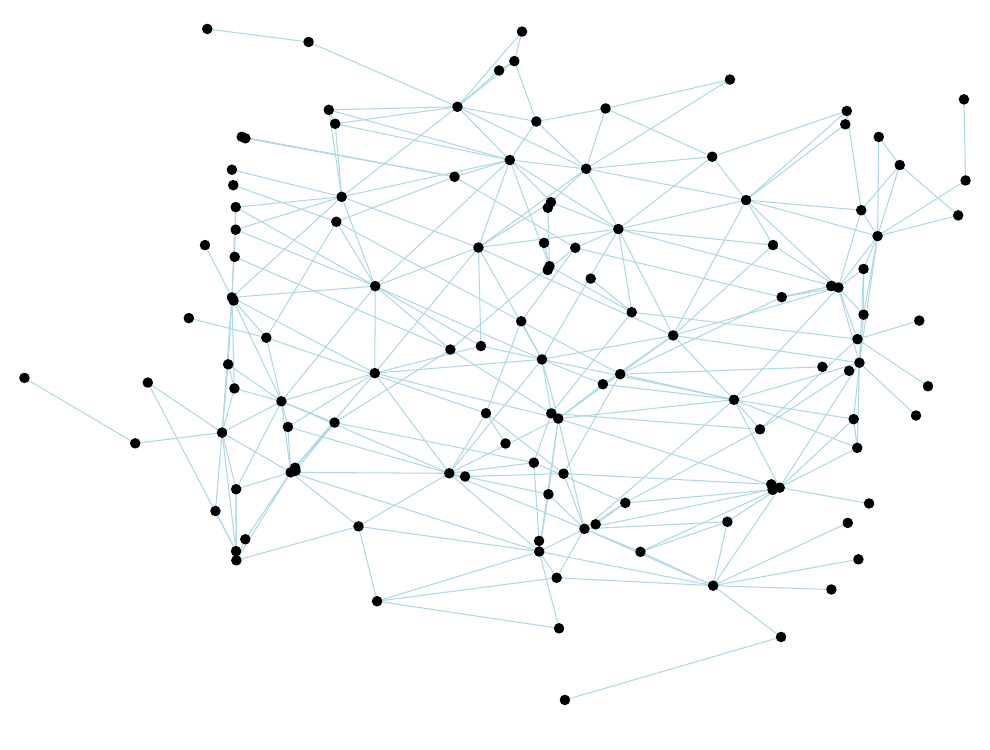} &
  \includegraphics[width=9.5mm]{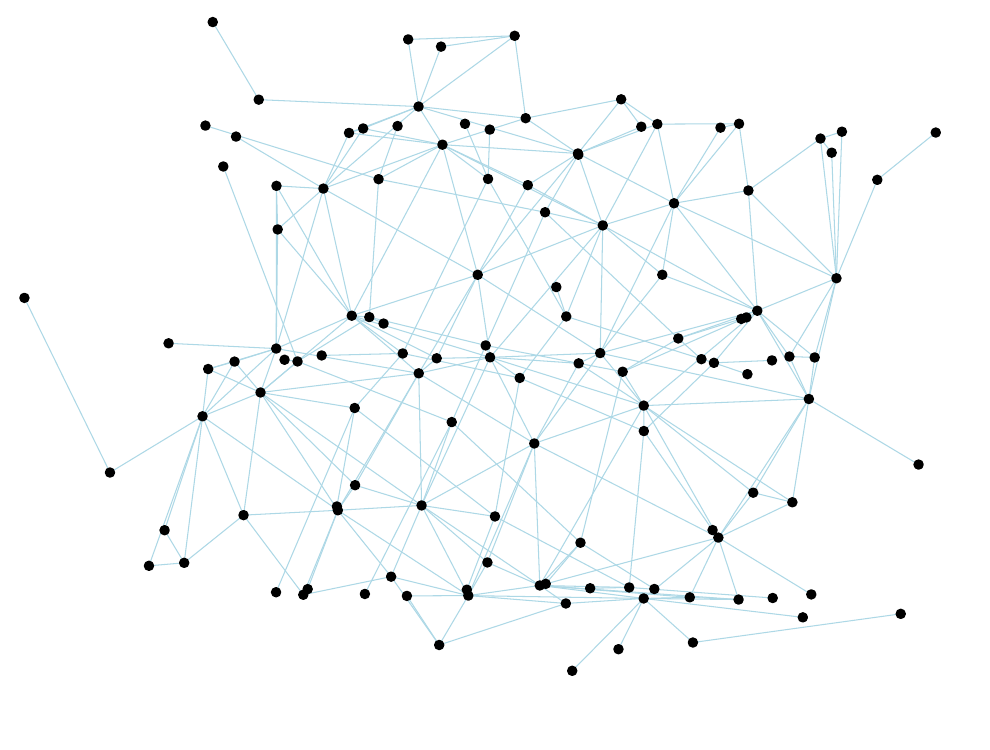} &
  \includegraphics[width=9.5mm]{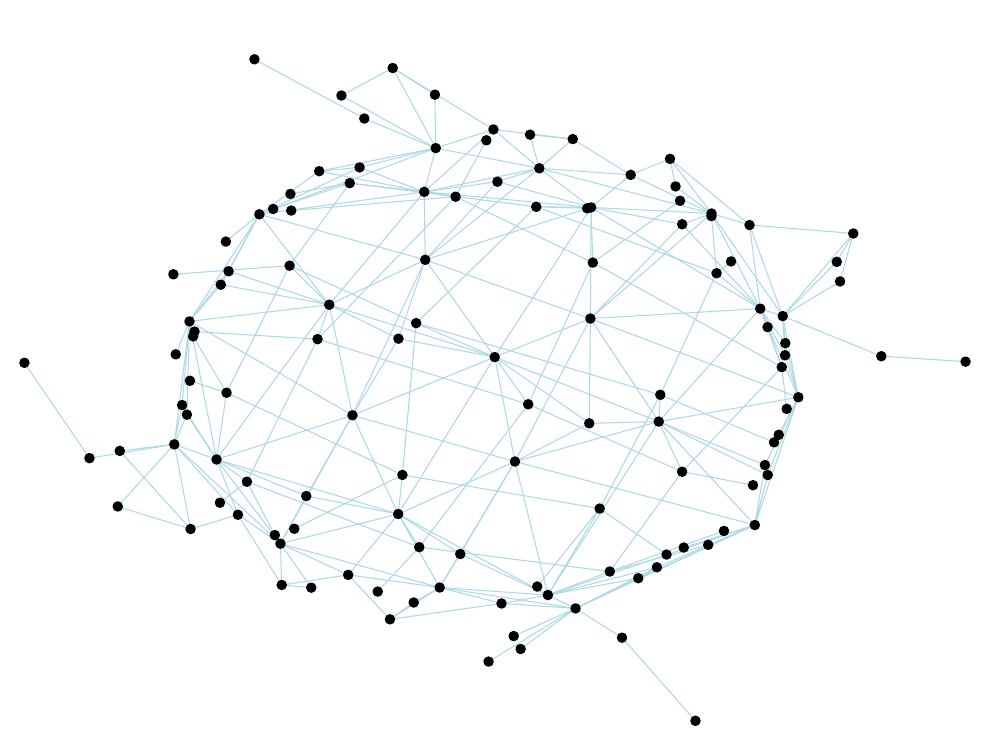} &
  \includegraphics[width=9.5mm]{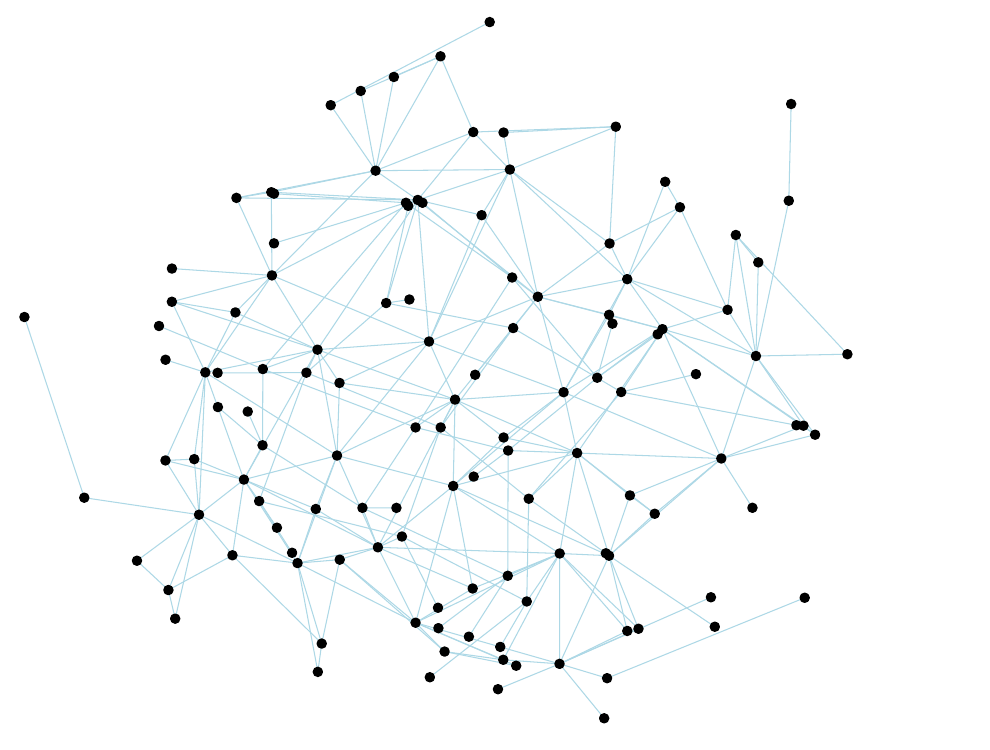} &
  \includegraphics[width=9.5mm]{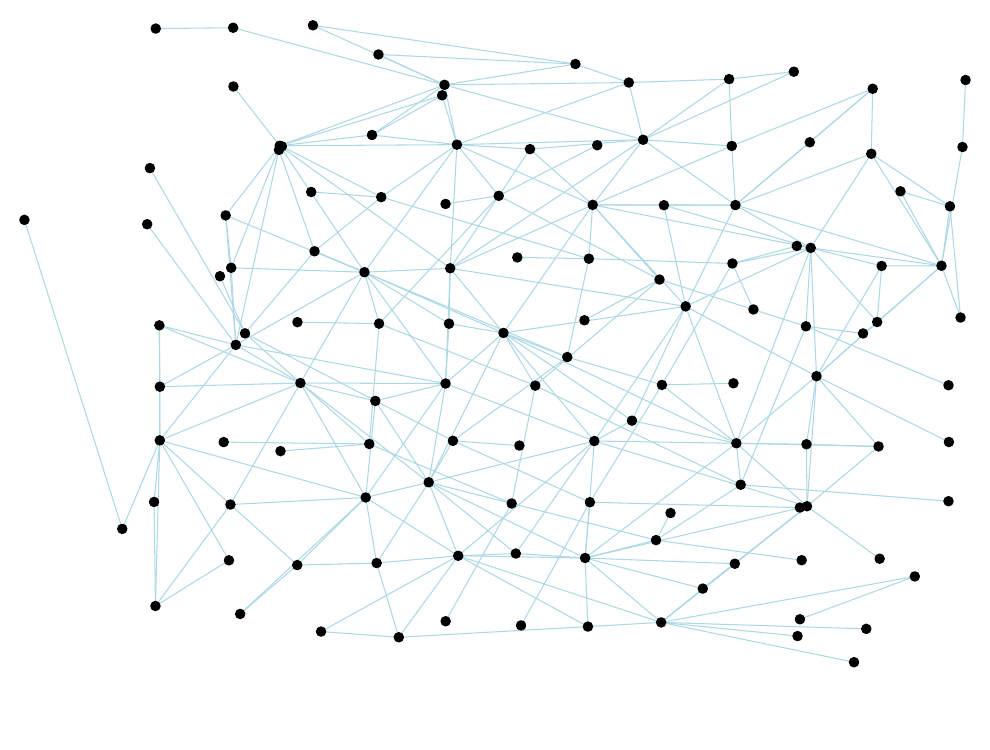}
  \\
      & \texttt{ST-AR} & & \includegraphics[width=9.5mm]{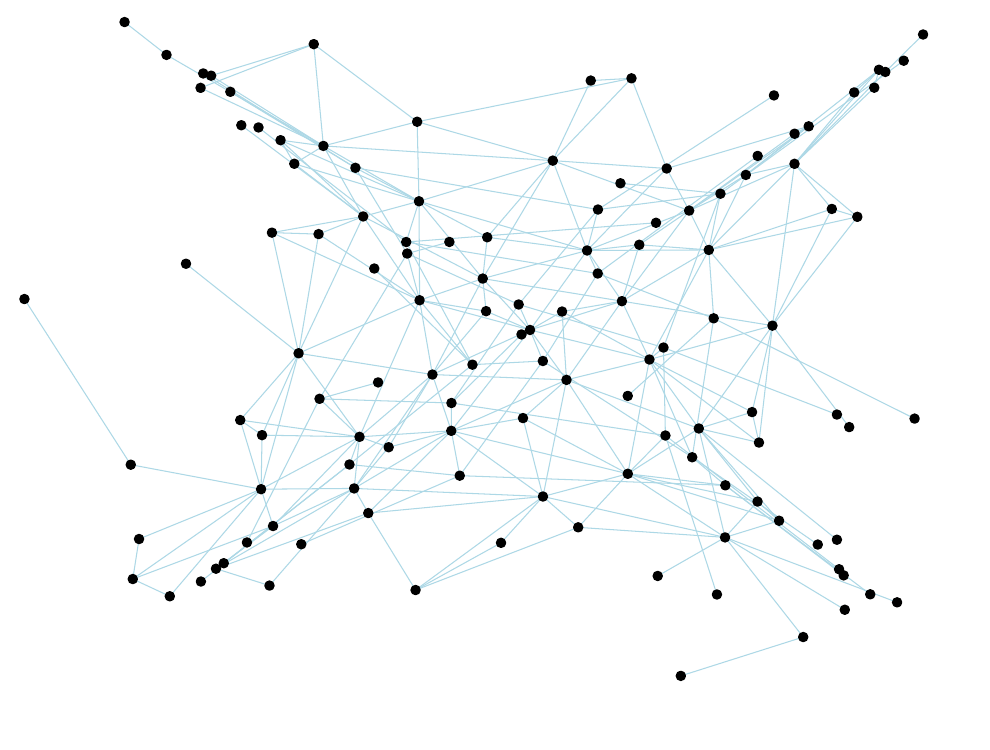} &
  \includegraphics[width=9.5mm]{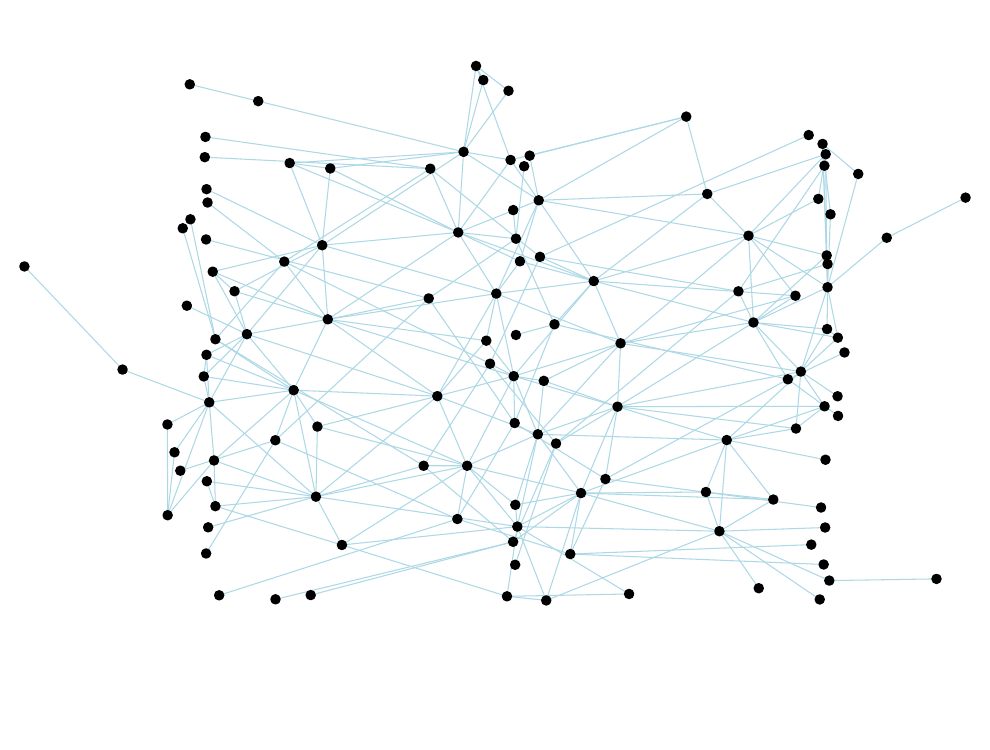} &
  \includegraphics[width=9.5mm]{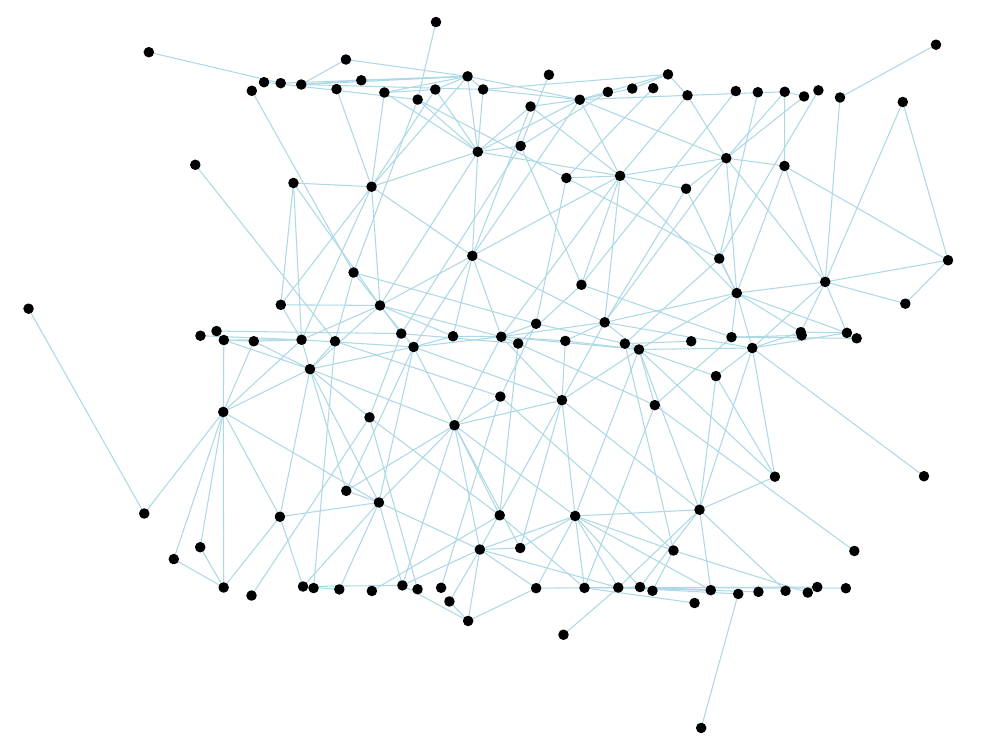} &
  \includegraphics[width=9.5mm]{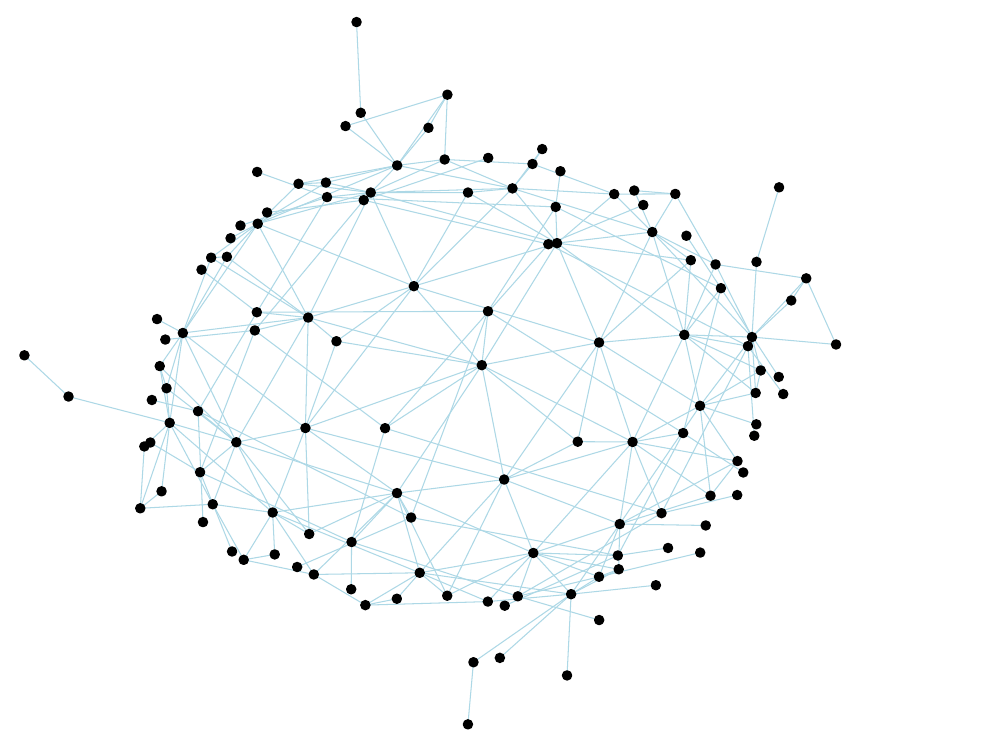} &
  \includegraphics[width=9.5mm]{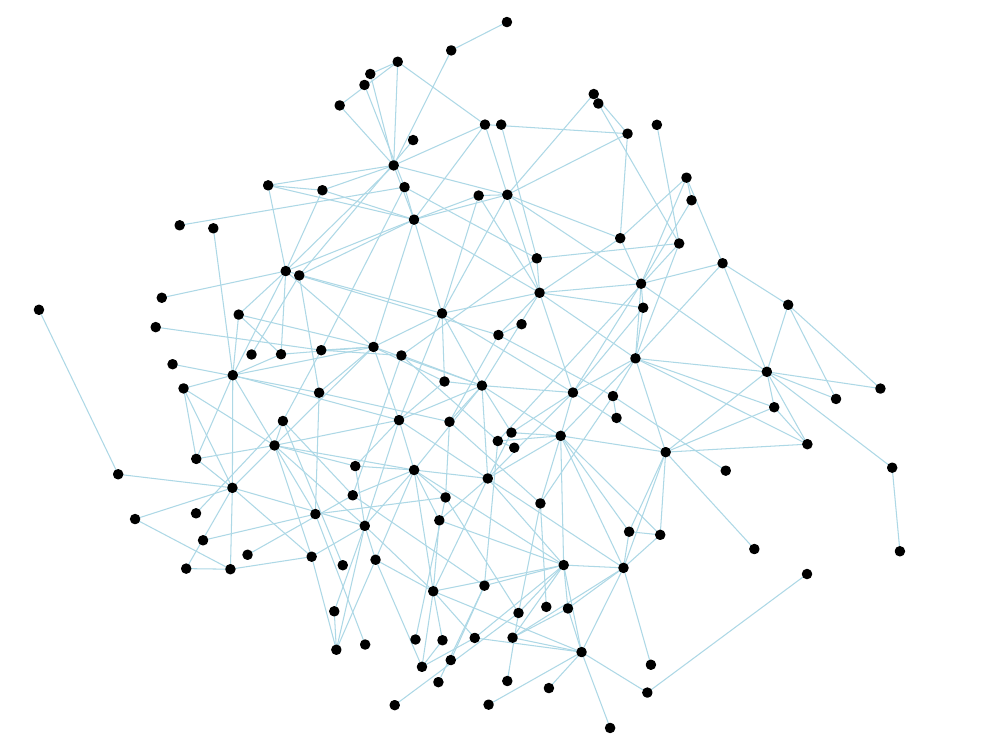} &
  \includegraphics[width=9.5mm]{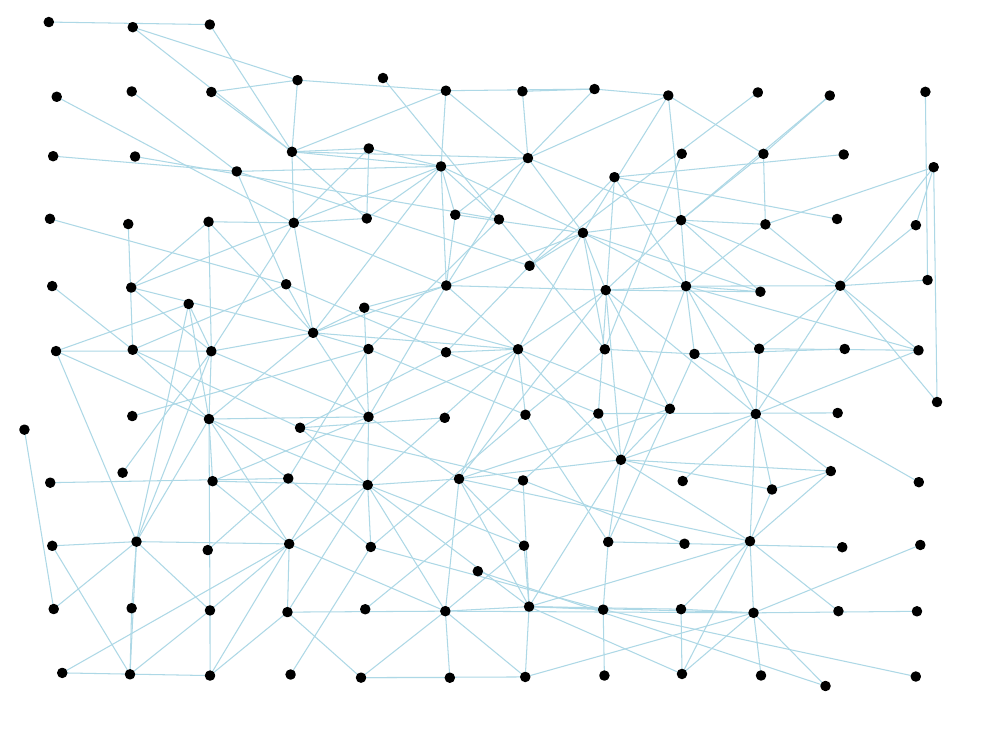}
  \\
     & \texttt{ELD-CN} & & \includegraphics[width=9.5mm]{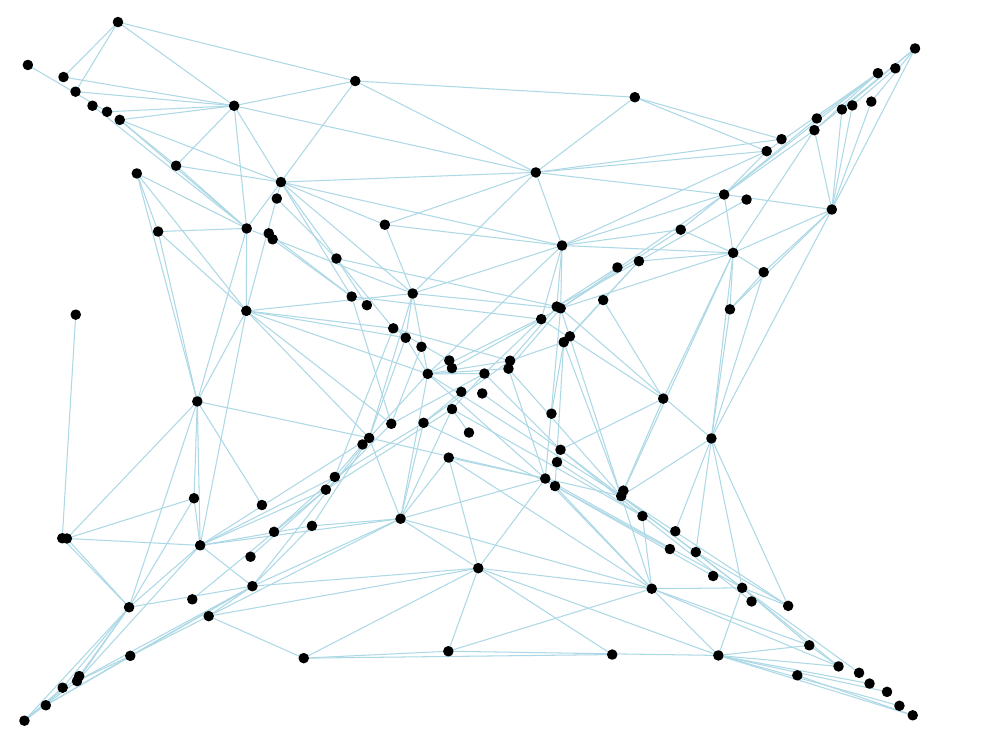} &
  \includegraphics[width=9.5mm]{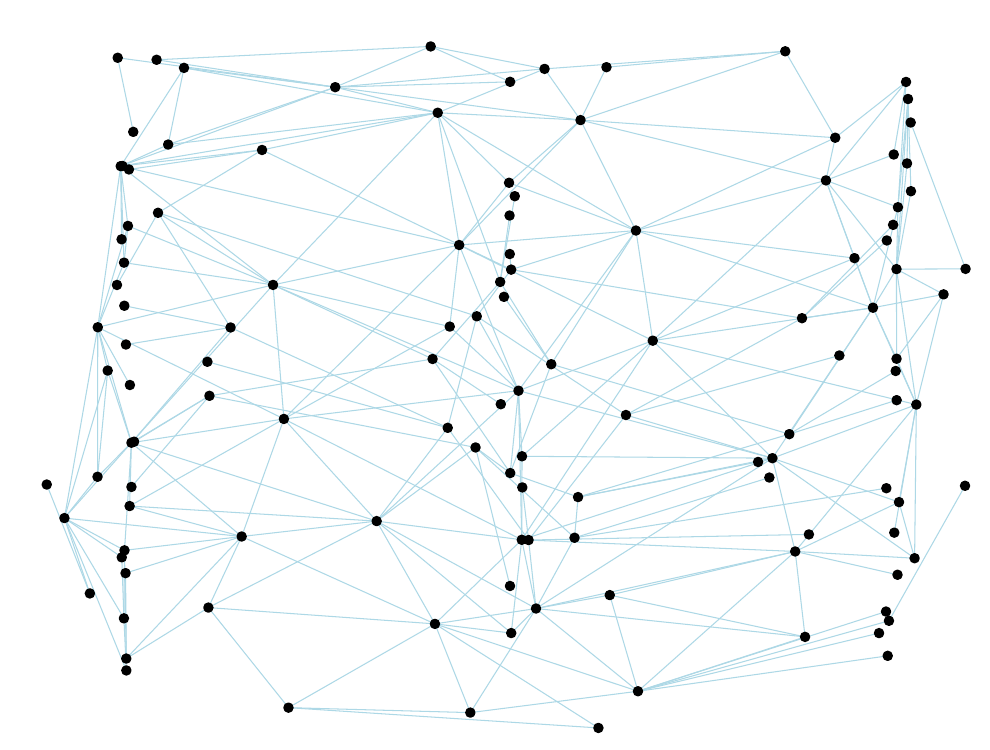} &
  \includegraphics[width=9.5mm]{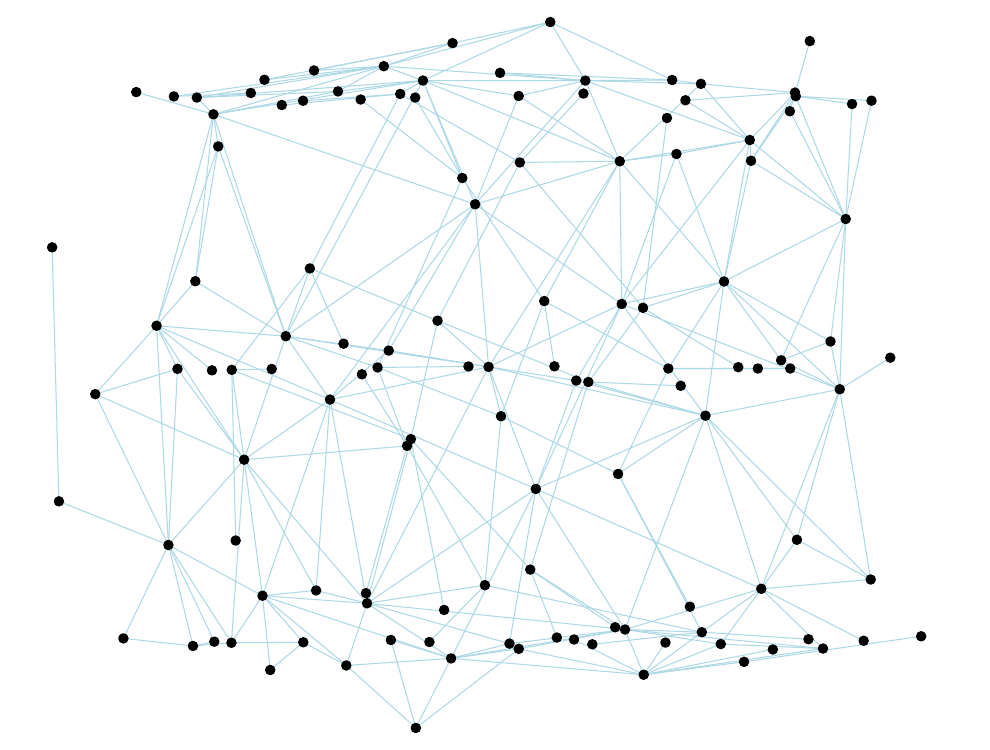} &
  \includegraphics[width=9.5mm]{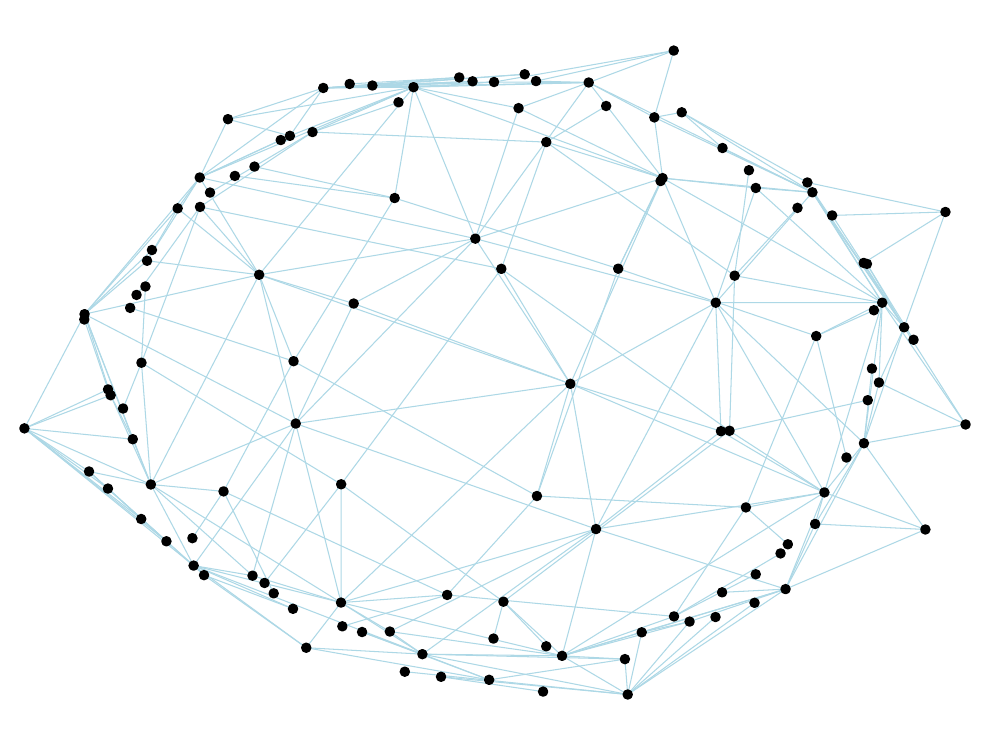} &
  \includegraphics[width=9.5mm]{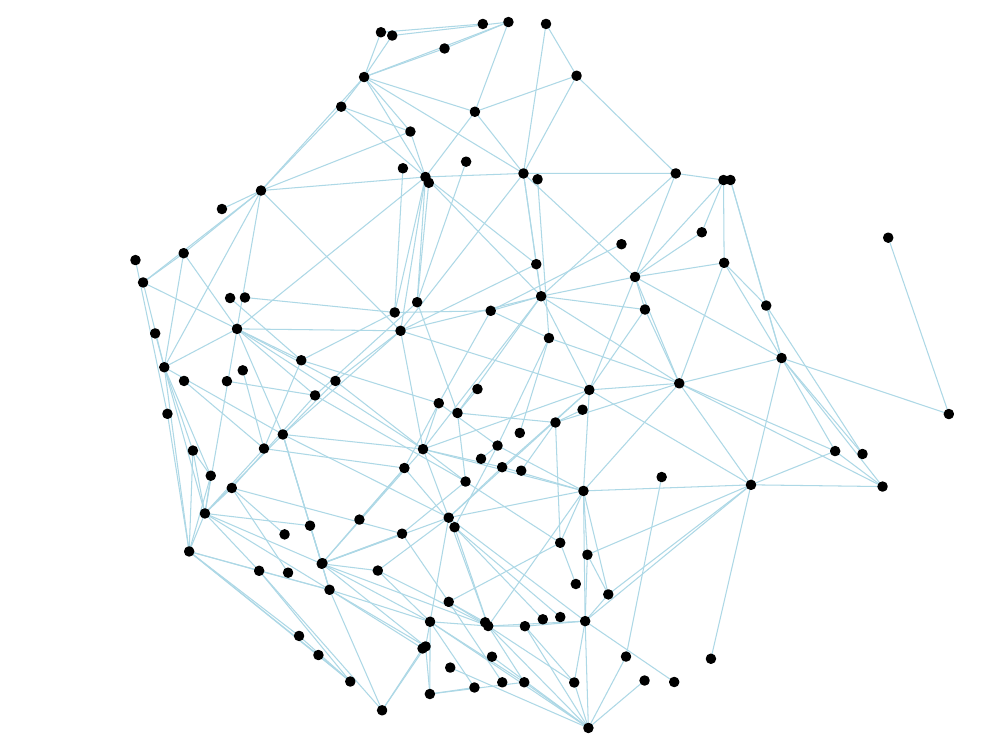} &
  \includegraphics[width=9.5mm]{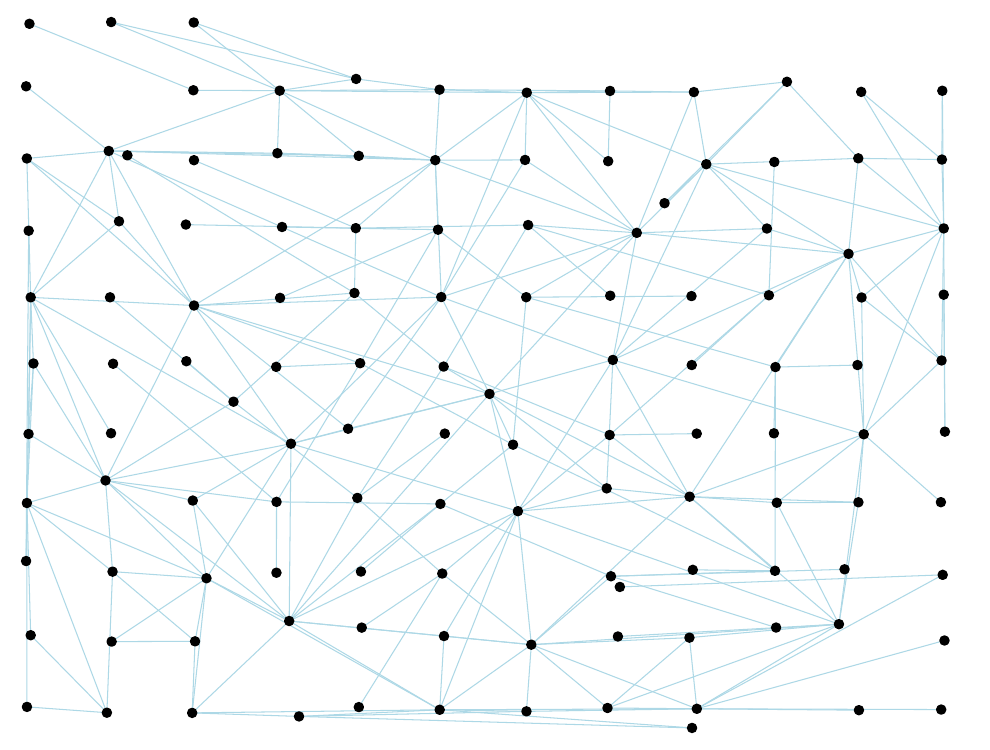}
  \\
       & \texttt{ELD-AR} & & \includegraphics[width=9.5mm]{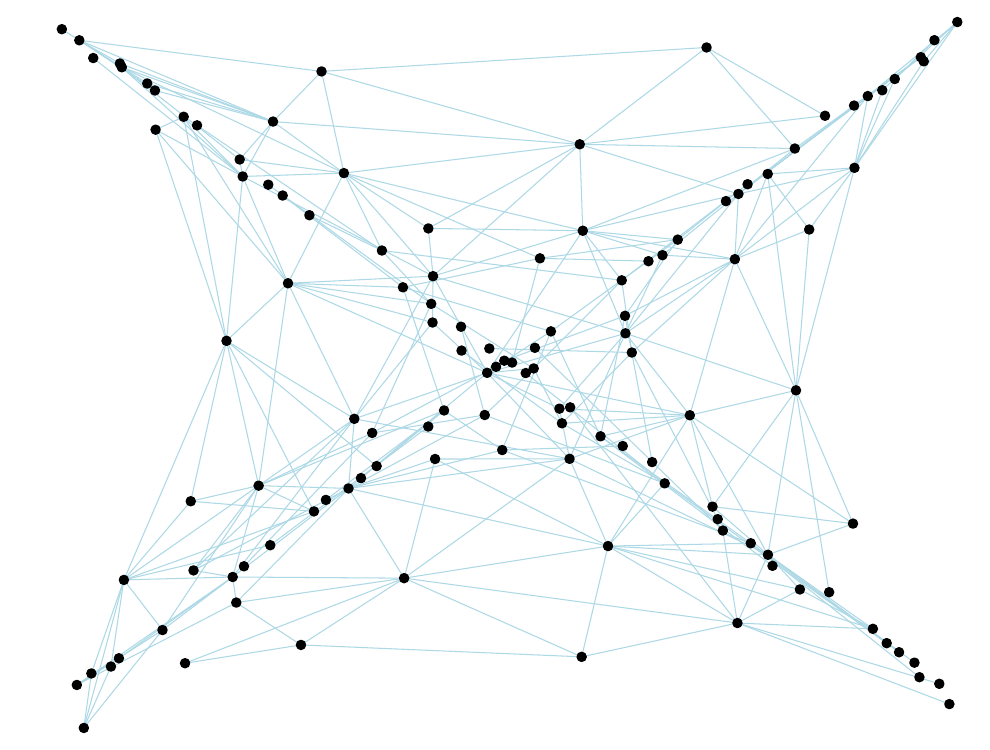} &
  \includegraphics[width=9.5mm]{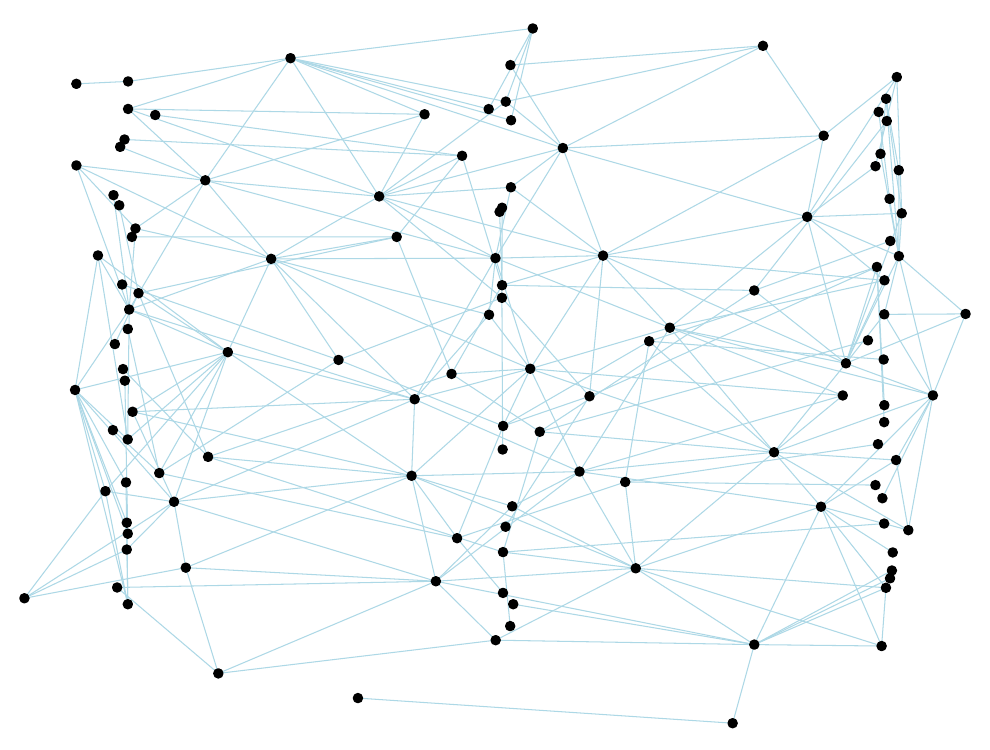} &
  \includegraphics[width=9.5mm]{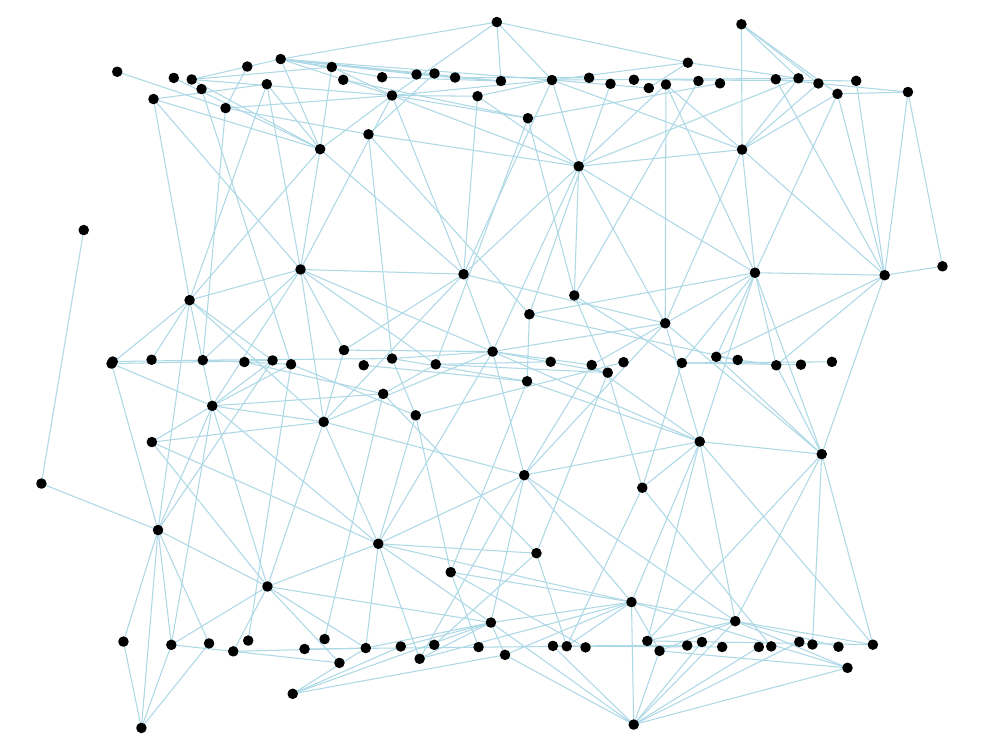} &
  \includegraphics[width=9.5mm]{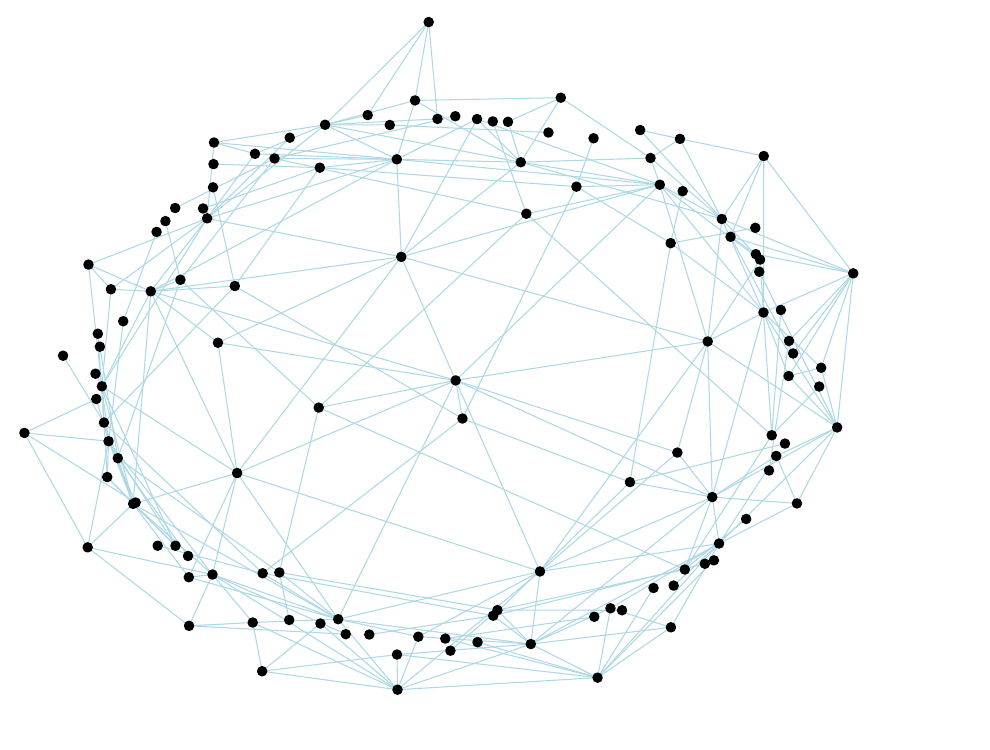} &
  \includegraphics[width=9.5mm]{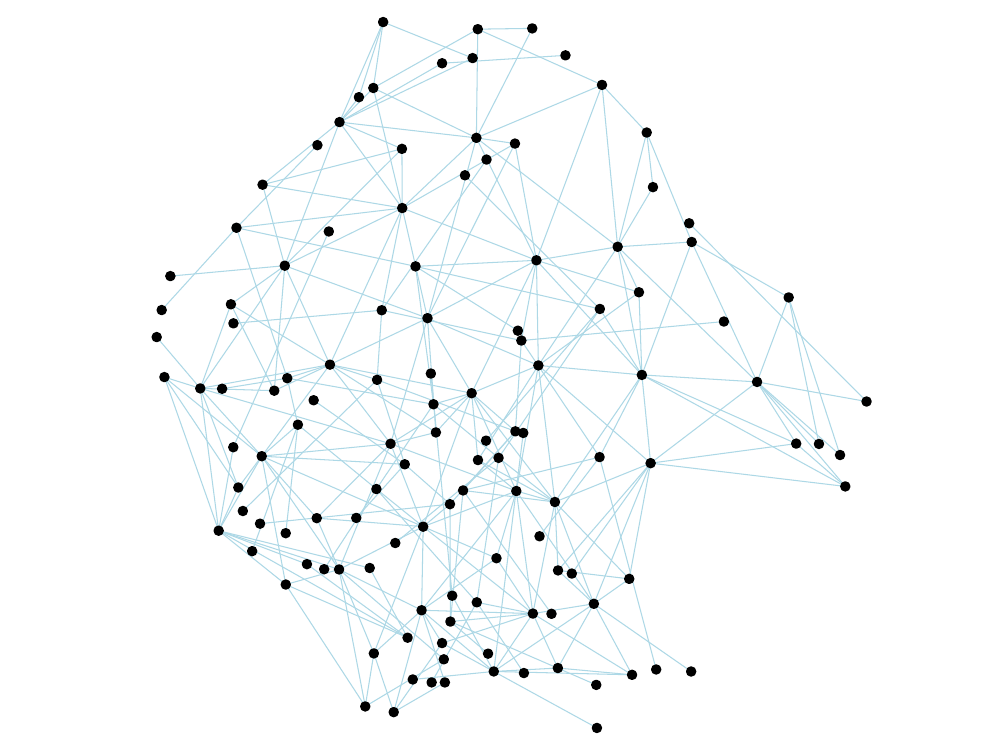} &
  \includegraphics[width=9.5mm]{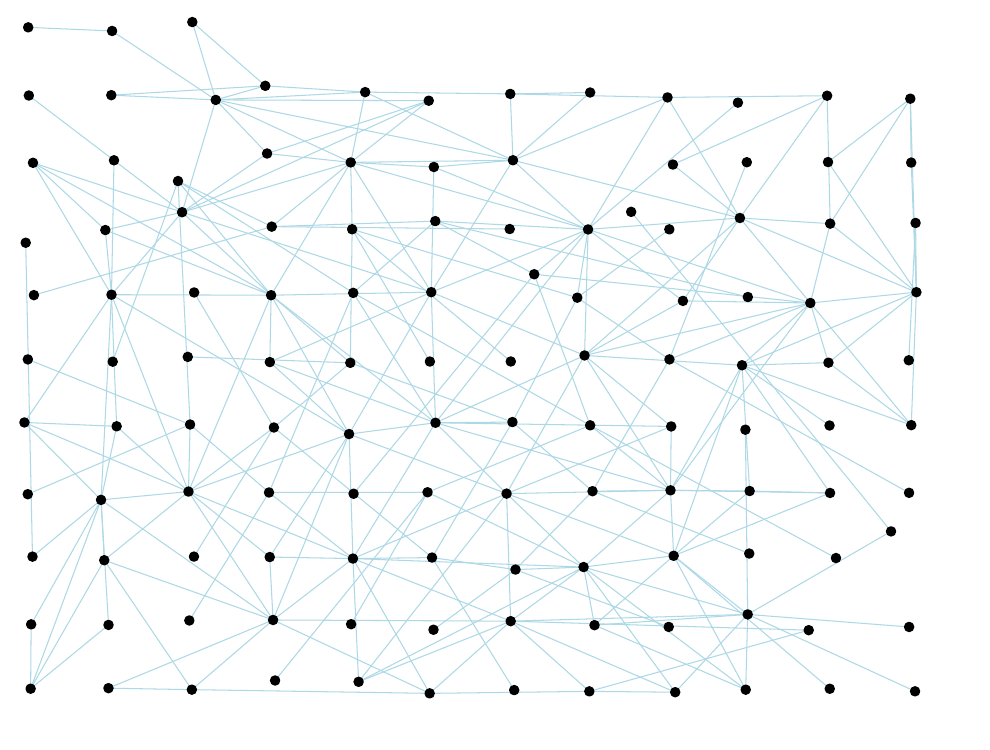}
  \\
       & \texttt{CN-AR} & & \includegraphics[width=9.5mm]{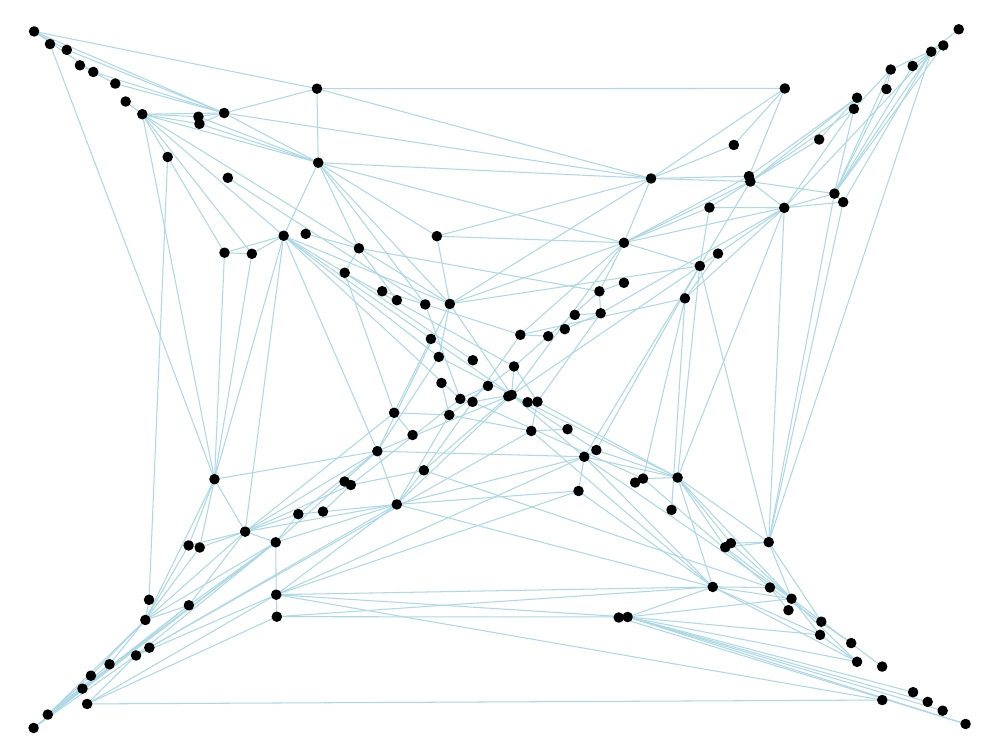} &
  \includegraphics[width=9.5mm]{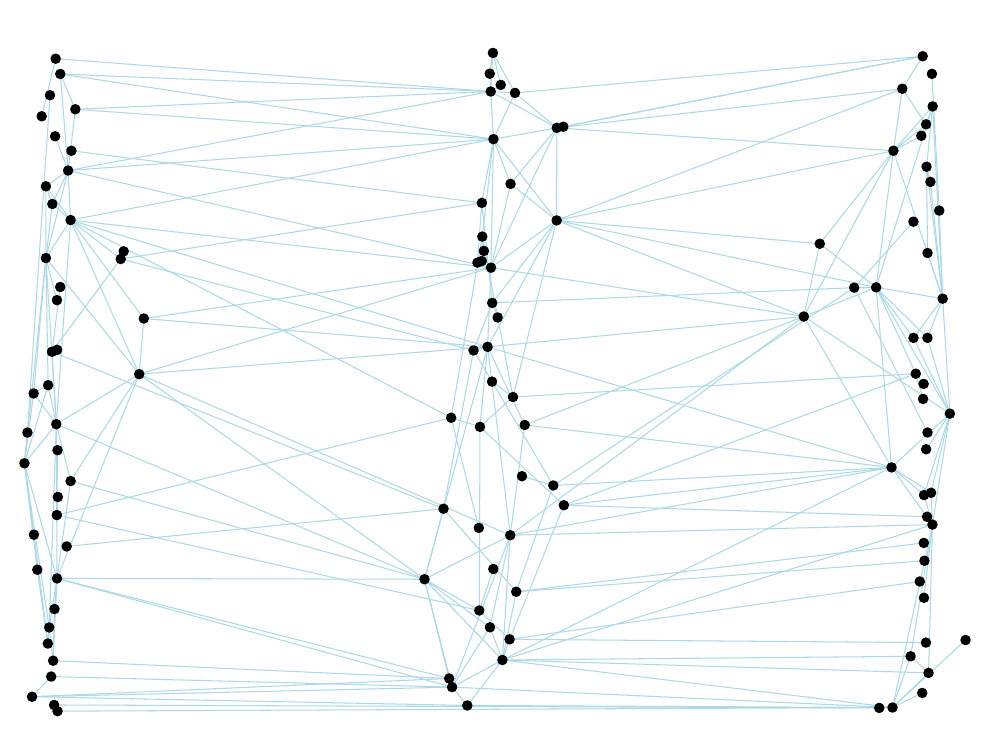} &
  \includegraphics[width=9.5mm]{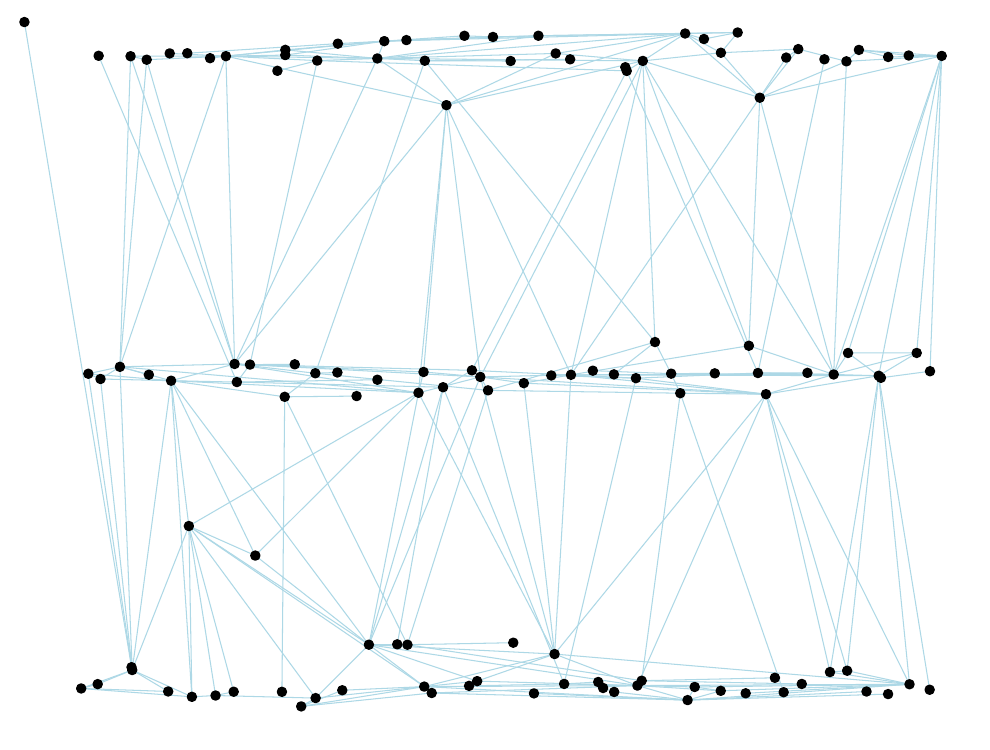} &
  \includegraphics[width=9.5mm]{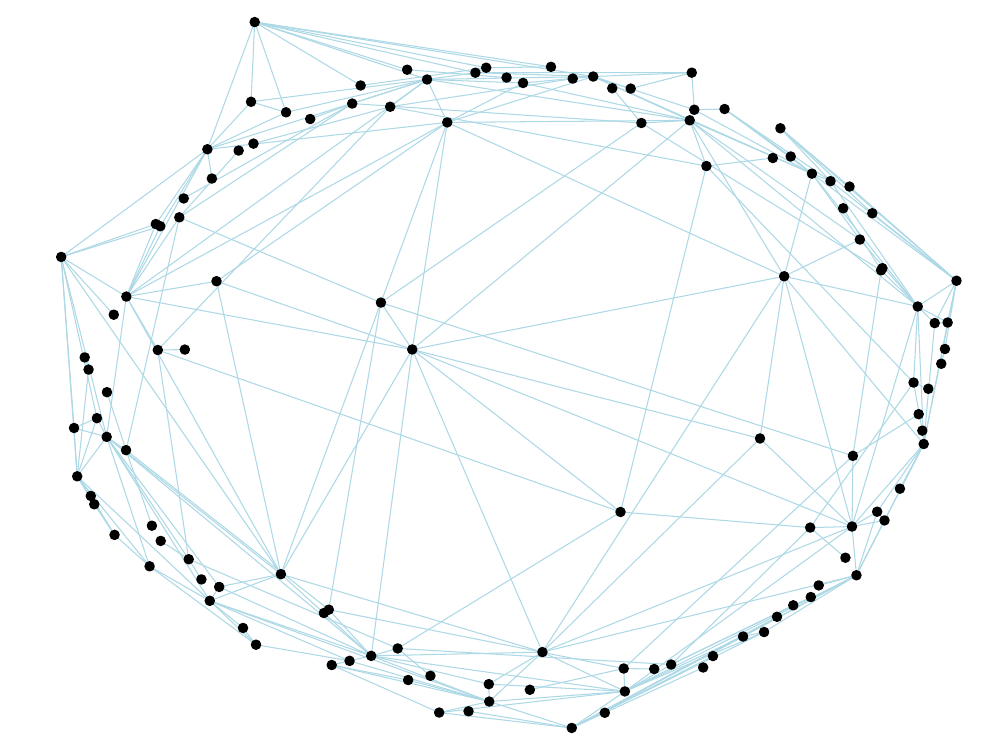} &
  \includegraphics[width=9.5mm]{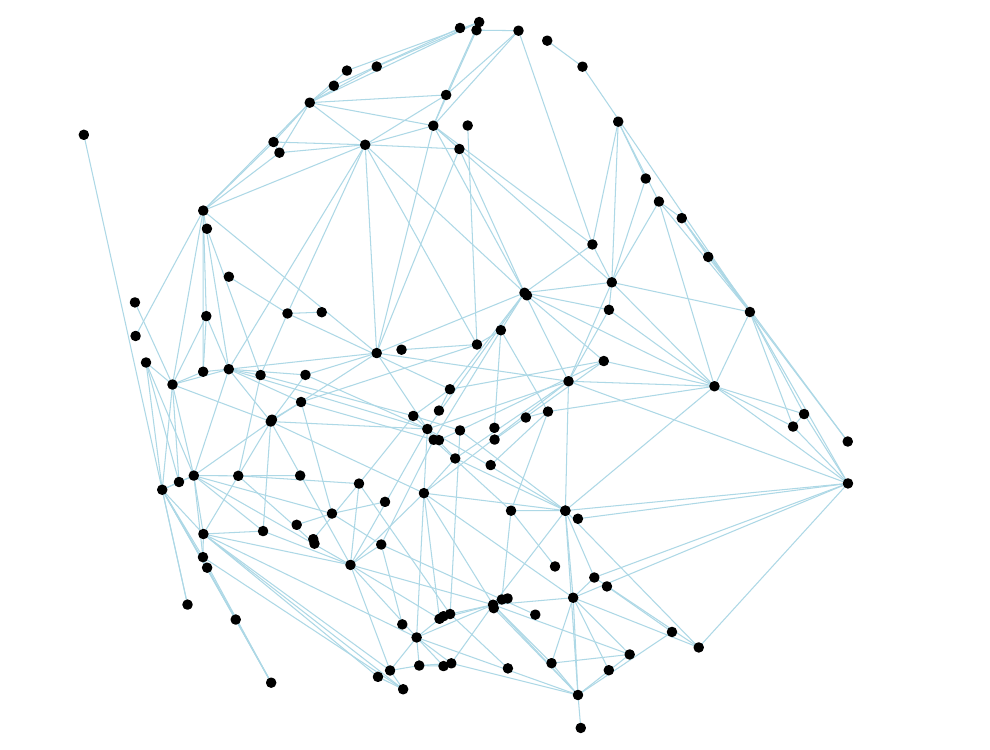} &
  \includegraphics[width=9.5mm]{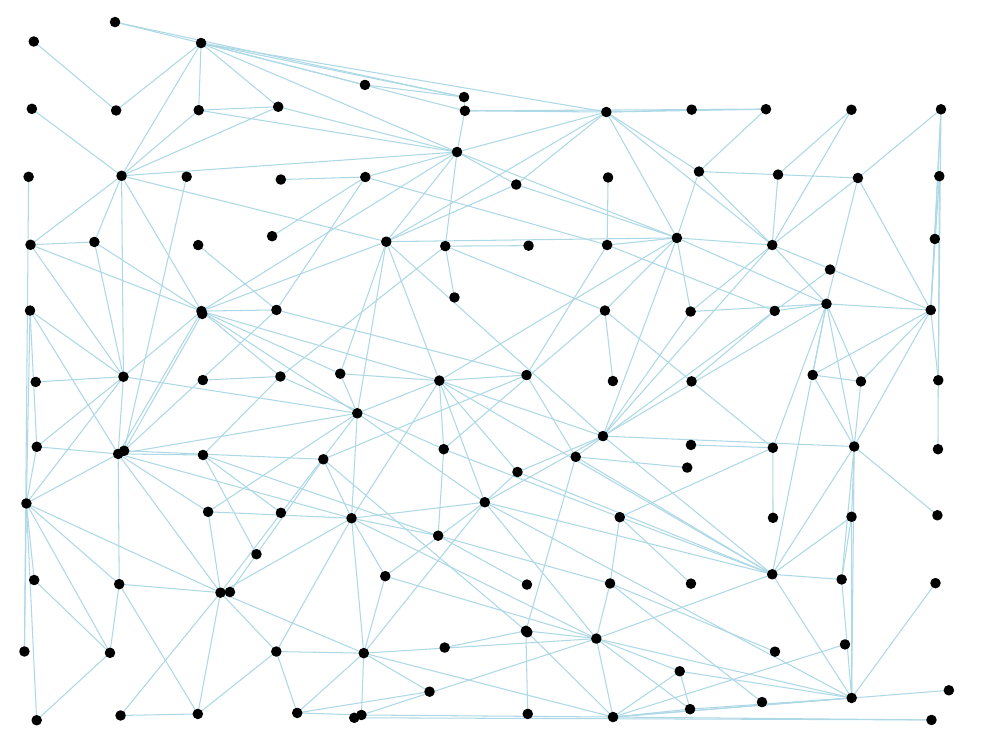}
  \\
       & \texttt{ST-ELD-CN} & &\includegraphics[width=9.5mm]{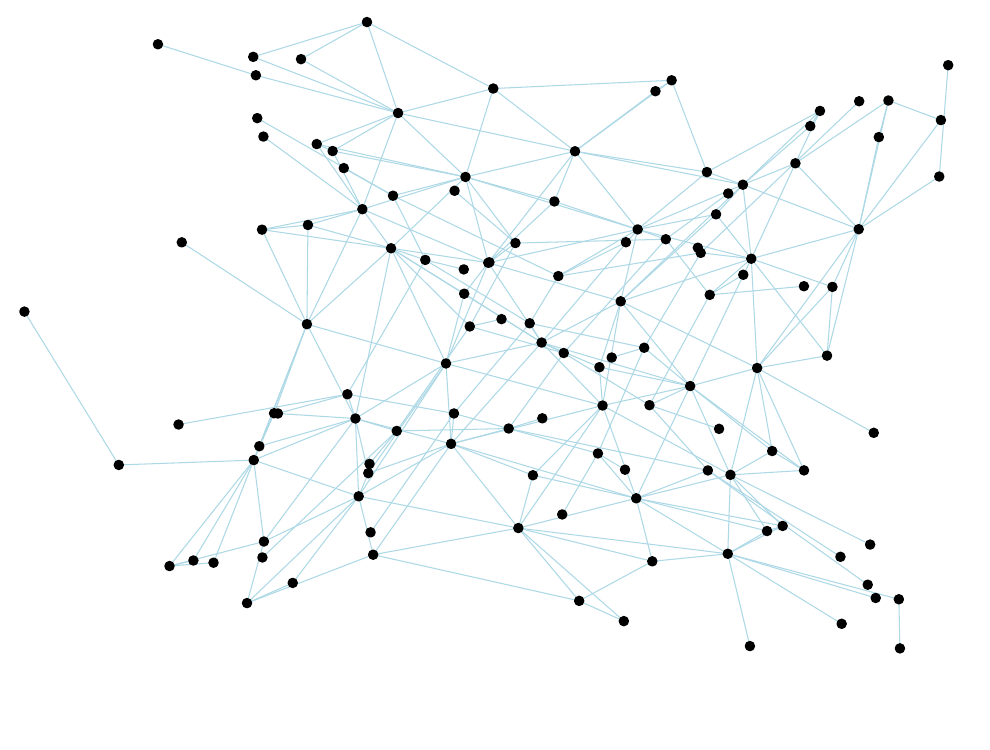} &
  \includegraphics[width=9.5mm]{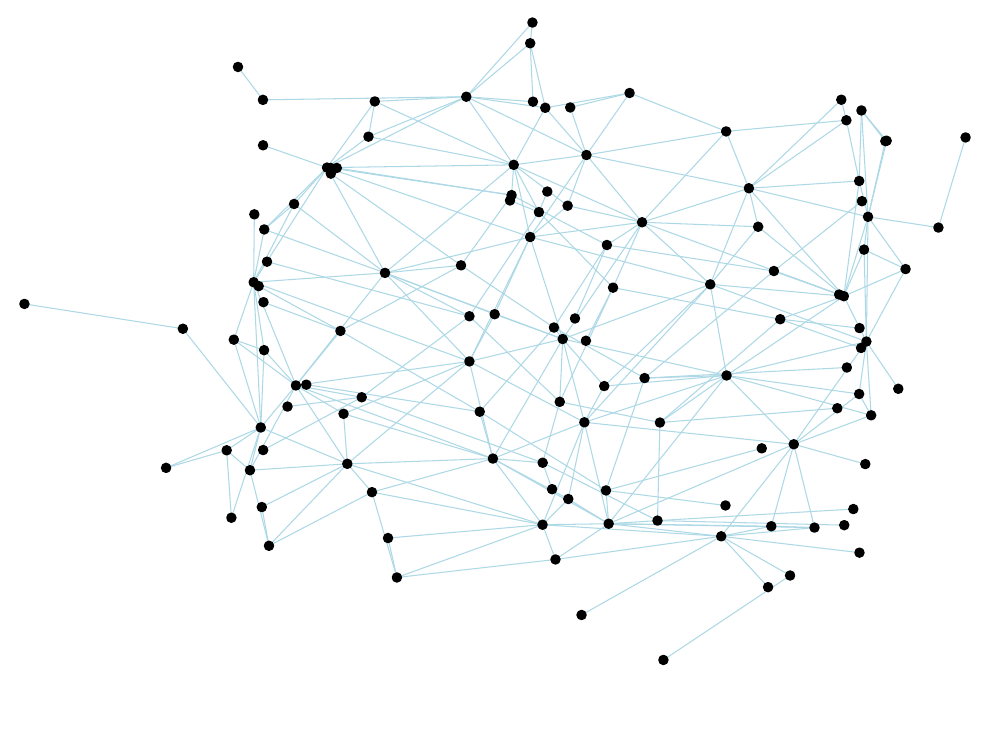} &
  \includegraphics[width=9.5mm]{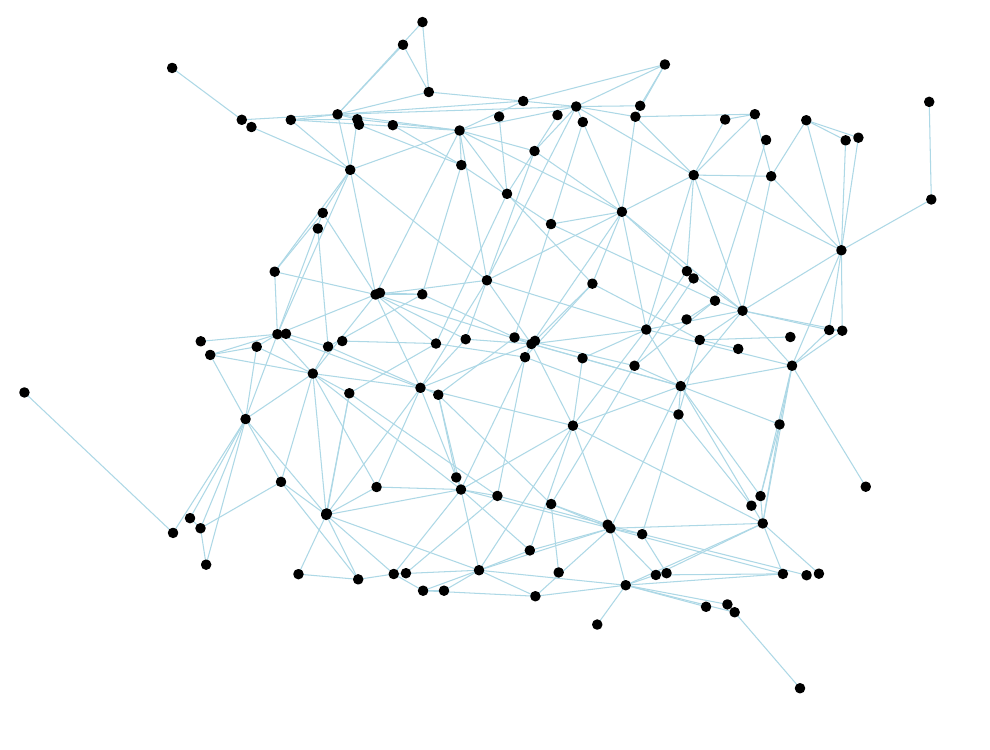} &
  \includegraphics[width=9.5mm]{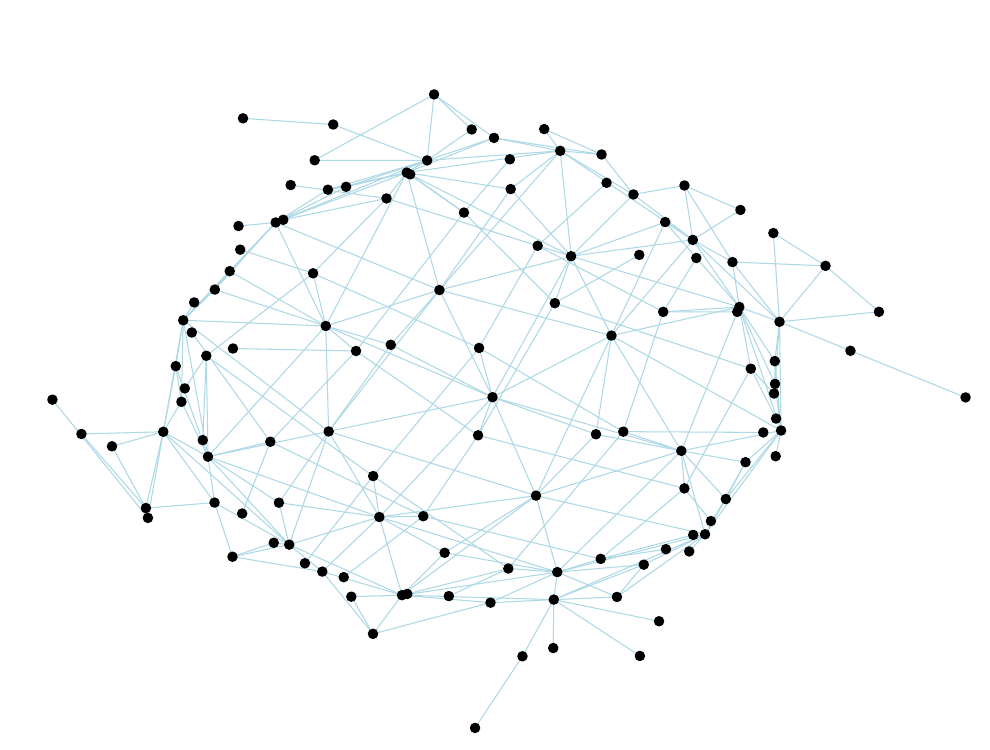} &
  \includegraphics[width=9.5mm]{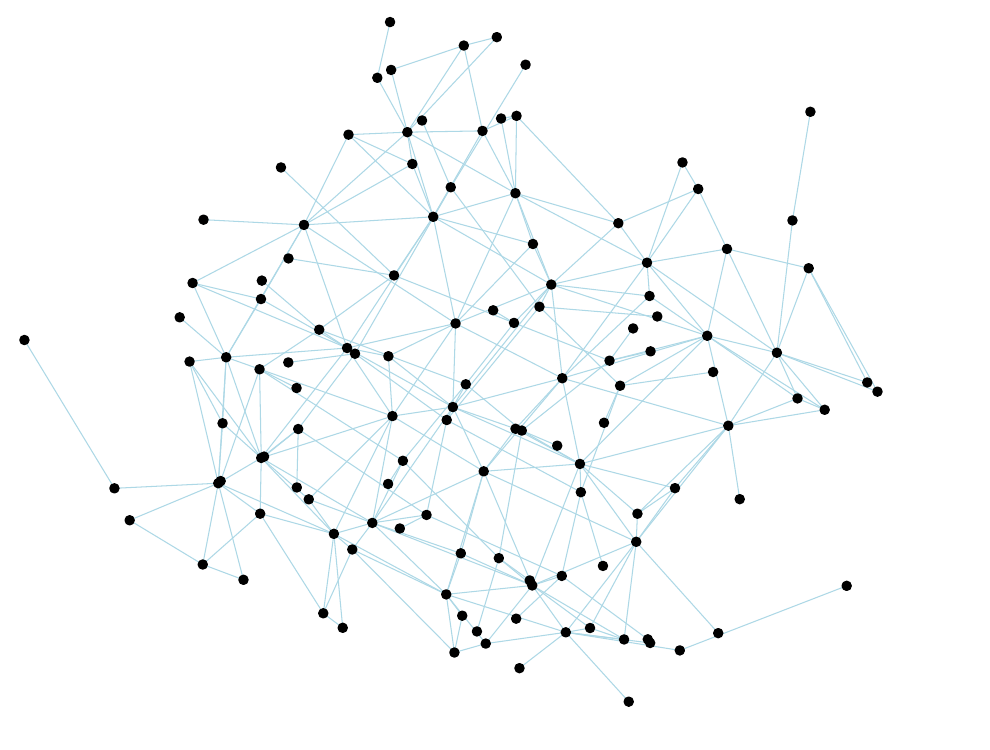} &
  \includegraphics[width=9.5mm]{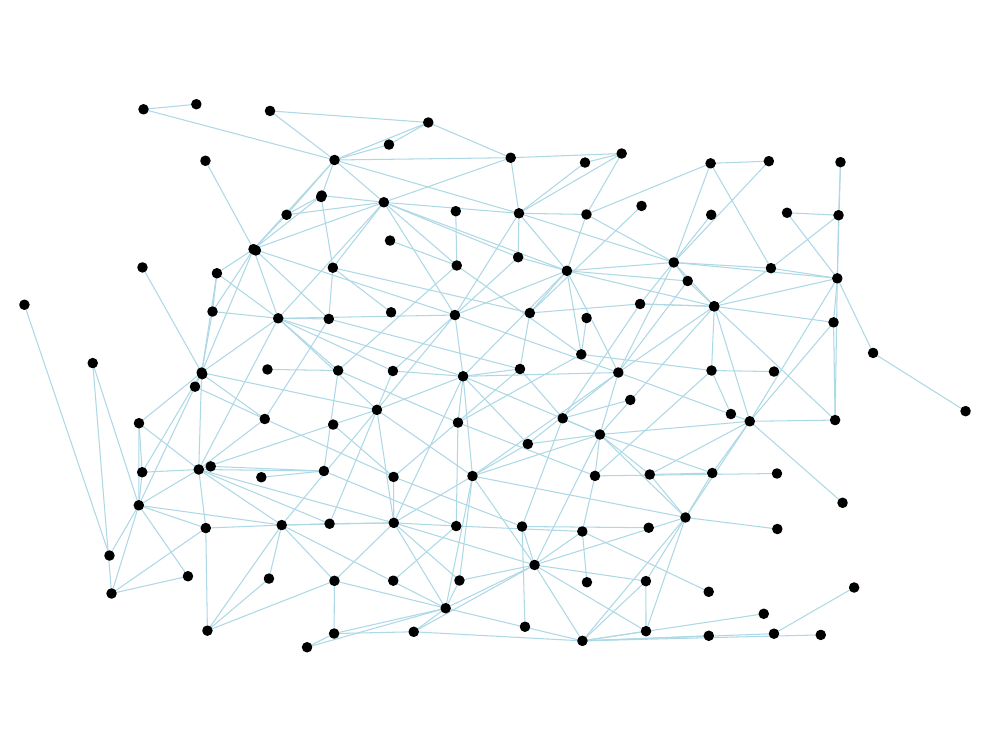}
  \\
         & \texttt{ST-ELD-AR} & &\includegraphics[width=9.5mm]{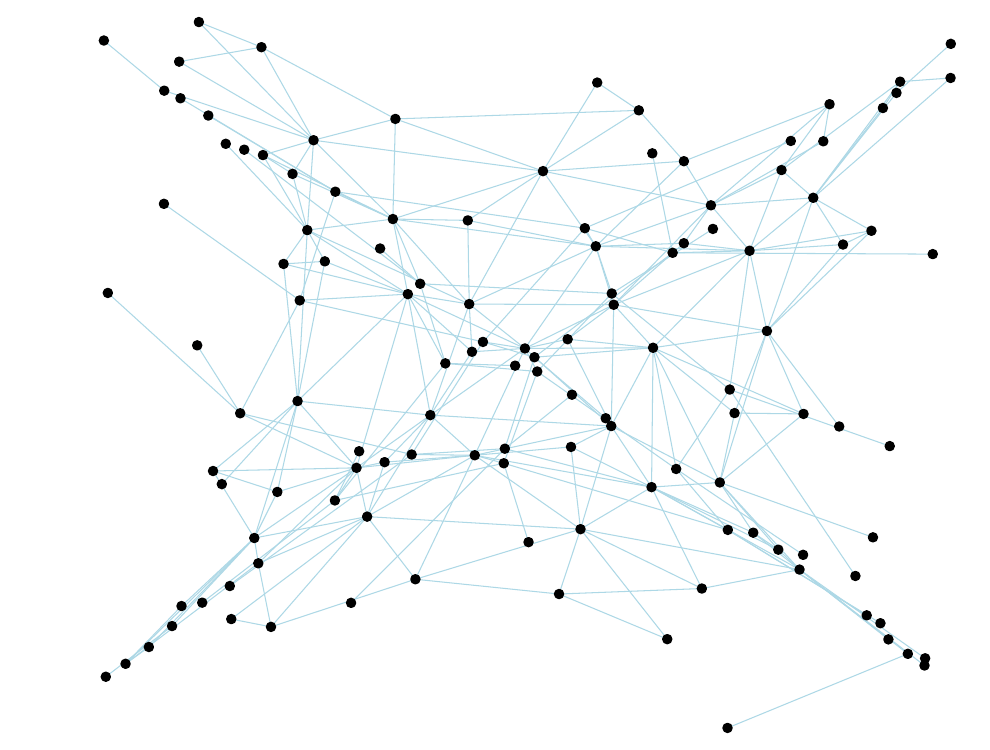} &
  \includegraphics[width=9.5mm]{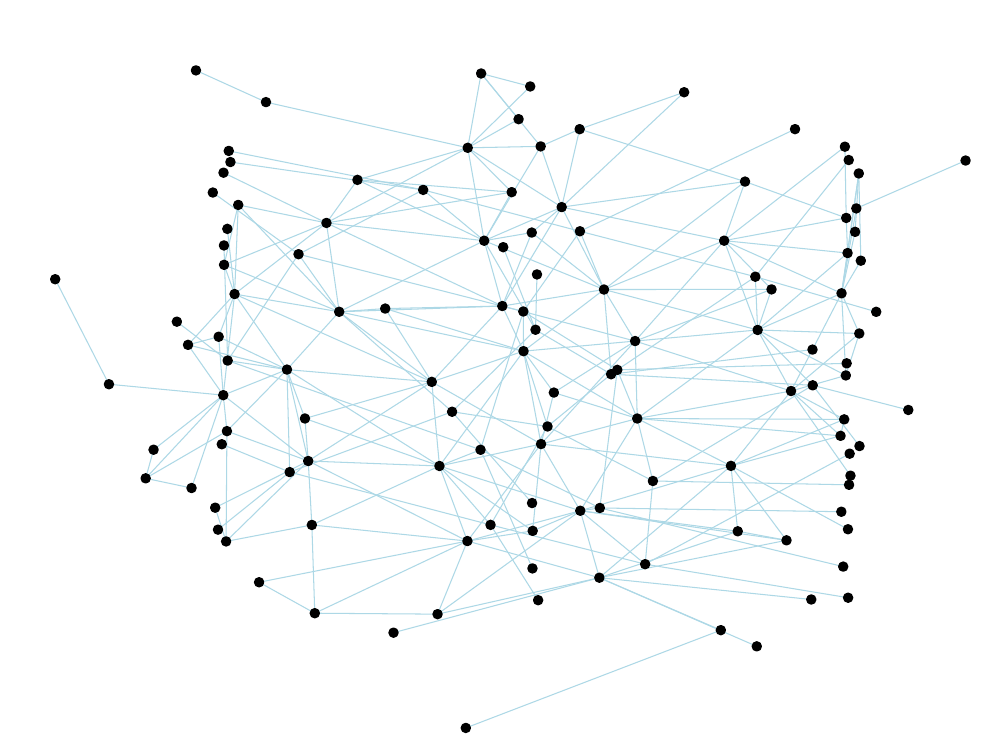} &
  \includegraphics[width=9.5mm]{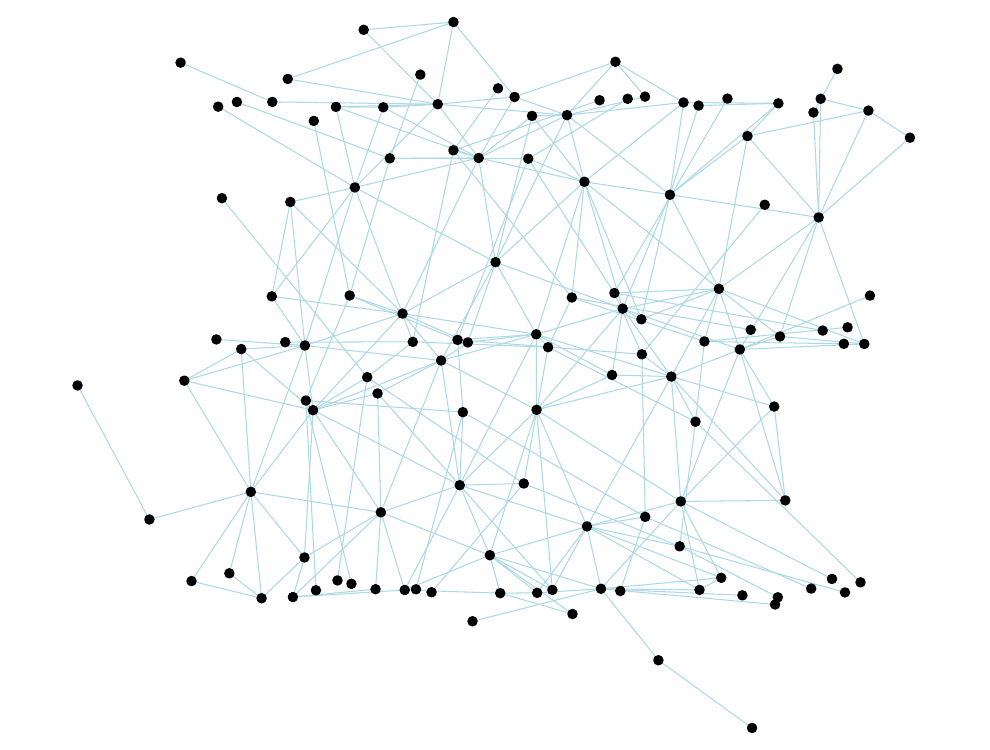} &
  \includegraphics[width=9.5mm]{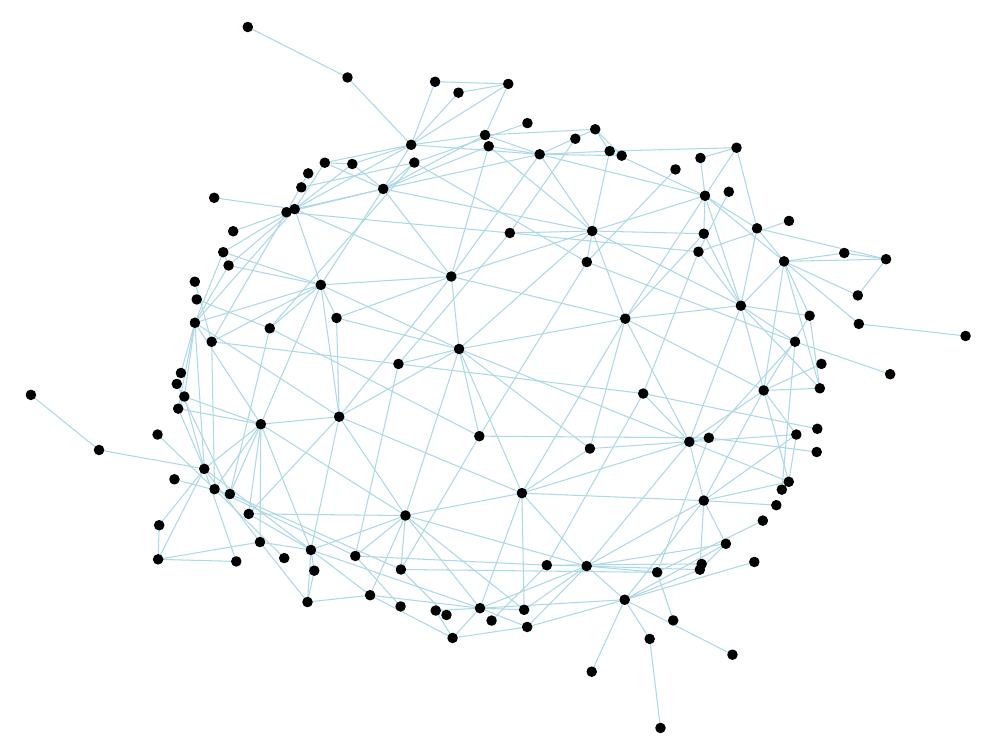} &
  \includegraphics[width=9.5mm]{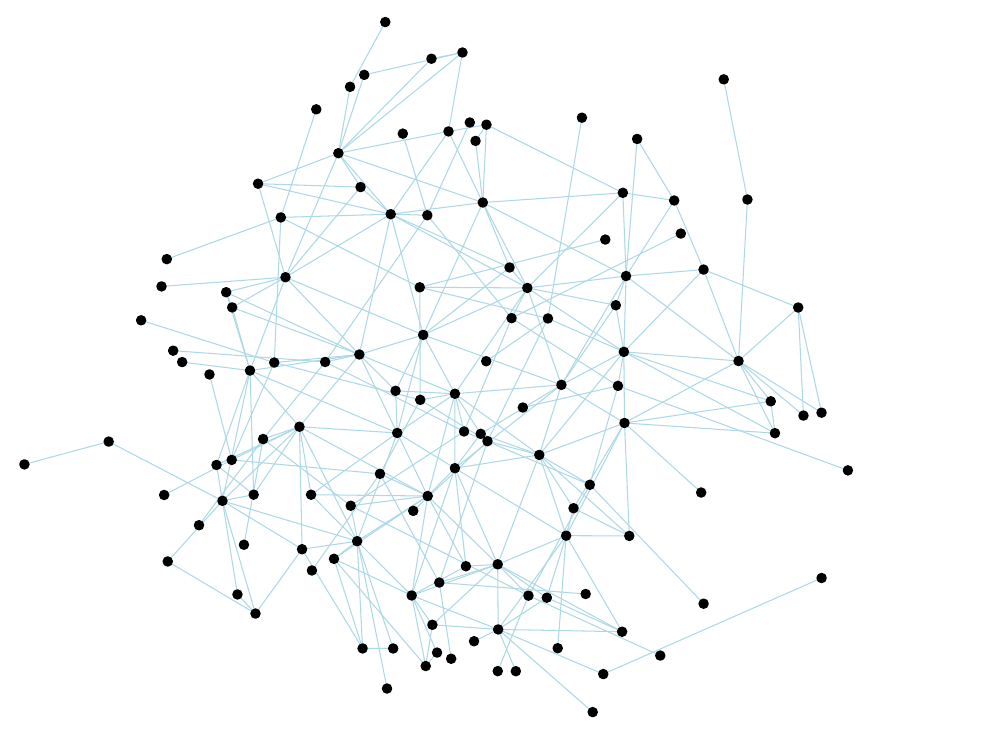} &
  \includegraphics[width=9.5mm]{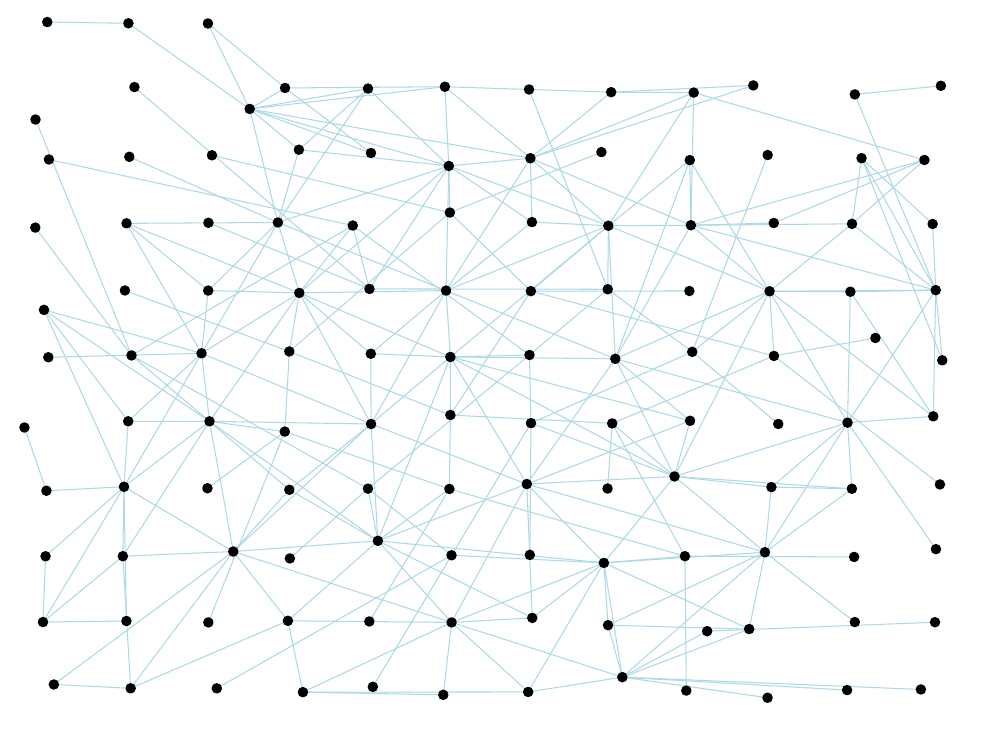}
  \\
         & \texttt{ST-CN-AR} & &\includegraphics[width=9.5mm]{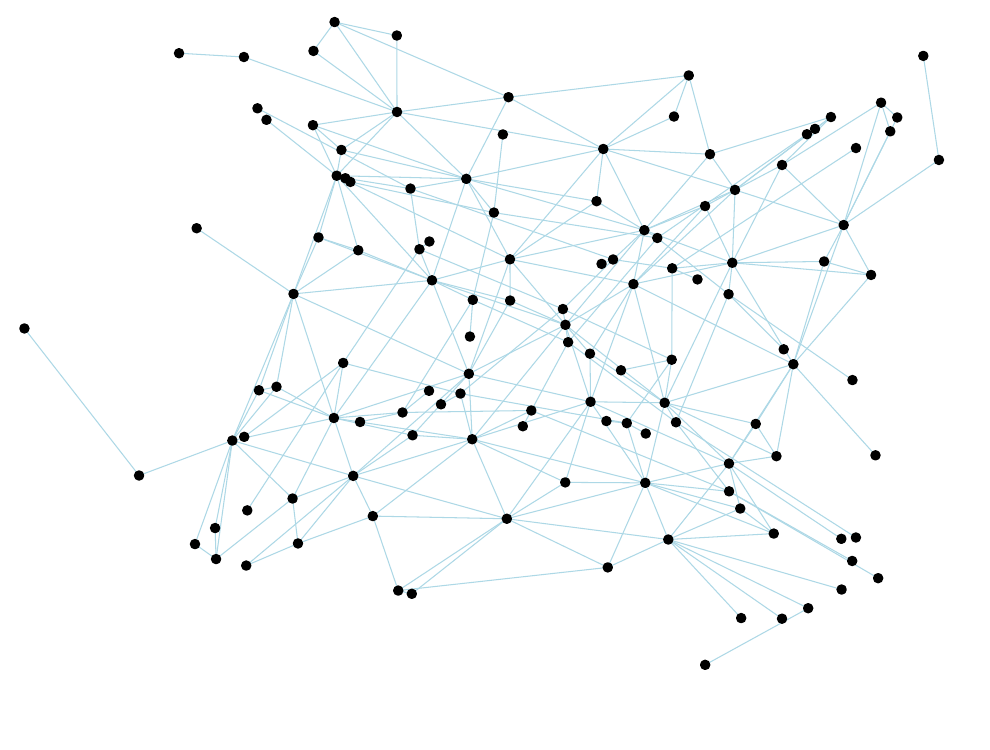} &
  \includegraphics[width=9.5mm]{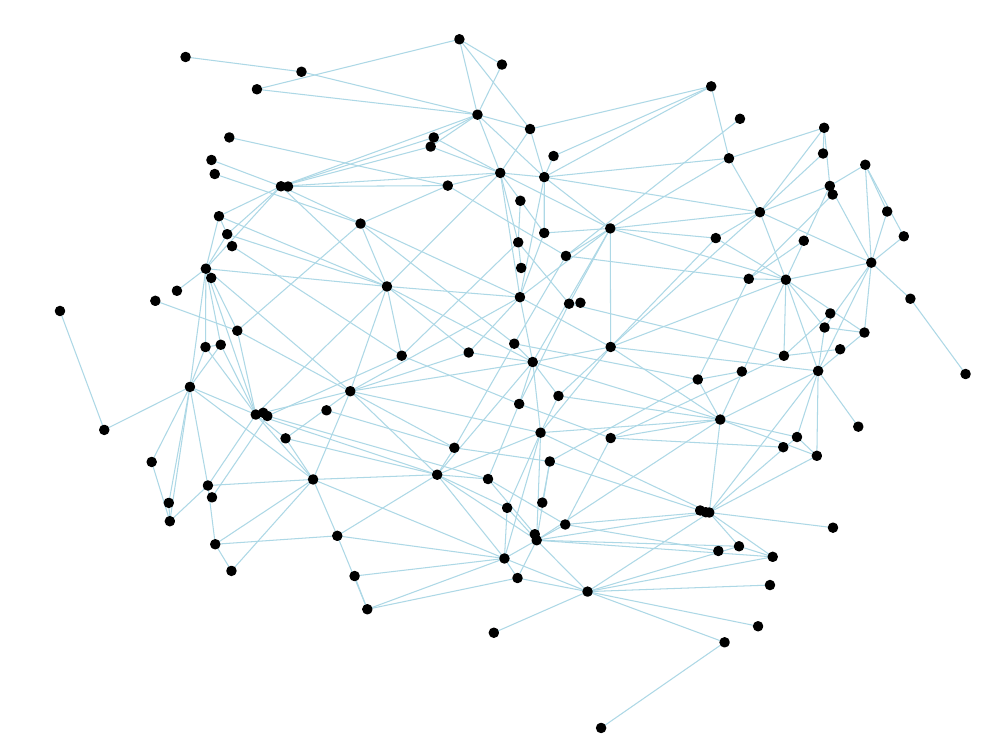} &
  \includegraphics[width=9.5mm]{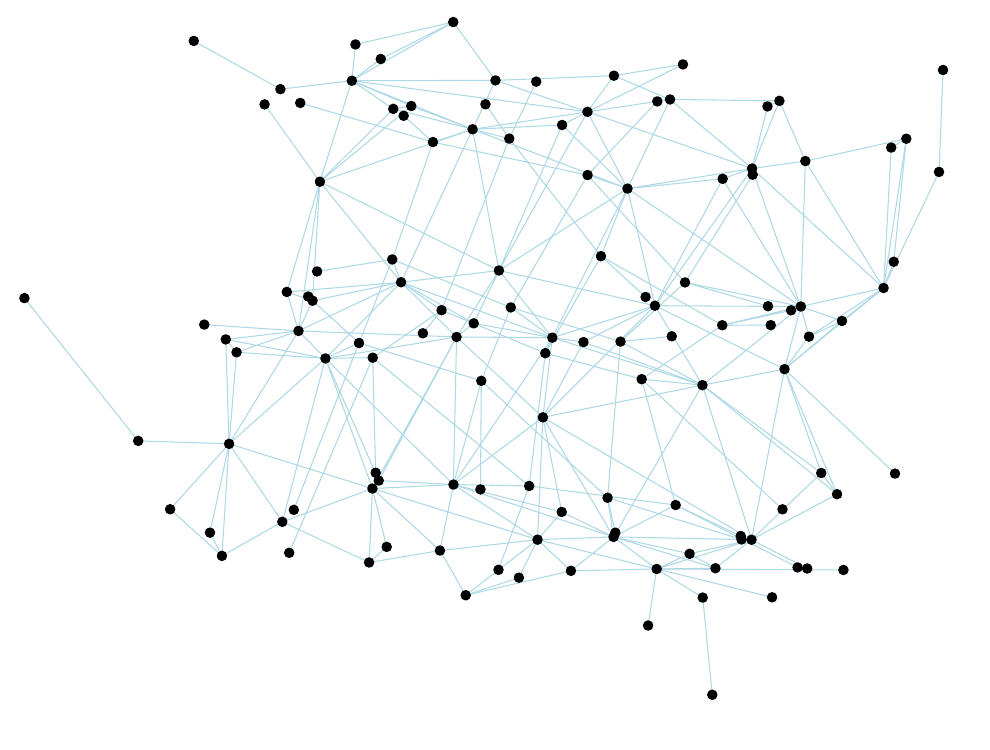} &
  \includegraphics[width=9.5mm]{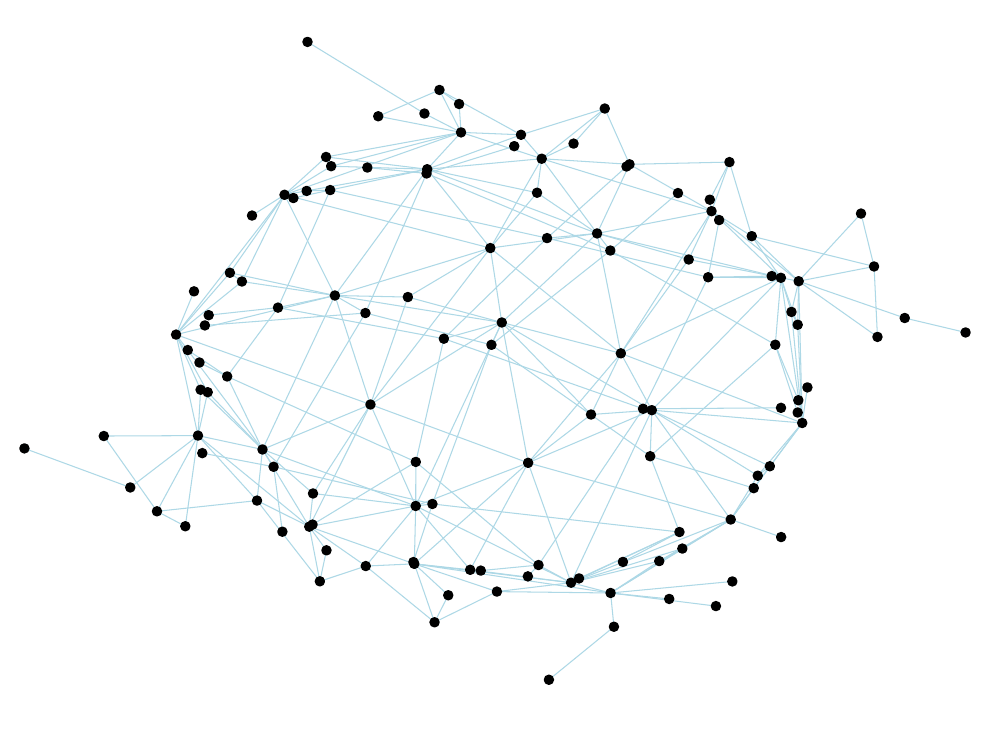} &
  \includegraphics[width=9.5mm]{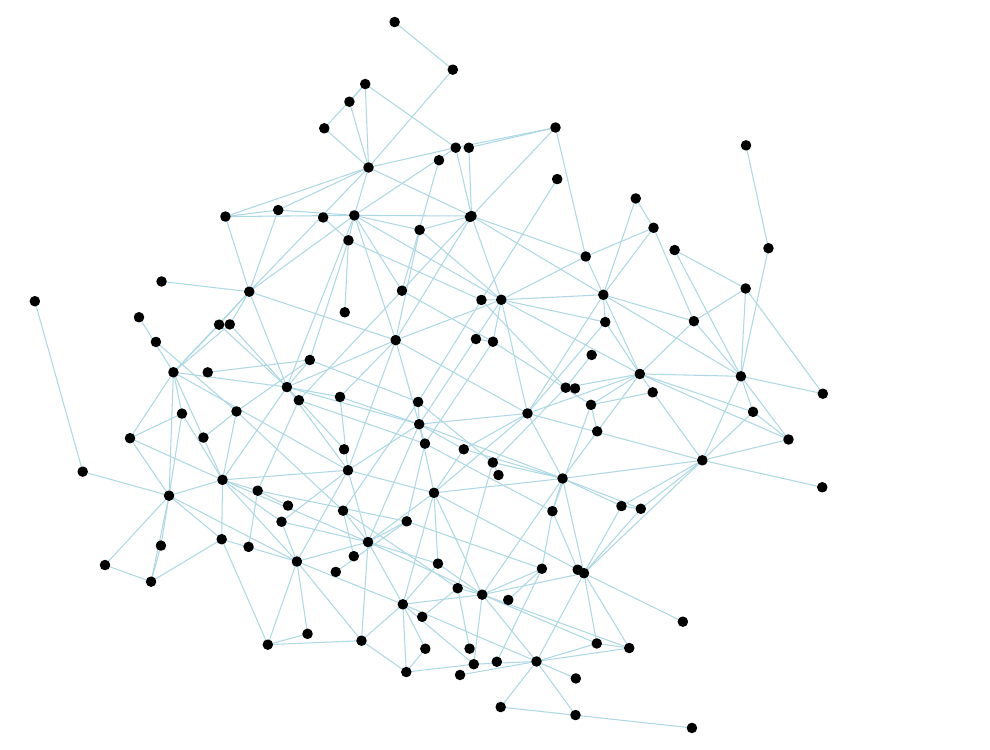} &
  \includegraphics[width=9.5mm]{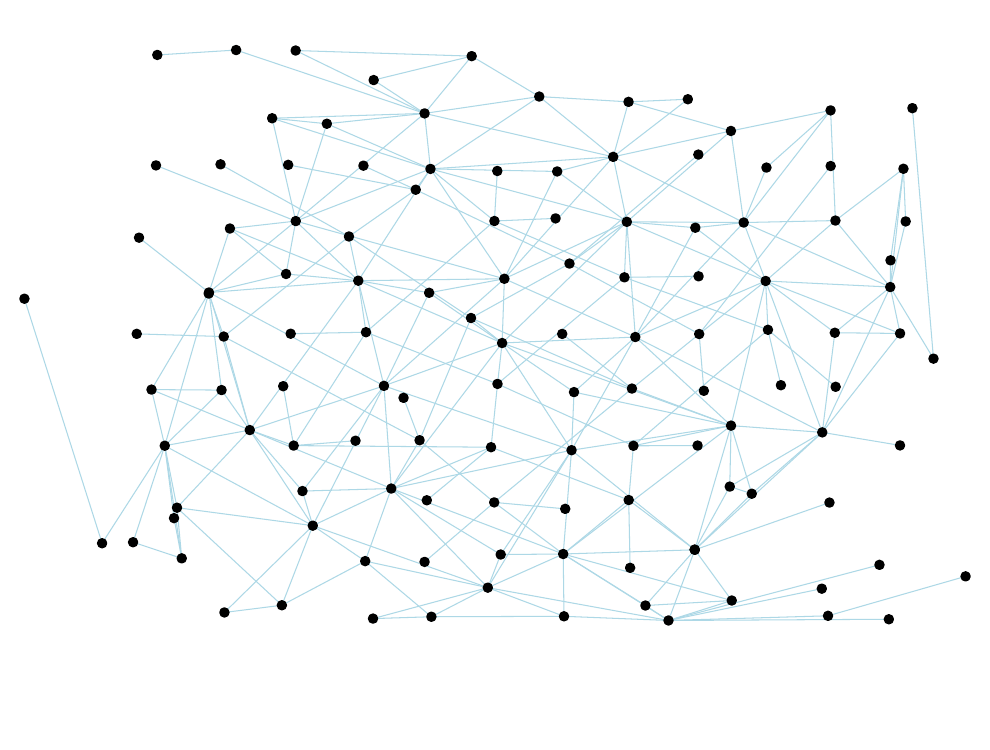}
  \\
         & \texttt{ELD-CN-AR} & & \includegraphics[width=9.5mm]{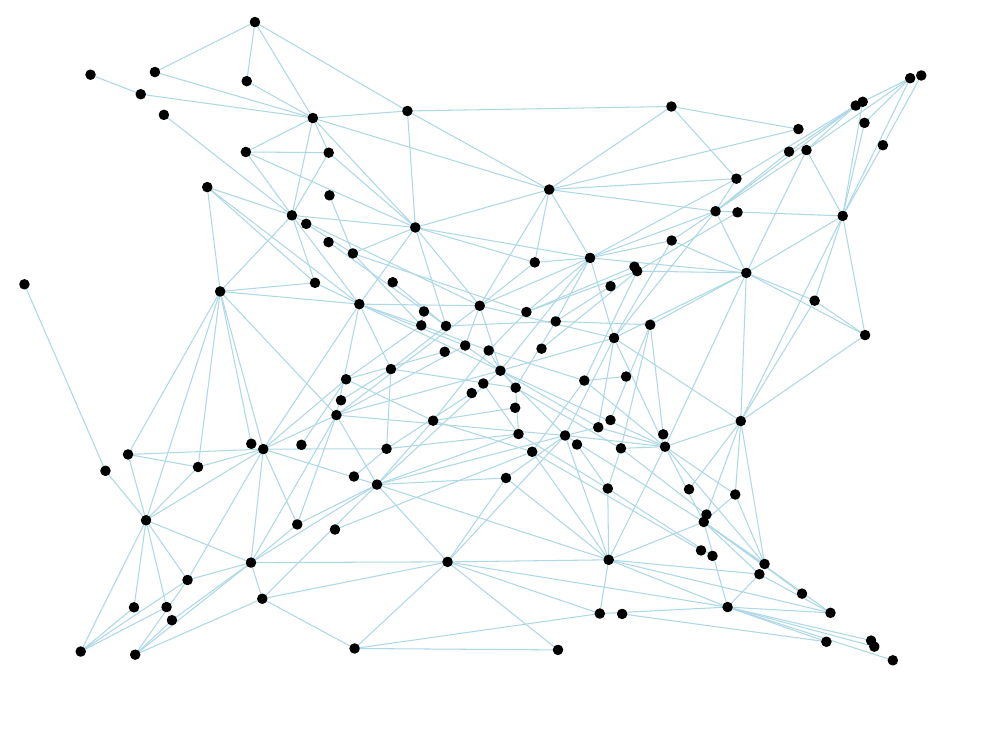} &
  \includegraphics[width=9.5mm]{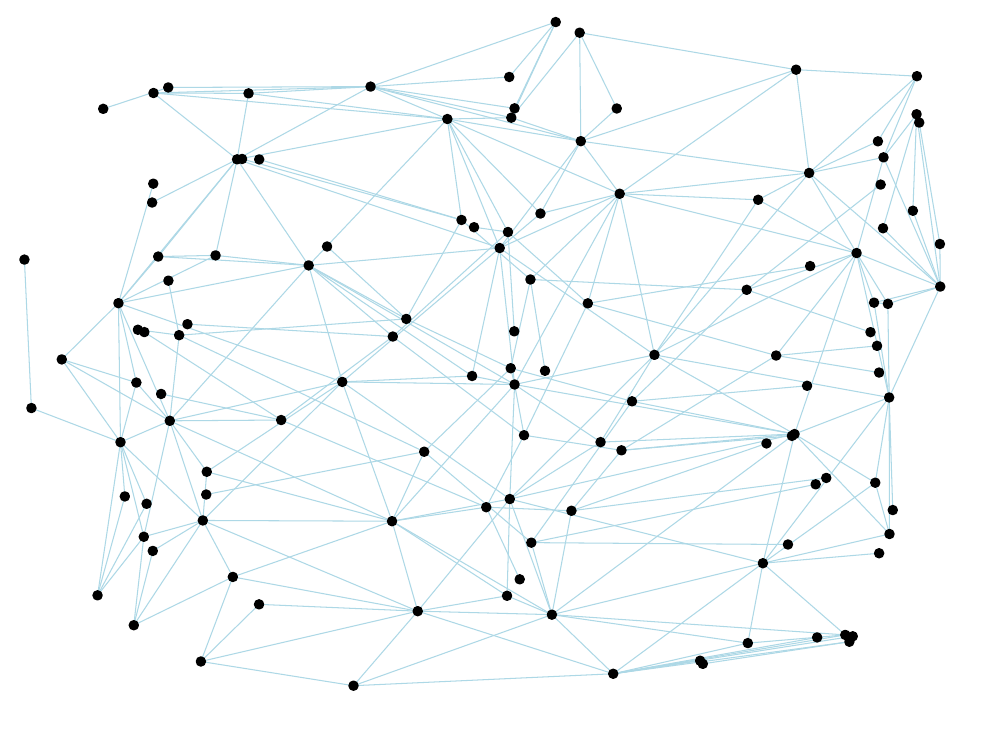} &
  \includegraphics[width=9.5mm]{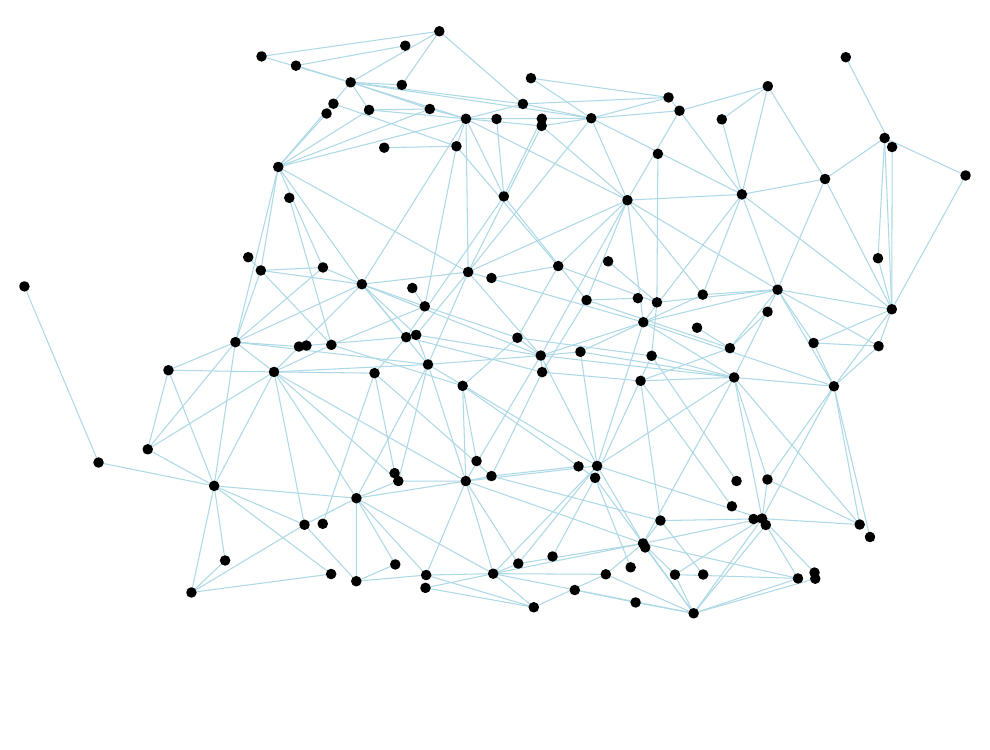} &
  \includegraphics[width=9.5mm]{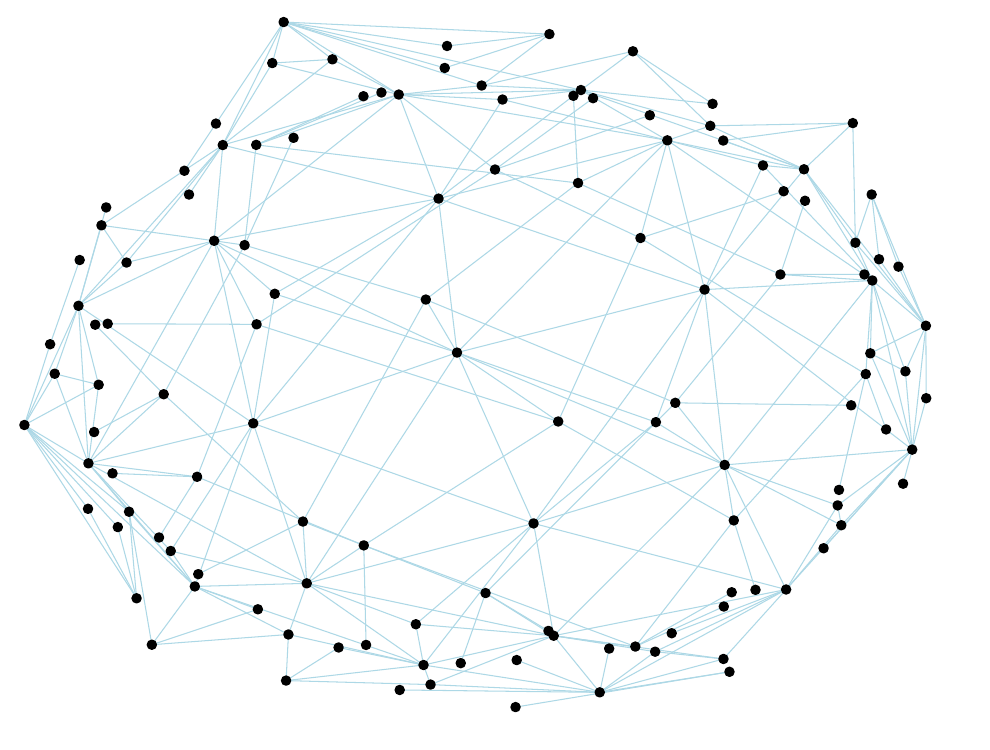} &
  \includegraphics[width=9.5mm]{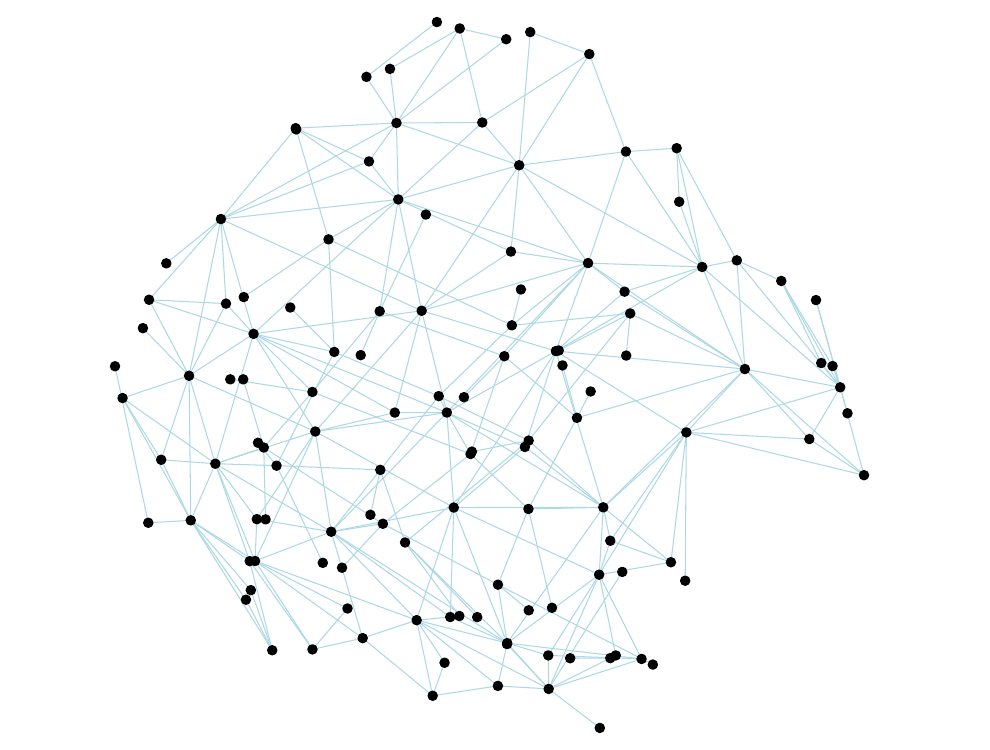} &
  \includegraphics[width=9.5mm]{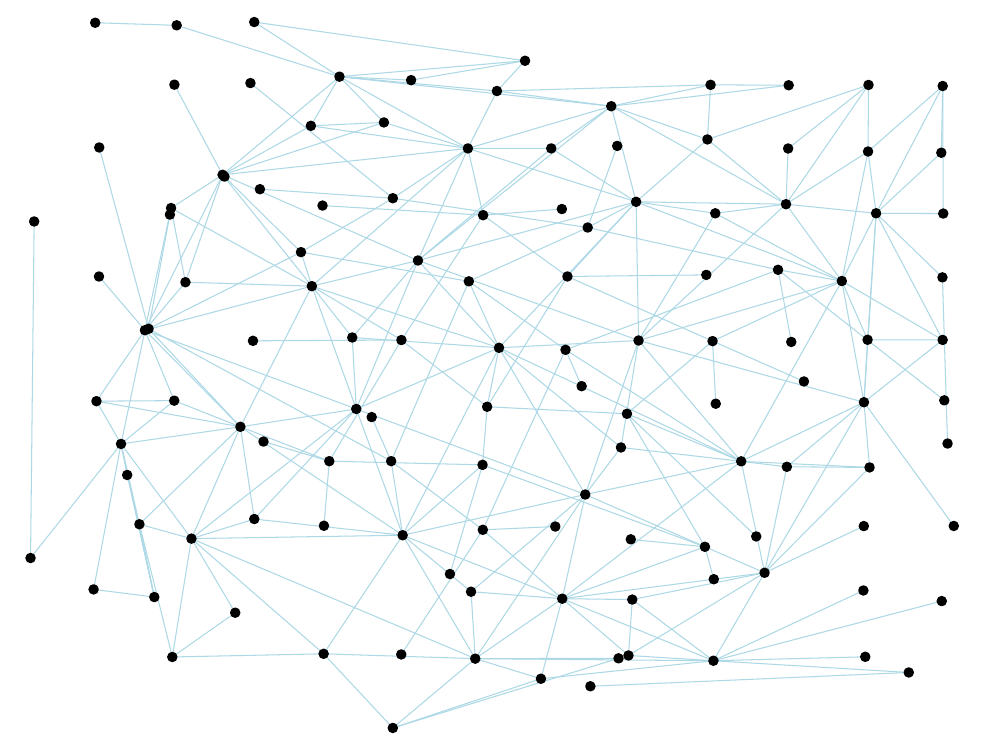}
  \\
         & \texttt{ST-ELD-CN-AR} & & \includegraphics[width=9.5mm]{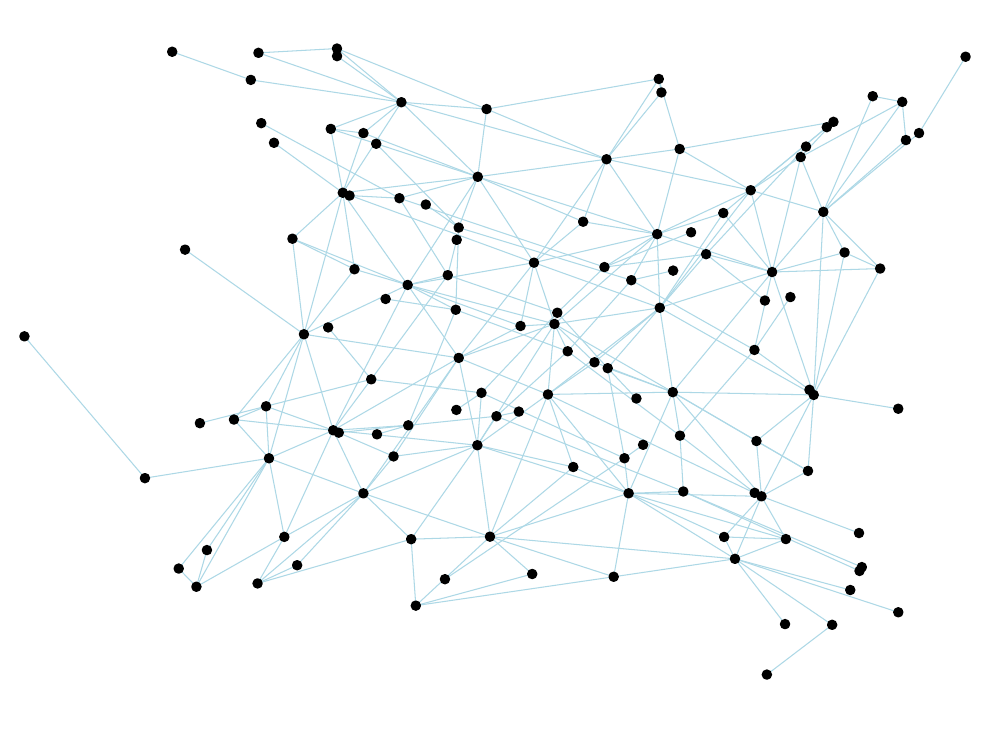} &
  \includegraphics[width=9.5mm]{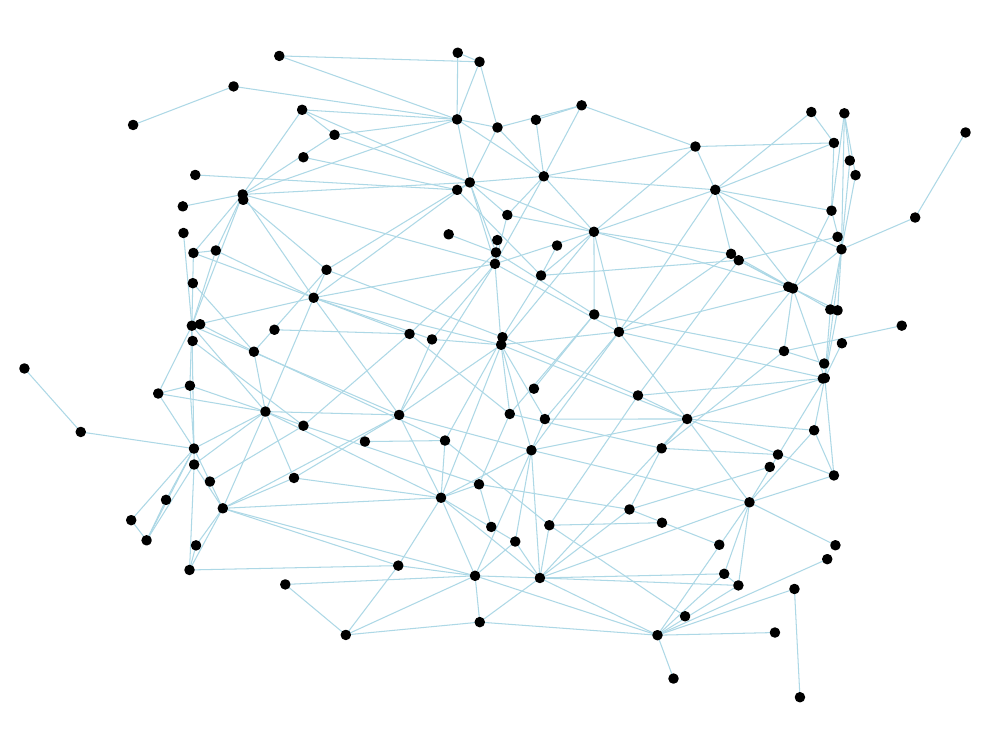} &
  \includegraphics[width=9.5mm]{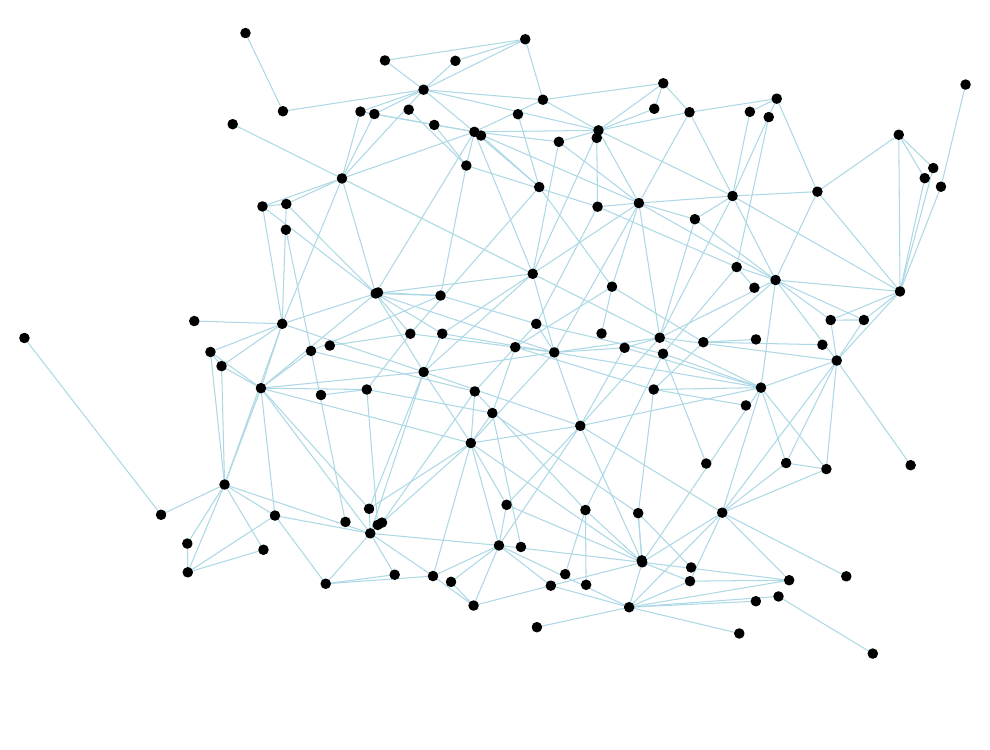} &
  \includegraphics[width=9.5mm]{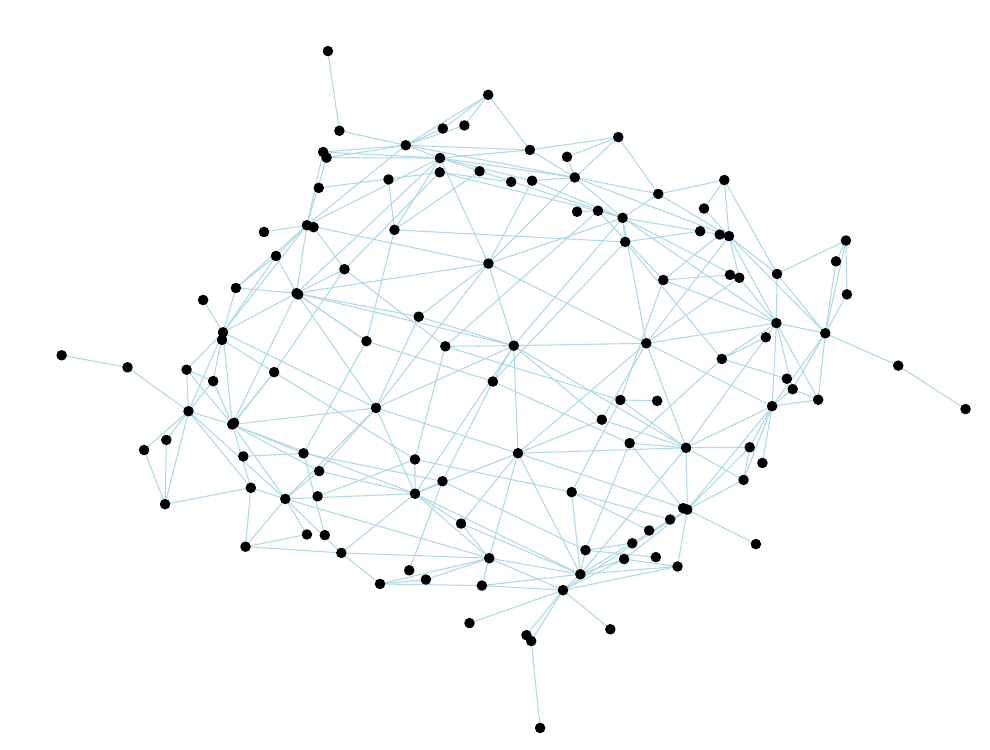} &
  \includegraphics[width=9.5mm]{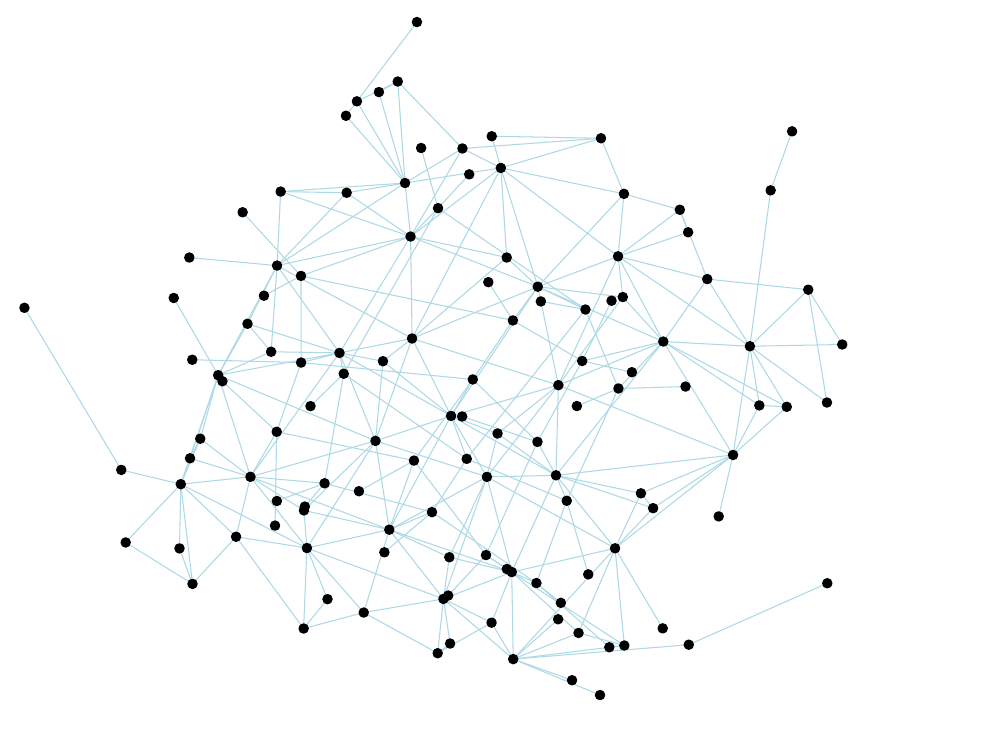} &
  \includegraphics[width=9.5mm]{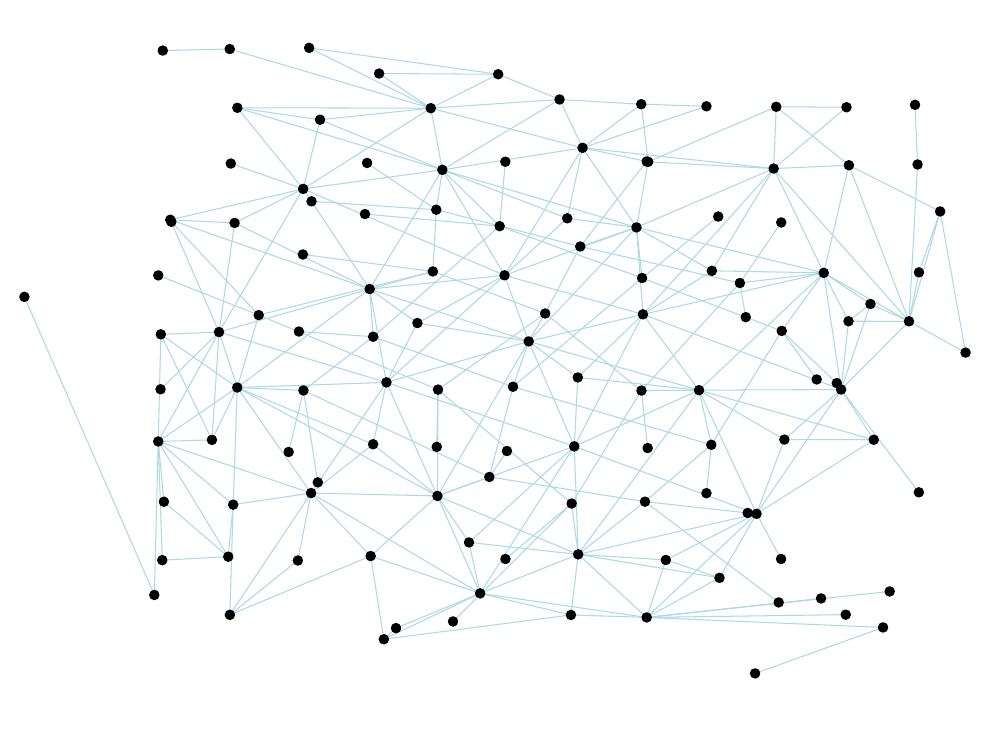}
  \\
  \hline

    \emph{gams10am} & \texttt{ST-ELD} & \includegraphics[width=9.5mm]{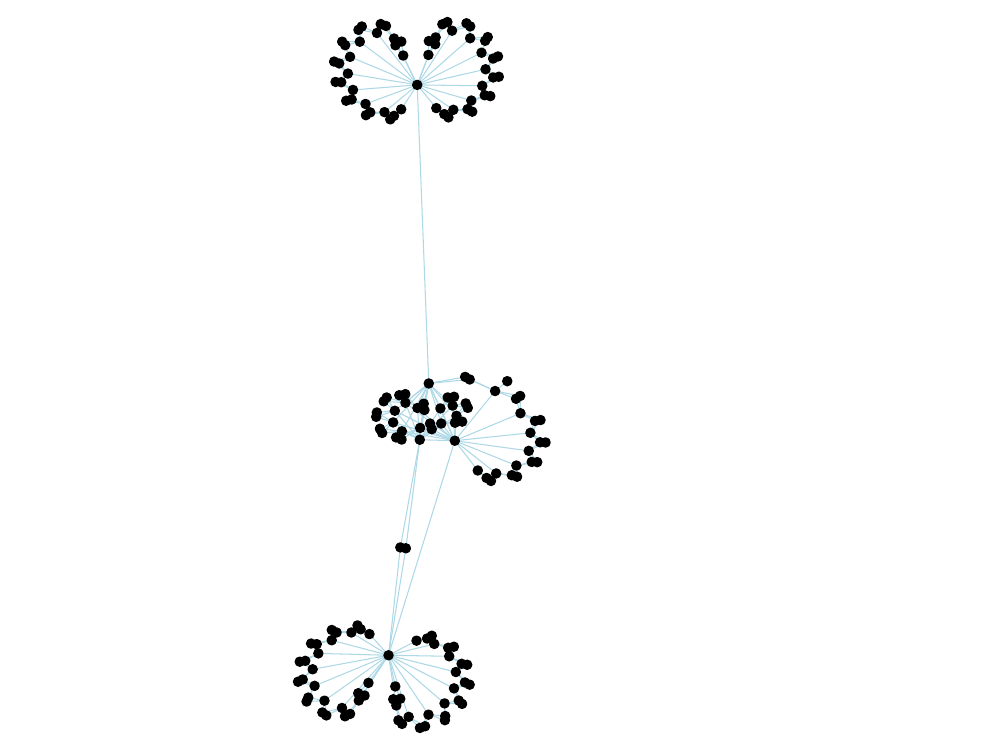} & \includegraphics[width=9.5mm]{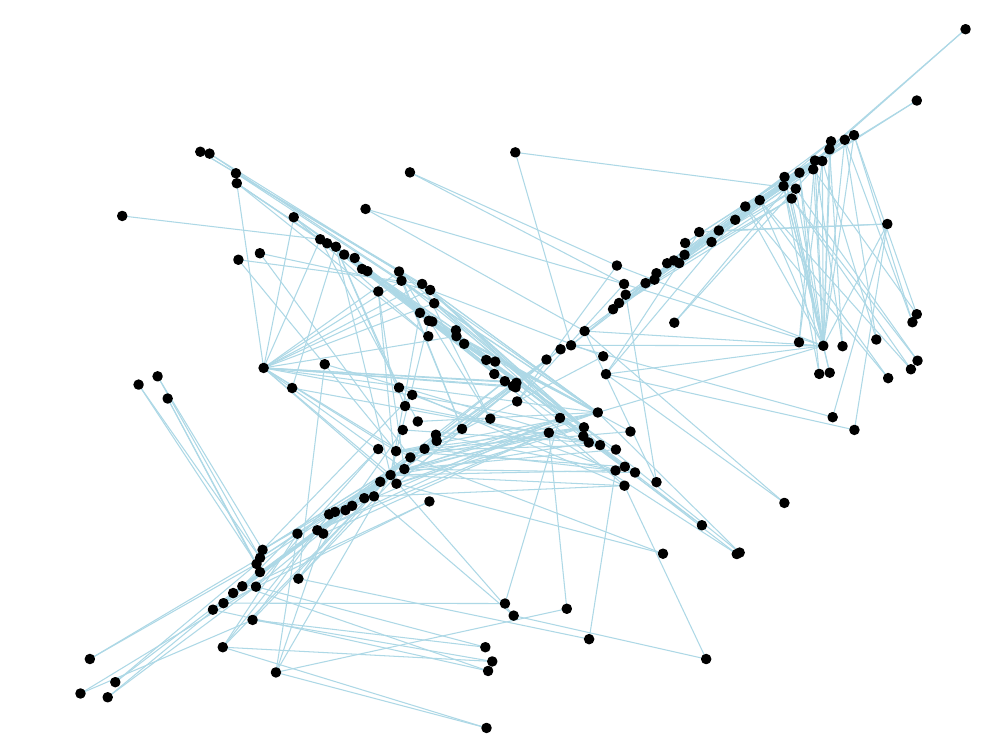} &
  \includegraphics[width=9.5mm]{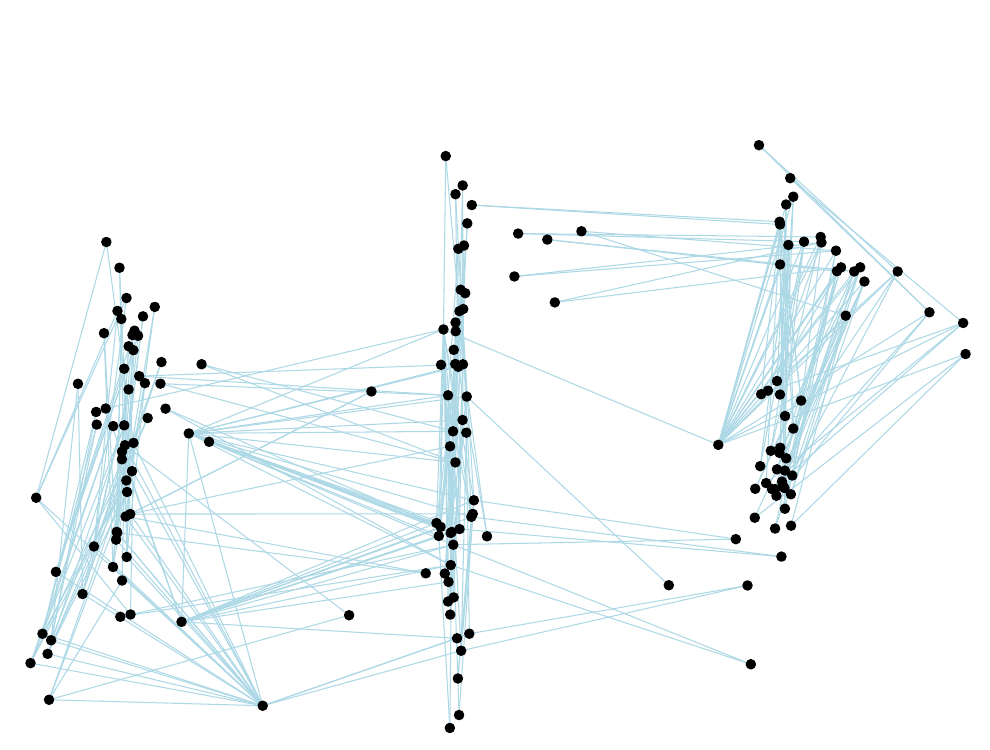} &
  \includegraphics[width=9.5mm]{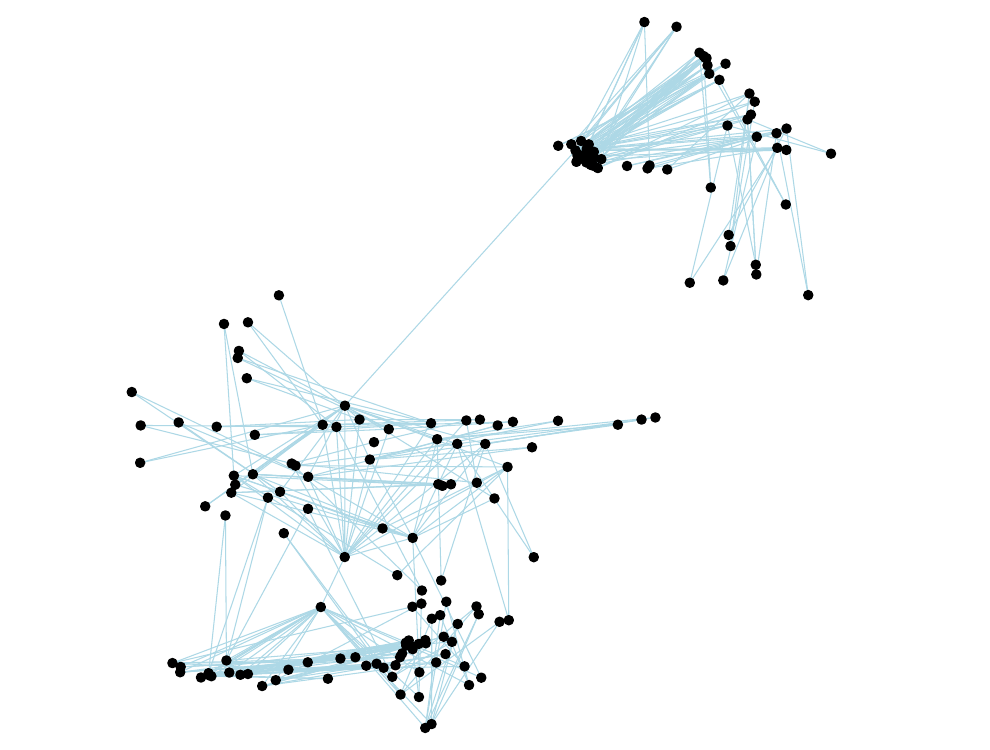} &
  \includegraphics[width=9.5mm]{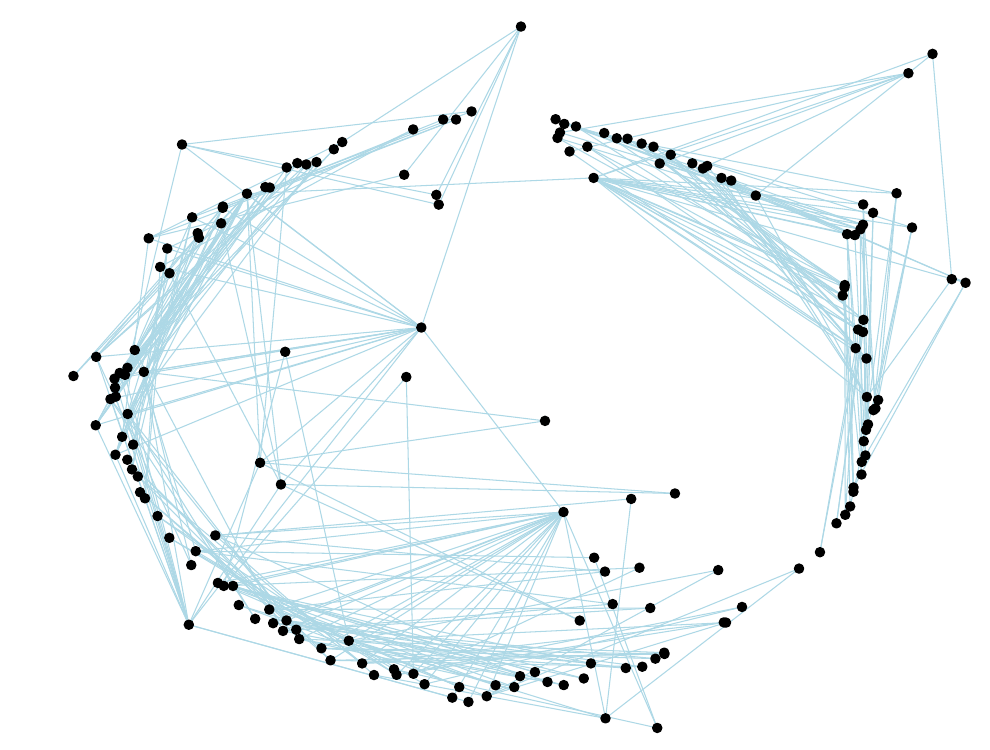} &
  \includegraphics[width=9.5mm]{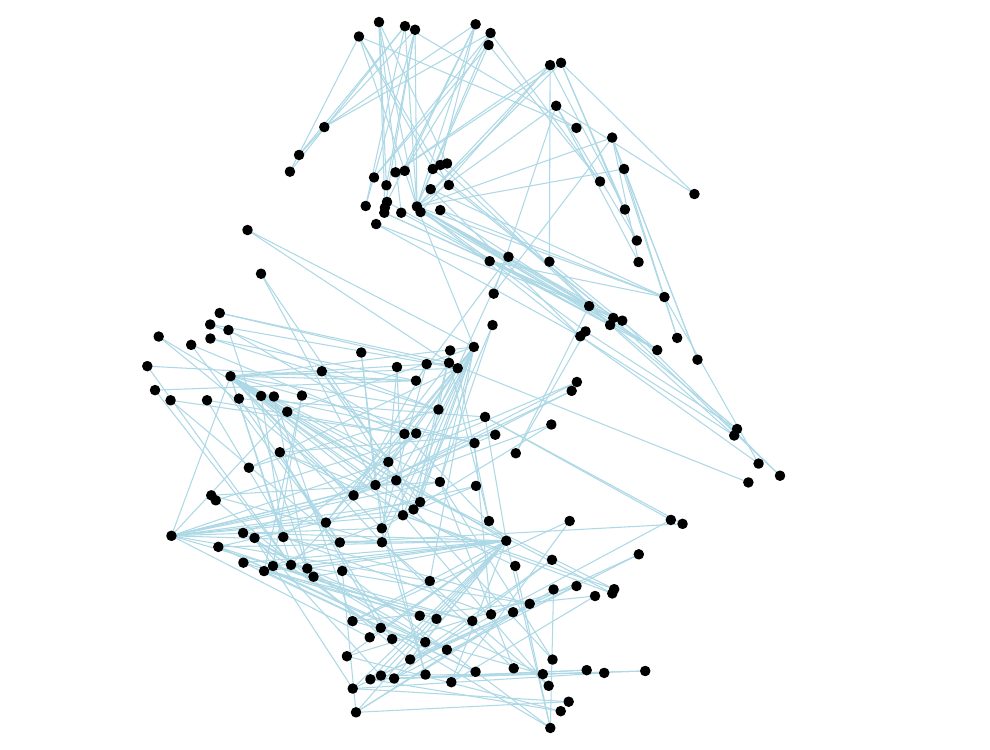} &
  \includegraphics[width=9.5mm]{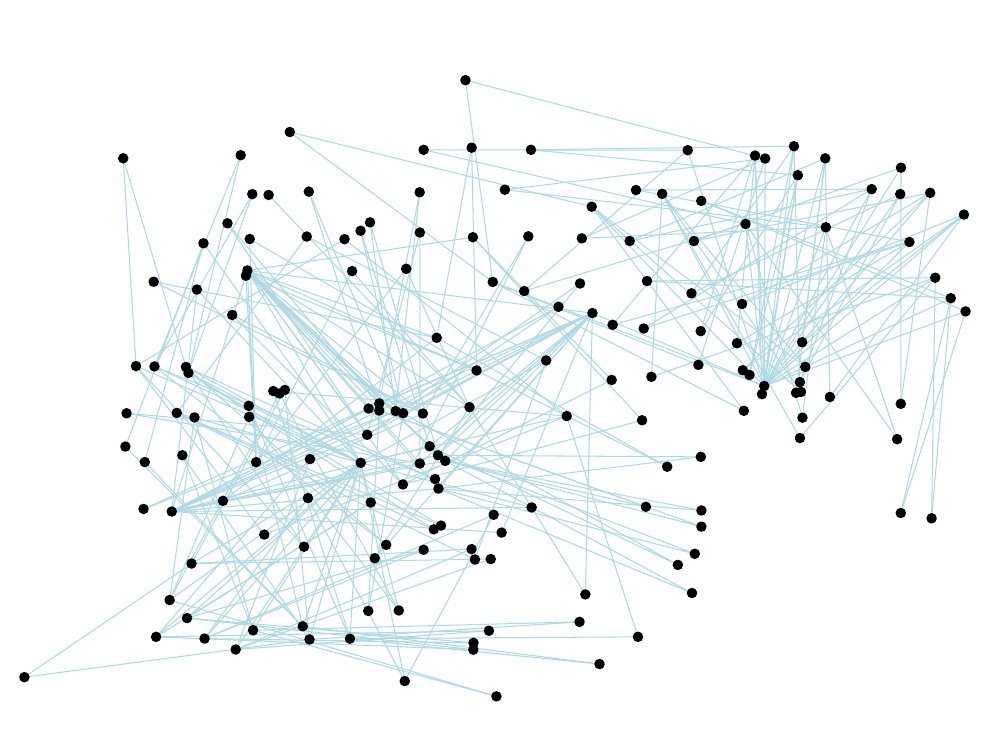}
  \\
     & \texttt{ST-CN} & & \includegraphics[width=9.5mm]{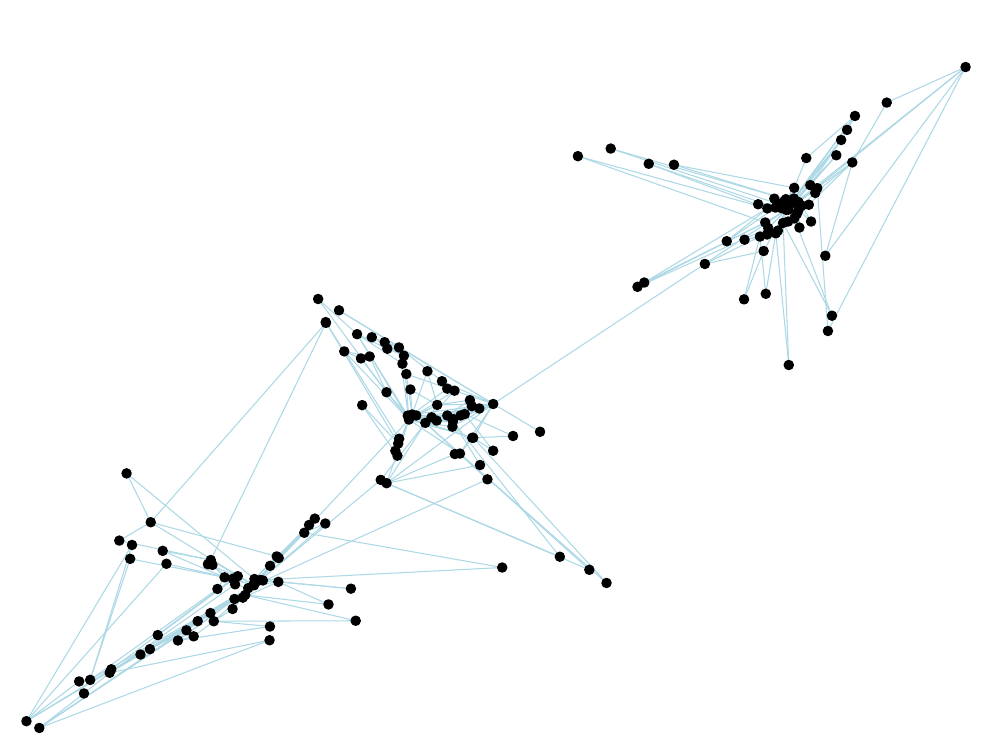} &
  \includegraphics[width=9.5mm]{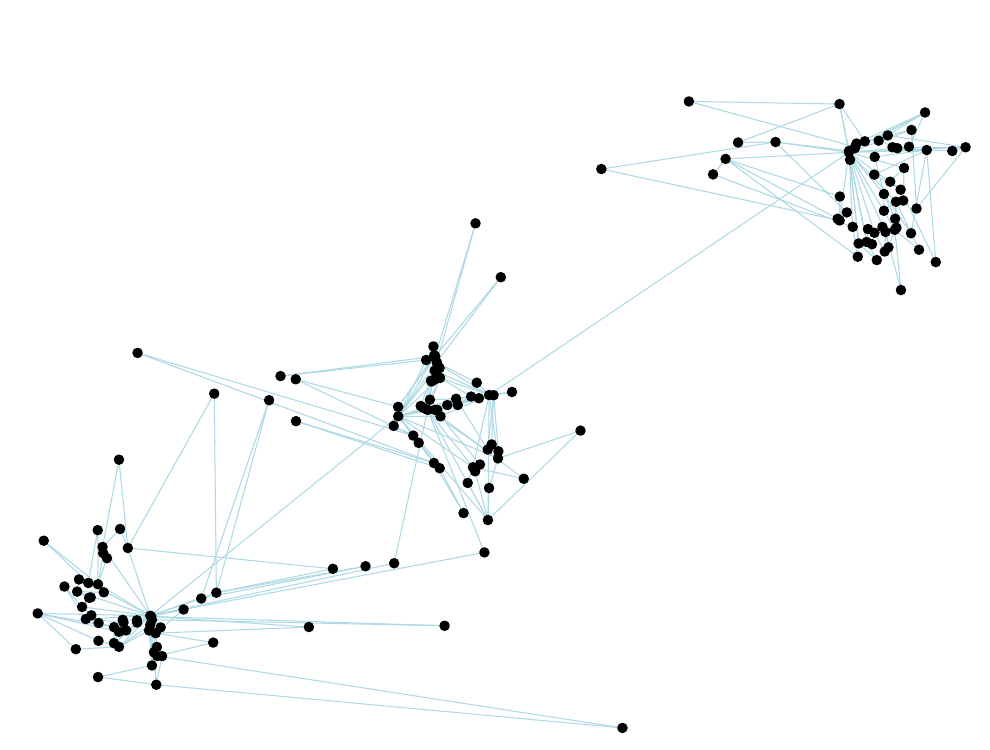} &
  \includegraphics[width=9.5mm]{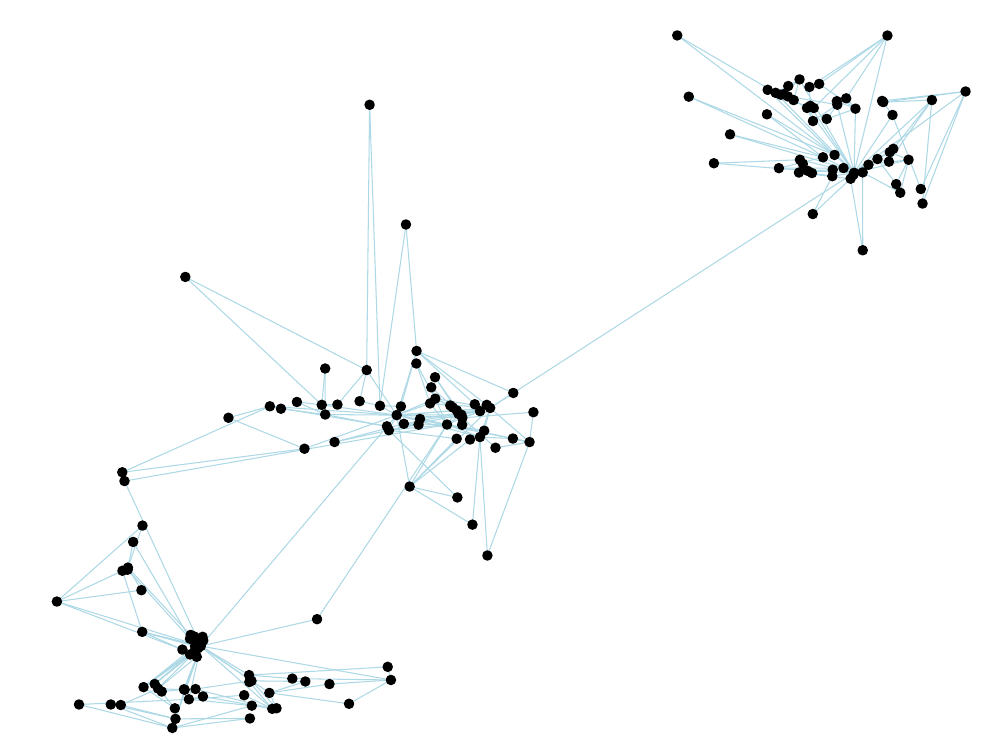} &
  \includegraphics[width=9.5mm]{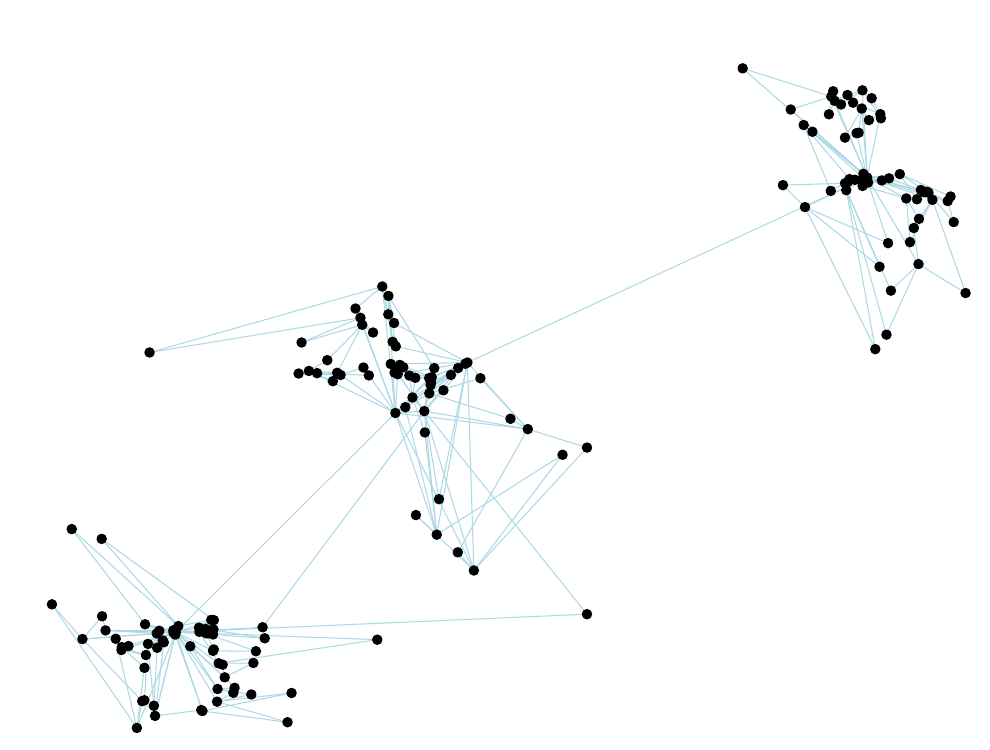} &
  \includegraphics[width=9.5mm]{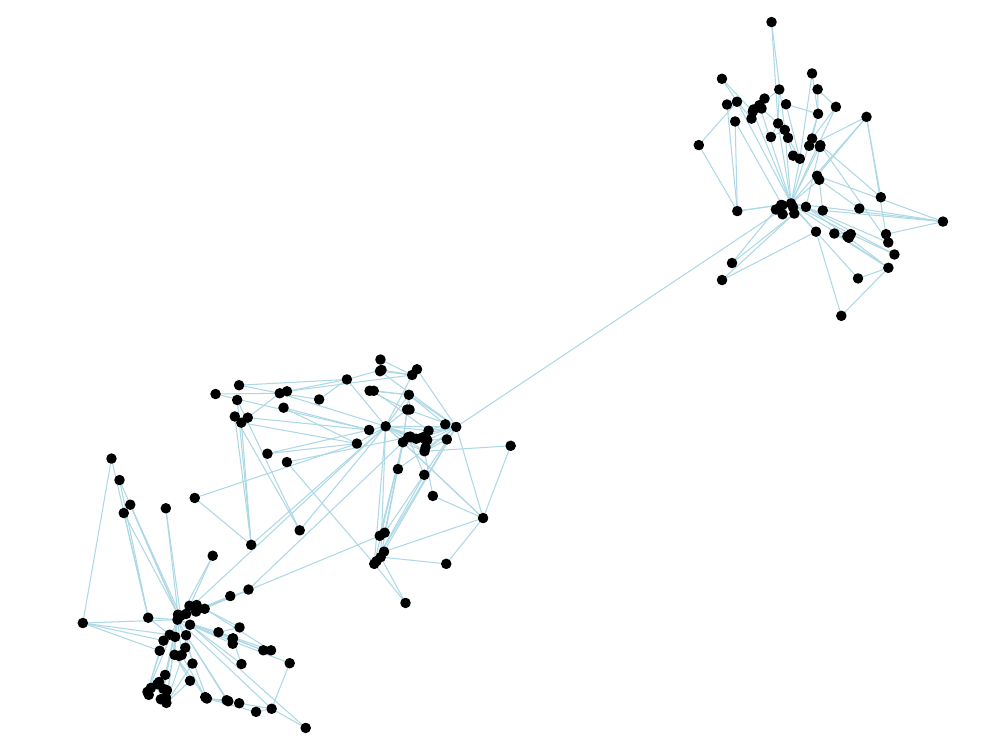} &
  \includegraphics[width=9.5mm]{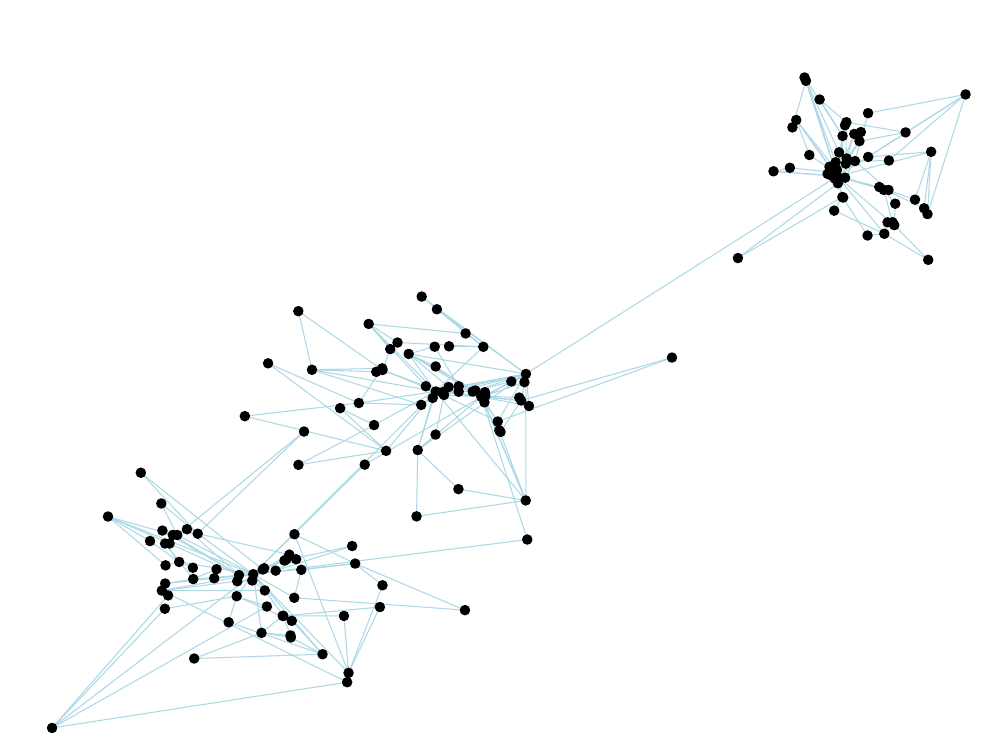}
  \\
        & \texttt{ST-AR} &
   & \includegraphics[width=9.5mm]{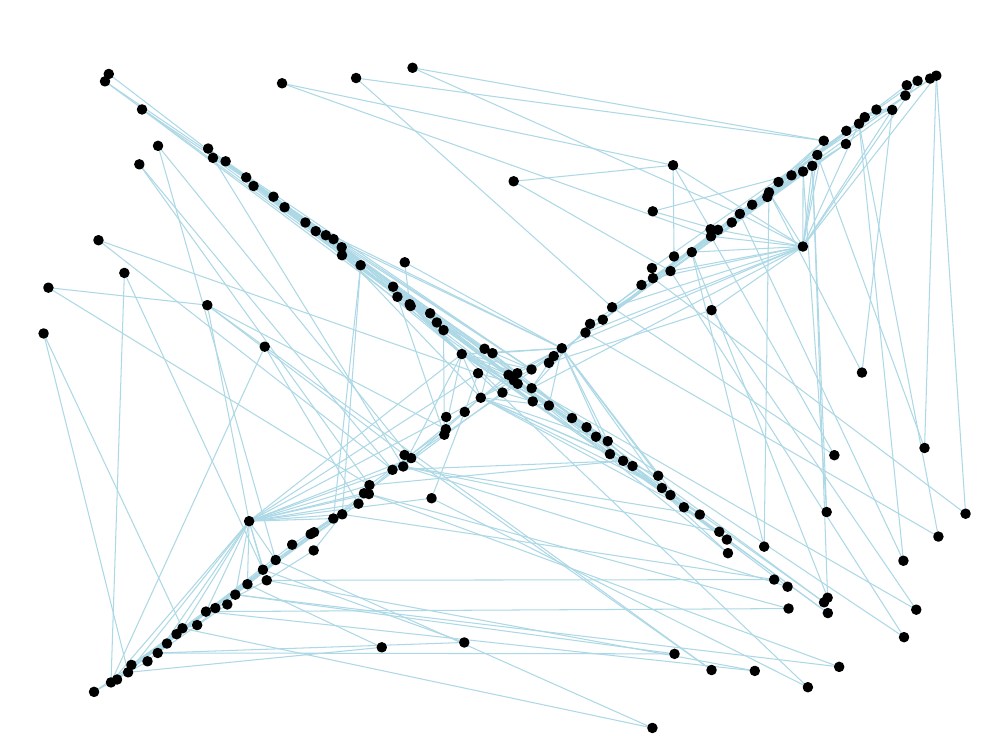} &
  \includegraphics[width=9.5mm]{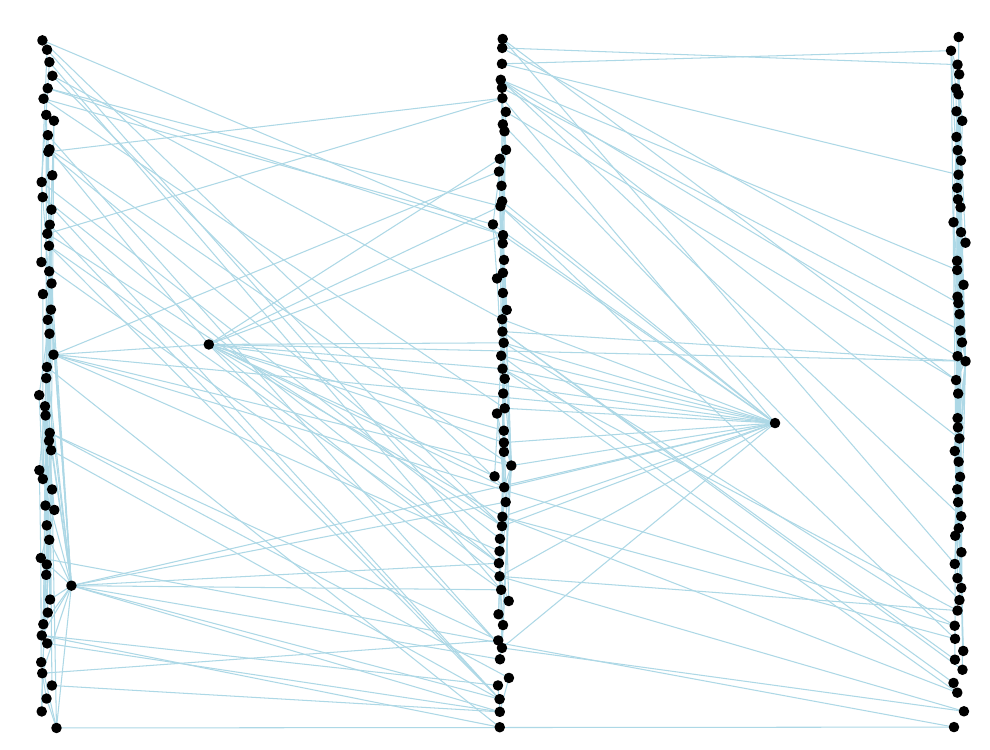} &
  \includegraphics[width=9.5mm]{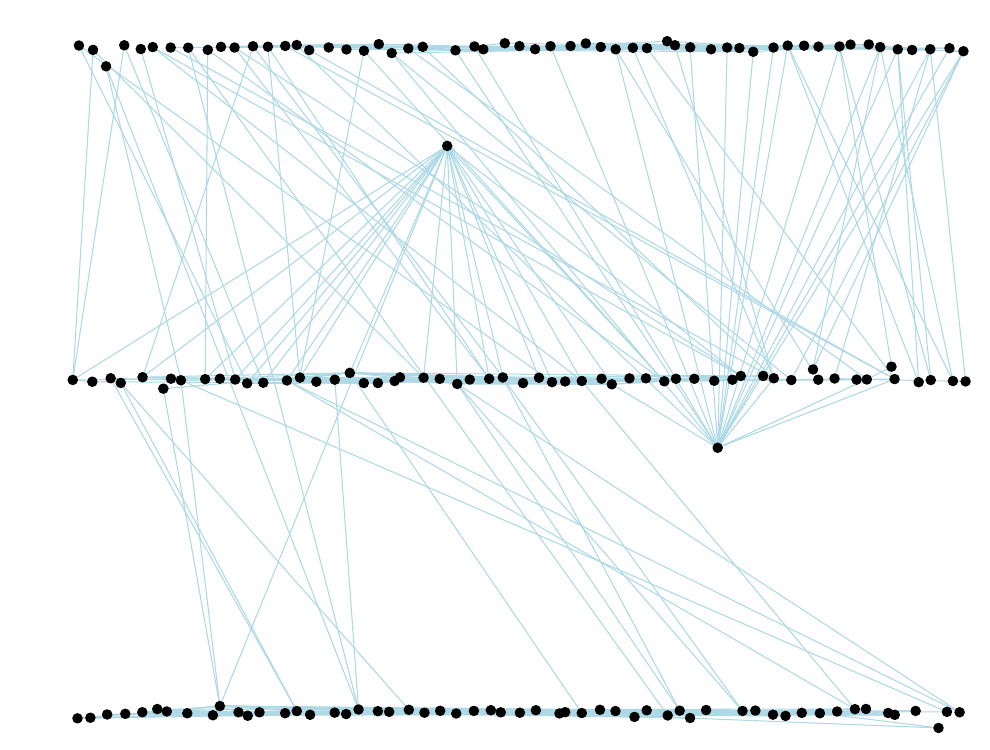} &
  \includegraphics[width=9.5mm]{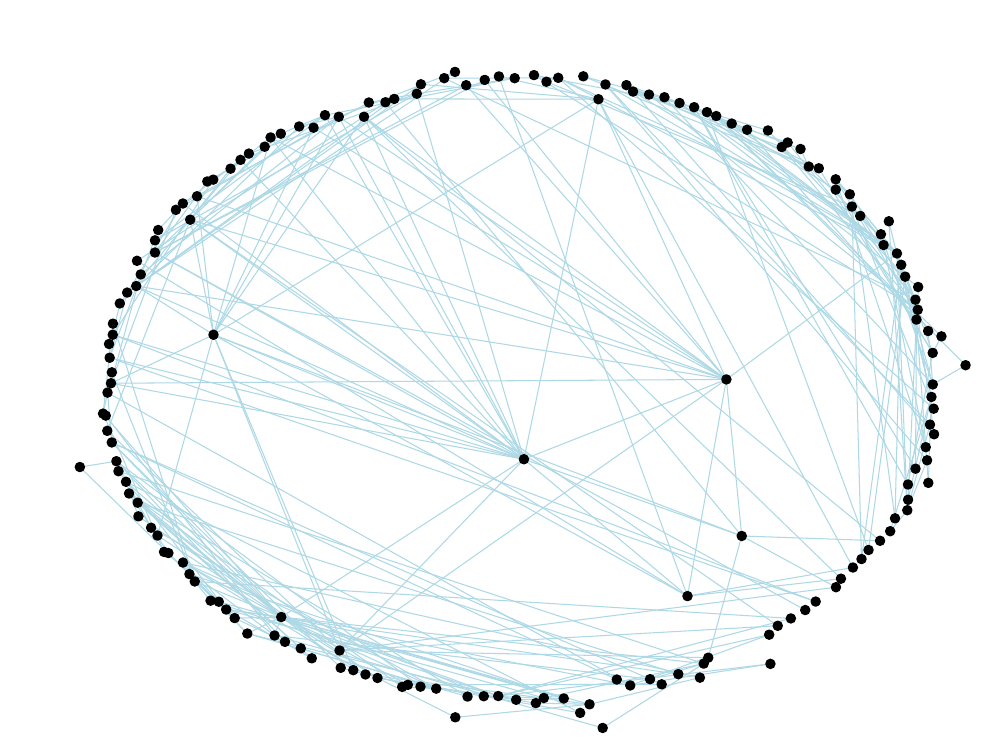} &
  \includegraphics[width=9.5mm]{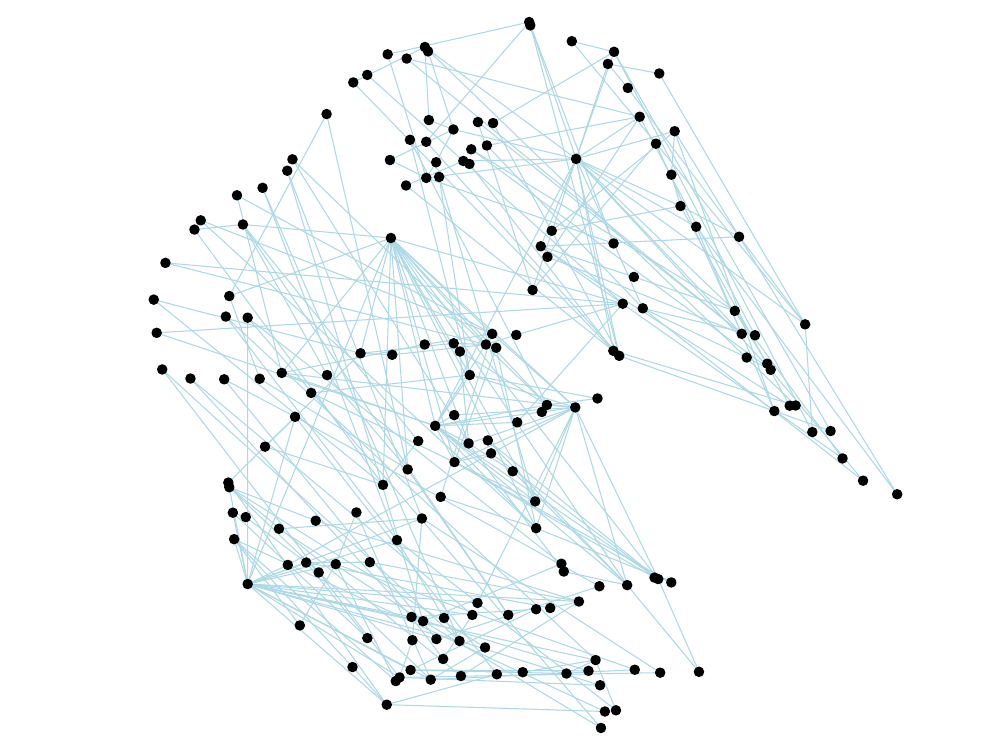} &
  \includegraphics[width=9.5mm]{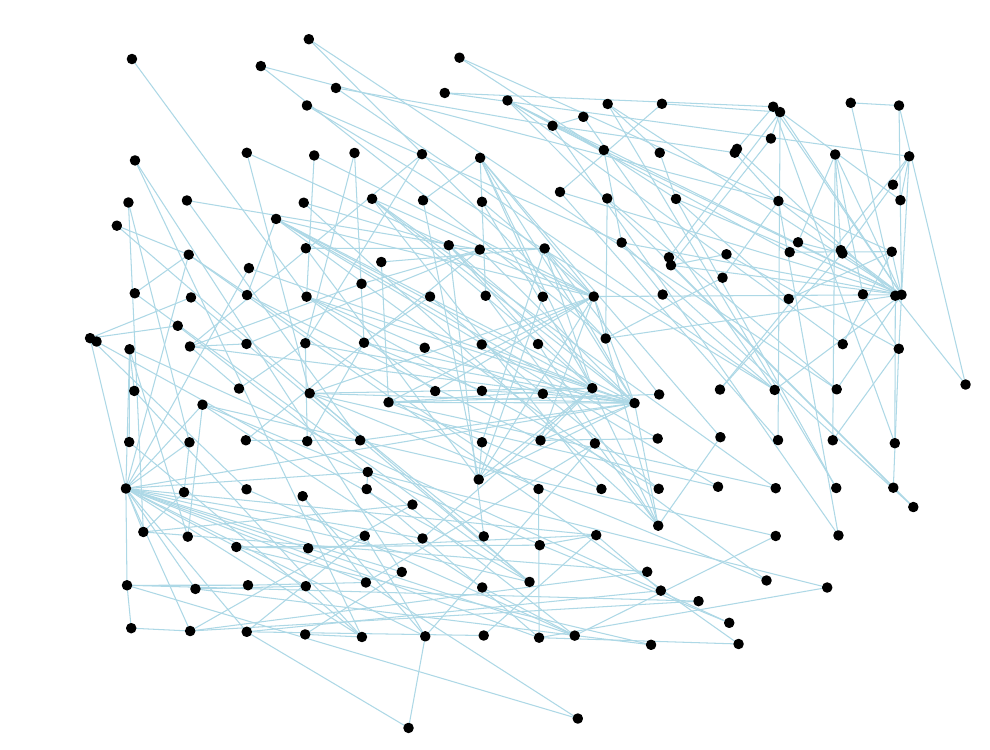}
  \\
     & \texttt{ELD-CN} & & \includegraphics[width=9.5mm]{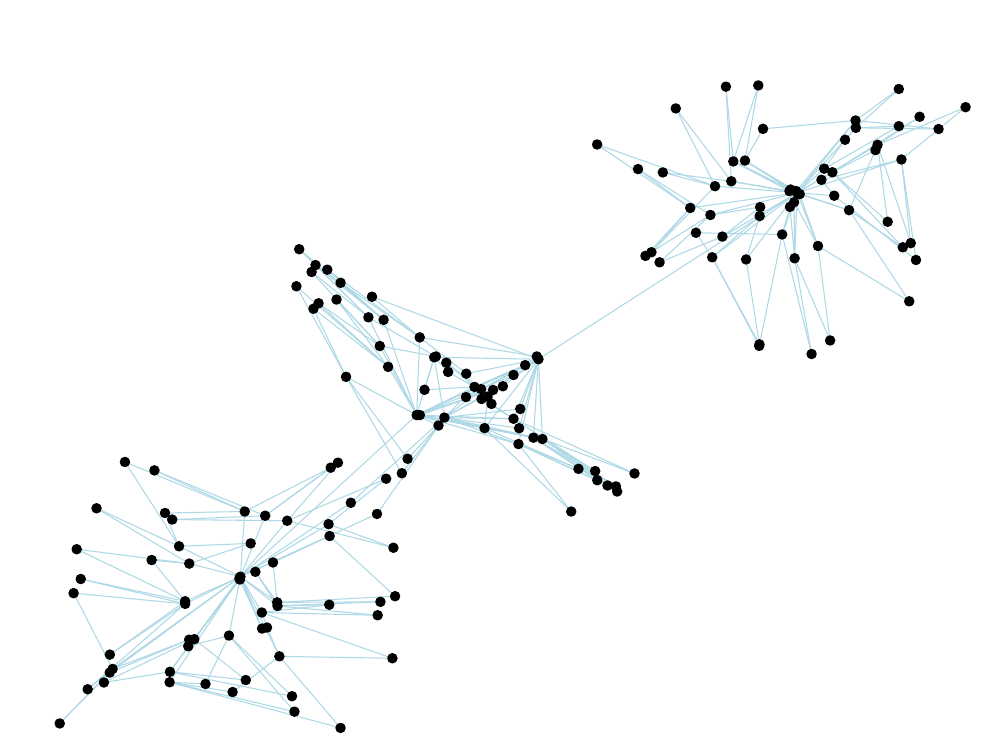} &
  \includegraphics[width=9.5mm]{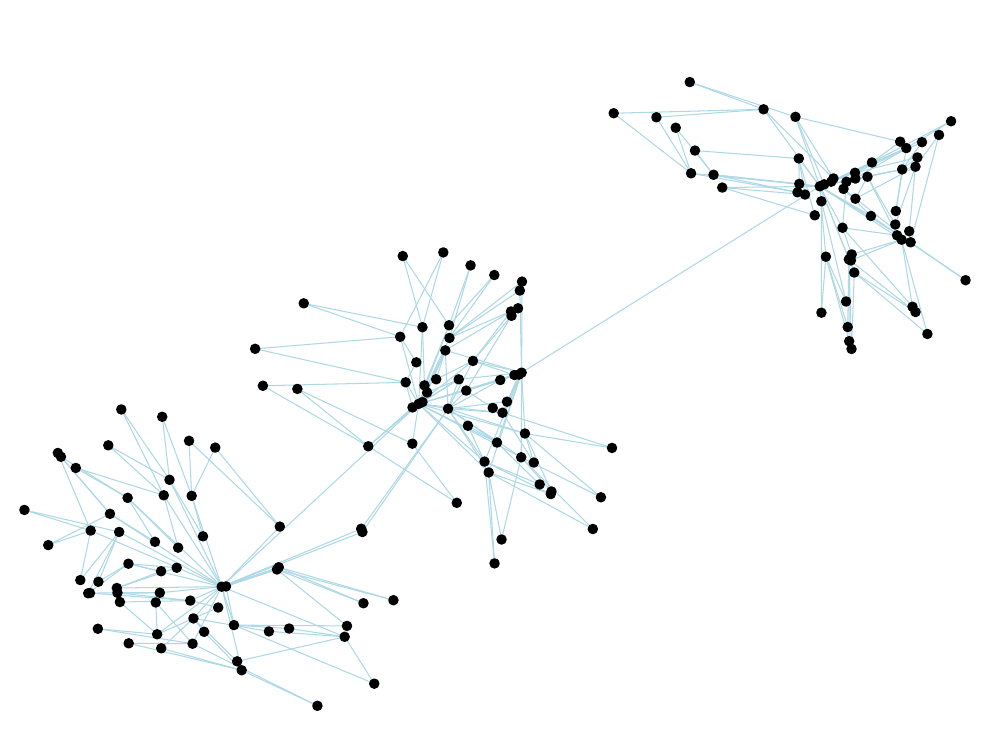} &
  \includegraphics[width=9.5mm]{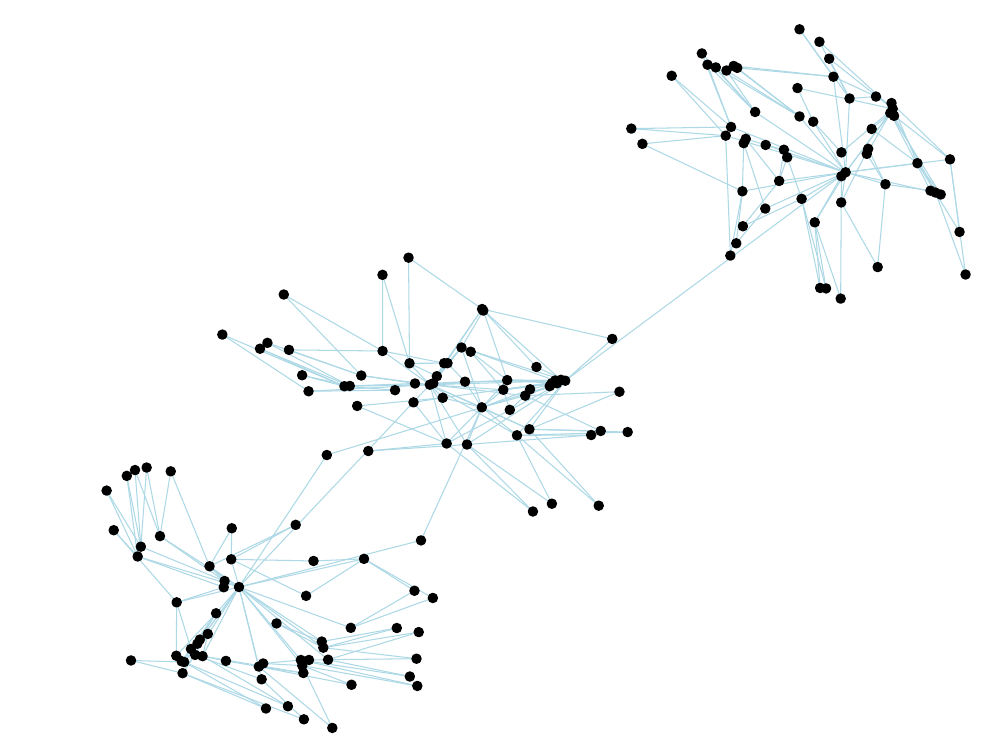} &
  \includegraphics[width=9.5mm]{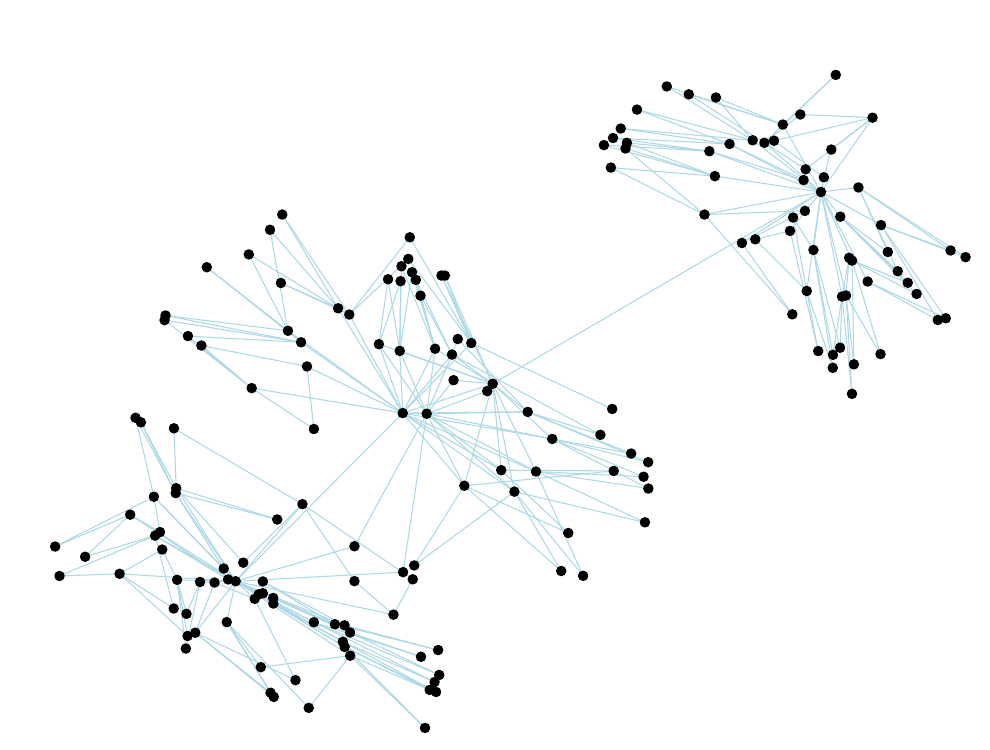} &
  \includegraphics[width=9.5mm]{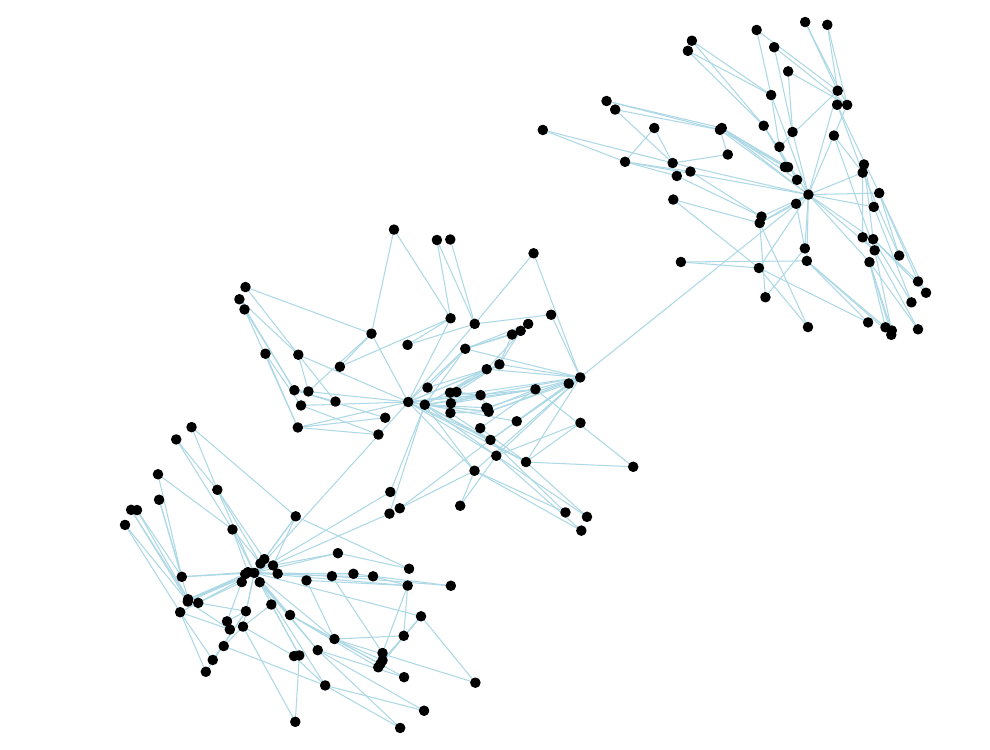} &
  \includegraphics[width=9.5mm]{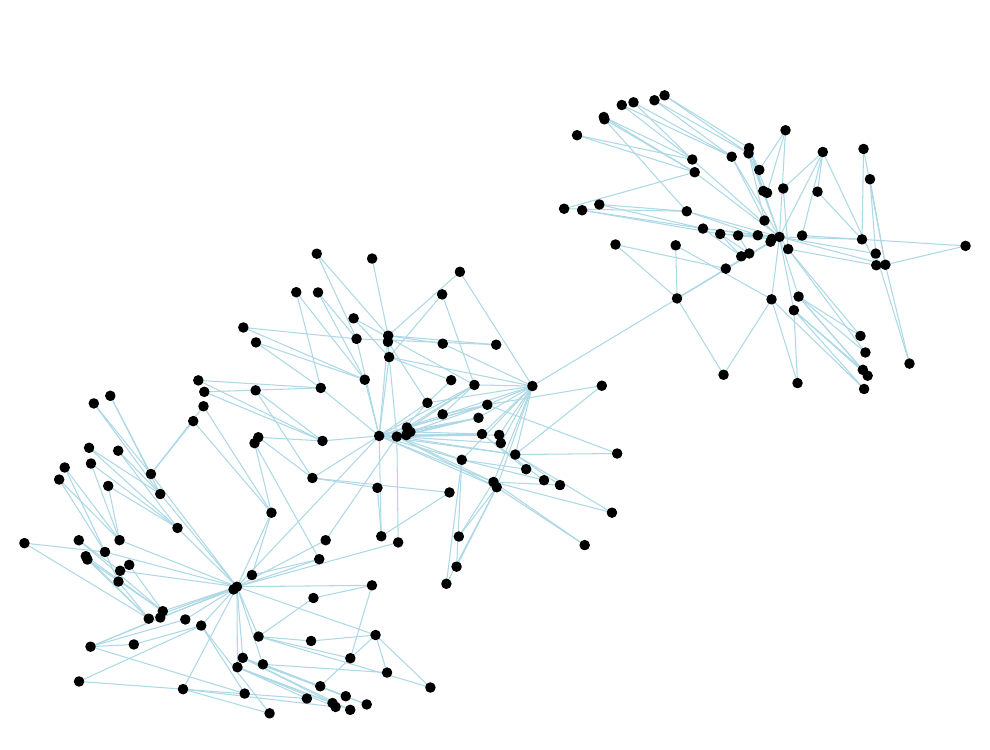}
  \\
       & \texttt{ELD-AR} & & \includegraphics[width=9.5mm]{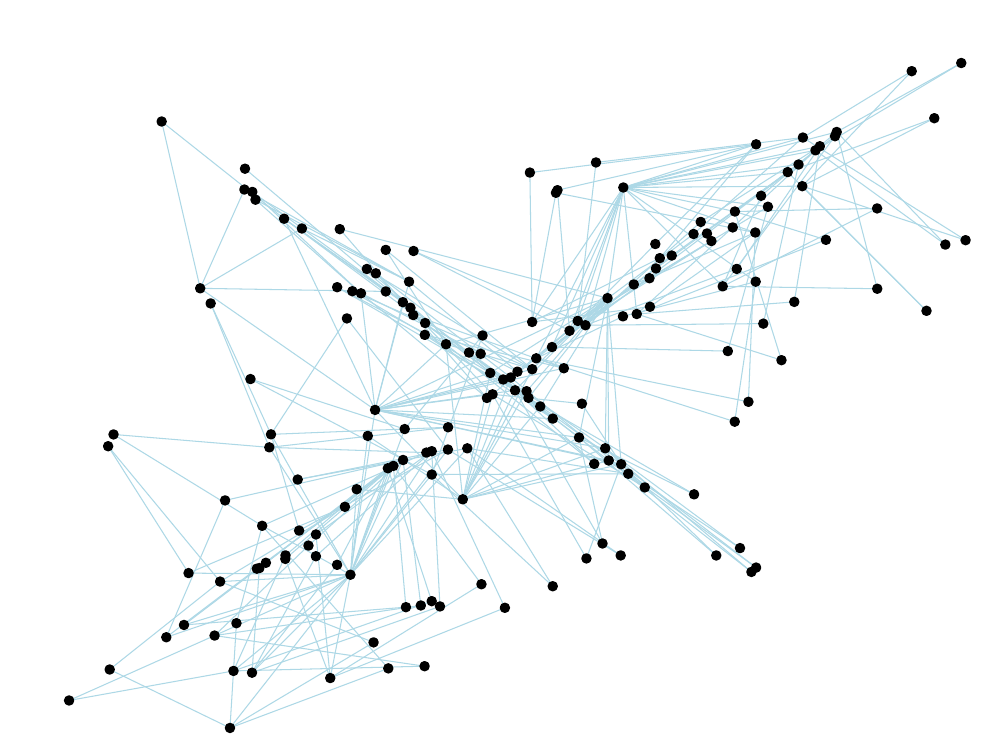} &
  \includegraphics[width=9.5mm]{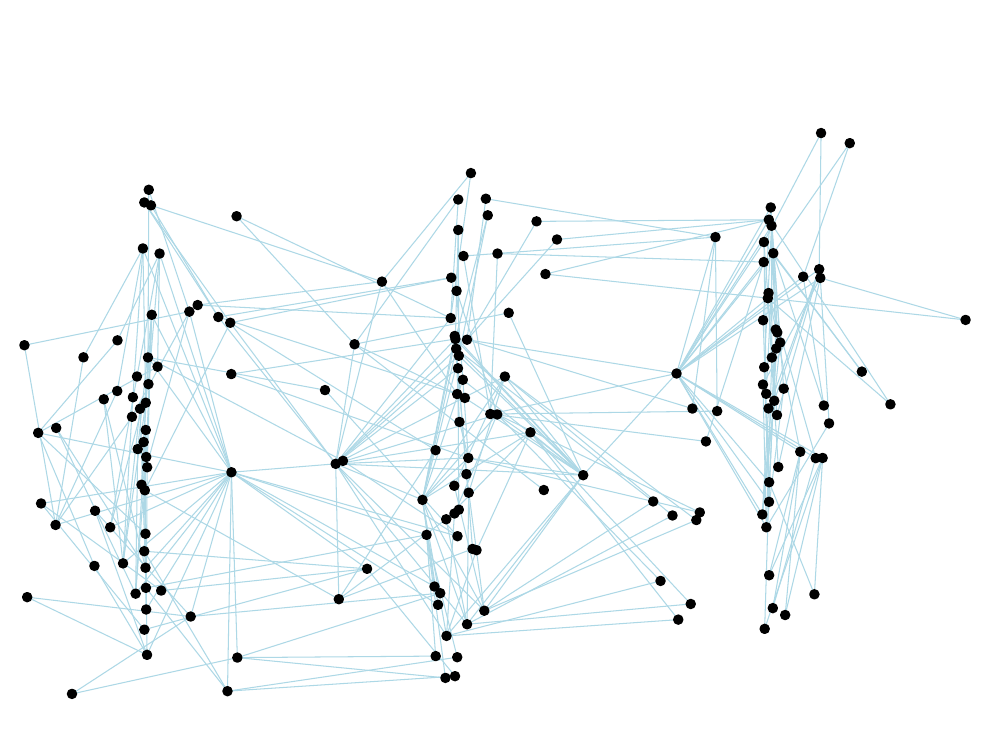} &
  \includegraphics[width=9.5mm]{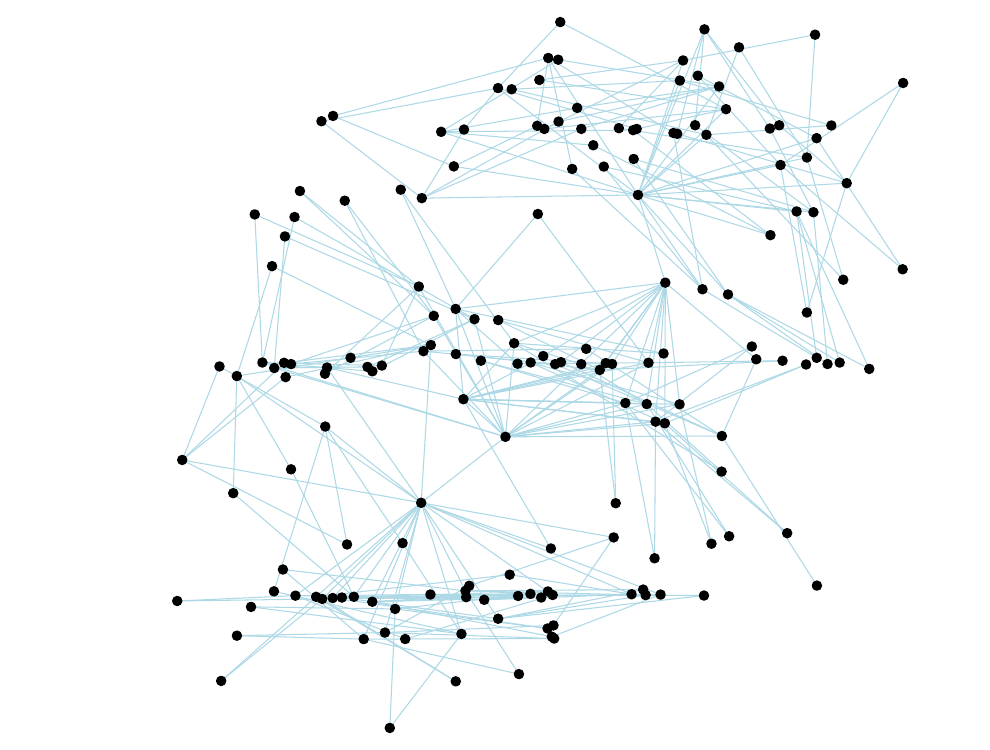} &
  \includegraphics[width=9.5mm]{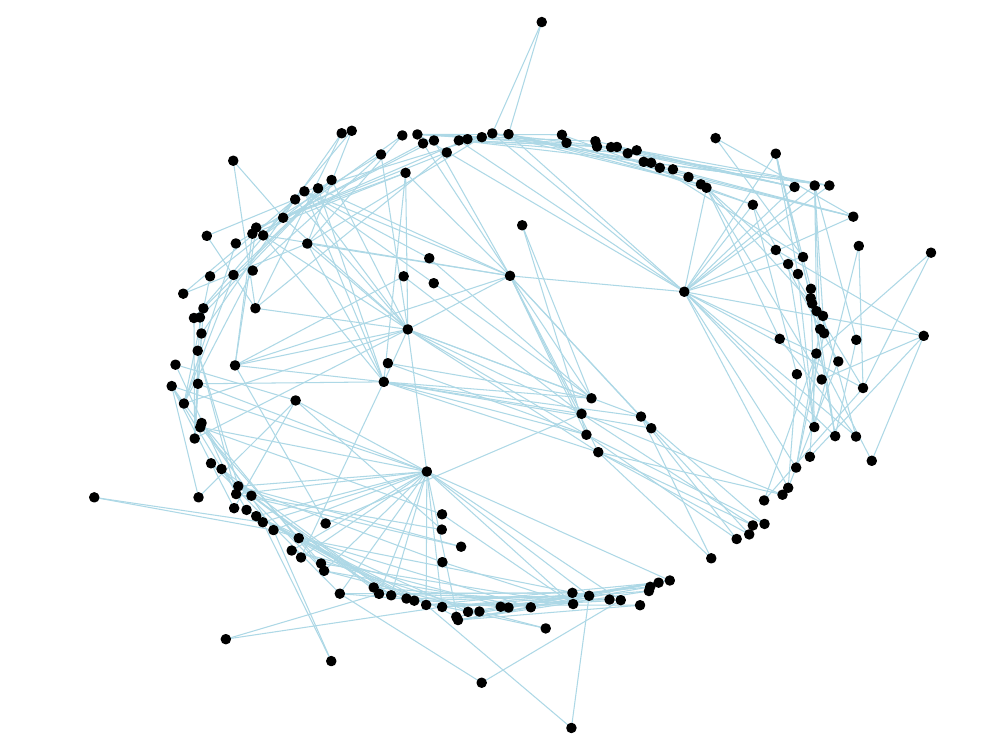} &
  \includegraphics[width=9.5mm]{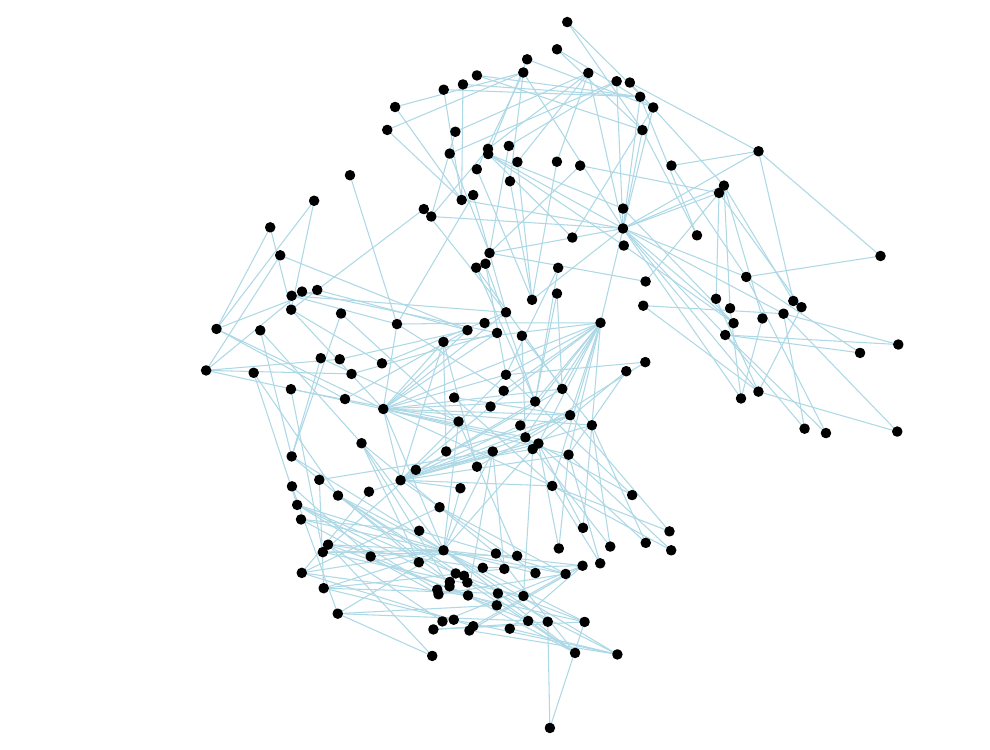} &
  \includegraphics[width=9.5mm]{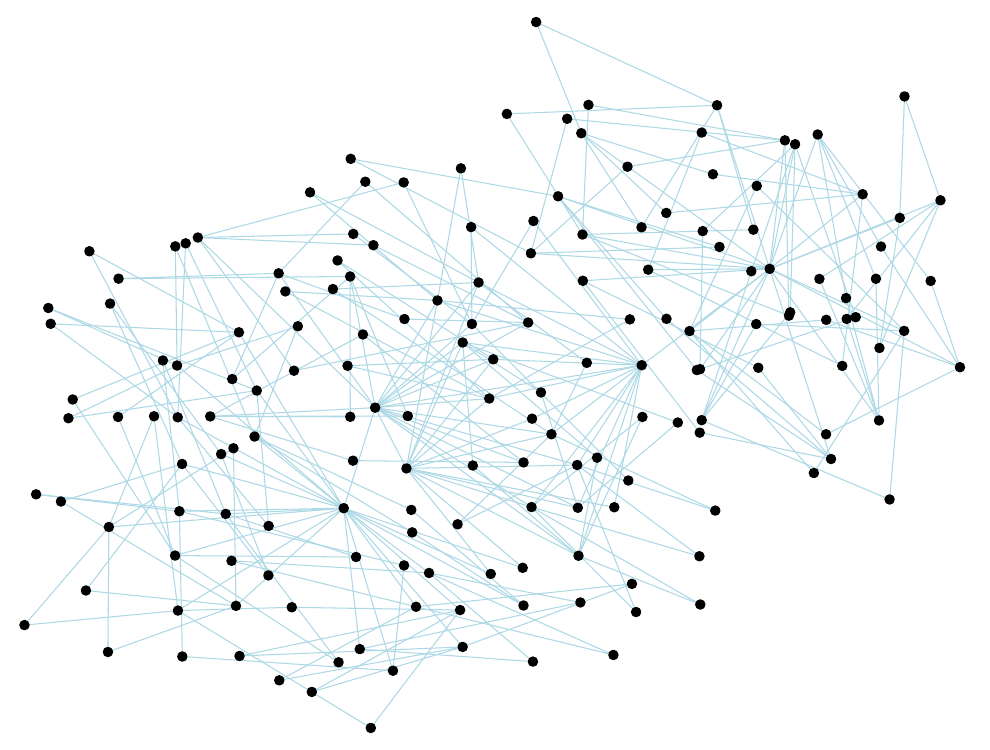}
  \\
       & \texttt{CN-AR} & & \includegraphics[width=9.5mm]{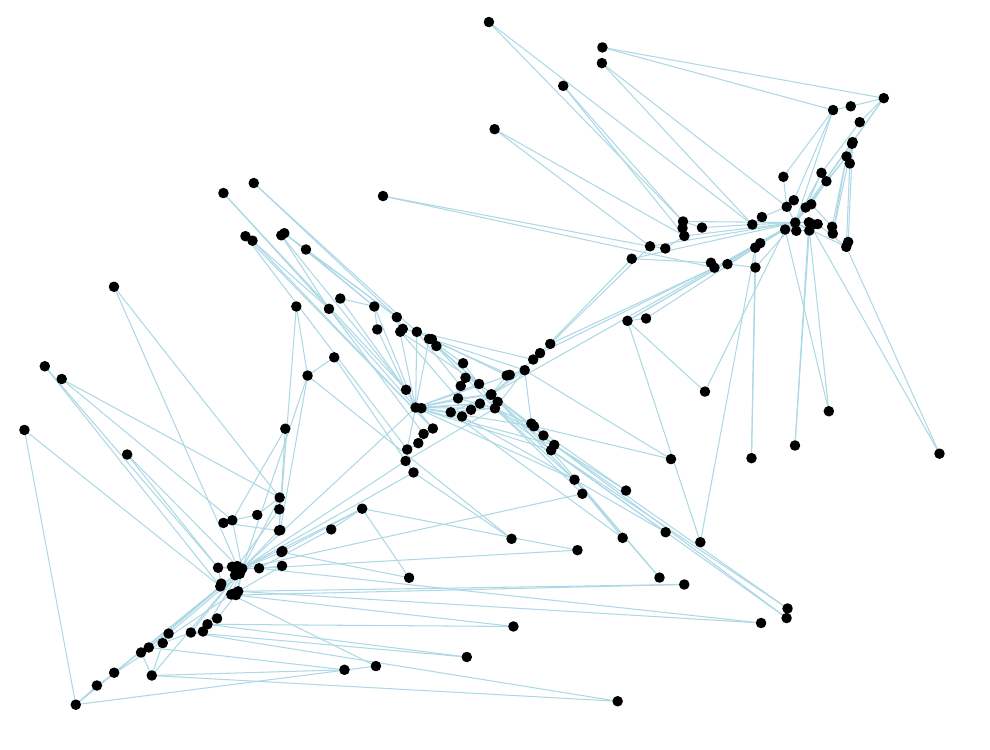} &
  \includegraphics[width=9.5mm]{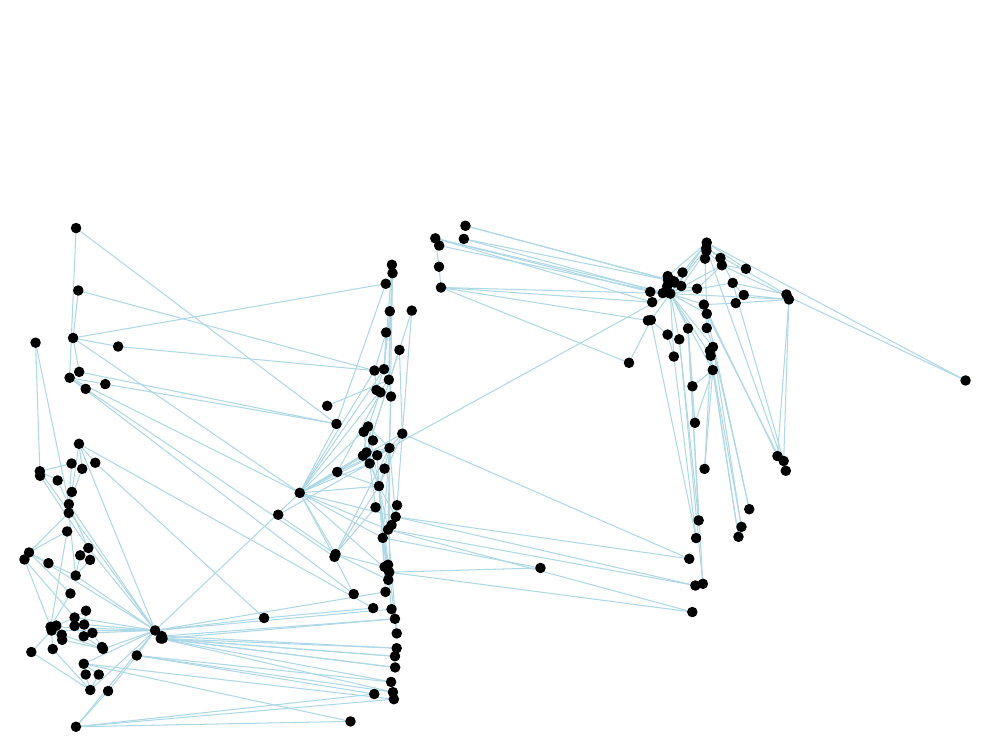} &
  \includegraphics[width=9.5mm]{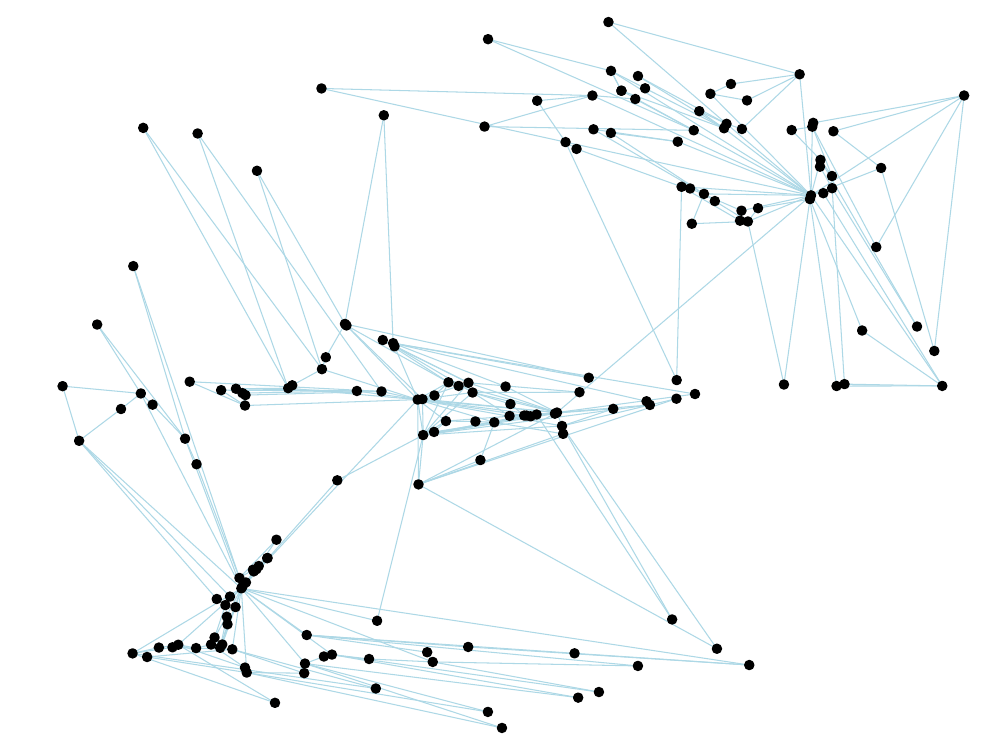} &
  \includegraphics[width=9.5mm]{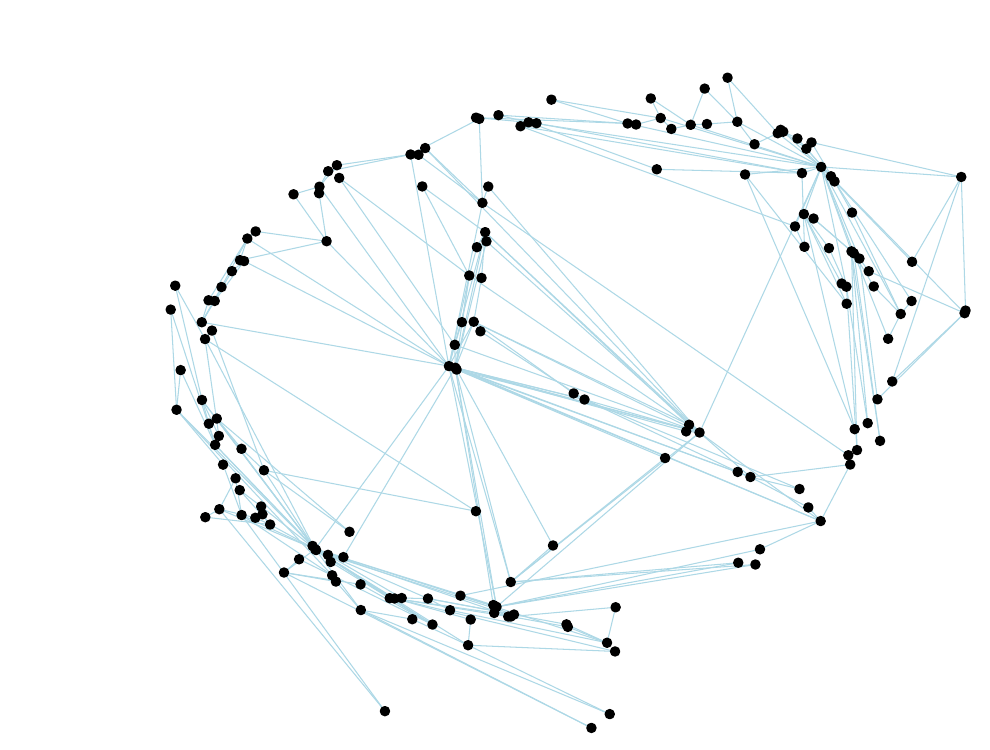} &
  \includegraphics[width=9.5mm]{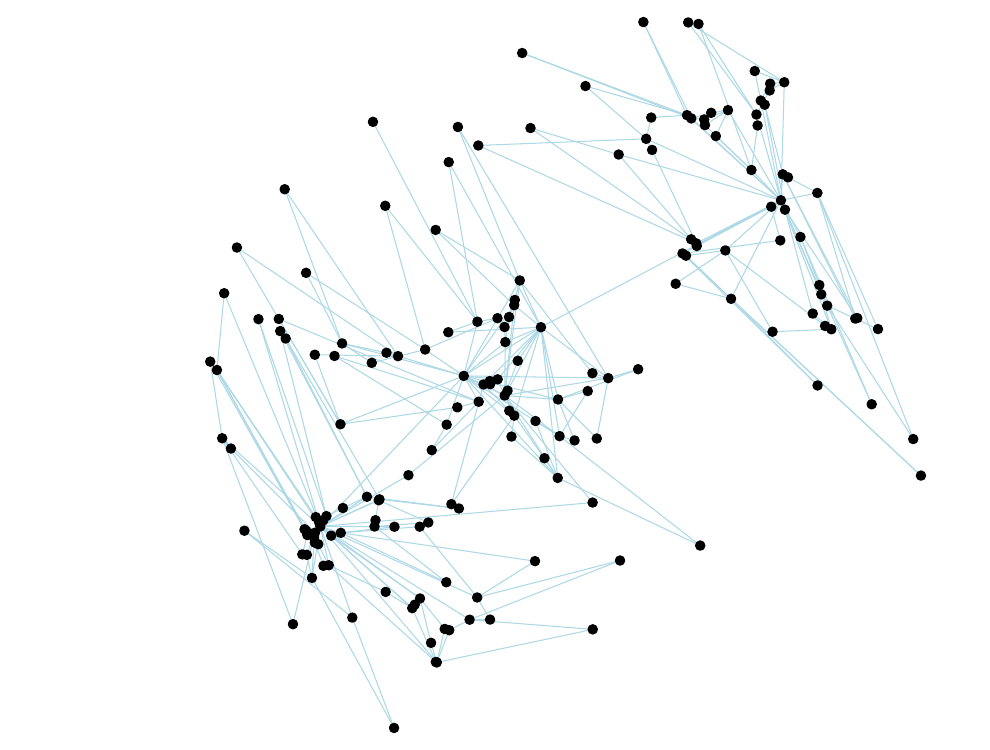} &
  \includegraphics[width=9.5mm]{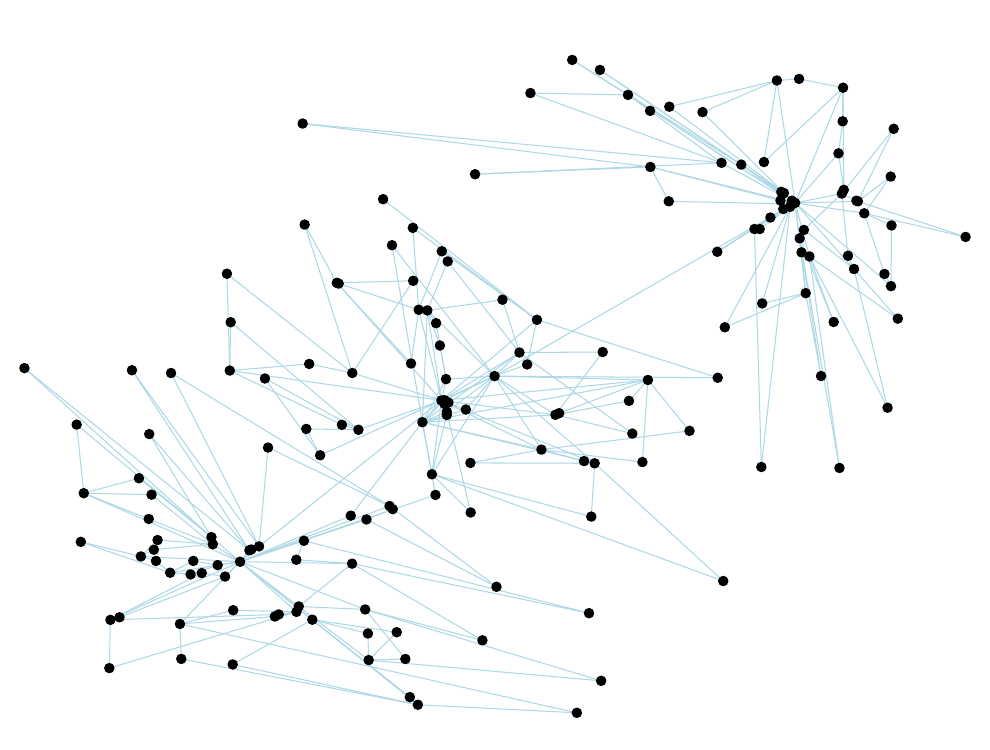}
  \\
        & \texttt{ST-ELD-CN} & &\includegraphics[width=9.5mm]{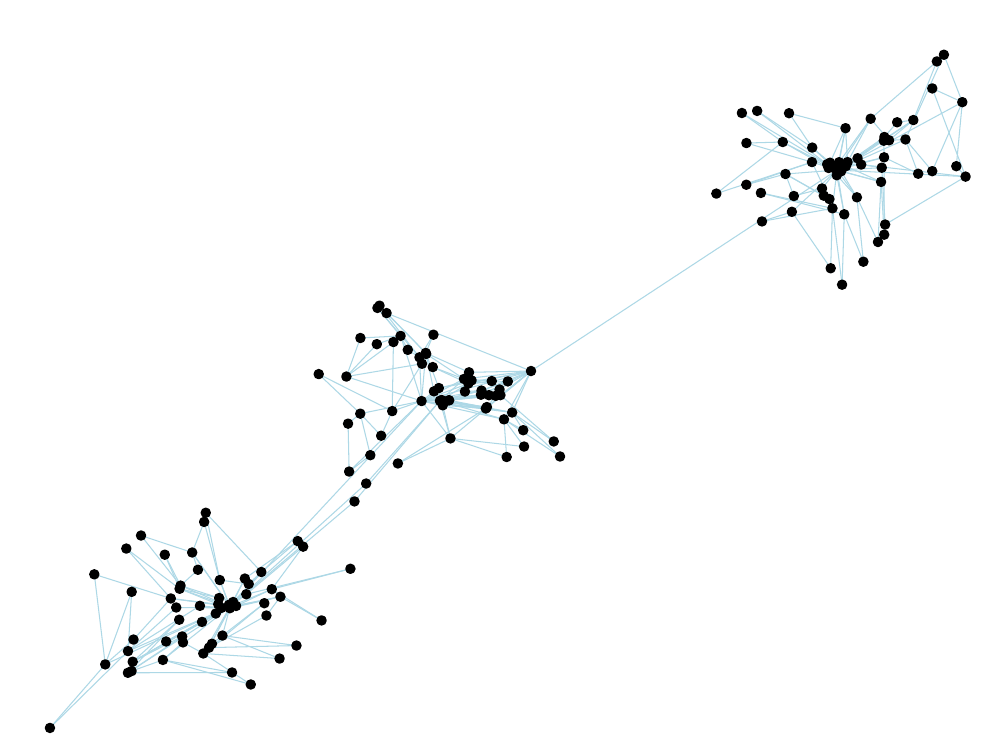} &
  \includegraphics[width=9.5mm]{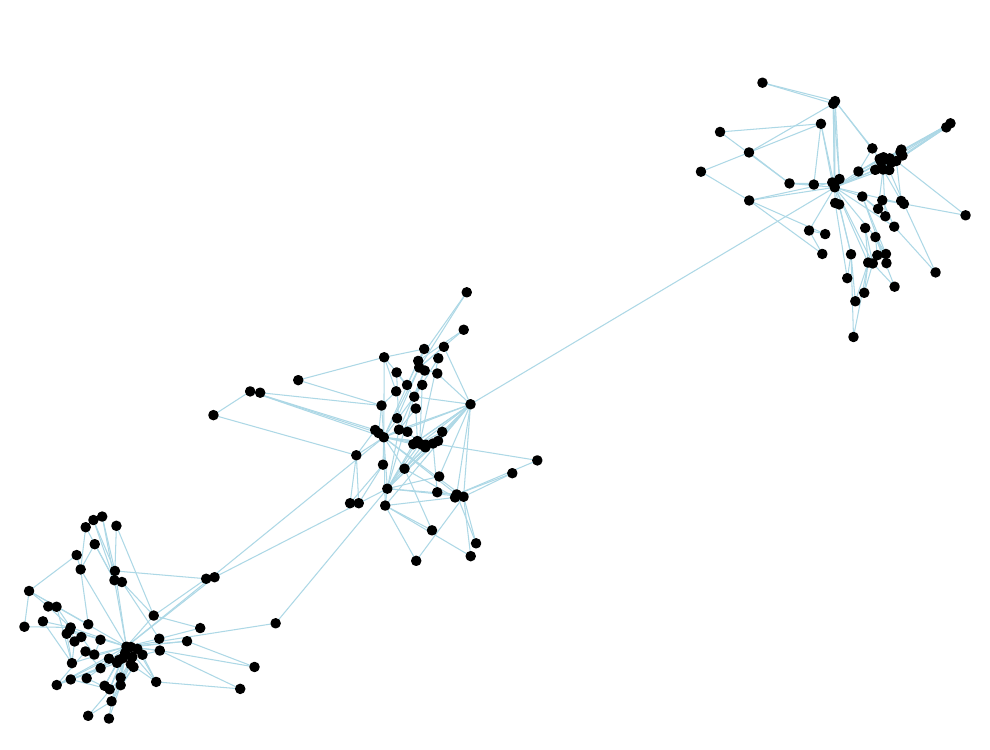} &
  \includegraphics[width=9.5mm]{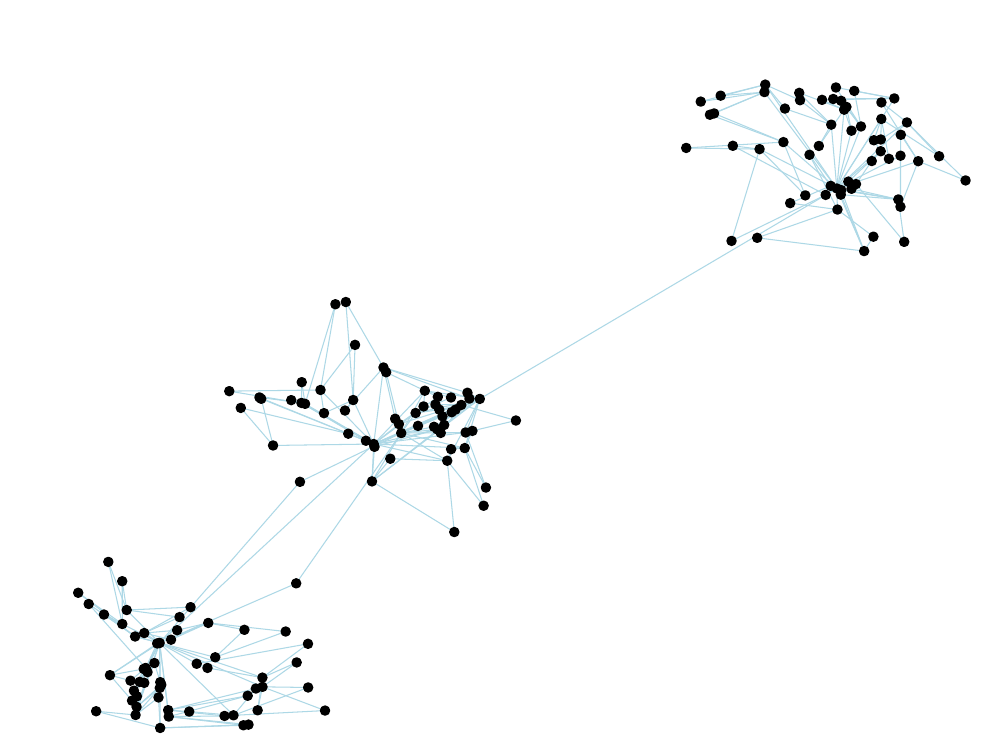} &
  \includegraphics[width=9.5mm]{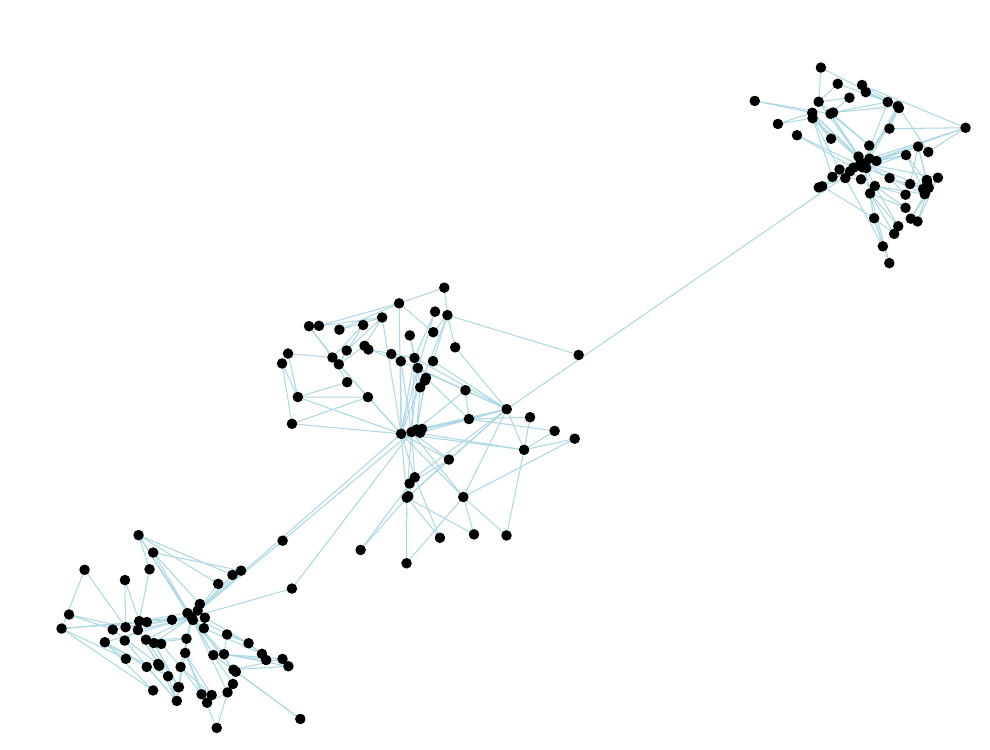} &
  \includegraphics[width=9.5mm]{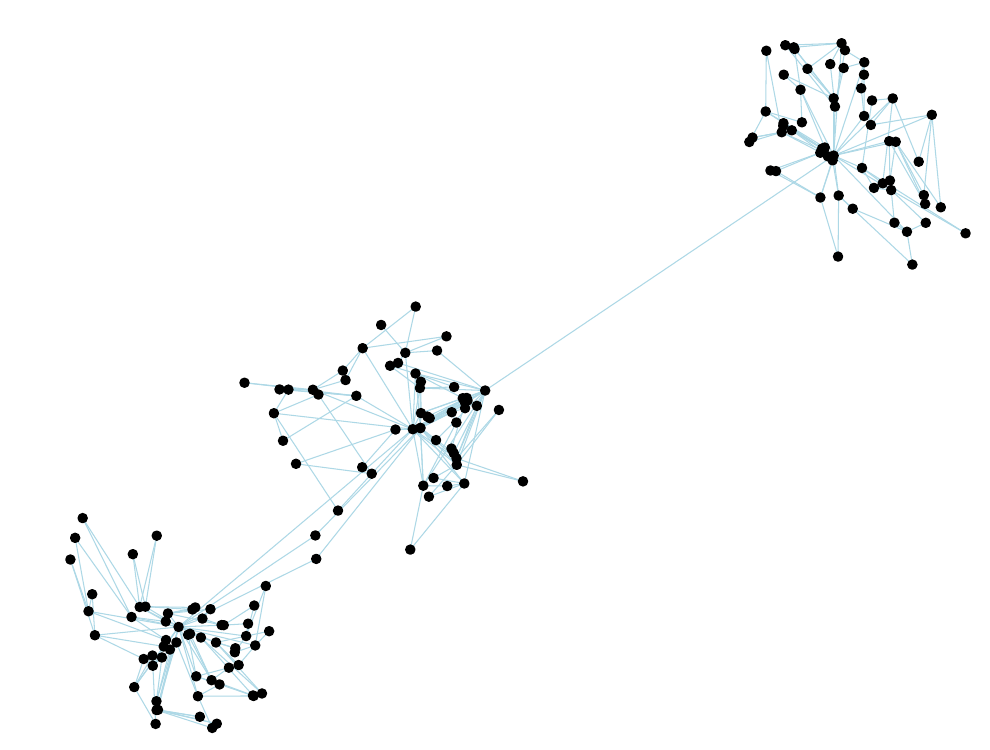} &
  \includegraphics[width=9.5mm]{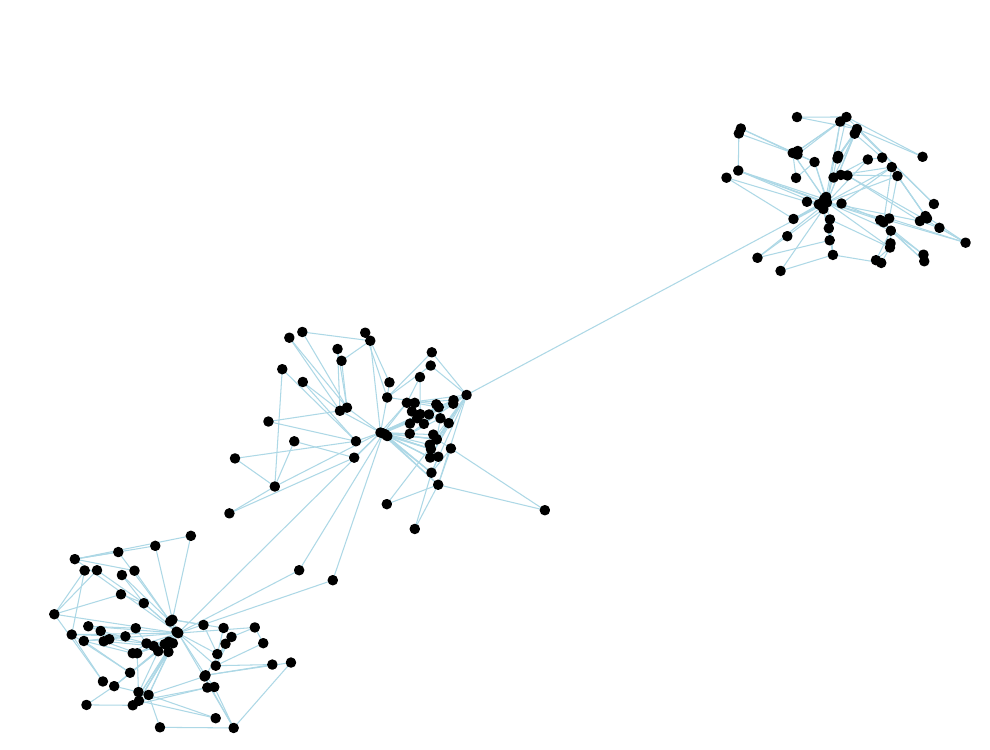}
  \\
         & \texttt{ST-ELD-AR} & &\includegraphics[width=9.5mm]{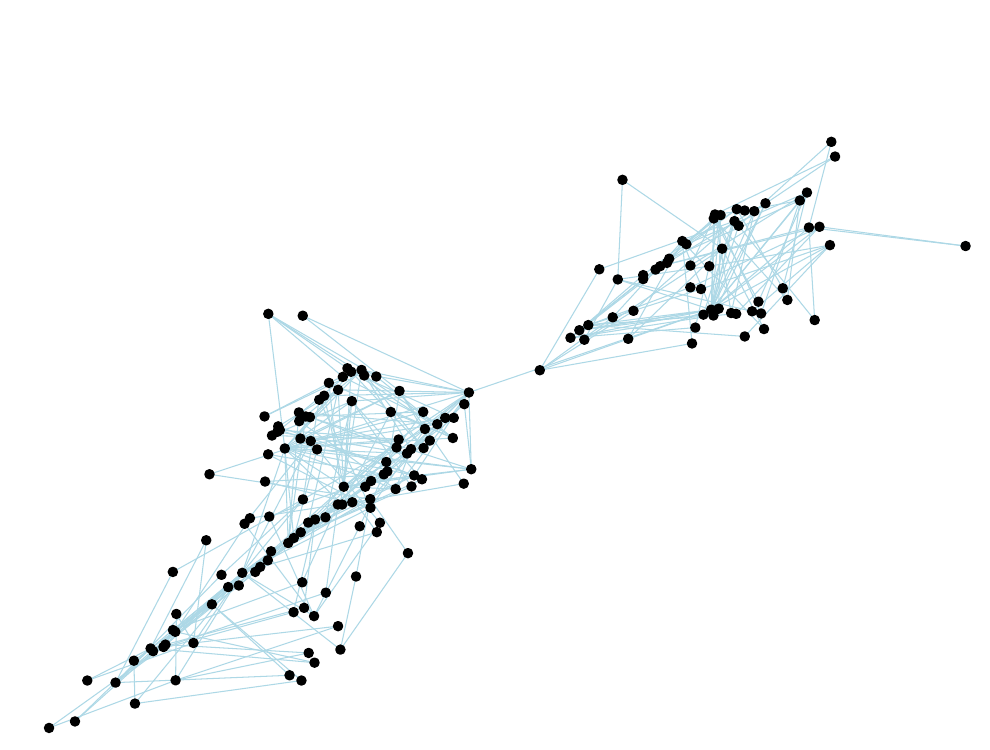} &
  \includegraphics[width=9.5mm]{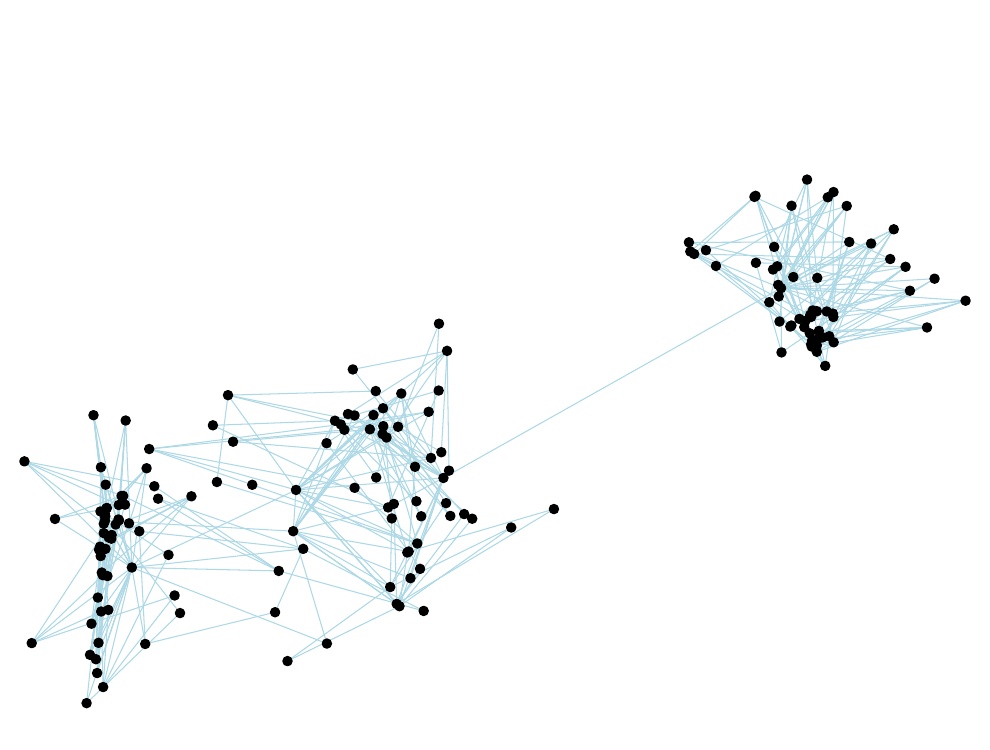} &
  \includegraphics[width=9.5mm]{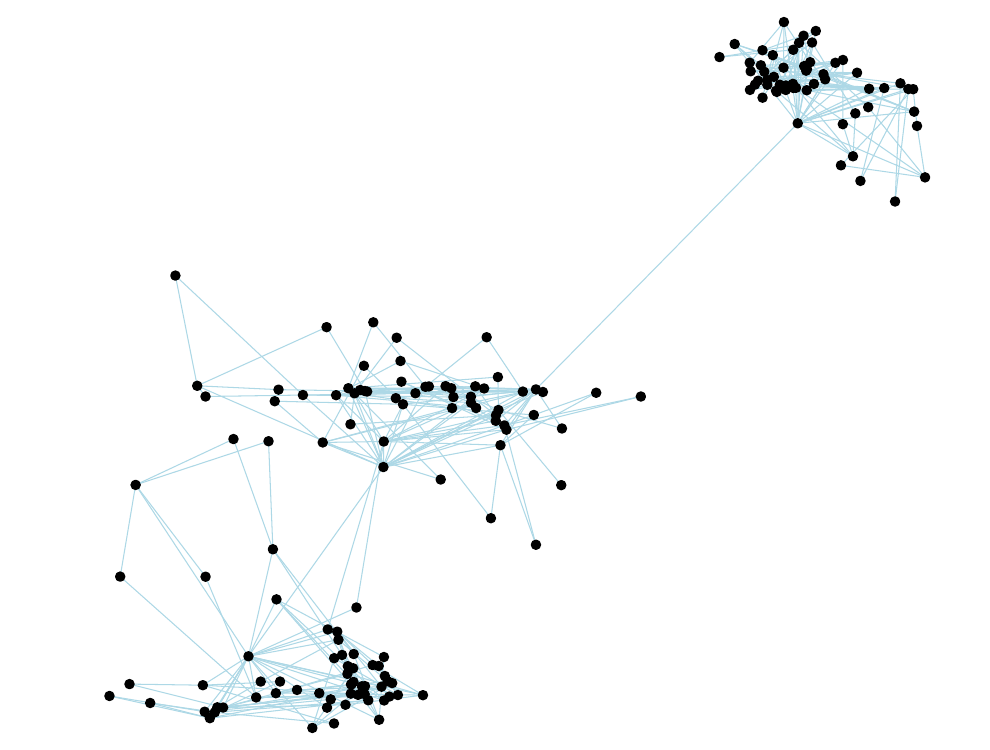} &
  \includegraphics[width=9.5mm]{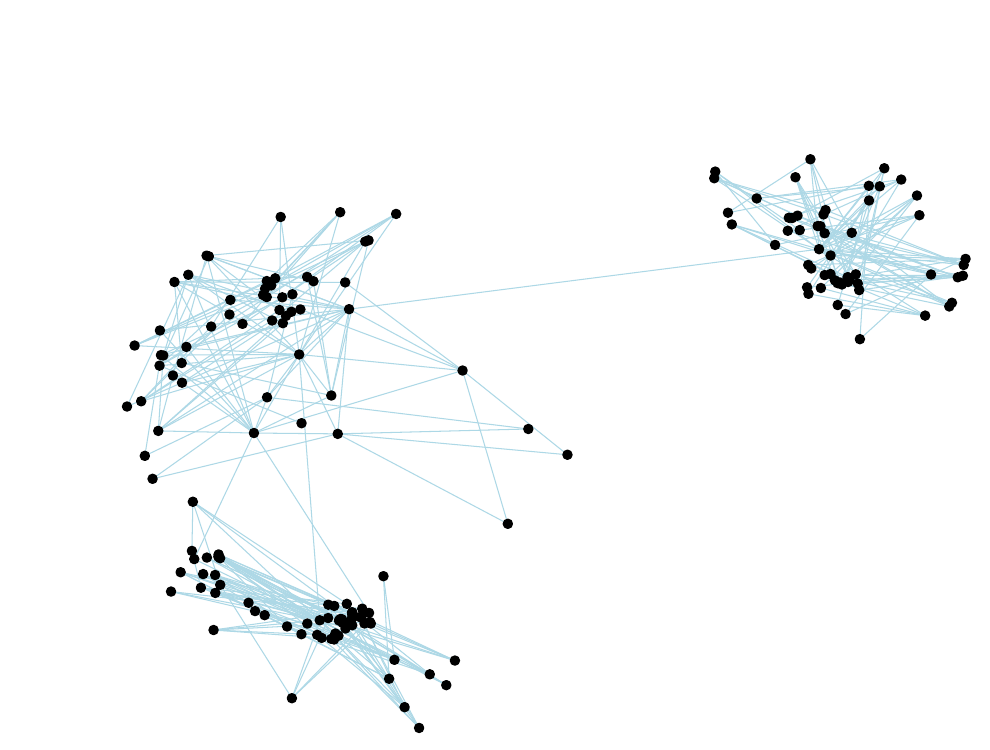} &
  \includegraphics[width=9.5mm]{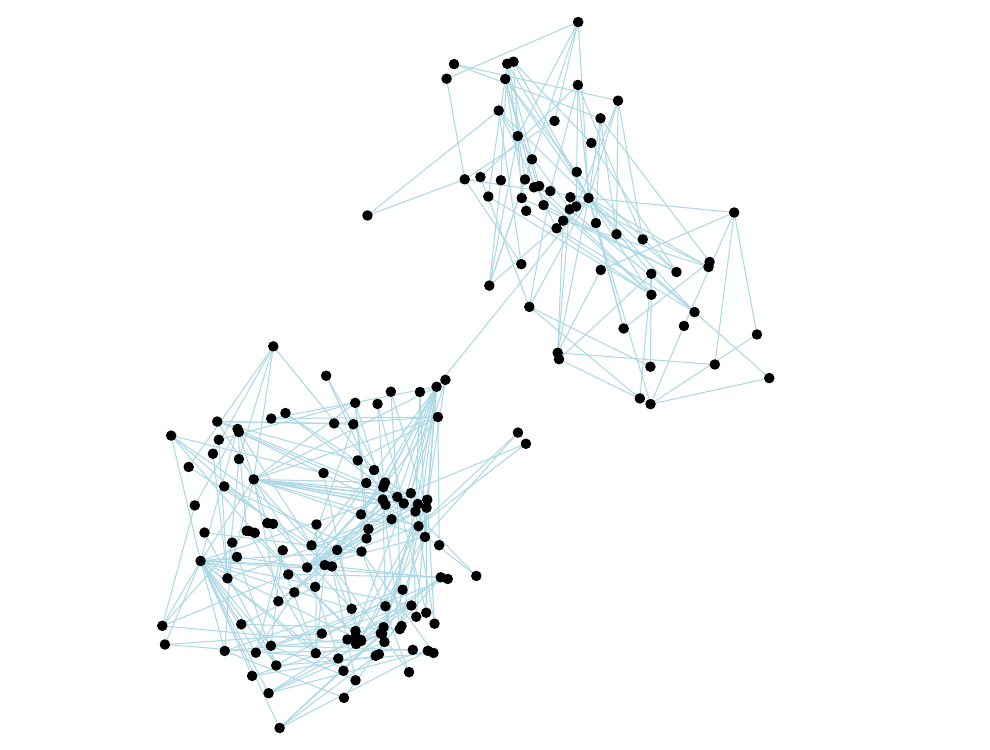} &
  \includegraphics[width=9.5mm]{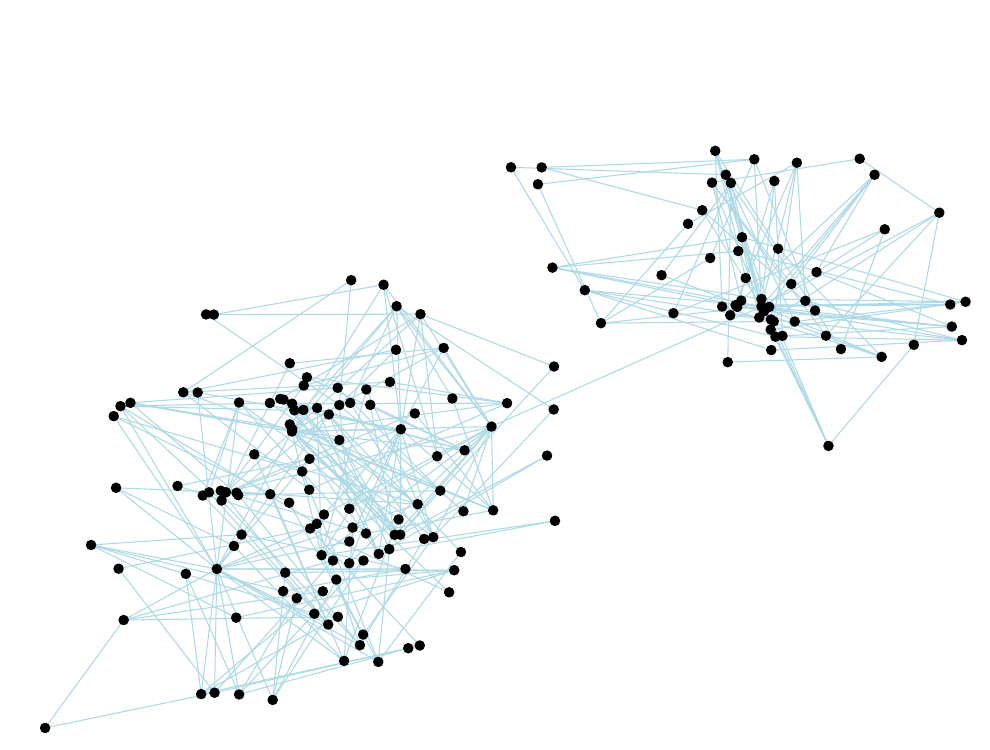}
  \\
         & \texttt{ST-CN-AR} & &\includegraphics[width=9.5mm]{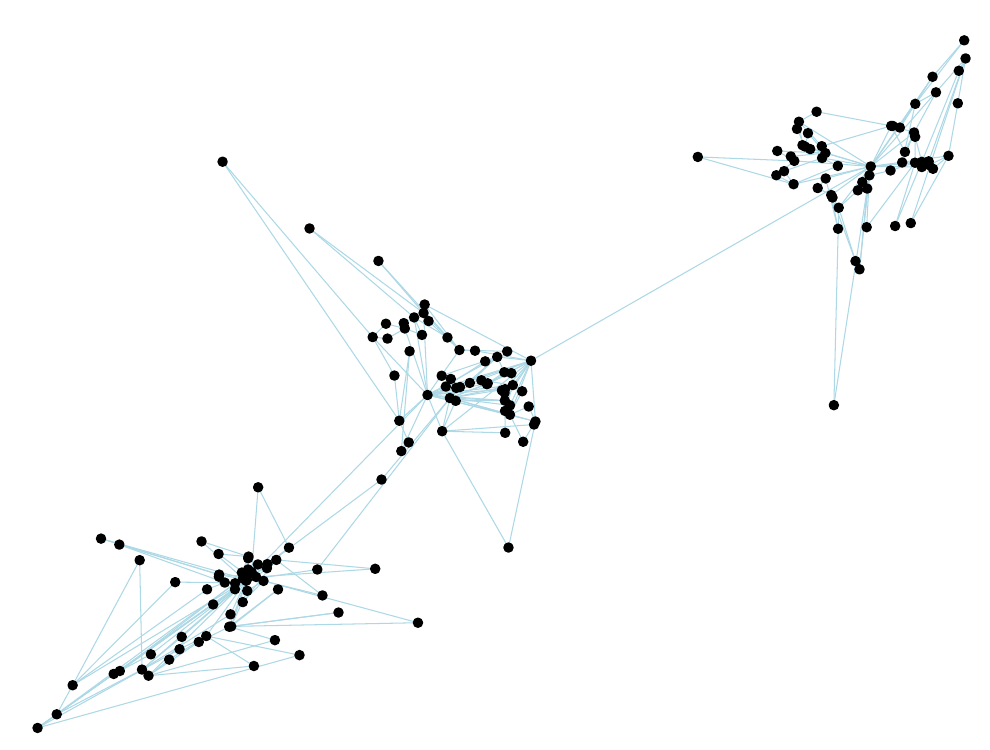} &
  \includegraphics[width=9.5mm]{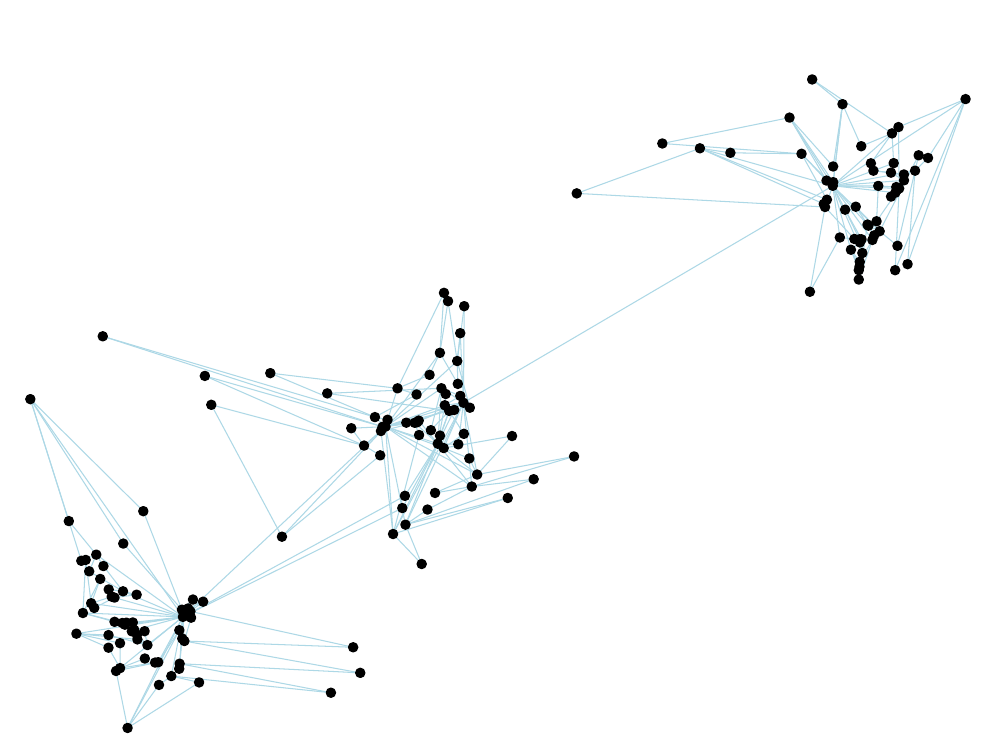} &
  \includegraphics[width=9.5mm]{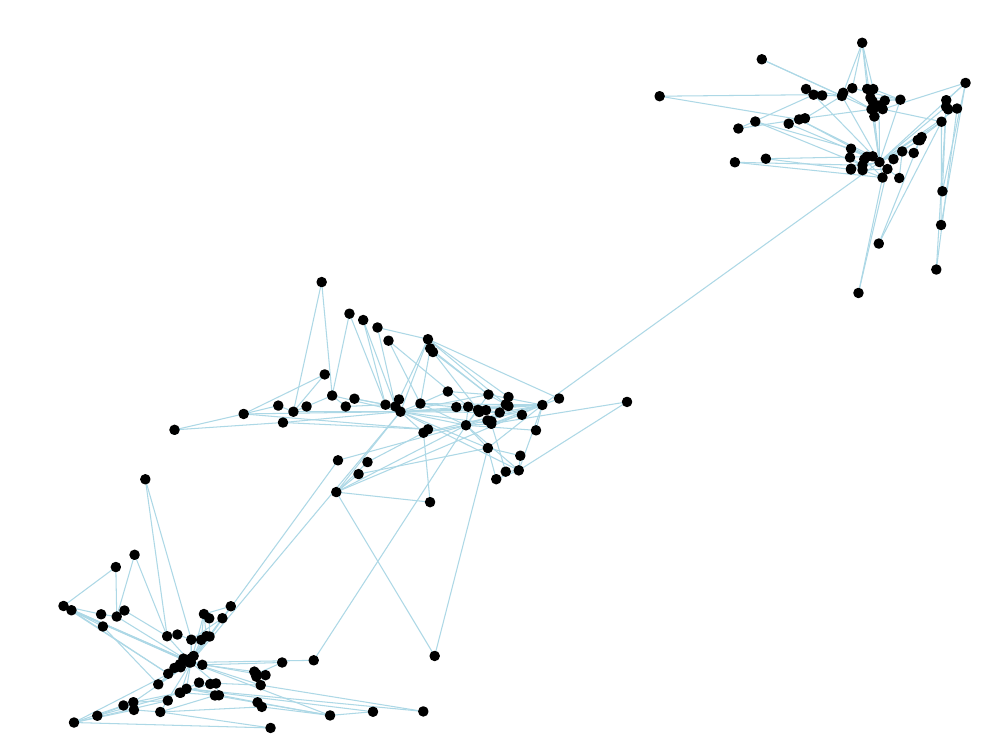} &
  \includegraphics[width=9.5mm]{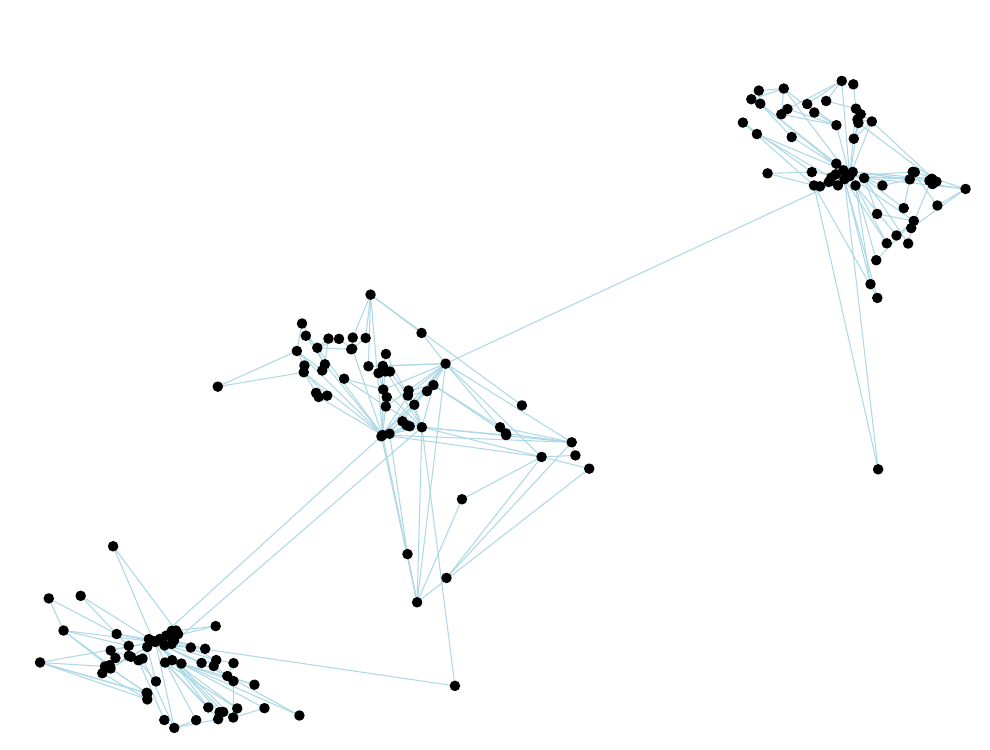} &
  \includegraphics[width=9.5mm]{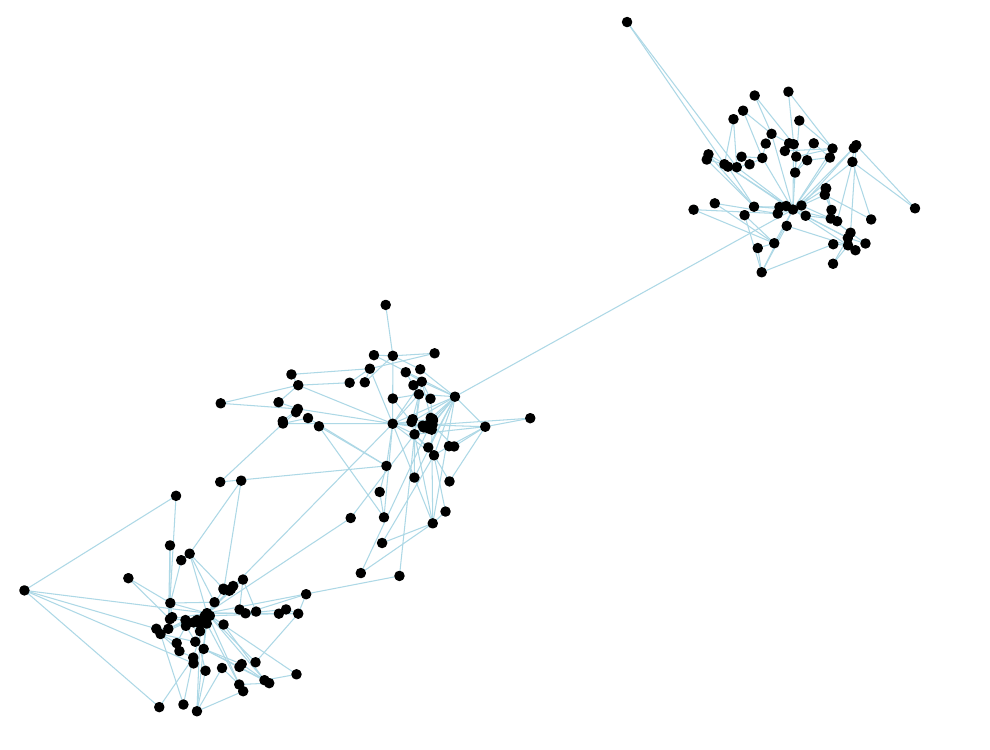} &
  \includegraphics[width=9.5mm]{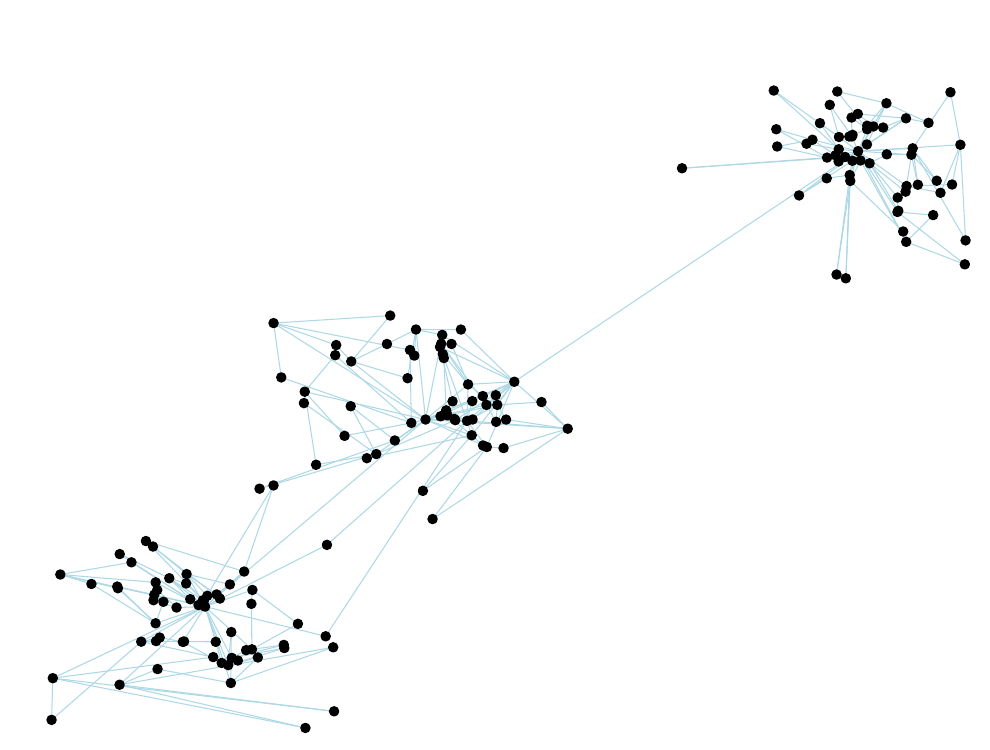}
  \\
         & \texttt{ELD-CN-AR} & & \includegraphics[width=9.5mm]{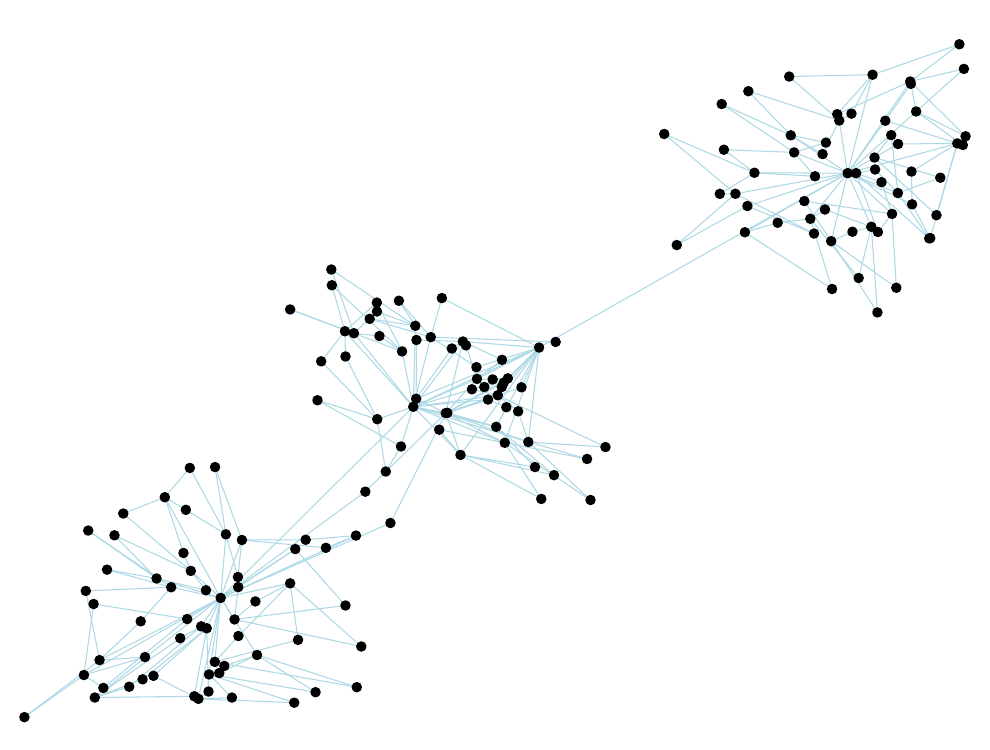} &
  \includegraphics[width=9.5mm]{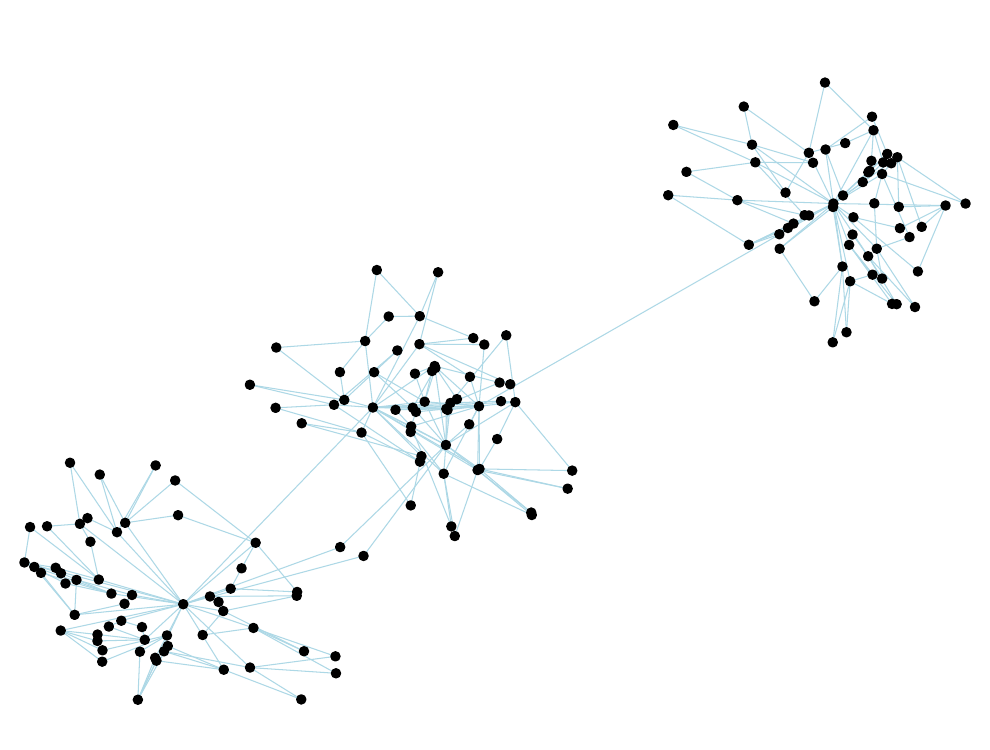} &
  \includegraphics[width=9.5mm]{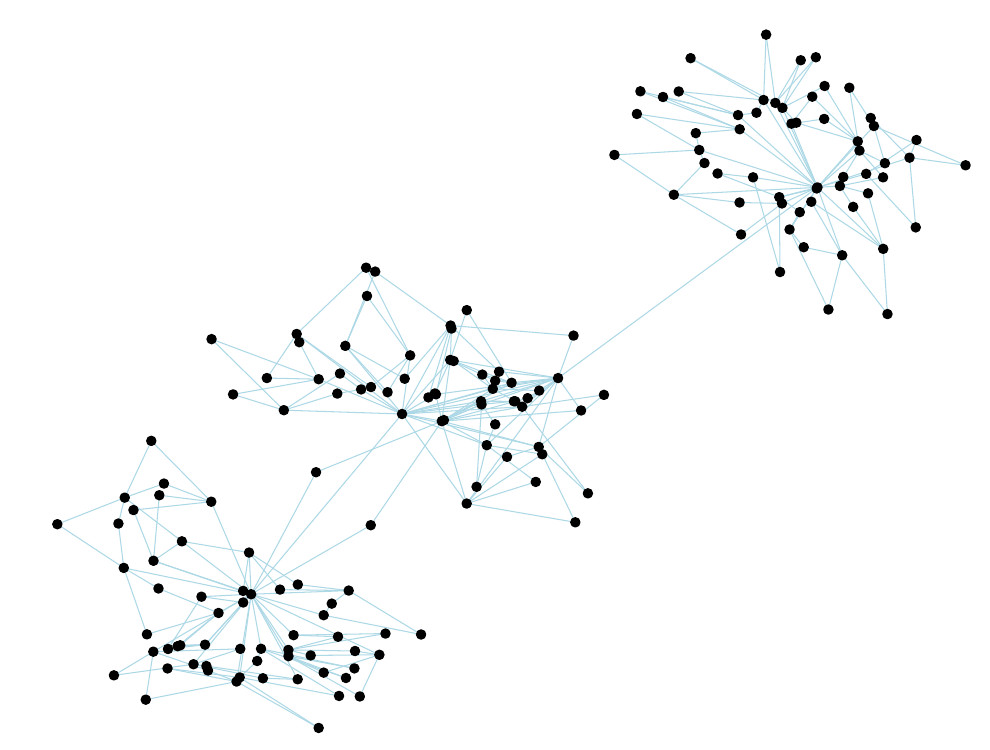} &
  \includegraphics[width=9.5mm]{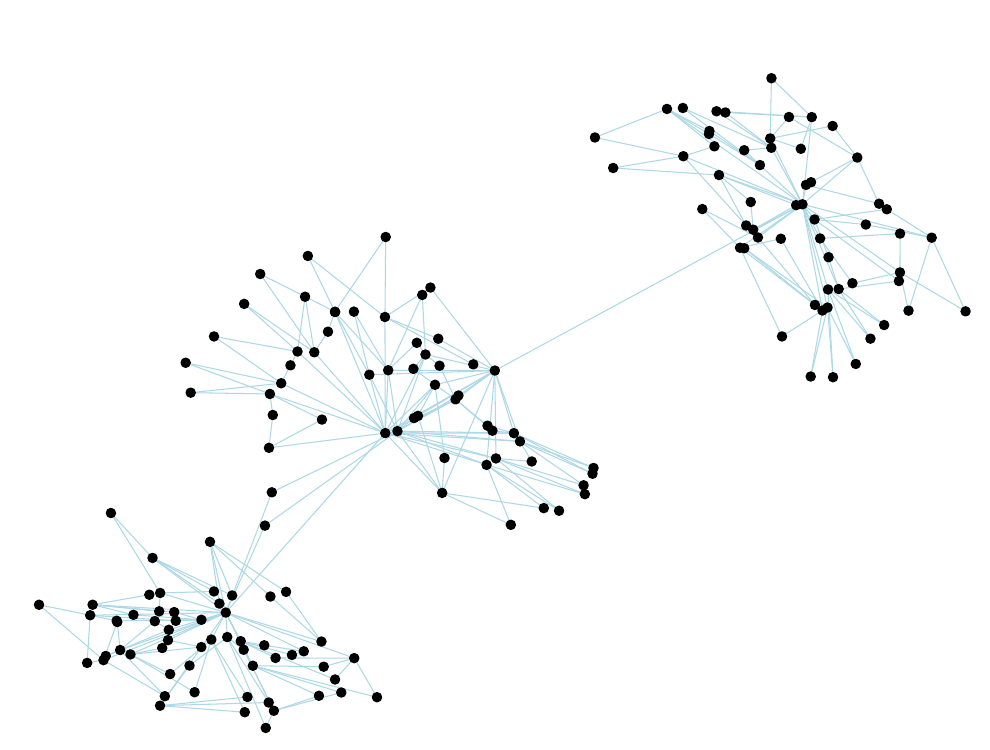} &
  \includegraphics[width=9.5mm]{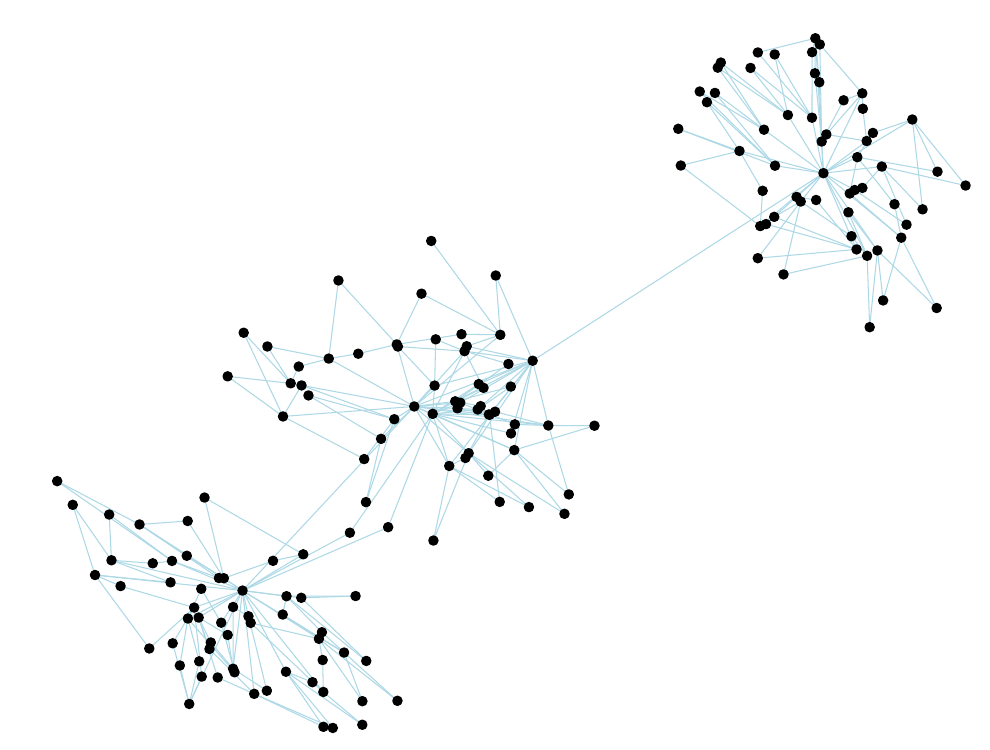} &
  \includegraphics[width=9.5mm]{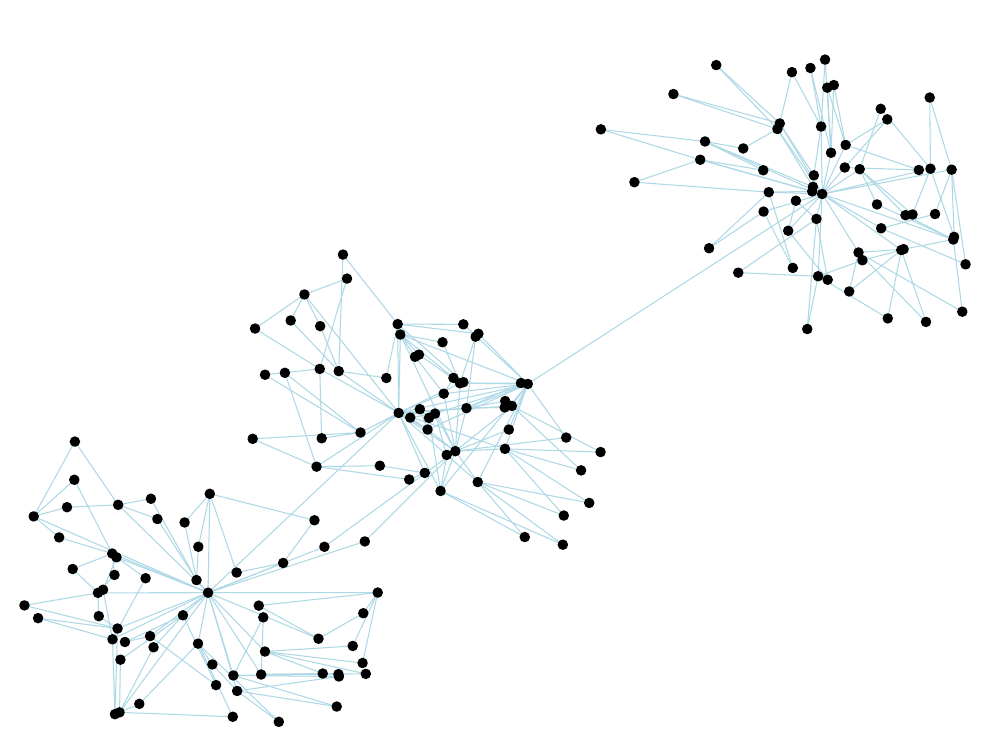}
  \\
         & \texttt{ST-ELD-CN-AR} & & \includegraphics[width=9.5mm]{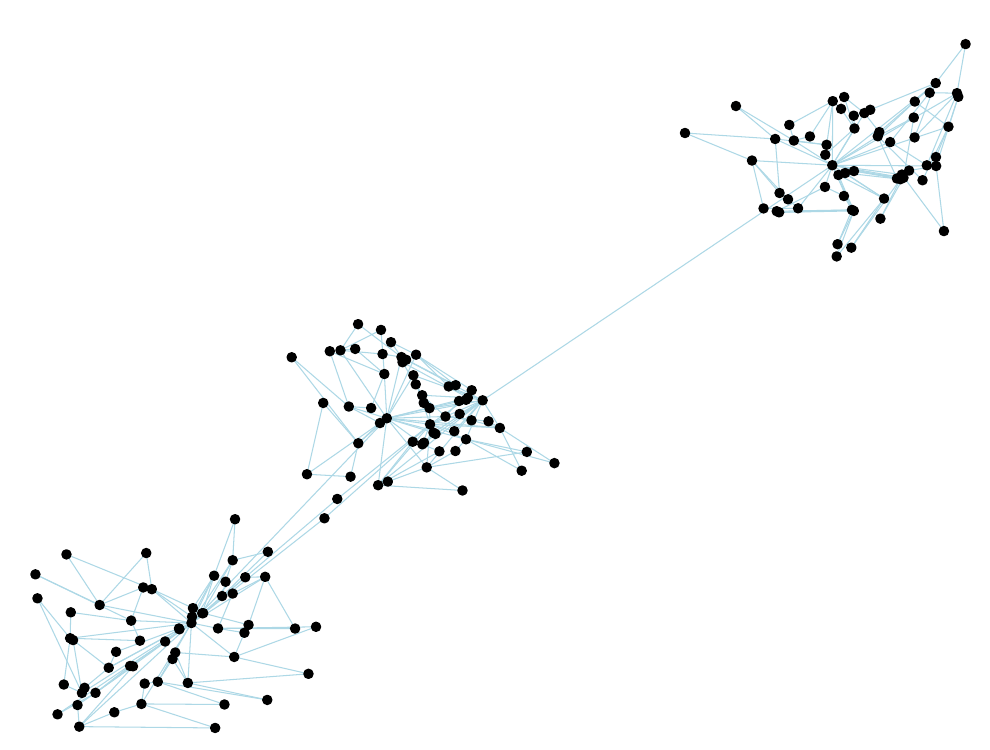} &
  \includegraphics[width=9.5mm]{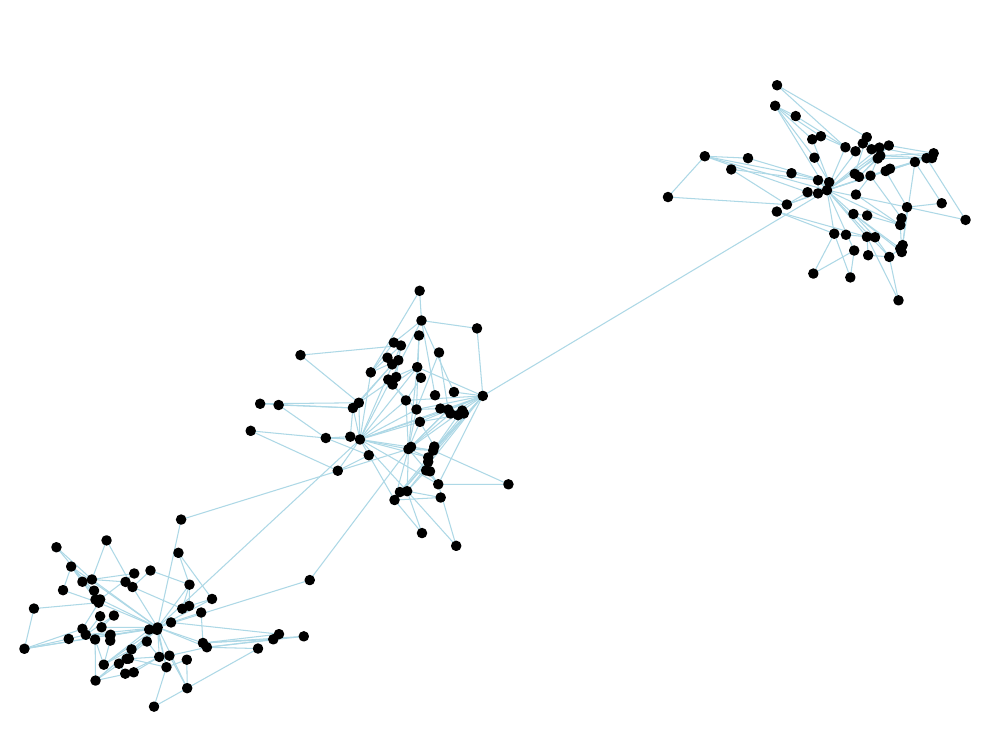} &
  \includegraphics[width=9.5mm]{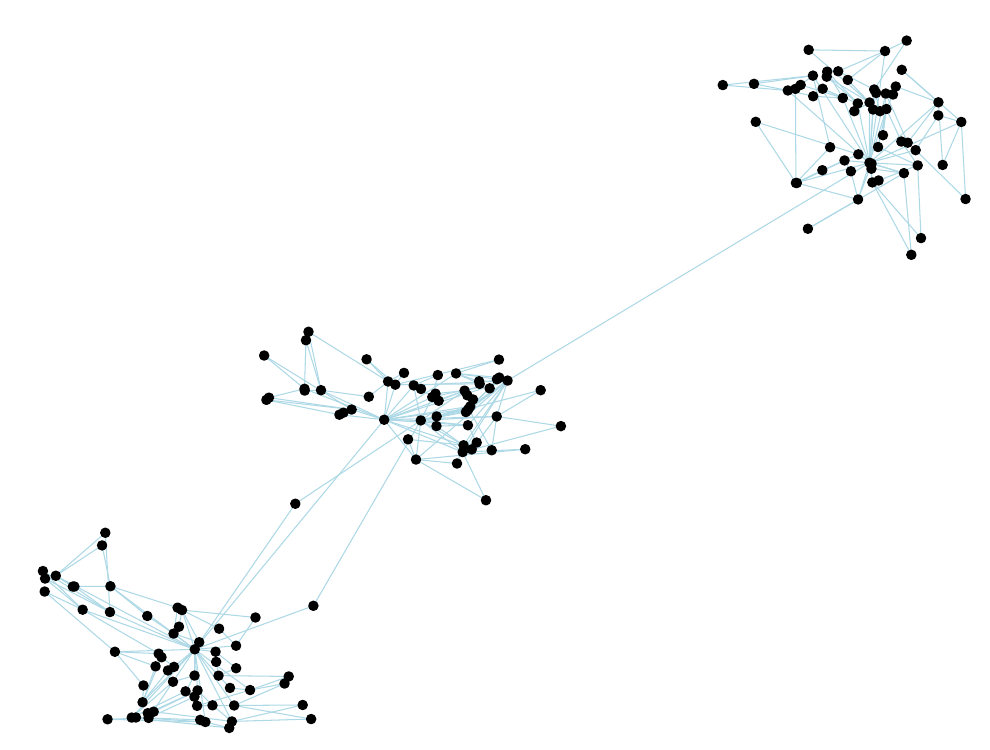} &
  \includegraphics[width=9.5mm]{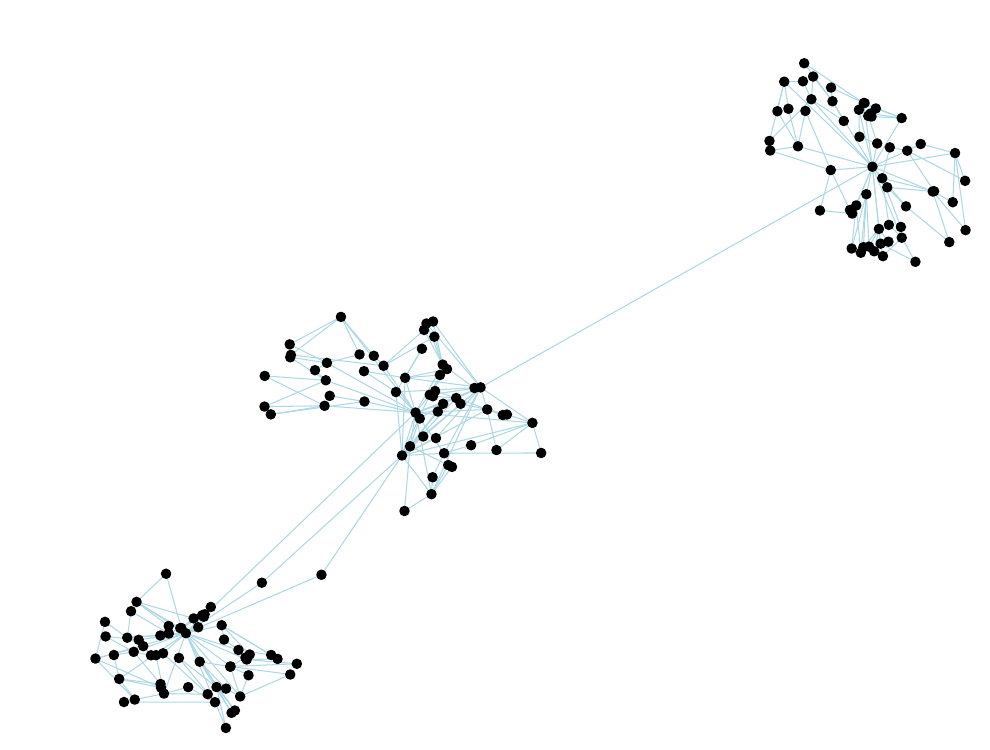} &
  \includegraphics[width=9.5mm]{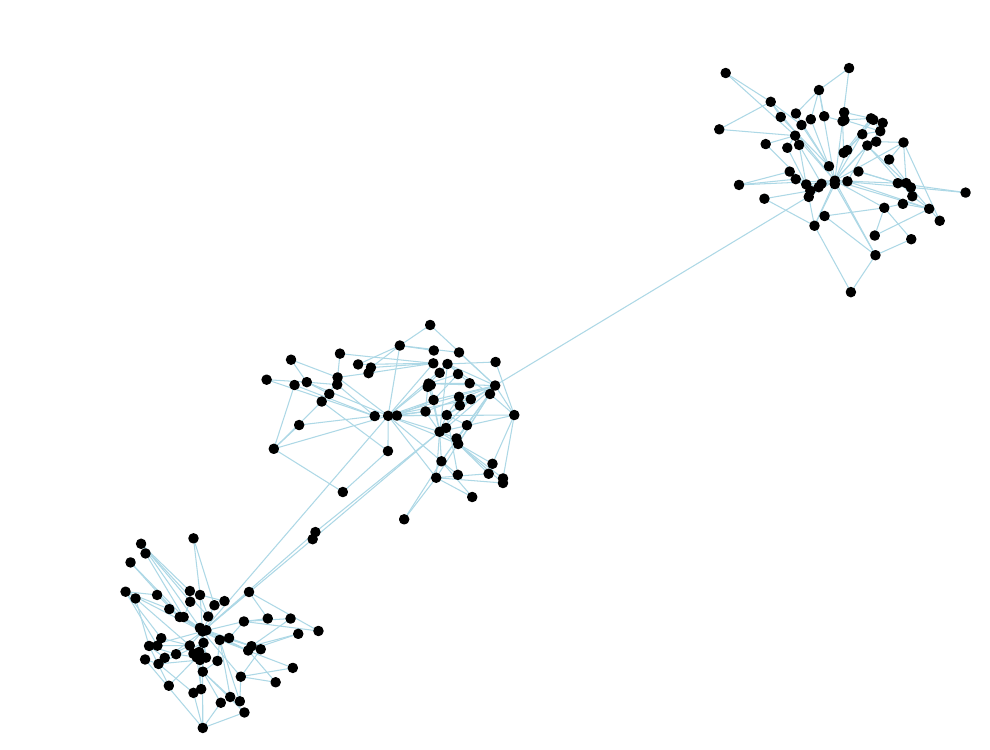} &
  \includegraphics[width=9.5mm]{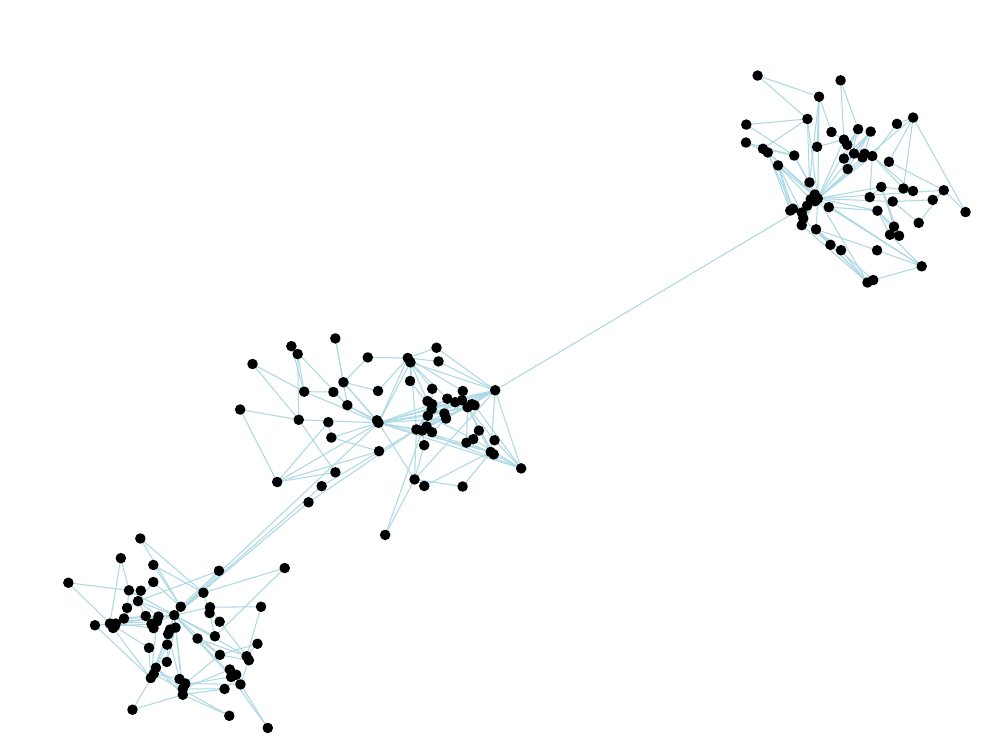}
  \\
  \hline
\end{tabular}
\caption{Collection of different graph drawings. The \texttt{START} column indicates the starting drawing of various graphs. The results of applying Alg.~\ref{alg:sim_anneal}  to \texttt{START} to six target shapes (\texttt{X}, \texttt{VERT}, \texttt{HOR}, \texttt{O}, \texttt{DINO}, \texttt{GRID}) are shown in their respective columns. The rows indicate the metrics that have $\pm\epsilon=0.0025$ for combinations of \texttt{ST,ELD,AR} and $\pm\epsilon=\texttt{CN}(\Gamma)*0.05$ for \texttt{CN}.}
\label{fig:results_combs_lnsp_gams}
\end{figure*}

\clearpage
\subsection{Similarity percentages}
\begin{figure*}[!ht]
\centering
\includegraphics[width=\linewidth]{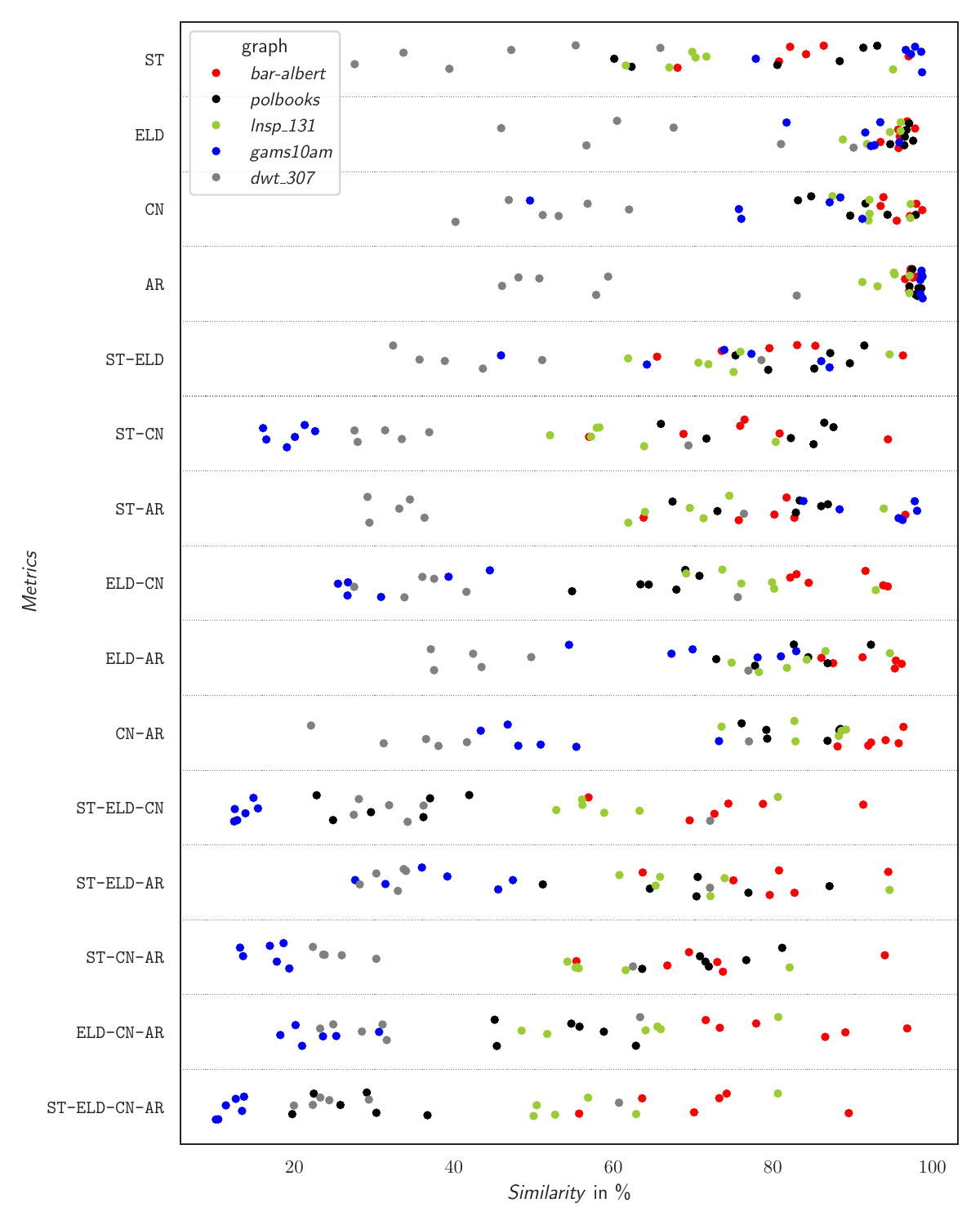}
\caption{Distributions of how similar the resulting drawings are to the six target shapes.}
\label{fig:jitter_all}
\end{figure*}

Figure~\ref{fig:jitter_all}$\star$ visualizes the distributions of the similarity percentages. Here, each dot represents how similar the fooled drawing is to its target shape. 

\subsubsection{Metric comparison}
To test significant differences between the non-normally distributed results of the fooled metrics we first employ the Friedman test. We find that there are significant differences ($\chi^2 =314$, $p < 0.05$) between the results of metrics. Since our data is paired, we then perform pairwise metric comparisons using the Wilcoxon signed-rank test. Additionally, we correct the p-values using the  Bonferroni adjustments. 

\begin{figure*}[!ht]
\centering
\includegraphics[width=\linewidth]{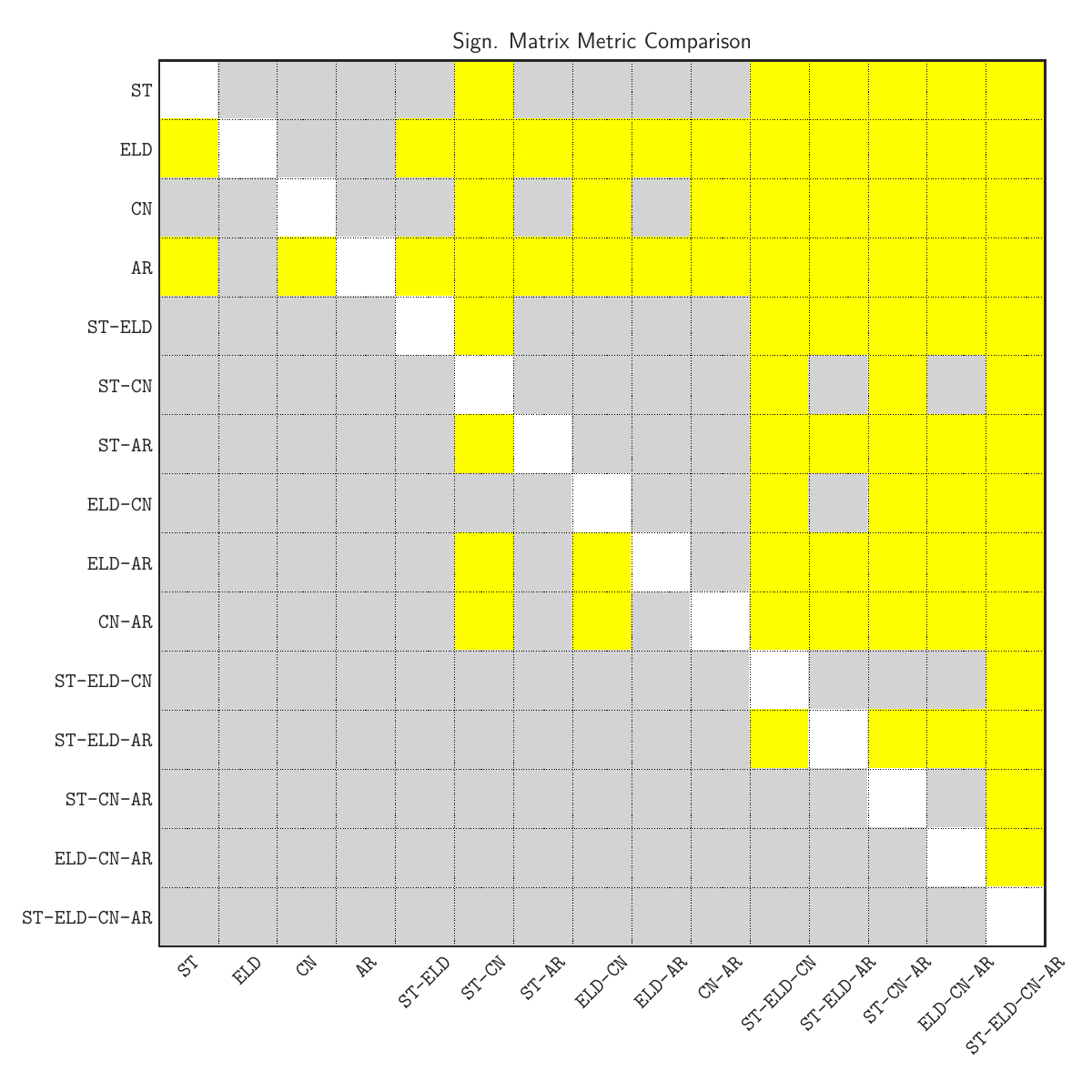}
\caption{Comparing the distributions of each metric's results with each other metric. Each box indicates that the resulting distribution of the results of the row metric minus the results of the column metric are non-significantly (gray) and significantly (yellow) greater (easier to fool) than a distribution symmetric about zero.}
\label{fig:sig_tests_metrics}
\end{figure*}

In Figure~\ref{fig:sig_tests_metrics}$\star$ we display the results of the significance tests with $p < 0.05$. Here, the metric of a row is compared with the metrics of all columns and vice-versa. A yellow (resp., gray) color indicates that the resulting distribution of the row metric minus the column metric is significantly (resp., or non-significantly) greater than a distribution symmetric about zero. As we are using a \emph{greater than} alternative hypothesis, the pattern in Figure~\ref{fig:sig_tests_metrics}$\star$ is not symmetric. For example, for \texttt{ELD} and \texttt{ST} we see that there is a significant difference, meaning that the difference between the distributions of \texttt{ELD} and \texttt{ST} is greater than a distribution around zero, i.e. \texttt{ELD} is easier to fool than \texttt{ST}. However, that means that the difference between the distributions of \texttt{ST} and \texttt{ELD} is smaller than a distribution around zero which is why this pair is not significant for the ``greater than'' test.

From this pair-wise metric comparison, we observe that fooling individual metrics is easier than fooling combinations, with \texttt{ELD} and \texttt{AR} being the simplest. For pairs of quality metrics, the \texttt{ST-CN} combination is more resistant to fooling than others. We see the same for any other larger combination that includes \texttt{ST} and \texttt{CN}, such as \texttt{ST-ELD-CN} and \texttt{ST-CN-AR} which are the most difficult to fool.

\clearpage
\subsubsection{Target shape comparison}
\begin{figure*}[!ht]
\centering
\includegraphics[scale=0.75]{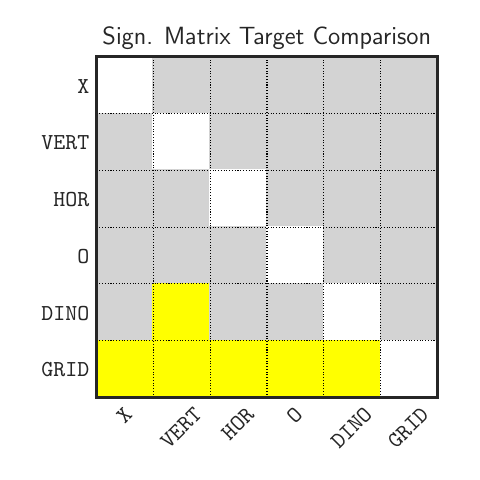}
\caption{Comparing the distributions of each target's results with each other target. Each box indicates that the resulting distribution of the results of the row metric target the results of the column target are non-significantly (gray) and significantly (yellow) greater (easier to fool) than a distribution symmetric about zero..  }
\label{fig:sig_tests_target}
\end{figure*}

We test the differences between graphs in exactly the same manner as for the metric comparison, with now the dependent variable being the target shape. The Friedman test shows that there are significant differences between the results of the target shapes ($\chi^2 = 42.9$, $p < 0.05$). The  results of the pairwise Wilcoxon signed-rank tests are visualized in Figure~\ref{fig:sig_tests_target}$\star$. From these results we observe that the \texttt{GRID} shape is significantly easier to morph into than any other shape. All the other shapes are more or less equally difficult to morph any drawing into, with the exception that the \texttt{DINO} shape is significantly easier than \texttt{VERT}.

\clearpage
\subsubsection{Graph comparison}

\begin{figure*}[!ht]
\centering
\includegraphics[scale=0.7]{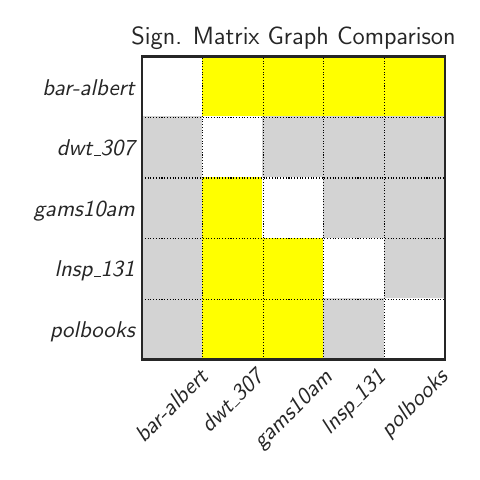}
\caption{Comparing the distributions of each graph's results with each other graph. Each box indicates that the resulting distribution of the results of the row graph minus the results of the column graph are non-significantly (gray) and significantly (yellow) greater (easier to fool) than a distribution symmetric about zero.}
\label{fig:sig_tests_graphs}
\end{figure*}

We test the differences between graphs in a similar way to the metric comparison, with now the dependent variable being the graph. However, due to the graph data being independent samples we use the Mann-Whitney U test to determine significant differences between pairs of graphs. These results are visualized in Figure~\ref{fig:sig_tests_graphs}$\star$. From these results we observe that for the \emph{bar-albert} graph it significantly easier to fool any metric combination than it is for any other graph. Conversely, fooling metrics for the \emph{dwt\_307} graph is significantly more difficult than for any other graph

\end{document}